\documentclass[journal]{IEEEtran}
\usepackage{amsmath,amssymb,amsthm,lipsum} %
\usepackage{cite}
\usepackage{algorithmic}
\usepackage{algorithm}
\usepackage{graphicx}
\usepackage{textcomp}
\usepackage{stfloats}
\usepackage{booktabs}
\usepackage{bm}
\usepackage{subfigure}
\usepackage{setspace}
\usepackage{color}
\usepackage[T1]{fontenc}
\usepackage{cases}
\usepackage{indentfirst}
\allowdisplaybreaks[4]
\ifCLASSINFOpdf
\else
\fi
\theoremstyle{Theorem}
\newtheorem{theo}{Theorem}
\newtheorem{theoremmaindescription1}[theo]{Theorem}
\newtheorem{theoremmaindescription2}[theo]{Theorem}
\newtheorem{theoremmaindescription3}[theo]{Theorem}
\newtheorem{theoremmaindescription4}[theo]{Theorem}
\newtheorem{theoremmaindescription5}[theo]{Theorem}
\newtheorem{theoremmaindescription6}[theo]{Theorem}
\newtheorem{theoremmaindescription7}[theo]{Theorem}
\newtheorem{theoremmaindescription8}[theo]{Theorem}

\theoremstyle{remark}
\newtheorem{rmk}{Remark}
\newtheorem{rmk2}[rmk]{Remark}
\newtheorem{rmk3}[rmk]{Remark}
\newtheorem{rmk4}[rmk]{Remark}
\newtheorem{rmk5}[rmk]{Remark}
\newtheorem{rmk6}[rmk]{Remark}
\newtheorem{rmk7}[rmk]{Remark}
\newtheorem{rmk8}[rmk]{Remark}
\newtheorem{rmk9}[rmk]{Remark}
\newtheorem{rmk10}[rmk]{Remark}
\newtheorem{rmk11}[rmk]{Remark}

\theoremstyle{Definition}
\newtheorem{def1}{Definition}
\newtheorem{def2}[def1]{Definition}
\newtheorem{def3}[def1]{Definition}
\newtheorem{def4}[def1]{Definition}
\newtheorem{def5}[def1]{Definition}
\newtheorem{def6}[def1]{Definition}
\newtheorem{def7}[def1]{Definition}
\newtheorem{def8}[def1]{Definition}

\theoremstyle{Lemma}
\newtheorem{lemma1}{Lemma}

\newtheorem{lemma4}[lemma1]{Lemma}
\newtheorem{lemma5}[lemma1]{Lemma}

\theoremstyle{Corollary}
\newtheorem{Corollary1}{Corollary}

\newtheorem{Corollary4}[Corollary1]{Corollary}

\newtheorem{Corollary6}[Corollary1]{Corollary}
\newtheorem{Corollary7}[Corollary1]{Corollary}
\newtheorem{Corollary8}[Corollary1]{Corollary}

\theoremstyle{Proposition}
\newtheorem{prop}{Proposition}
\newtheorem{proposition1}[prop]{Proposition}
\newtheorem{proposition2}[prop]{Proposition}

\begin{document}

\title{Block-Sparse Tensor Recovery}

\author{Liyang~Lu,~\IEEEmembership{Member,~IEEE,}
	      Zhaocheng~Wang,~\IEEEmembership{Fellow,~IEEE,}
	      Zhen~Gao,~\IEEEmembership{Member,~IEEE,}\\
	      Sheng~Chen,~\IEEEmembership{Life Fellow,~IEEE,}
	      H.~Vincent~Poor,~\IEEEmembership{Life Fellow,~IEEE}
\thanks{This work was supported in part by the National Natural Science Foundation of
	China under Grant U22B2057, in part by the Postdoctoral Fellowship Program of
	CPSF under Grant Number GZC20231374, in part by the China Postdoctoral
	Science Foundation under Grant Number 2024M751680, and in part by the grant from the Princeton Language and Intelligence. (\textit{Corresponding author:
		Zhaocheng Wang}.)}
\thanks{L.~Lu and Z.~Wang are with the Beijing National Research Center for Information Science and Technology, Department of Electronic Engineering, Tsinghua University, Beijing 100084, China (e-mails: luliyang@mail.tsinghua.edu.cn, zcwang@tsinghua.edu.cn), Z.~Wang is also with the Tsinghua Shenzhen International Graduate School, Shenzhen 518055, China.} %
\thanks{Z. Gao is with the Advanced Research Institute of Multidisciplinary Science (ARIMS), Beijing Institute of Technology, Beijing 100081, China (e-mail: gaozhen16@bit.edu.cn).} %
\thanks{S.~Chen is with the School of Electronics and Computer Science, University of Southampton, Southampton SO17 1BJ, U.K. (e-mail: sqc@ecs.soton.ac.uk).} %
\thanks{H. Vincent Poor is with the Department of Electrical and Computer Engineering, Princeton University, Princeton, NJ 08544 USA (e-mail: poor@princeton.edu).} %
\vspace*{-5mm}
}

\markboth{}%
{Shell \MakeLowercase{\textit{et al.}}: Bare Demo of IEEEtran.cls for IEEE Journals}

\maketitle

\begin{abstract}
This work explores the fundamental problem of the recoverability of a sparse tensor being reconstructed from its compressed embodiment. We present a generalized model of block-sparse tensor recovery as a theoretical foundation, where concepts measuring holistic mutual incoherence property (MIP) of the measurement matrix set are defined. A representative algorithm based on the orthogonal matching pursuit (OMP) framework, called tensor generalized block OMP (T-GBOMP), is applied to the theoretical framework elaborated  for analyzing both noiseless and noisy recovery conditions. Specifically, we present the exact recovery condition (ERC) and sufficient conditions for establishing it with consideration of different degrees of restriction. Reliable reconstruction conditions, in terms of the residual convergence, the estimated error and the signal-to-noise ratio bound, are established to reveal the computable theoretical interpretability based on the newly defined MIP, which we introduce. The flexibility of tensor recovery is highlighted, i.e., the reliable recovery can be guaranteed by optimizing MIP of the measurement matrix set. Analytical comparisons demonstrate that the theoretical results developed are tighter and less restrictive than the existing ones (if any). Further discussions provide tensor extensions for several classic greedy algorithms, indicating that the sophisticated results derived are universal and applicable to all these tensorized variants.
\end{abstract}


\IEEEpeerreviewmaketitle

%\newpage

\section{Introduction}\label{introduction} % S1

\IEEEPARstart{R}{ecovering} an $n$-mode sparse tensor $\mathcal{X}\in\mathbb{C}^{N_1\times N_2\times\cdots\times N_n}$ from  linear measurements \cite{Hoang2022,Changxiao2021,Changxiao2023}
\begin{equation}\label{tensorCSmodel} % eq.1
	\mathcal{Y}=\mathcal{X}\times_1\mathbf{D}_1\times_2\mathbf{D}_2\times_3\cdots\times_n\mathbf{D}_n+\mathcal{N},
\end{equation}
occurs in many applications, such as channel estimation \cite{ch2022} and direction-of-arrival estimation \cite{hzheng2023}, where $\mathcal{Y}\in\mathbb{C}^{M_1\times M_2\times\cdots\times M_n}$ is the given measurement tensor, $\mathbf{D}_t\in\mathbb{C}^{M_t\times N_t}$ $(t\in\{1,2,\cdots,n\})$ are measurement matrices with $M_t\ll N_t$, $\mathcal{N}$ denotes the additive noise, and $\times_i$  $(i\in\{1,2,\cdots,n\})$ denotes the $i$th mode product of a tensor and a matrix. More specifically, the tensor product $\mathcal{X}\times_i\mathbf{D}_i$ $(i=1,\cdots,n)$ is equivalent to first multiplying each slice of the $i$th mode of the tensor with the matrix $\mathbf{D}_i$, and then arranging the resulting matrices in sequence to form a new tensor. This model is called the generalized sparse recovery model since the conventional ones, e.g., single measurement vector (SMV) \cite{liyang2022} and multiple measurement vector (MMV) \cite{Slawski2022} models, are one-mode tensor recovery models, i.e., the parameter $n$ in (\ref{tensorCSmodel}) is equal to 1. As the exploitation of $n$ modes provides instrumental correlation information from various domains and the confirmed potential to filter out noise inference, high-dimensional tensors exhibit more desirable performance  than ones with low modes. Moreover, compared with the MMV problem, the intrinsic structure of the nonzero tensor supports is more flexible, since MMV assumes that the sparse patterns of each sparse signal are the same. These advantages, in terms of recoverability and flexibility, have sparked much interest in tensor recovery.

Representative tensor recovery algorithms, e.g., tensor orthogonal matching pursuit (T-OMP) \cite{AraUjo2019,ch2022} and tensor block OMP (T-BOMP) \cite{Zubair2014}, are promising due to their fast implementation through the greedy iterative mechanism. They gradually construct an estimated support of the sparse tensors by adding one index that is most strongly correlated with the residual into it per iteration, and then calculate the sparse approximation over the enlarged support. Different from T-OMP, T-BOMP considers the reconstruction of the sparse tensors that exhibit additional structure in the form of the nonzero atoms appearing in clusters, which are referred to as block-sparse \cite{Eldar2010,liyang2022}. Making explicit use of block sparsity obtains better provable recovery characteristic than the schemes which treat the signals being randomly sparse \cite{liyang2022}. In the applications, e.g., muti-band signal processing \cite{liyang2022tccn,Azghani2019tvt,liyang2023tcom}, the block sparsity occurs naturally.

A fundamental question in the analysis of tensor recovery is the characterization of its recoverability. The restricted isometry property (RIP) \cite{ripWen2017,ripKim2021,ripHerzet2013} is one of the main tools for measuring the recoverability of greedy algorithms in conventional SMV and MMV models. It indicates that if a measurement matrix satisfies the RIP with some suitable restricted isometry constants (RICs), then the sparse signals can be recovered reliably. However, calculating the RIC of a matrix is an NP-hard problem. By contrast, the mutual incoherence property (MIP) \cite{greed2004,Herzet2016,liyang2022} is computable, and it provides a stronger condition than that of the RIP, i.e., meeting MIP implies that RIP holds but the converse is not true \cite{cai2011}. However, the existing MIP framework is not applicable to the tensor recovery scenario since it only represents the characteristic of a single measurement matrix and cannot materialize the cross coherence within a given matrix set.

Moreover, developing less restrictive theoretical guarantees is useful for practical decision-making, which is highly dependent on an enabling tensor model. The block-sparse tensor is regarded as a generalized case with further consideration of the internal structure characteristic \cite{Boyer2016,Zubair2014}. The authors of \cite{Boyer2016} consider the case in which one of the nonzero supports of a tensor consists of multiples indices, resulting in block sparsity, and the authors then reformulate the tensor recovery problem into the MMV model-based problem. The work \cite{Zubair2014} proposes to tensorize data in a block-sparse manner via utilizing its spatial distributions. In practical scenarios, the need for block-sparse tensor recovery arises in various applications, including spectrum sensing \cite{Zhao2019blockstr,liyang2022tccn,liyang2023tcom}, channel estimation \cite{Pizzo2020holo,Wei2022multiuser,Pizzo2022,Guo2024wang,Luwang2024,Wanghduanh2024,ch2022,AraUjo2019} and sparse representation \cite{Caiafa2012Cichocki,Pfeiferliu2017,ZhangZhao2014,ZelnikRosenblum2012,WangTang2019} across different modes. The signals of interest not only possess a multidimensional structure with each mode representing a specific physical attribute like time, frequency or space, leading to the tensor structure, but also exhibit block sparsity due to factors such as multiple-band spectrum utilization and  energy spread effect \cite{Fundamentals2005,Caiafa2012Cichocki,tutorialWang2024}. By leveraging both the tensor structure and block sparsity, block-sparse tensor recovery holds promise in terms of enhancing recovery accuracy and simplifying implementation complexity. Despite these advancements, a comprehensive block-sparse tensor model elucidated through an associated MIP framework is yet to be formulated for computably feasible interpretability.

The focus of the present paper is on revealing theoretical recovery guarantees by exploiting the generalized tensor recovery model. The main contributions are summarized as follows.

\begin{enumerate}
\item We define concepts revealing the MIP of a given matrix set, as a foundation for the theoretical illumination of the tensor recovery problem. They are intended as indicators of whether the matrix set is sufficiently rich to allow for reliable reconstruction. The flexibility of tensor recovery is further highlighted, since these newly defined concepts can be optimized by adjusting the internal measurement matrices in the set. Analytical comparisons evince the newly defined concepts' stronger ability of tensor structured representation compared with conventional ones.
 	
\item A generalized tensor recovery model with block sparsity is formulated. We model a new block structure inside the sparse tensor as shadow block sparsity. It quantifies the number of related blocks between each measurement matrix and the nonzero support blocks of the sparse tensor. Meanwhile, the cases in which the tensor has regular or irregular block structure, i.e., in which the block lengths in certain modes are the same or different, are considered. Correspondingly, we present analytical solutions for these distinct scenarios with the redefined MIP concepts.
 	
\item As a generalized variant, a tensor generalized BOMP (T-GBOMP) stemming from the OMP framework that contains multiple tensor block selection mechanism, similar to those of compressive sampling matching pursuit (CoSaMP) \cite{cosamp2009} and multiple orthogonal least squares (MOLS) \cite{jwang2017mols}, is organized and summarized for the first time. The algorithm is exploited as an important tool to derive reliable reconstruction conditions of block-sparse tensor recovery. We further summarize and present several tensorized algorithm variants based on the existing greedy algorithms, and point out that the theoretical results derived are universal and applicable to all of these algorithm variants.

\item In-depth analyses are conducted based on the aforementioned preparations. The theoretical results contain two parts, i.e., the analyses on exact recovery and reliable reconstruction in the noiseless and noisy scenarios, respectively. Firstly, we present less restrictive exact recovery condition (ERC) for T-GBOMP (\textbf{Theorem~\ref{theo4}} and \emph{Remark~\ref{rmk1}}), and provide corresponding sufficient conditions, in terms of higher reconstructible sparsity level (\textbf{Theorem~\ref{theo8}}, \textbf{Corollary~\ref{Corollary8}}, \emph{Remark~\ref{R3}}, \emph{Remark~\ref{remarkremark4}}, and \emph{Remark~\ref{rmk5rmk5}}). These results guarantee that T-GBOMP reconstructs the sparsest tensor from the measurements exactly if the sparsity level falls within an upper bound. Meanwhile, the exponential convergence of residual tensor has also been demonstrated (\textbf{Theorem \ref{theo5}}). We have presented the lower and upper bounds of shadow block sparsity in \textbf{Proposition \ref{prop1}}, along with specific tensor structures corresponding to these bounds, which facilitate a more intuitive analysis of the results presented in this study. By leveraging the concept of shadow block sparsity, we have obtained analytical results pertaining to crucial parameters, such as different coherences and tensor block length. These results showcase how a general tensor can align with the tensor structure that reaches the lower bound of shadow block sparsity (\emph{Remark \ref{rmk7rmk7}}). Secondly, noisy recovery conditions are derived based on MIP, which involves upper bounds on the error between the original tensor and the estimated one under various assumptions (\textbf{Theorem~\ref{theo1}}, \textbf{Theorem~\ref{theo3}} and \textbf{Theorem~\ref{theo7}}), the lower bound of SNR required for reliable recovery (\textbf{Theorem~\ref{theo2}}), and the norm of the residual tensor at specific iterations (\textbf{Theorem~\ref{theo6}}). It is worth mentioning that the theoretical results derived are related to the number of tensor modes. Based on these findings and asymptotic analysis, it has been revealed that a larger tensor mode allows for a lower bound on the required SNR for reliable recovery (\emph{Remark~\ref{rmkrmk8}}). Additionally, block orthogonality of measurement matrices can reduce the upper bound of reconstruction error, wherein a larger block length further increases the performance gap between block orthogonal and non block-orthogonal measurement matrices (\emph{Remark~\ref{rmk9rmk9}}). The aforementioned results derived are tighter and less restrictive than the existing ones (if any and comparable), demonstrating the superiority of block-sparse tensor recovery over other methodologies.
\end{enumerate}

\section{Preliminaries}\label{S2}

\subsection{Notations}\label{S2.1}

We briefly summarize the notations used in this paper. Matrices are denoted by boldface uppercase letters, e.g., $\mathbf{D}$, and tensors are denoted by calligraphic uppercase letters, e.g., $\mathcal{X}$. Vectorization of tensor $\mathcal{X}$ is denoted by ${\rm vec}(\mathcal{X})$. The element in the $i$-th row and $j$-th column of matrix $\mathbf{D}$ is denoted by $\mathbf{D}_{(i,j)}$, and $\mathbf{D}_{(:,j)}$ denotes the $j$-th column of $\mathbf{D}$. $\mathbf{D}_{\mathbf{\Theta}}$ is a submatrix of $\mathbf{D}$ that contains the columns indexed by the set $\mathbf{\Theta}$. $\mathbf{D}\backslash \mathbf{D}_{\mathbf{\Theta}}$ is the residual matrix after all the columns indexed by $\mathbf{\Theta}$ are removed from $\mathbf{D}$. $\mathbf{D}^{\rm T}$ represents the transpose of $\mathbf{D}$, and $\mathbf{W}^{\rm H}$ denotes the conjugate transpose of $\mathbf{W}$. Given a block length, the $j$-th column-block submatrix of matrix $\mathbf{D}$ is denoted as $\mathbf{D}_{[j]}$. The set $\mathbf{\Theta}_B$ consists of the block indices of the set $\mathbf{\Theta}$ based on a given block length. $\langle\cdot,\cdot\rangle$ denotes the inner product. ${\rm inf}\{f(x)\}$ denotes the infimum of the function $f(x)$  with respect to the variable $x$. $\mathbf{I}$ stands for the identity matrix. The operation ${\rm supp}(\mathcal{X})$ returns the index set of the nonzero support tensor blocks of $\mathcal{X}$. If $\mathbf{D}_{\mathbf{\Theta}}$ has full column rank, $\mathbf{D}_{\mathbf{\Theta}}^\dag\! =\! \big(\mathbf{D}_{\mathbf{\Theta}}^{\rm H}\mathbf{D}_{\mathbf{\Theta}}\big)^{-1}\mathbf{D}^{\rm H}_{\mathbf{\Theta}}$ represents the pseudoinverse of $\mathbf{D}_{\mathbf{\Theta}}$. $\text{span}(\mathbf{D}_{\mathbf{\Theta}})$ denotes the space spanned by the columns of $\mathbf{D}_{\mathbf{\Theta}}$, and $\mathbf{P}_{\mathbf{D}_{\mathbf{\Theta}}}\! =\! \mathbf{D}_{\mathbf{\Theta}}\mathbf{D}_{\mathbf{\Theta}}^\dag$ is the projection onto $\text{span}(\mathbf{D}_{\mathbf{\Theta}})$, while $\mathbf{P}_{\mathbf{D}_{\mathbf{\Theta}}}^\bot\! =\! \mathbf{I}-\mathbf{P}_{\mathbf{D}_{\mathbf{\Theta}}}$ is the projection onto the orthogonal complement of span$(\mathbf{D}_{\mathbf{\Theta}})$. For a vector $\mathbf{r}$, ${\rm diag}(\mathbf{r})$ denotes the diagonal matrix whose diagonal elements are the entries of $\mathbf{r}$. The all zero vector/matrix/tensor is uniformly denoted by $\mathbf{0}$. 
$|\mathbf{\Theta}|$ stands for the cardinality or the block cardinality that allows repetition of set $\mathbf{\Theta}$, and $|c|$ is the absolute value of scalar $c$. Give an $n$-mode tensor $\mathcal{X}$, denote its $i$th $(i\! \in\! \{1,2,\cdots,n\})$ mode tensor index set as $\mathbf{\Theta}_i$ with $|\mathbf{\Theta}_i|\! =\! k$. The corresponding tensor index set for $\mathcal{X}$ can be represented as 
\begin{align}
	\mathbf{\Theta}\! =\! \{&(\mathbf{\Theta}_{1_1},\mathbf{\Theta}_{2_1},\cdots,\mathbf{\Theta}_{n_1}),(\mathbf{\Theta}_{1_2},\mathbf{\Theta}_{2_2},\cdots,\mathbf{\Theta}_{n_2}),\nonumber\\
	&\cdots,(\mathbf{\Theta}_{1_k},\mathbf{\Theta}_{2_k},\cdots,\mathbf{\Theta}_{n_k})\}\nonumber
\end{align}
 with $|\mathbf{\Theta}_i|\! =\! |\mathbf{\Theta}|\! =\! k$ $(i\! \in\! \{1,2,\cdots,n)\}$, where $\mathbf{\Theta}_{i_j}$ denotes the $j$th element of the set $\mathbf{\Theta}_i$ $(j\! \in\! \{1,2,\cdots,k\})$. The $(i_1,\cdots,i_n)$th tensor block of $\mathcal{X}$ with respect to $\mathbf{\Theta}$ is denoted by $\mathcal{X}_{[\mathbf{\Theta}_{1_{i_1}},\cdots,\mathbf{\Theta}_{n_{i_n}}]}\! \in\! \mathbb{C}^{d_1\times\cdots\times d_n}$. The Kronecker product is represented by $\otimes$. Based on a matrix set $\mathbf{\Upsilon}=\big\{\mathbf{D}_i\in\mathbb{C}^{M_i\times s_i d_i}, 1\le i\le n\big\}$ $(n \ge 1)$ and the tensor index set $\mathbf{\Theta}$, we denote the cascading matrix:
\begin{align}
 	\ddot{\mathbf{D}}_{\mathbf{\Theta}} = \Big[&\mathbf{D}_{n_{\mathbf{\Theta}_{{n_{[1]}}}}}\otimes \cdots \otimes \mathbf{D}_{1_{\mathbf{\Theta}_{{1_{[1]}}}}}, 
 		\mathbf{D}_{n_{\mathbf{\Theta}_{{n_{[2]}}}}}\otimes \cdots \otimes \mathbf{D}_{1_{\mathbf{\Theta}_{{1_{[2]}}}}}, \nonumber\\
 		&\cdots, \mathbf{D}_{n_{\mathbf{\Theta}_{{n_{[|\mathbf{\Theta}_n|]}}}}} \otimes \cdots \otimes \mathbf{D}_{1_{\mathbf{\Theta}_{{1_{[|\mathbf{\Theta}_1|]}}}}}\Big],\nonumber
\end{align}
where $\mathbf{D}_{t_{\mathbf{\Theta}_{{t_{[j]}}}}}$, $t\in\{1,\cdots,n\}$ and $j\in\{1,\cdots,|\mathbf{\Theta}_t|\})$, is a column-block submatrix of $\mathbf{D}_t$ indexed by $\mathbf{\Theta}_{{t_{[j]}}}$. The operation $|\cdot|_u$ denotes the non-repeating cardinality of its argument. 

\begin{figure}[!tp]
\vspace*{-1mm}
\begin{center}
\includegraphics[width=.5\textwidth]{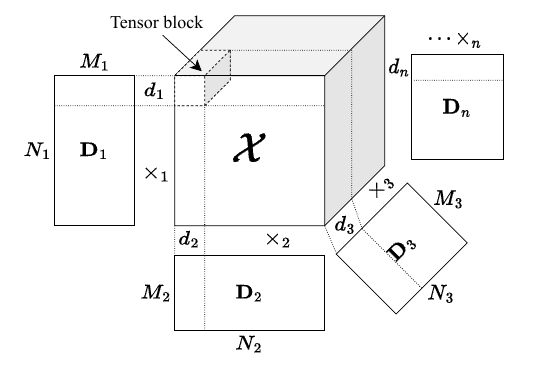}
\end{center}
\vspace*{-8mm}
\caption{Illustration of the block-sparse tensor recovery model.}
\label{blocksparsemodel} % Fig.1
\vspace*{2mm}
\end{figure}

\vspace*{-2mm}

\subsection{Block-Sparse Tensor Model and Useful Definitions}\label{S2.2}

As illustrated in Fig.~\ref{blocksparsemodel}, the block-sparse tensor $\mathcal{X}\in\mathbb{C}^{N_1 \times\cdots\times N_n}$ is compressed by the $n$ measurement matrices in the set $\mathbf{\Upsilon}=\big\{\mathbf{D}_i\in\mathbb{C}^{M_i\times N_i}, 1\le i\le n\big\}$ ($n\ge 1$), where $N_n=s_n d_n$.
 If there exist $k$ support tensor blocks with nonzero F-norms, $\mathcal{X}$ is called $k$ block-sparse tensor, and the corresponding indices constitute the set $\mathbf{\Xi}$ with $|\mathbf{\Xi}|=k$. The measurement matrices $\mathbf{D}_i$ for $i\in\{1,\cdots,n\}$ can be rewritten as a concatenation of the $s_i$ column-block submatrices, i.e.,
\begin{align} % eq.3
	\mathbf{D}_i = \Big[&\underbrace{\mathbf{D}_{i_{(:,1)}}\cdots  \mathbf{D}_{i_{(:,d)}} }_{\mathbf{D}_{i_{[1]}}} \underbrace{\mathbf{D}_{i_{(:,d+1)}}\cdots \mathbf{D}_{i_{(:,2d)}}}_{\mathbf{D}_{i_{[2]}}}\nonumber\\
	&\cdots \underbrace{\mathbf{D}_{i_{(:,N-d+1)}}\cdots \mathbf{D}_{i_{(:,N)}}}_{\mathbf{D}_{i_{[s_i]}}} \Big] ,\nonumber
\end{align}
where $\mathbf{D}_{i_{[j]}}\in \mathbb{R}^{M_i\times d_i}$ is the $j$-th column-block submatrix of $\mathbf{D}_i$. Throughout the paper, the columns in the measurement matrix are normalized to have the unit $\ell_2$-norm.

Based on the model presented, we provide the MIP framework, which is an important tool to analyze the theoretical guarantees of compressed sensing (CS) approaches \cite{cai2011}. We first give the conventional MIP concept of a matrix $\mathbf{D}$, including the matrix coherence $\underline{\mu}_D$, and the block-structure coherence $\mu_D$ and $\nu_D$. Then, we propose the new concepts of mutual block coherence and mutual sub-coherence, which can be regarded as the generalized high-dimensional extensions of the conventional MIP.

\begin{def3}\label{definitionofcoherence} % def.1
(Matrix coherence \cite{greed2004}) The coherence of a matrix $\mathbf{D}$, which represents the similarity of its elements, is defined as
\begin{equation}
	\underline{\mu}_{\mathbf{D}} = \max\limits_{\forall i,j\neq i} \big|<\mathbf{D}_{(:,i)},\mathbf{D}_{(:,j)}>\big| .\nonumber
\end{equation}
\end{def3}

\begin{def4}\label{definitionofsubcoherence} % def.2
(Block-structure coherence \cite{Eldar2010}) This block-structure coherence of a matrix $\mathbf{D}$ consists of two individual concepts, i.e., block coherence and sub-coherence. Let $d$ be the block length. The block coherence of $\mathbf{D}$ is defined as
\begin{equation} % eq.5
	\mu_{\mathbf{D}} = \max\limits_{\forall i,j\neq i} \frac{\big\|\mathbf{M}_{[i,j]}\big\|_2}{d},\nonumber
\end{equation}
where $\mathbf{M}_{[i,j]}=\mathbf{D}^{\rm H}_{[i]}\mathbf{D}_{[j]}$, and $\mathbf{D}_{[i]}$ is the $i$-th column-block submatrix of $\mathbf{D}$ with block length $d$. The sub-coherence of $\mathbf{D}$ is defined as
\begin{equation} % eq.6
	\nu_{\mathbf{D}} = \max\limits_{\forall l}\max\limits_{\forall i,j\neq i} \big| <\mathbf{D}_{{[l]}_{(:,i)}},\mathbf{D}_{{[l]}_{(:,j)}} > \big| ,\nonumber
\end{equation}
where $\mathbf{D}_{{[l]}_{(:,i)}}$ is the $i$-th column of $\mathbf{D}_{[l]}$.
\end{def4}

The block-structure coherence is a generalized extension of the matrix coherence, which is more capable of exploring the MIP-based performance guarantees offered by the block-structure characteristic of a matrix. Theoretical bounds based on the block-structure coherence are usually less restrictive than those based on conventional matrix coherence under the same assumptions \cite{Eldar2010}. When the block length $d=1$, then block-structure coherence reduces to the matrix coherence, i.e., $\mu_D=\underline{\mu}_D$ and $\nu_D=0$ \cite{Eldar2010}. For the tensor recovery case, there exists a matrix set, consisting of $n$ measurement matrices. Neither the matrix coherence nor the block-structure coherence, which are both applicable only to single matrix, can accurately measure the cross coherence among the given multiple matrices. To represent the degree of coherence among multiple block-structured matrices, we propose new coherence concepts that provide the MIP among $n$ measurement matrices as follows.

\begin{def7}\label{crossblockcoherece} % def.3
(Mutual block coherence) Given a matrix set $\mathbf{\Upsilon}=\big\{\mathbf{D}_i\in\mathbb{C}^{M_i\times s_i d_i}, 1\le i\le n\big\}$ $(n\geq 1)$, the mutual block coherence of $\mathbf{\Upsilon}$ is defined by the following two formulations.

Firstly, considering that $\forall (i_1,i_2,\cdots,i_n),(j_1\neq i_1,j_2\neq i_2,\cdots,j_n\neq i_n)$, mutual block coherence $	\varpi_{\mathbf{\Upsilon}} $ is given by 
\begin{align}\label{eqMBC} % eq.7
	\varpi_{\mathbf{\Upsilon}} 
	= &\max_{\forall (i_1,\cdots,i_n),(j_1\neq i_1,\cdots,j_n\neq i_n)}\nonumber\\
	& \bigg(\frac{1}{\prod_{i=1}^n d_i} \Big\|\big(\mathbf{D}_{n_{[i_n]}}\otimes\cdots\otimes\mathbf{D}_{1_{[i_1]}}\big)^{\rm H}\nonumber\\
	&\qquad\qquad\quad\times\big(\mathbf{D}_{n_{[j_n]}}\otimes\cdots\otimes\mathbf{D}_{1_{[j_1]}}\big)\Big\|_2\bigg)^{\frac{1}{n}},
\end{align}
where $d_i$ is the block length of the matrix $\mathbf{D}_i$.

Secondly, without loss of generality, $\forall (i_1,i_2,\cdots,i_n),(j_1=i_1,j_2=i_2,\cdots,j_t=i_t,j_{t+1}\neq i_{t+1},\cdots,j_n\neq i_n)$ $(t\geq1)$, mutual block coherence is defined as
\begin{align}\label{defofmutualblockcoherence2} % eq.8
	\varpi_{\mathbf{\Upsilon}} =& \max_{\forall (i_1,\cdots,i_n),(j_1= i_1,\cdots,j_t=i_t,j_{t+1}\neq i_{t+1},\cdots, j_n\neq i_n)}\nonumber\\
	& \bigg(\frac{1}{\prod_{i=1}^n d_i} \Big\|\big(\mathbf{D}_{n_{[i_n]}}\otimes\cdots\otimes\mathbf{D}_{t+1_{[i_{t+1}]}}\nonumber\\
	&\qquad\qquad\quad\otimes\mathbf{D}_{t_{[i_t]}}\otimes\cdots\otimes\mathbf{D}_{1_{[i_1]}}\big)^{\rm H}\nonumber\\
	&\qquad\qquad\enspace\times\big(\mathbf{D}_{n_{[j_n]}}\otimes\cdots\otimes\mathbf{D}_{t+1_{[j_{t+1}]}}\nonumber\\
	&\qquad\qquad\quad\otimes\mathbf{D}_{t_{[i_t]}}\otimes\cdots\otimes\mathbf{D}_{1_{[i_1]}}\big)\Big\|_2\bigg)^{\frac{1}{n}}.
\end{align}
\end{def7}

The definition in (\ref{eqMBC}) is more similar to the conventional sparse recovery case as defined in Definition~\ref{definitionofsubcoherence}, wherein the indices of the measurement matrix blocks used for coherence calculation differ. However, in high-dimensional tensor recovery scenarios, the measurement matrix labels corresponding to different tensor blocks may be partially the same, as long as these labels are not identical, which results in the definition in (\ref{defofmutualblockcoherence2}).

\begin{def8}\label{crosssubcoherence} % def.4
(Mutual sub-coherence) Given a matrix set $\mathbf{\Upsilon}=\big\{\mathbf{D}_i\in\mathbb{C}^{M_i\times s_i d_i}, 1\le i\le n\big\}$ $(n\geq 1)$, the mutual sub-coherence of $\mathbf{\Upsilon}$ is given by the following two definitions.

Firstly, considering that $\forall (i_1,i_2,\cdots,i_n),(j_1\neq i_1,j_2\neq i_2,\cdots,j_n\neq i_n)$, mutual sub-coherence $\tau_{\mathbf{\Upsilon}}$ is given by 
\begin{align}\label{eqMSC} % eq.9
	\tau_{\mathbf{\Upsilon}} = &\max_{\forall (i_1,\cdots,i_n),(j_1\neq i_1,\cdots,j_n\neq i_n)}\nonumber\\
	&\bigg(\Big|\big(\mathbf{D}_{n_{(:,i_n)}}\otimes\cdots\otimes\mathbf{D}_{1_{(:,i_1)}}\big)^{\rm H}\nonumber\\
&\quad\times\big(\mathbf{D}_{n_{(:,j_n)}}\otimes\cdots\otimes\mathbf{D}_{1_{(:,j_1)}}\big)\Big|\bigg)^{\frac{1}{n}},
\end{align}
where $d_i$ is the block length of the matrix $\mathbf{D}_i$, and $\mathbf{D}_{l_{(:,i_l)}}, \mathbf{D}_{l_{(:,j_l)}}\in\mathbf{D}_{l_{[t_l]}}$, $t_l\in\{1,\cdots,s_l\}$, $l\in\{1,\cdots,n\}$.

Secondly, without loss of generality, $\forall (i_1,i_2,\cdots,i_n),(j_1=i_1,j_2=i_2,\cdots,j_t=i_t,j_{t+1}\neq i_{t+1},\cdots,j_n\neq i_n)$ $(t\geq1)$, mutual sub-coherence is defined as 
\begin{align}\label{eqsubcoherencedef2} % eq.10
	\tau_{\mathbf{\Upsilon}} = &\max_{\forall (i_1,\cdots,i_n),(j_1= i_1,\cdots,j_t=i_t,j_{t+1}\neq i_{t+1},\cdots, j_n\neq i_n)}\nonumber\\
	&\bigg(\Big|\big(\mathbf{D}_{n_{(:,i_n)}}\otimes\cdots\otimes\mathbf{D}_{t+1_{(:,i_{t+1})}}\nonumber\\
	&\quad\otimes\mathbf{D}_{t_{(:,i_t)}}\otimes\cdots\otimes\mathbf{D}_{1_{(:,i_1)}}\big)^{\rm H}\nonumber\\
	&\quad\times\big(\mathbf{D}_{n_{(:,j_n)}}\otimes\cdots\otimes\mathbf{D}_{t+1_{(:,j_{t+1})}}\nonumber\\
	&\quad\otimes\mathbf{D}_{t_{(:,i_t)}}\otimes\cdots\otimes\mathbf{D}_{1_{(:,i_1)}}\big)\Big|\bigg)^{\frac{1}{n-t}}.
\end{align}
\end{def8}

More specially, the definition in (\ref{eqsubcoherencedef2}) satisfies
\begin{align}
	\tau_{\mathbf{\Upsilon}} =&\max_{\forall (i_{t+1},\cdots,i_n),(j_{t+1}\neq i_{t+1},\cdots, j_n\neq i_n)}\nonumber\\
	&\Big\|\big(\mathbf{D}^{\rm H}_{n_{(:,i_n)}}\mathbf{D}_{n_{(:,j_n)}}\big)\otimes\cdots\otimes\big(\mathbf{D}^{\rm H}_{t_{(:,i_{t+1})}}\mathbf{D}_{t_{(:,j_{t+1})}}\big)\Big\|^{\frac{1}{n-t}}_2,\nonumber
\end{align}
since the measurement matrices are normalized to have unit column norms.

Compared with the conventional and block-structure coherence, a significant advantage of the representativeness of the mutual coherence is that it can be optimized by adding a matrix with satisfactory coherence or replacing a matrix in the given set with a more desirable matrix. This indicates that the mutual coherence can provide the overall coherence variation within the matrix set.
When $n=1$ and $d\neq1$, the mutual block coherence and mutual sub-coherence degenerate into the block coherence and sub-coherence of \textbf{Definition~\ref{definitionofsubcoherence}}, respectively. On the other hand, if $n=1$ and $d=1$, they naturally reduce to the matrix coherence of \textbf{Definition~\ref{definitionofcoherence}}. Moreover, for the definition in (\ref{eqMBC}), we have
\begin{align} % eqs.12-16
  \varpi_{\mathbf{\Upsilon}} &= \max_{\forall (i_1,\cdots,i_n),(j_1\neq i_1,\cdots,j_n\neq i_n)} \nonumber\\
  &\quad\bigg(\frac{1}{\prod_{i=1}^n d_i} \Big\|\big(\mathbf{D}^{\rm H}_{n_{[i_n]}}\mathbf{D}_{n_{[j_n]}}\big)\otimes\cdots\otimes\big(\mathbf{D}^{\rm H}_{1_{[i_1]}}\mathbf{D}_{1_{[j_1]}}\big) \Big\|_2 \bigg)^{\frac{1}{n}} \label{miud1} \\
	&= \max_{\forall (i_1,\cdots,i_n),(j_1\neq i_1,\cdots,j_n\neq i_n)} \Bigg( \prod_{l=1}^n \frac{\big\|\mathbf{D}^{\rm H}_{l_{[i_l]}}\mathbf{D}_{l_{[j_l]}}\big\|_2}{d_l} \Bigg)^{\frac{1}{n}} \label{miud2} \\
  &= \bigg(\prod_{l=1}^n \mu_{\mathbf{D}_l}\bigg)^{\frac{1}{n}} \label{miud3} \\
  &\leq \bigg(\prod_{l=1}^n \underline{\mu}_{\mathbf{D}_l}\bigg)^{\frac{1}{n}} \label{miud4} \\
  &\leq 1 \label{miud5} ,
\end{align}
where (\ref{miud1}) is from the definition of mutual block coherence, (\ref{miud2}) is based on the property of the norm of Kronecker product, (\ref{miud3}) is due to the definition of block coherence, (\ref{miud4}) is from \cite[Eq. (10)]{Eldar2010} which states that given a matrix $\mathbf{D}$, $\mu_{\mathbf{D}}\leq \underline{\mu}_{\mathbf{D}}$, and (\ref{miud5}) is because $\underline{\mu}_{\mathbf{D}}\leq1$. If the matrices in the matrix set consist of orthonormal blocks, i.e., $\mathbf{D}^{\rm H}_{t_{[j]}}\mathbf{D}_{t_{[j]}}=\mathbf{I}$ for $t\in\{1,\cdots,n\}$ and $j\in\{1,\cdots,s_t\}$, we have
\begin{align}
	\varpi_{\mathbf{\Upsilon}}\leq\bigg(\prod_{l=1}^n\frac{1}{d_l}\bigg)^{\frac{1}{n}},\nonumber
\end{align}
which is because given a matrix $\mathbf{D}$ with orthonormal blocks and block length $d$, $\mu_{\mathbf{D}}\leq\frac{1}{d}$ \cite{Eldar2010}. As for the definition in (\ref{defofmutualblockcoherence2}), we have
\begin{align} % eqs.18,19
	\varpi_{\mathbf{\Upsilon}} & = \max_{\forall (i_1,\cdots,i_n),(j_1= i_1,\cdots,j_t=i_t,j_{t+1}\neq i_{t+1},\cdots, j_n\neq i_n)}\nonumber\\ &\quad\Bigg(\prod_{l=1}^{t} \frac{\big\|\mathbf{D}^{\rm H}_{l_{[i_l]}}\mathbf{D}_{l_{[i_l]}}\big\|_2}{d_l} \prod_{l=t+1}^n \frac{\big\|\mathbf{D}^{\rm H}_{l_{[i_l]}}\mathbf{D}_{l_{[j_l]}}\big\|_2}{d_l} \Bigg)^{\frac{1}{n}}\nonumber\\
	&\leq\bigg(\prod_{l=1}^{t} \frac{1+(d_l-1)\nu_{\mathbf{D}_l}}{d_l}\prod_{l=t+1}^n \mu_{\mathbf{D}_l}\bigg)^{\frac{1}{n}}\label{defnition222}\\
	&\leq1,\label{equto1}
\end{align}
where (\ref{defnition222}) is from the Ger\v{s}gorin's disc theorem \cite{matrixanalysis}, and (\ref{equto1}) is because $\frac{1+(d_l-1)\nu_{\mathbf{D}_l}}{d_l}=\frac{1}{d_l}(1-\nu_{\mathbf{D}_l})+\nu_{\mathbf{D}_l}\leq1$ and $\mu_{\mathbf{D}_l}\leq 1$.

Similarly, for the mutual sub-coherence defined in (\ref{eqMSC}), we have
\begin{align} % eqs.20-23
	\tau_{\mathbf{\Upsilon}} &= \max_{\forall (i_1,\cdots,i_n),(j_1\neq i_1,\cdots,j_n\neq i_n)} \nonumber\\
	&\quad \bigg( \Big\|\big(\mathbf{D}^{\rm H}_{n_{(:,i_n)}}\mathbf{D}_{n_{(:,j_n)}}\big)\otimes\cdots\otimes\big(\mathbf{D}^{\rm H}_{1_{(:,i_1)}}\mathbf{D}_{1_{(:,j_1)}}\big) \Big\|_2 \bigg)^{\frac{1}{n}} \label{nu1} \\
	&= \max_{\forall (i_1,\cdots,i_n),(j_1\neq i_1,\cdots,j_n\neq i_n)} \bigg( \prod_{l=1}^n \Big\| \mathbf{D}^{\rm H}_{l_{(:,i_l)}}\mathbf{D}_{l_{(:,j_l)}}\Big\|_2 \bigg)^{\frac{1}{n}} \label{nu2} \\
	&= \bigg(\prod_{l=1}^n\nu_{\mathbf{D}_l}\bigg)^{\frac{1}{n}} \label{nu3} \\
	&\leq \bigg(\prod_{l=1}^n \underline{\mu}_{\mathbf{D}_l}\bigg)^{\frac{1}{n}} , \label{nu4}
\end{align}
where (\ref{nu4}) is because given a matrix $\mathbf{D}$, $\nu_{\mathbf{D}}\leq \underline{\mu}_{\mathbf{D}}$. Based on the similar derivations in (\ref{nu1})-(\ref{nu4}), for the definition in (\ref{eqsubcoherencedef2}), we have
\begin{align}
	\tau_{\mathbf{\Upsilon}} \leq \bigg(\prod_{l=t+1}^n \underline{\mu}_{\mathbf{D}_l}\bigg)^{\frac{1}{n-t}}.\nonumber
\end{align}

\begin{rmk6}\label{R1} % Remark 1
\emph{The above definitions apply to the scenario where the block structure of the sparse tensor is regular, i.e., the block length of each block in each mode is the same. For the case in which the block lengths within the $t$th $(t\in\{1,\cdots,n\})$ mode are different, we present two useful solutions. The first one considers that irregular block lengths do not lead to a block structure, which results in a block length 1 for the modes where the block lengths are different. The second one considers this irregular block lengths case as a generalized block structure, and we provide extended MIP concepts serving for this generalized structural characteristic. Note that we consider the definitions in (\ref{eqMBC}) and (\ref{eqMSC}) as examples, and similar analyses can be easily extended to those of the definitions in (\ref{defofmutualblockcoherence2}) and (\ref{eqsubcoherencedef2}).}

\emph{We now present the first solution. The case in which the block lengths of the $t$th mode are different is provided as a concise example, which can be directly extended to the scenario where the block lengths of multiple modes are different. To this end, let the block length of the $t$th $(t\in\{1,\cdots,n\})$ mode be 1, i.e., $d_t=1$. From \textbf{Definitions~\ref{crossblockcoherece}} and \textbf{\ref{crosssubcoherence}}, we have
\begin{align} % eqs.25,26
	\varpi_{\mathbf{\Upsilon}} =& \max_{(i_1,\cdots,i_t,\cdots,i_n),(j_1\neq i_1,\cdots,j_t\neq i_t,\cdots,j_n\neq i_n)}\nonumber\\
	&\bigg(\frac{1}{\big(\prod_{l=1}^{t-1} d_l\big) d_t \big(\prod_{l=t+1}^n d_l\big)} \nonumber \\
	& \times \Big\|\big(\mathbf{D}_{n_{[i_n]}}\otimes\cdots\otimes\mathbf{D}_{t_{[i_t]}}\otimes\cdots\otimes\mathbf{D}_{1_{[i_1]}}\big)^{\rm H}\nonumber\\
	&\quad\enspace\times\big(\mathbf{D}_{n_{[j_n]}}\otimes\cdots\otimes\mathbf{D}_{t_{[j_t]}}\otimes\cdots\otimes\mathbf{D}_{1_{[j_1]}}\big)\Big\|_2\bigg)^{\frac{1}{n}} \nonumber \\
	=& \max_{(i_1,\cdots,i_n),(j_1\neq i_1,\cdots,j_n\neq i_n)}\nonumber\\
	&\bigg(\frac{1}{\big(\prod_{l=1}^{t-1} d_l\big) \big(\prod_{l=t+1}^n d_l\big)} \nonumber \\
	& \times \Big\|\big(\mathbf{D}_{n_{[i_n]}}\otimes\cdots\otimes\mathbf{D}_{t_{(:,i_t)}}\otimes\cdots\otimes\mathbf{D}_{1_{[i_1]}}\big)^{\rm H}\nonumber\\
	&\quad\enspace\times\big(\mathbf{D}_{n_{[j_n]}}\otimes\cdots\otimes\mathbf{D}_{t_{(:,j_t)}}\otimes\cdots\otimes\mathbf{D}_{1_{[j_1]}}\big)\Big\|_2\bigg)^{\frac{1}{n}} , \nonumber \\
	\tau_{\mathbf{\Upsilon}} =& \max_{(i_1,\cdots,i_n),(j_1\neq i_1,\cdots,j_n\neq i_n)}\nonumber\\ &\bigg(\Big\|\big(\mathbf{D}_{n_{(:,i_n)}}\otimes\cdots\otimes\mathbf{D}_{t_{(:,i_t)}}\otimes\cdots\otimes\mathbf{D}_{1_{(:,i_1)}}\big)^{\rm H} \nonumber \\
	& \quad\times\big(\mathbf{D}_{n_{(:,j_n)}}\otimes\cdots\otimes\mathbf{D}_{t_{(:,j_t)}}\otimes\cdots\otimes\mathbf{D}_{1_{(:,j_1)}}b\big)\Big\|_2\bigg)^{\frac{1}{n}} . \nonumber
\end{align}
Thus, similar to (\ref{miud3}) and (\ref{nu3}), we have
\begin{align} % eqs.27,28
	\varpi_{\mathbf{\Upsilon}} =& \underline{\mu}_{\mathbf{D}_t}^{\frac{1}{n}} \bigg(\prod_{l=1}^{t-1} \mu_{\mathbf{D}_l} \prod_{l=t+1}^{n}\mu_{\mathbf{D}_{l}}\bigg)^{\frac{1}{n}} , \nonumber \\
	\tau_{\mathbf{\Upsilon}} =& \underline{\mu}_{\mathbf{D}_t}^{\frac{1}{n}}\bigg(\prod_{l=1}^{t-1}\nu_{\mathbf{D}_l} \prod_{l=t+1}^{n} \nu_{\mathbf{D}_{l}} \bigg)^{\frac{1}{n}} . \nonumber
\end{align}
Under a special case where $d_1=\cdots=d_n=1 $, we define $\tau_{\mathbf{\Upsilon}}=0$.
In a second case, suppose that the block lengths of the $t$th $(t\in\{1,\cdots,n\})$ mode are different, and there are $s_t$ tensor blocks in the $t$th mode with the block lengths being $d_{t_{i}}$, $i\in\{1,\cdots,s_t\}$. Then, we define
\begin{align} % eqs.29,30
	\varpi_{\mathbf{\Upsilon}} &=\! \max_{(i_1,\cdots,i_n),(j_1\neq i_1,\cdots,j_n\neq i_n)}\nonumber\\
	& \quad\bigg(\frac{1}{\prod_{t=1}^{n}\! \sqrt{d_{t_{i_n}} d_{t_{j_n}}}}  
	  \Big\|\big(\mathbf{D}_{n_{[i_n]}}\otimes\cdots\otimes\mathbf{D}_{1_{[i_1]}}\big)^{\rm H}\nonumber\\
	  &\qquad\qquad\qquad\qquad\times\big(\mathbf{D}_{n_{[j_n]}}\otimes\cdots\otimes\mathbf{D}_{1_{[j_1]}}\big)\Big\|_2\bigg)^{\frac{1}{n}}\! ,\! \nonumber \\
	\tau_{\mathbf{\Upsilon}} &= \max_{\forall (i_1,\cdots,i_n),(j_1\neq i_1,\cdots,j_n\neq i_n)}\nonumber\\
	&\quad\bigg(\Big\|\big(\mathbf{D}_{n_{(:,i_n)}}\otimes\cdots\otimes\mathbf{D}_{1_{(:,i_1)}}\big)^{\rm H}\nonumber\\
&\qquad\times\big(\mathbf{D}_{n_{(:,j_n)}}\otimes\cdots\otimes\mathbf{D}_{1_{(:,j_1)}}\big)\Big\|_2\bigg)^{\frac{1}{n}}\! .\! \label{notneatau22}
\end{align}
It can be seen that $\tau_{\mathbf{\Upsilon}}$ defined in (\ref{notneatau22}) is the same as that in (\ref{eqMSC}) with indices $i_t$ and $j_t$ $(t\in\{1,\cdots,n\})$ being within the same block of $\mathbf{D}_{t}$.
It can be observed that the first solution does not require any additional input of structural information for reliable recovery analysis as the block length is regarded as 1, while the second solution relies on specific block length information of the irregular structure. For concise symbol representation, in the sequel, we consider the theoretical derivation of the regular block structure scenario, which can be directly extended to the irregular block length case by using the two solutions presented in this remark.}
\end{rmk6}

For symbol simplicity, we denote the mutual block coherence and mutual sub-coherence as $\varpi_{\mathbf{\Upsilon},t}$ and $\tau_{\mathbf{\Upsilon},t}$, respectively, when $t$ indices $(0\leq t\leq n)$ are the same in their definitions. Furthermore, when $t=0$, the mutual block coherence and mutual sub-coherence can be simplified to be represented as $\varpi_{\mathbf{\Upsilon}}$ and $\tau_{\mathbf{\Upsilon}}$, respectively.
In the following, a new structural concept of the block-sparse tensor is presented, which can figuratively display the number of measurement matrix blocks that are related to the support tensor blocks in tensor recovery.

\begin{figure}[!t]
%\vspace*{-1mm}
\begin{center}
\includegraphics[width=.47\textwidth]{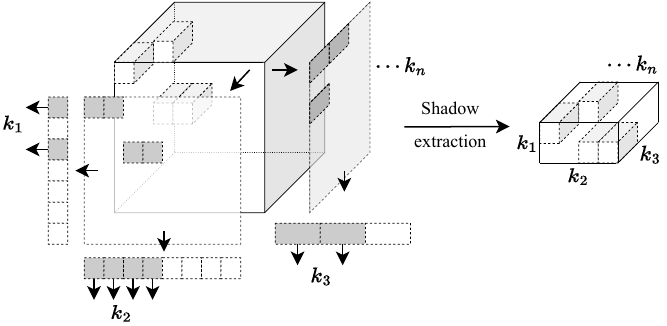}
\end{center}
\caption{Illustration of shadow block sparsity and shadow extraction.}
\label{shadow-block-MODEL} % Fig.2
\vspace*{2mm}
\end{figure}

\begin{def5}\label{definitionofshadowblocksparsity} % def.5
(Shadow block sparsity) As illustrated in Fig.~\ref{shadow-block-MODEL}, the $i$th $(i=1,2\cdots,n)$ mode shadow block sparsity $k_i$ of a tensor $\mathcal{X}\in\mathbb{C}^{N_1\times\cdots\times N_n}$ is defined as the number of nonzero blocks blocking the field of view in the $i$th mode direction.
\end{def5}

Denoting the index set of nonzero tensor blocks in $\mathcal{X}$ as $\mathbf{\Xi}\! =\! \{(\mathbf{\Xi}_{1_1},\mathbf{\Xi}_{2_1},\cdots,\mathbf{\Xi}_{n_1}),(\mathbf{\Xi}_{1_2},\mathbf{\Xi}_{2_2},\linebreak\cdots,\mathbf{\Xi}_{n_2}),\cdots,(\mathbf{\Xi}_{1_k},\mathbf{\Xi}_{2_k},\cdots,\mathbf{\Xi}_{n_k})\}$, we have $k_i\! =\! |\mathbf{\Xi}_i|_u$ and $k_i\! \in\! [1,\,k]$ $(i\in[1,2,\cdots,n])$, since there exist duplicate indices in $\mathbf{\Xi}_i$, where the elements in $\mathbf{\Xi}$ are $n$-dimensional.
 We further define a shadow extraction operation, which returns a neat subtensor consisting of the nonzero supports of the original tensor but with sufficient gaps as shown in Fig. \ref{shadow-block-MODEL}. In other words, the shadow extraction operation removes all of the zero slices of the tensor and returns the obtained subtensor, which is sized as $k_1\times k_2\times\cdots\times k_n$. Mathematically, given an index set $\mathbf{\Theta}$, the extraction operation on $\mathcal{X}_{\mathbf{\Theta}}$ returns the neat subtensor consisting of the tensor blocks indexed by $\mathbf{\Theta}$ and with sufficient gaps.

To aid the derivation of both exact and reliable recovery conditions in noiseless and noisy scenarios, we further present the definitions of block mixed norm, SNR and an extension of minimum-to-average ratio (MAR) \cite{jwang2017mols} on block-sparse tensors as follows.

\begin{def6}\label{defofmixednorm} % def.6
(Block mixed norm \cite{Eldar2010}) For a matrix $\mathbf{D}\in\mathbb{C}^{M\times N}$ with $M=m d$ and $N=n d$, denote $\mathbf{D}_{[i,j]}$ as the $(i,j)$th $d\times d$ block of $\mathbf{D}$. The two components of the block mixed norm of $\mathbf{D}$ are given by
\begin{align} % eqs.31,32
	\rho_r(\mathbf{D}) =& \max_i\sum_j\|\mathbf{D}_{[i,j]}\|_2, \nonumber \\
	\rho_c(\mathbf{D}) =& \max_j\sum_i\|\mathbf{D}_{[i,j]}\|_2. \nonumber
\end{align}
\end{def6}

\begin{def2}\label{defofSNR} % def.7
(SNR) The SNR is defined as the ratio of the signal component to the noise component, i.e.,
\begin{align}
	{\rm SNR} =& \frac{\|\mathcal{X}\times_1\mathbf{D}_1\times_2\mathbf{D}_2\times_3\cdots\times_n\mathbf{D}_n\|^2_F}{\|\mathcal{N}\|^2_F}.\nonumber
\end{align}
\end{def2}

\begin{def7}\label{defofmar} % def.8
(MAR on block-sparse tensors) Given a $k$ block-sparse tensor $\mathcal{X}$ whose nonzero support tensor blocks are indexed by the set $\mathbf{\Xi}$, the MAR on $\mathcal{X}$ is defined as
\begin{align}
	{\rm MAR}_* =&\frac{\sqrt{k}\min\limits_{(i_1,\cdots,i_n)\in\mathbf{\Xi}} \big\|\mathcal{X}_{[i_1,\cdots,i_n]}\|_F}{\|\mathcal{X}\big\|_F}.\nonumber
	\end{align}
\end{def7}

\section{Useful Lemmas and Block-Sparse Tensor Recovery Algorithms}\label{3}

\subsection{Some Useful Lemmas}\label{3.1}

Analyses related to the extreme eigenvalues of measurement matrices are useful, and they have been exploited in many works, such as \cite{liyang2022,cai2011}. Before presenting the theoretical results bounding the eigenvalues, a useful proposition is provided first for an intuitive expression of the eigenvalue bounds.

\begin{proposition2}\label{prop2} % Propo1
Consider a matrix set $\mathbf{\Upsilon}=\{\mathbf{D}_i\in\mathbb{C}^{M_i\times s_i d_i} , 1\le i\le n\}$ $(n\geq1)$. Denote $G_i = s_i-1$, and define the following generating function:
\begin{align}
	g(x) &= (x+G_1)(x+G_2)\cdots(x+G_n) \nonumber \\
	&=C_nx^n+C_{n-1}x^{n-1}+\cdots+C_0x^0.\nonumber
\end{align}
For the block matrix $\mathbf{T}^{\rm H}\mathbf{T}$ whose blocks are $(\mathbf{D}_{1_{[i_1]}}\otimes\mathbf{D}_{2_{[i_2]}}\otimes\cdots\otimes\mathbf{D}_{n_{[i_t]}}\otimes\mathbf{D}_{n_{[i_{t+1}]}}\otimes\cdots\otimes\mathbf{D}_{n_{[i_n]}})^{\rm H}(\mathbf{D}_{1_{[j_1]}}\otimes\mathbf{D}_{2_{[j_2]}}\otimes\cdots\otimes\mathbf{D}_{n_{[j_t]}}\otimes\mathbf{D}_{n_{[j_{t+1}]}}\otimes\cdots\otimes\mathbf{D}_{n_{[j_n]}})$ sized as $\prod_{i=1}^nd_i\times\prod_{i=1}^nd_i$, where $\mathbf{T}=\mathbf{D}_1\otimes\mathbf{D}_2\otimes\cdots\otimes\mathbf{D}_n$, the number of blocks in the block matrix $\mathbf{T}^{\rm H}\mathbf{T}$ satisfying $i_1=j_1$, $i_2=j_2$, $\cdots$, $i_t=j_t$ is equal to $C_t\prod_{l=1}^ns_l$ $(1\leq t\leq n)$. Specifically, if there is no identical index, the corresponding number is equal to $C_0\prod_{l=1}^ns_l$. If considering block matrices on a certain column or row, then the aforementioned numbers are equal to $C_t$ and $C_0$, respectively.
  
Similarly, denote $\underline{G}_i = d_i-1$, and define the following generating function:
\begin{align}
  \underline{g}(x) &= (x+\underline{G}_1)(x+\underline{G}_2)\cdots(x+\underline{G}_n)\nonumber\\
  &=\underline{C}_nx^n+\underline{C}_{n-1}x^{n-1}+\cdots+\underline{C}_0x^0.\nonumber
\end{align}
For the matrix $\underline{\mathbf{T}}^{\rm H}\underline{\mathbf{T}}=\big(\mathbf{D}_{1_{[i_1]}}\otimes\mathbf{D}_{2_{[i_2]}}\otimes\cdots\otimes\mathbf{D}_{t_{[i_{t}]}}\otimes\mathbf{D}_{t+1_{[i_{t+1}]}}\otimes\cdots\otimes\mathbf{D}_{n_{[i_n]}}\big)^{\rm H} \big(\mathbf{D}_{1_{[i_1]}}\otimes\mathbf{D}_{2_{[i_2]}}\otimes\cdots\otimes\mathbf{D}_{t_{[i_{t}]}}\otimes\mathbf{D}_{t+1_{[i_{t+1}]}}\otimes\cdots\otimes\mathbf{D}_{n_{[i_n]}}\big)$, where $\underline{\mathbf{T}}=\mathbf{D}_{1_{[i_1]}}\otimes\mathbf{D}_{2_{[i_2]}}\otimes\cdots\otimes\mathbf{D}_{t_{[i_{t}]}}\otimes\mathbf{D}_{t+1_{[i_{t+1}]}}\otimes\cdots\otimes\mathbf{D}_{n_{[i_n]}}$, whose elements are $\big(\mathbf{D}_{1_{[i_1]_{(:,j_1)}}}\otimes\mathbf{D}_{2_{[i_2]_{(:,j_2)}}}\otimes\cdots\otimes\mathbf{D}_{t_{[i_{t}]_{(:,j_t)}}}\otimes\mathbf{D}_{t+1_{[i_{t+1}]_{(:,j_{t+1})}}}\otimes\cdots\otimes\mathbf{D}_{n_{[i_n]_{(:,j_n)}}}\big)^{\rm H} \big(\mathbf{D}_{1_{[i_1]_{(:,q_1)}}}\otimes\mathbf{D}_{2_{[i_2]_{(:,q_2)}}}\otimes\cdots\otimes\mathbf{D}_{t_{[i_{t}]_{(:,q_t)}}}\otimes\mathbf{D}_{t+1_{[i_{t+1}]_{(:,q_{t+1})}}}\otimes\cdots\otimes\mathbf{D}_{n_{[i_n]_{(:,q_n)}}}\big)$, the number of elements in $\underline{\mathbf{T}}^{\rm H}\underline{\mathbf{T}}$ satisfying $j_1=q_1$, $j_2=q_2$, $\cdots$, $j_t=q_t$ is equal to $\underline{C}_t\prod_{l=1}^ns_l$ $(1\leq t\leq n)$. Specifically, if there is no identical index, the corresponding number is equal to $\underline{C}_0\prod_{l=1}^ns_l$. If considering elements on a certain column or row, then the aforementioned numbers are equal to $\underline{C}_t$ and $\underline{C}_0$, respectively.
\end{proposition2}

Note that $\sum\limits_{l=0}^nC_l=\prod\limits_{l=1}^ns_l$ and $\sum\limits_{l=0}^n\underline{C}_l=\prod\limits_{l=1}^nd_l$. Meanwhile, $C_t$ and $\underline{C}_t$ can be formulated concretely as  
\begin{align} % eqs.37,38
	C_t &= \sum_{1\leq j_1<j_2<\cdots<j_{n-t}<\cdots\leq n}G_{j_1}G_{j_2}\cdots G_{j_n-t}\nonumber\\
	&=\sum_{1\leq j_1<j_2<\cdots<j_{n-t}<\cdots\leq n}\prod_{i=1}^{n-t}G_{ji}, \nonumber \\
	\underline{C}_t &= \sum_{1\leq j_1<j_2<\cdots<j_{n-t}<\cdots\leq n}\underline{G}_{j_1}\underline{G}_{j_2}\cdots \underline{G}_{j_n-t} \nonumber \\
	&=\sum_{1\leq j_1<j_2<\cdots<j_{n-t}<\cdots\leq n}\prod_{i=1}^{n-t}\underline{G}_{ji}, \nonumber
\end{align}
respectively. For the sake of simplicity in notation, in the sequel, we denote $C_t$ satisfying $\sum\limits_{l=0}^nC_l=\alpha$ and $\underline{C}_t$ satisfying $\sum\limits_{l=0}^n\underline{C}_l=\beta$ as $C_t^{\alpha}$ and $\underline{C}_t^{\beta}$, respectively. Based on \textbf{Proposition \ref{prop2}}, the following lemma reveals that the eigenvalues of the Kronecker product of the measurement matrices are bounded in terms of mutual block coherence and mutual sub-coherence.

\begin{lemma5}\label{lemma5} % L1
Consider a matrix set $\mathbf{\Upsilon}=\{\mathbf{D}_i\in\mathbb{C}^{M_i\times s_i d_i} , 1\le i\le n\}$ $(n\geq1)$. Let $\mathbf{T}=\mathbf{D}_1\otimes\cdots\otimes\mathbf{D}_n$. Define $\lambda_{\min}$ and $\lambda_{\max}$ as the minimum and maximum eigenvalues of the matrix $\mathbf{T}^{\rm H}\mathbf{T}$. If $\underline{W}_{\mathbf{\Upsilon},\prod_{t=1}^ns_t}>0$, then the following inequality holds:
\begin{align}
	& \underline{W}_{\mathbf{\Upsilon},\prod_{t=1}^ns_t}\leq \lambda_{\min}
		\leq \lambda_{\max}\leq \overline{W}_{\mathbf{\Upsilon},\prod_{t=1}^ns_t} ,\nonumber
\end{align}
where
\begin{align} % eqs.40,41
	\underline{W}_{\mathbf{\Upsilon},\prod_{t=1}^{n}s_t} =& 1\!\!-\!\!\sum_{t=0}^{n-1}\underline{C}^{\prod_{t=1}^{n}d_t}_t\tau_{\mathbf{\Upsilon},t}^{n-t}
	\!\!-\!\!\sum_{t=0}^{n-1}C^{\prod_{t=1}^{n}s_t}_t\varpi_{\mathbf{\Upsilon},t}^n\prod^{n}_{i=1}d_i, \nonumber \\
	\overline{W}_{\mathbf{\Upsilon},\prod_{t=1}^{n}s_t} =& 1\!\!+\!\!\sum_{t=0}^{n-1}\underline{C}^{\prod_{t=1}^{n}d_t}_t\tau_{\mathbf{\Upsilon},t}^{n-t}
	\!\!+\!\!\sum_{t=0}^{n-1}C^{\prod_{t=1}^{n}s_t}_t\varpi_{\mathbf{\Upsilon},t}^n\prod^{n}_{i=1}d_i, \nonumber
\end{align}
in which $\varpi_{\Upsilon}$ and $\tau_{\Upsilon}$ are the mutual block coherence and mutual sub-coherence of the matrix set $\mathbf{\Upsilon}$, respectively.
\end{lemma5}

\begin{IEEEproof}
See Appendix~\ref{proofoflemma5}.
\end{IEEEproof}

It can be observed from \textbf{Lemma~\ref{lemma5}} that as the mutual block coherence $\varpi_{\Upsilon}$ and the mutual sub-coherence $\tau_{\Upsilon}$ decrease, $\underline{W}_{\mathbf{\Upsilon},\prod_{t=1}^{n}s_t}$ and $\overline{W}_{\mathbf{\Upsilon},\prod_{t=1}^{n}s_t}$ approach 1, indicating tighter eigenvalue bounds. Generally speaking, to recover the block-sparse tensor accurately, the degree of linear dependency among the entries of the corresponding measurement matrices needs to be small. This reveals that there is a significant amount of effective information in the measurement matrices. Assume that the measurement matrices in the set $\mathbf{\Upsilon}$ are block orthogonal, which results in the mutual sub-coherence $\tau_{\mathbf{\Upsilon}}$ being equal to 0, and we have
\begin{align}\label{lamdanewbounds} % eq.42
	&1 - \sum_{t=0}^{n-1}C^{\prod_{t=1}^{n}s_t}_t\varpi_{\mathbf{\Upsilon},t}^n\prod^{n}_{i=1}d_i\leq\lambda_{\min}\nonumber\\
		&\qquad\qquad\leq\lambda_{\max}\leq1+\sum_{t=0}^{n-1}C^{\prod_{t=1}^{n}s_t}_t\varpi_{\mathbf{\Upsilon},t}^n\prod^{n}_{i=1}d_i.
\end{align}
The bounds in (\ref{lamdanewbounds}) becomes tighter than those in \textbf{Lemma~\ref{lemma5}}, which indicates that a block orthogonal structure can bring significant improvements in the eigenvalue boundary. Due to the orthogonality of the measurement matrix blocks, the atomic information contained within these blocks is relatively high, leading to a more reliable recovery performance.

\textbf{Lemma~\ref{lemma5}} is applicable to symmetric matrices, while, in the general case, the matrices to be analyzed are not symmetric matrices. Therefore, we present the following lemma in which an upper bound on the matrix spectral norm is provided, which is suitable for any given matrix.

\begin{lemma4}\label{lemma4} % L2
Given a matrix $\mathbf{D}\in\mathbb{C}^{M\times N}$ with $M=md$ and $N=nd$, we have
\begin{align} % eqs.43,44
	\|\mathbf{D}\|_2&\leq\sqrt{\rho_c(\mathbf{D})\rho_r(\mathbf{D})}\label{gersthey1}\\
	&\leq\max\{\rho_r(\mathbf{D}),\rho_c(\mathbf{D})\}, \label{gersthey22}
\end{align}
where $\mathbf{D}_{[i,j]}$ is the $(i,j)$th $d\times d$ block of $\mathbf{D}$.
\end{lemma4}

\begin{IEEEproof}
See Appendix \ref{proofoflemmager}.
\end{IEEEproof}

It is interesting that (\ref{gersthey1}) can be seen as an extension of the Ger\v{s}gorin's disc theorem \cite{matrixanalysis} for any block matrix and does not require the restriction that matrices need to be square. Moreover, the inequality (\ref{gersthey22}) introduces a scaling, which is useful as it allows to bound the matrix norm based on just one of the block mixed norms. In certain particular cases where the rows and columns of a matrix have different meanings, such as the rows of a measurement matrix represent coefficients from different candidate entries and the columns represent individual entries, the inequality (\ref{gersthey22}) can provide intuitive bounds that reflect the characteristics of either the rows or the columns of the matrix.

Based on \textbf{Lemma~\ref{lemma4}}, for the case in which the matrix is square, the following corollary holds.

\begin{Corollary1}\label{Corollary1} % C1
Given a symmetric matrix $\mathbf{D}\in\mathbb{C}^{M\times M}$ with $M=m d$, we have
\begin{align} 
  \|\mathbf{D}\|_2\leq\min\{\rho_r(\mathbf{D}),\rho_c(\mathbf{D})\},\nonumber
\end{align}
where $\mathbf{D}_{[i,j]}$ is the $(i,j)$th $d\times d$ block of $\mathbf{D}$.
\end{Corollary1}

The proof of Corollary \ref{Corollary1} is direct since the spectral radius of a symmetric matrix is equal to its spectral norm, and the spectral radius of a square matrix is less than or equal to any other matrix norm. Now consider a symmetric matrix $\mathbf{D}_*$. Ger\v{s}gorin's disc theorem on $\mathbf{D}_*$ is given by
\begin{align}\label{Gerbounds} % eq.46
	\|\mathbf{D}_*\|_2\leq\min\{\|\mathbf{D}_*\|_{\infty},\|\mathbf{D}_*\|_{1}\}.
\end{align}
Since
\begin{align} % eqs.47,48
	\|\mathbf{D}_*\|_{\infty} &= \max_l\sum_r\big|\mathbf{D}_{*_{(l,r)}}\big| \nonumber\\
	&\ge
	  \max_i\sum_j\big\|\mathbf{D}_{*_{[i,j]}}\big\|_2 =
	  \rho_r(\mathbf{D}_*), \nonumber \\
	\|\mathbf{D}_*\|_{1} &= \max_r\sum_l\big|\mathbf{D}_{*_{(l,r)}}\big| \nonumber\\
	&\ge
	  \max_j\sum_i\big\|\mathbf{D}_{*_{[i,j]}}\big\|_2 =
	  \rho_c(\mathbf{D}_*),\nonumber
\end{align}
we have $\min\{\rho_r(\mathbf{D}_*),\rho_c(\mathbf{D}_*)\}\leq\min\{\|\mathbf{D}_*\|_{\infty},\|\mathbf{D}_*\|_{1}\}$, which indicates that the bound in \textbf{Corollary~\ref{Corollary1}} is tighter than (\ref{Gerbounds}).

Exploiting \textbf{Lemmas~\ref{lemma5}} and \textbf{\ref{lemma4}}, we present the following two corollaries. They are the building blocks for the feasible derivation of both exact and reliable recovery conditions in noiseless and noisy scenarios.

\begin{Corollary4}\label{Corollary4} % C2
Consider a matrix set $\mathbf{\Upsilon}=\{\mathbf{D}_i\in\mathbb{C}^{M_i\times s_i d_i} , 1\le i\le n\}$ $(n\geq1)$, and $n$ pairs of column-block submatrices, $\mathbf{D}_{1_{\mathbf{\Xi}_1}}$ and $\mathbf{D}_{1_{\mathbf{\Xi}^*_1}}$ of $\mathbf{D}_1$, $\cdots$, $\mathbf{D}_{n_{\mathbf{\Xi}_n}}$ and $\mathbf{D}_{n_{\mathbf{\Xi}^*_n}}$ of $\mathbf{D}_n$ with $\mathbf{\Xi}_1\cap\mathbf{\Xi}^*_1=\mathbf{\emptyset}$, $\cdots$, $\mathbf{\Xi}_n\cap\mathbf{\Xi}^*_n=\mathbf{\emptyset}$, $|\mathbf{\Xi}_1|=l_1$,  $|\mathbf{\Xi}^*_1|=l_1^*$, $\cdots$, $|\mathbf{\Xi}_n|=l_n$ and $|\mathbf{\Xi}_n^*|=l_n^*$. Let $\mathbf{T}=\mathbf{D}_{1_{\mathbf{\Xi}_1}}\otimes\cdots\otimes\mathbf{D}_{n_{\mathbf{\Xi}_n}}$ and $\mathbf{T}_*=\mathbf{D}_{1_{\mathbf{\Xi}^*_1}}\otimes\cdots\otimes\mathbf{D}_{n_{\mathbf{\Xi}^*_n}}$. Define $\sigma_{\min}$ and $\sigma_{\max}$ as the minimum and maximum singular values of $\mathbf{T}^{\rm H}\mathbf{T}_*$. If $\underline{W}_{\mathbf{\Upsilon},\prod_{t=1}^{n}l_t}>0$ and $\underline{W}_{\mathbf{\Upsilon},\prod_{t=1}^{n}l^*_t}>0$, then the following inequality holds:
\begin{align}
	&\underline{W}_{\mathbf{\Upsilon},\prod_{t=1}^{n}l_t}^{\frac{1}{2}}\underline{W}_{\mathbf{\Upsilon},\prod_{t=1}^{n}l^*_t}^{\frac{1}{2}} \leq \sigma_{\min}\nonumber\\
		&\qquad\leq \sigma_{\max}\leq \max\Big\{\varpi_{\mathbf{\Upsilon}}^n\prod_{t=1}^{n}l_td_t,\varpi_{\mathbf{\Upsilon}}^n\prod_{t=1}^{n}l^*_td_t\Big\},\nonumber
\end{align}
where $\underline{W}_{\mathbf{\Upsilon},l_n}$ and $\underline{W}_{\mathbf{\Upsilon},l^*_n}$ can be obtained by \textbf{Lemma \ref{lemma5}}, and $d_t$ $(t\in\{1,2,\cdots,n\})$ is the block length of $\mathbf{D}_t$.
\end{Corollary4}

\begin{IEEEproof}
See Appendix~\ref{proofofCorollary4}.
\end{IEEEproof}

\begin{Corollary6}\label{Corollary6} % C3
Consider an $n$-mode $k$ block-sparse tensor $\mathcal{X}\in\mathbb{C}^{N_1\times N_2\times\cdots\times N_n}$ with block length and block shadow sparsity of the $n$th mode equal to $d_n$ and $k_n$ respectively, and a measurement matrix set $\mathbf{\Upsilon}=\{\mathbf{D}_i\in\mathbb{C}^{M_i\times s_i d_i} , 1\le i\le n\}$ $(n\geq1)$. Then we have
\begin{align} % eqs.50,51
	 \underline{W}_{\mathbf{\Upsilon},\prod_{t=1}^{n}k_t}^{\frac{1}{2}}\|\mathcal{X}\|_F\leq&\|\mathcal{X}\times_1\mathbf{D}_1\times_2\cdots\times_n\mathbf{D}_n\|_F\nonumber\\
		\leq& \overline{W}_{\mathbf{\Upsilon},\prod_{t=1}^{n}k_t}^{\frac{1}{2}}\|\mathcal{X}\|_F, \nonumber \\
	 \underline{W}_{\mathbf{\Upsilon},\prod_{t=1}^{n}k_t}\|\mathcal{X}\|_F
		\leq &\|\mathcal{X}\times_1(\mathbf{D}_1^{\rm H}\mathbf{D}_1)\times_2(\mathbf{D}_2^{\rm H}\mathbf{D}_2)\nonumber\\
		&\;\times_3\cdots\times_n(\mathbf{D}_n^{\rm H}\mathbf{D}_n)\|_F\nonumber\\
		\leq &\overline{W}_{\mathbf{\Upsilon},\prod_{t=1}^{n}k_t}\|\mathcal{X}\|_F,\label{Corollary6main2}
\end{align}
where $\underline{W}_{\mathbf{\Upsilon},k_n}$ and $\overline{W}_{\mathbf{\Upsilon},k_n}$ can be obtained by \textbf{Lemma~\ref{lemma5}}.
		
For $n$ pairs of column-block submatrices $\mathbf{D}_{1_{\mathbf{\Xi}_1}}$ and $\mathbf{D}_{1_{\mathbf{\Xi}^*_1}}$ of $\mathbf{D}_1$, $\cdots$, $\mathbf{D}_{n_{\mathbf{\Xi}_n}}$ and $\mathbf{D}_{n_{\mathbf{\Xi}^*_n}}$ of $\mathbf{D}_n$ with $\mathbf{\Xi}_1\cap\mathbf{\Xi}^*_1=\mathbf{\emptyset}$, $\cdots$, $\mathbf{\Xi}_n\cap\mathbf{\Xi}^*_n=\mathbf{\emptyset}$, $|\mathbf{\Xi}_1|=l_1$,  $|\mathbf{\Xi}^*_1|=l_1^*$, $\cdots$, $|\mathbf{\Xi}_n|=l_n$ and $|\mathbf{\Xi}_n^*|=l_n^*$, we have
\begin{align}\label{proposition3MAIN} % eq.52
	&\underline{W}_{\mathbf{\Upsilon},\prod_{t=1}^{n}l_t}^{\frac{1}{2}}\underline{W}_{\mathbf{\Upsilon},\prod_{t=1}^{n}l^*_t}^{\frac{1}{2}}\|\mathcal{X}\|_F\nonumber\\
	  & \leq \|\mathcal{X}\times_1(\mathbf{D}_{1_{\mathbf{\Xi}_1}}^H\mathbf{D}_{1_{\mathbf{\Xi}^*_1}})\times_2\cdots\times_n(\mathbf{D}_{n_{\mathbf{\Xi}_n}}^H\mathbf{D}_{n_{\mathbf{\Xi}^*_n}})\|_F \nonumber \\
		& \leq \max\bigg\{\varpi_{\mathbf{\Upsilon}}^n\prod_{t=1}^{n}l_td_t,\varpi_{\mathbf{\Upsilon}}^n\prod_{t=1}^{n}l^*_td_t\bigg\}\|\mathcal{X}\|_F.
\end{align}	
	
For any vector $\mathbf{x}\in\mathbb{C}^{s_n\prod_{t=1}^nd_t}$, we have
\begin{align}\label{proposition4MAIN} % eq.53
	\underline{W}^{\frac{1}{2}}_{\mathbf{\Upsilon},s_n}\|(\mathbf{T}^{\dagger})^{\rm H}\mathbf{x}\|_F \leq \|\mathbf{x}\|_F \leq \overline{W}^{\frac{1}{2}}_{\mathbf{\Upsilon},s_n}\|(\mathbf{T}^{\dagger})^{\rm H}\mathbf{x}\|_F,
\end{align}
where $\mathbf{T}=\Big[\mathbf{D}_{n_{\mathbf{\Xi}_{n_{[1]}}}}\otimes\cdots\otimes\mathbf{D}_{1_{\mathbf{\Xi}_{1_{[1]}}}}\mid \cdots\mid\mathbf{D}_{n_{\mathbf{\Xi}_{n_{[s_n]}}}}\otimes\cdots\otimes\mathbf{D}_{1_{\mathbf{\Xi}_{1_{[s_n]}}}}\Big]$.
\end{Corollary6}

\begin{IEEEproof}
See Appendix \ref{proofofCorollary6}.
\end{IEEEproof}

\begin{algorithm}[!t]
	%\doublespacing
	\renewcommand{\algorithmicrequire}{\textbf{Input:}}
	\renewcommand{\algorithmicensure}{\textbf{Output:}}
	\caption{T-GBOMP}
	\label{alg:T-GBOMP} % Alg.1
	\begin{algorithmic}[1]
		\REQUIRE $\mathbf{D}_1\cdots\mathbf{D}_n$, $\mathcal{Y}$, block sparsity level $k$, block length $d_1,d_2\cdots,d_n$, residual tolerant $\epsilon$ and selection parameter $s\leq k$.
		\ENSURE $\hat{\mathcal{X}}$, $\hat{\mathbf{\Xi}}$.
		\STATE $\mathbf{Initialization}$: $l=0$, $\mathcal{R}^0=\mathcal{Y}$, $\mathbf{\Xi}^0=\emptyset$, $\mathcal{X}^0=\mathbf{0}$.
		\WHILE {$l< k$ and $\|\mathcal{R}^l\|_2>\epsilon$}
		\STATE Identify the support by $\mathbf{\Theta}^{l+1}=\arg\max\limits_{\mathbf{\Theta}:|\mathbf{\Theta}|=s} \sum\limits_{(i_1,i_2,\cdots,i_n)\in\mathbf{\Theta}}
		\|\mathcal{R}^l\times_1\mathbf{D}^{\rm H}_{1_{[i_1]}}\times_2\cdots\times_n\mathbf{D}^{\rm H}_{n_{[i_n]}}\|_F$.
		\STATE Augment $\mathbf{\Xi}^{l+1}=\mathbf{\Xi}^l\cup\mathbf{\Theta}^{l+1}$.
		\STATE Estimate $\mathcal{X}^{l+1}=\arg\min\limits_{\mathcal{X}:{\rm supp}(\mathcal{X})=\mathbf{\Xi}^{l+1}}\|{\rm vec}(\mathcal{Y})-\sum\limits_{(i_1,\cdots,i_n)\in\mathbf{\Xi}^{l+1}}(\mathbf{D}_{n_{[i_n]}}\otimes\cdots\otimes\mathbf{D}_{1_{[i_1]}}){\rm vec}(\mathcal{X}_{[i_1,\cdots,i_n]})\|_F$.
		\STATE Update $\mathcal{R}^{l+1}=\mathcal{Y}-\mathcal{X}^{l+1}\times_1\mathbf{D}_1\times_2\cdots\times_n\mathbf{D}_n$.
		\STATE Update $l=l+1$.
		\ENDWHILE
		\STATE \textbf{return} $\hat{\mathbf{\Xi}}=\arg\min\limits_{\mathbf{\Theta}:|\mathbf{\Theta}|=k}\|\mathcal{X}^l-\mathcal{X}^l_{\mathbf{\Theta}}\|_F$ and $\hat{\mathcal{X}}$ with $\hat{\mathcal{X}}_{\mathbf{\Omega}\backslash\hat{\mathbf{\Xi}}}=\mathbf{0}$.
	\end{algorithmic}
\end{algorithm}

\subsection{T-GBOMP and Relations with Other Existing Algorithms}\label{tgbomp} % S3.2

As illustrated in Algorithm~\ref{alg:T-GBOMP}, the T-GBOMP algorithm chooses $s$ tensor blocks that are most strongly correlated with the residual tensor, and adds the selected indices to the support list per iteration. Then, it estimates the block-sparse tensor over the enlarged support matrix. As T-GBOMP incorporates multiple useful mechanisms from existing greedy algorithms, i.e., the multiple support selection \cite{jwang2017mols}, the structure calculation \cite{liyang2022}, the correlation discriminant \cite{Eldar2010}, and the high-dimensional spatial mapping \cite{ch2022} mechanisms, T-GBOMP can be seen as a generalization of these earlier algorithms. Theoretical results derived for T-GBOMP can thus be applied/extended straightforwardly to the other algorithms. To lay the groundwork for this generalization, this section lists essential characteristics of several representative greedy algorithms, and provides the basic ideas for extending them. Note that OMP and
OLS are the two classic frameworks for sparse recovery algorithms, and T-GBOMP is developed based on the OMP framework. We first analyze the relationship between T-GBOMP and other OMP-framework-based algorithms.

\subsubsection{OMP framework-based algorithms} 

Sparse recovery in vector form, i.e., the SMV model, is regarded as the 1-mode sparse tensor reconstruction. OMP is one of the most classical 1-mode recovery algorithms, which selects only one most correlated index per iteration \cite{cai2011}. Thus, T-GBOMP degenerates into OMP directly by setting the block lengths, the number of multiple supports selected and the number of modes as 1. The block version variant of OMP, i.e., BOMP \cite{liyang2023tcom}, can be obtained by setting the number of multiple supports selected and the number of the modes in T-GBOMP to 1. CoSaMP is another representative algorithm that selects multiple entries per iteration, but only retains the most correlated nonzero atoms, the number of which is equal to the sparsity \cite{cosamp2009}. This mechanism is equivalent to that of T-GBOMP by setting the selection parameter as sparsity and the block length as 1. Furthermore, T-GBOMP reduces to the two high-dimensional tensor recovery algorithms, T-OMP \cite{AraUjo2019} and T-BOMP \cite{Zubair2014}, by setting the block length and the selection parameter to 1, and only setting the selection parameter to 1, respectively. Note that the sparse recovery based on the MMV model can also be regarded as the 1-mode tensor recovery problem. The mechanisms of other OMP-framework-based algorithms are similar and will not be elaborated here.

\subsubsection{OLS framework-based algorithms} 

The OLS and the OMP frameworks differ in the way of selecting the new support entry \cite{Herzet2016,Soussen2013}. More specifically, OLS chooses a column that minimizes the power of the new residual. It can be shown that compared with the OMP framework, the OLS framework exhibits better convergence behavior, at the expense of imposing higher computational complexity \cite{jwang2017mols}. Mathematically, the main difference between the OMP-framework-based algorithm and the OLS-framework-based algorithm is that there exists a normalization factor in the selection mechanism of the OLS-framework-based algorithm \cite{liyang2022}. For example, the following is the selection mechanism of OLS \cite{liyang2022}:
\begin{align} % eq.54
	i^{l+1} &= \arg\max\limits_{j\in\{1,..,N\}\backslash\mathbf{\Omega}^{l}} \bigg| \bigg\langle\frac{\mathbf{D}_j}{\|\mathbf{P}^\bot_{\mathbf{\Omega}^{l}}\mathbf{D}_j\|_2},\mathbf{r}^l \bigg\rangle\bigg|, \label{normalizationfactor}
\end{align}
where $\|\mathbf{P}^\bot_{\mathbf{S}^{l}}\mathbf{D}_j\|_2$ is a normalization factor. As illustrated in existing works, e.g., \cite{liyang2022} and \cite{Herzet2016}, (\ref{normalizationfactor}) can be lower and upper bounded, providing a means of applying the theoretical results derived in the next section to OLS-framework-based algorithms. Similarly, popular OLS framework-based approaches, such as MOLS \cite{jwang2017mols} and BOLS \cite{liyang2022}, have incorporated useful iterative mechanisms, resulting in desirable performance and complexity.

In a nutshell, OMP- and OLS-framework-based algorithms can be extended into the tensor case, and T-GBOMP acts as a generalized framework for these greedy algorithms. The analyses based on T-GBOMP in the sequel can thus serve as a theoretical basis for tensor greedy algorithms in general, and can be easily applied to the many existing algorithms by setting certain parameters to specific values as mentioned above.

\section{Main Results}\label{S4}

We now present our main results, namely, conditions that guarantee T-GBOMP can recover a block-sparse tensor from measurements.  This section is divided into two parts. In the first part, an analysis under the noiseless case is conducted, including development of the ERC and reconstructible sparsity. In the second part, we present conditions for reliable recovery in the noisy scenario, including discussions of the difference between the original tensor and the estimated tensor, the residual convergence with respect to both the measurements and noise, and a lower bound of the SNR required for reliable recovery.

\subsection{Noiseless Recovery Conditions}\label{S4.1}

An exact recovery condition was first developed for OMP in the work \cite{greed2004}, after which several ERCs applicable to various other algorithms, such as BOMP \cite{Eldar2010} and BOLS \cite{liyang2022}, were derived. These ERCs reveal that, if the measurement matrix satisfies certain conditions, then the corresponding algorithms will select all the correct supports during their sparsity iterations. The conditions apply only to $1$-mode tensor recovery, and they are not suitable for the generalized $n$-mode tensor reconstruction scenarios, as discussed in Subsection~\ref{tgbomp}. Our first result is an ERC for an $n$-mode tensor recovery algorithm T-GBOMP to solve the exact sparse problem, which is a sufficient condition for T-GBOMP to identify the representation of the input tensor that exploits the least number of tensor blocks, i.e., the sparsest one.

\begin{theoremmaindescription4}\label{theo4} % T1
Let $\mathcal{X}$ be a $k$ block-sparse tensor, and for a given measurement matrix set $\mathbf{\Upsilon}=\{\mathbf{D}_i\in\mathbb{C}^{M_i\times s_i d_i} , 1\le i\le n\}$ $(n\geq1)$, let $\mathcal{Y}=\mathcal{X}\times_1\mathbf{D}_1\times_2\mathbf{D}_2\times_3\cdots\times_n\mathbf{D}_n$. A sufficient condition for T-GBOMP to reconstruct $\mathcal{X}$ is
\begin{align} % eq.55
	\frac{Z_{\mathcal{R}^l}}{\sqrt{s}}\big\|\ddot{\mathbf{D}}^{\dagger}_{\mathbf{\Xi}}\ddot{\mathbf{D}}_{\mathbf{\Psi}}\big\|_2<1, \label{theo4main}
\end{align}
where
\begin{align} % eq.56
	Z_{\mathcal{R}^l} =& \frac{\big\|\ddot{\mathbf{D}}^{\rm H}_{\mathbf{\Xi}}\, {\rm vec}(\mathcal{R}^l)\big\|_F}{\big\|\big(\mathbf{D}^{\rm H}_{n_{[i^*_n]}}\otimes\cdots\otimes\mathbf{D}^{\rm H}_{1_{[i^*_1]}}\big) {\rm vec}(\mathcal{R}^l)\big\|_F}, \label{zrldef}
\end{align}
$\mathbf{\Xi}$ is the tensor index set corresponding to the nonzero supports of $\mathcal{X}$, $\mathbf{\Psi}$ denotes the index set containing $s$ largest elements in  $\big\{\big\|\mathcal{R}^l\times_1\mathbf{D}^{\rm H}_{1_{[i_1]}}\times_2\cdots\times_n\mathbf{D}^{\rm H}_{n_{[i_n]}}\big\|_F\big\}_{(i_1,i_2,\cdots,i_n)\in\mathbf{\Omega}\backslash\mathbf{\Xi}}$, and $(i_1^*,\cdots,i_n^*)$ is the tensor index that leads to the largest element in the set $\big\{\big\|\mathcal{R}^l\times_1\mathbf{D}^{\rm H}_{1_{[i_1]}}\times_2\cdots\times_n\mathbf{D}^{\rm H}_{n_{[i_n]}}\big\|_F\big\}_{(i_1,i_2,\cdots,i_n)\in\mathbf{\Xi}}$.
\end{theoremmaindescription4}

\begin{IEEEproof}
See Appendix \ref{proofoftheo4}.
\end{IEEEproof}

\begin{rmk}\label{rmk1} % R2
\emph{Note that $Z_{\mathcal{R}^l}$ in (\ref{zrldef}) is bounded as $1\leq Z_{\mathcal{R}^l}\leq\sqrt{k}$. The lower bound can be obtained by assuming that $\ddot{\mathbf{D}}^{\rm H}_{\mathbf{\Xi}}$ is an all zero matrix except for the submatrix indexed by $(i_1^*,\cdots,i_n^*)$, while the upper bound is derived by letting $\ddot{\mathbf{D}}^{\rm H}_{\mathbf{\Xi}}$ be a cascade of $k$ $\mathbf{D}^{\rm H}_{n_{[i^*_n]}}\otimes\cdots\otimes\mathbf{D}^{\rm H}_{1_{[i^*_1]}}$ matrices. The ERC in (\ref{theo4main}) can thus be reformulated in terms of the least restricted variant and the most restricted variant, i.e.,
\begin{align} % eqs.57,58
	\frac{1}{\sqrt{s}}\big\|\ddot{\mathbf{D}}^{\dagger}_{\mathbf{\Xi}}\ddot{\mathbf{D}}_{\mathbf{\Psi}}\big\|_2 < & 1, \label{theo4main11} \\
	\big\|\ddot{\mathbf{D}}^{\dagger}_{\mathbf{\Xi}}\ddot{\mathbf{D}}_{\mathbf{\Psi}}\big\|_2 < & 1, \label{theo4main12}
\end{align}
respectively. Compared with the existing results in \cite{greed2004, Eldar2010, liyang2022}, (\ref{theo4main11}) provides progress due to the existing of $\frac{1}{\sqrt{s}}$. For instance, the ERC for BOMP with a 1-mode tensor is \cite{Eldar2010}
\begin{align} % eq.59
	\rho_c\big(\mathbf{D}^{\dagger}_{\mathbf{\Xi}}\mathbf{D}_{\mathbf{\Omega}\backslash\mathbf{\Xi}}\big)<1, \label{ercbomp}
\end{align}
where $\rho_c(\cdot)$ is the mixed matrix norm defined in \textbf{Definition~\ref{defofmixednorm}}, and $\mathbf{D}\in\mathbb{C}^{M\times N}$ is the 1-mode measurement matrix. It is worth noting that in most cases $|\mathbf{\Omega}\backslash\mathbf{\Xi}|>|\mathbf{\Psi}|=s$. Hence $\|\ddot{\mathbf{D}}^{\dagger}_{\mathbf{\Xi}}\ddot{\mathbf{D}}_{\mathbf{\Psi}}\|_2\leq	 \rho_c(\mathbf{D}^{\dagger}_{\mathbf{\Xi}}\mathbf{D}_{\mathbf{\Omega}\backslash\mathbf{\Xi}})$ under the reasonable assumption that the dimensions of measurement matrices are equivalent, i.e., $M=\prod_{t=1}^{n}M_t$ and $N=\prod_{t=1}^{n}N_t$. Since $\|\ddot{\mathbf{D}}^{\dagger}_{\mathbf{\Xi}}\ddot{\mathbf{D}}_{\mathbf{\Psi}}\|_2$ in (\ref{theo4main11}) is further multiplied by a parameter $\frac{1}{\sqrt{s}}\leq 1$, we obtain that the ERCs in (\ref{theo4main11}) and (\ref{theo4main12}) are less restrictive than the one in (\ref{ercbomp}). Comparisons to other existing ERCs are similar. Thus, we can conclude that the T-GBOMP algorithm is more capable of achieving exact sparse recovery than its  $1$-mode counterparts.}
\end{rmk}

Clearly (\ref{theo4main}) would not be very useful without the help of a method for checking if the ERC is satisfied. To this end, we present an intuitive result in terms of the sparsity bound required for the establishment of (\ref{theo4main}), named \textit{reconstructible sparsity}, by exploiting the MIP. Note that in this subsection up to Theorem \ref{theo5}, we assume that $\mathbf{\Xi}\cup\mathbf{\Psi}=\mathbf{\emptyset}$ for intuitive results, which facilitates clear comparisons with existing results. The outcomes when $\mathbf{\Xi}\cup\mathbf{\Psi}\neq\mathbf{\emptyset}$ can be easily obtained based on the analyses conducted in this subsection.

\begin{theoremmaindescription8}\label{theo8} % T2
A sufficient condition for the establishment of (\ref{theo4main}) is
\begin{align} % eq.60
	k < \frac{\sqrt{s}\Big(1-\big(\prod_{t=1}^nd_t-1\big)\tau_{\mathbf{\Upsilon}}^n+\varpi_{\mathbf{\Upsilon}}^n\prod_{t=1}^{n}d_t\Big)}{\big(Z_{\mathcal{R}^l}+\sqrt{s}\big)\varpi_{\mathbf{\Upsilon}}^n\prod_{t=1}^{n}d_t}. \label{theo8main}
\end{align}
\end{theoremmaindescription8}

\begin{IEEEproof}
See Appendix \ref{proofoftheo8}.
\end{IEEEproof}

It can be observed that \textbf{Theorem \ref{theo8}} provides an upper bound on the sparsity level, under which the T-GBOMP performs exact tensor recovery, i.e., the ERC in (\ref{theo4main}) holds. Based on \textit{Remark}~\ref{rmk1}, we obtain the least restricted reconstructible sparsity by setting $Z_{\mathcal{R}^l}=1$, as follows:
\begin{align}% eq.61
	k < \frac{\sqrt{s}\Big(1-\big(\prod_{t=1}^nd_t-1\big)\tau_{\mathbf{\Upsilon}}^n+\varpi_{\mathbf{\Upsilon}}^n\prod_{t=1}^{n}d_t\Big)}{\big(1+\sqrt{s}\big)\varpi_{\mathbf{\Upsilon}}^n\prod_{t=1}^{n}d_t}. \label{mostrestrictedsparsity}
\end{align}
In the subsequent remark, we provide some useful comparisons between (\ref{mostrestrictedsparsity}) and existing criteria. 

\begin{rmk2}\label{R3}
\emph{Since the existing reconstructible sparsity levels are developed for $1$-mode tensor recovery, we set $n=1$ in (\ref{mostrestrictedsparsity}) for fair comparison. Then, (\ref{mostrestrictedsparsity}) is reformulated into
\begin{align} % eq.62
	k d <\frac{\sqrt{s}d\big(1-(d-1)\nu_{\mathbf{D}}+\mu_{\mathbf{D}}d\big)}{(1+\sqrt{s})\mu_{\mathbf{D}}d}. \label{n=1sparsitylevel}
\end{align}
The classic reconstructible sparsity bounds in \cite{greed2004} and \cite{Eldar2010} are
\begin{align} % eqs.63,64
  K <& \frac{1}{2}\bigg(\frac{1}{\underline{\mu}_{\mathbf{D}}}+1\bigg), \label{conven1} \\
	k d <& \frac{1}{2}\bigg(\frac{1}{\mu_{\mathbf{D}}}+d-\frac{d\nu_{\mathbf{D}}}{\mu_{\mathbf{D}}}+\frac{\nu_{\mathbf{D}}}{\mu_{\mathbf{D}}}\bigg), \label{conven2}
\end{align}
respectively, where $K=k d$ represents the total sparsity. Firstly, let $d=1$ in (\ref{n=1sparsitylevel}), and (\ref{n=1sparsitylevel}) thus becomes
\begin{align} % eq.65
	K < \frac{\sqrt{s}}{1+\sqrt{s}}\bigg(\frac{1}{\underline{\mu}_{\mathbf{D}}}+1\bigg). \label{n=1sparsitylevel2}
\end{align}
Since $1\leq s\leq k$, we have $\frac{\sqrt{s}}{1+\sqrt{s}}\geq\frac{1}{2}$. This indicates that when $s=1$, (\ref{n=1sparsitylevel2}) converges to (\ref{conven1}), which is applicable for the conventional OMP, while in the case $s>1$, the upper bound in (\ref{n=1sparsitylevel2}) is better than that in (\ref{conven1}), 
which reveals that T-GBOMP is capable of exactly recovering tensors with larger sparsity levels in the scenarios where $Z_{\mathcal{R}^l}=1$. However, the condition $Z_{\mathcal{R}^l}=1$ is not achievable in practice, since the measurement matrix should be full rank for correct atom selection. Despite that the setting $Z_{\mathcal{R}^l}=1$ is impractical, the bound in (\ref{n=1sparsitylevel2}) provides a non-negligible improvement over the existing results, unveiling possible potential performance gain on tensor recovery by T-GBOMP.
Secondly, after some simplification operations, (\ref{n=1sparsitylevel}) becomes
\begin{align} % eq.66
	k d < \frac{\sqrt{s}}{1+\sqrt{s}}\bigg(\frac{1}{\mu_{\mathbf{D}}}+d-\frac{d\nu_{\mathbf{D}}}{\mu_{\mathbf{D}}}+\frac{\nu_{\mathbf{D}}}{\mu_{\mathbf{D}}}\bigg).\label{bompequ}
\end{align}
A similar conclusion can be made that the upper bound of reconstructible sparsity in (\ref{bompequ}) is equivalent to or better than that in (\ref{conven2}).}
\end{rmk2}

Now, considering the most restricted case of \textbf{Theorem \ref{theo8}}, i.e., $Z_{\mathcal{R}^l}=\sqrt{k}$, we provide the following corollary. Note that \textbf{Corollary~\ref{Corollary8}} replaces the uncertainty factor $Z_{\mathcal{R}^l}$ in (\ref{theo8main}) with a complex expression.

\begin{Corollary8}\label{Corollary8} % C4
If the following inequality
\begin{align} % eq.67
	k < \bigg( \sqrt[3]{-\frac{Q}{2}+\sqrt{\Delta}} + \sqrt[3]{-\frac{Q}{2}-\sqrt{\Delta}}-\frac{\sqrt{s}}{3}\bigg)^{2} \label{Corollary8main}
\end{align}
is satisfied, the ERC in (\ref{theo4main}) holds, where $P=-\frac{s}{3}$, $Q=\frac{27\delta+2s^{\frac{3}{2}}}{27}$, $\Delta=\big(\frac{Q}{2}\big)^2+\big(\frac{P}{3}\big)^3$, and
\begin{align}\label{eqC4delta} % eq.68
	\delta = -\frac{\sqrt{s}\Big(1-\big(\prod_{t=1}^nd_t-1\big)\tau_{\mathbf{\Upsilon}}^n+\varpi_{\mathbf{\Upsilon}}^n\prod_{t=1}^{n}d_t\Big)}{\varpi_{\mathbf{\Upsilon}}^n\prod_{t=1}^{n}d_t}.
\end{align}
\end{Corollary8}

\begin{IEEEproof}
See Appendix \ref{proofofcoro8}.
\end{IEEEproof}

\begin{rmk4}\label{remarkremark4} % Rem.4
\emph{We demonstrate that the most restricted reconstructible sparsity presented in \textbf{Corollary~\ref{Corollary8}} can be superior to the existing results (\ref{conven1}) and (\ref{conven2}). For reliable recovery in CS, the matrix formed by the measurement matrix blocks corresponding to the nonzero indices of the sparse tensor must be of full rank, which has been stated in} \emph{Remark~\ref{R3}}. \emph{This indicates that $Z_{\mathcal{R}^l}$ can be further bounded by $Z_{\mathcal{R}^l}\! <\! \sqrt{k}$. Since (\ref{Corollary8main}) that provides a complex reconstructible sparsity in closed-form is derived from (\ref{theo8main}), we come to the inequality (\ref{theo8main}) for an intuitive explanation. Let $Z_{\mathcal{R}^l}\! =\! \sqrt{k}-\theta$, where $\theta>0$ is a scalar, and we have
\begin{align} % eq.69
	k < \frac{\sqrt{s}\Big(1-\big(\prod_{t=1}^nd_t-1\big)\tau_{\mathbf{\Upsilon}}^n+\varpi_{\mathbf{\Upsilon}}^n\prod_{t=1}^{n}d_t\Big)}{(\sqrt{k}-\theta+\sqrt{s})\varpi_{\mathbf{\Upsilon}}^n\prod_{t=1}^{n}d_t}, \label{theo8mainresponse1}
\end{align}
which is the most restricted initial inequality for exporting the closed-form upper bound (\ref{Corollary8main}). By selecting a feasible $s$ value as $k$, (\ref{theo8mainresponse1}) changes into
\begin{align} % eq.70
	k < \frac{\Big(1-\big(\prod_{t=1}^nd_t-1\big)\tau_{\mathbf{\Upsilon}}^n+\varpi_{\mathbf{\Upsilon}}^n\prod_{t=1}^{n}d_t\Big)}{(2-\frac{\theta}{\sqrt{k}})\varpi_{\mathbf{\Upsilon}}^n\prod_{t=1}^{n}d_t}.\nonumber
\end{align}
For the case of $d=1$ and $n=1$, and for the case of $d>1$ and $n=1$, (\ref{theo8mainresponse1}) yields  respectively
\begin{align} % eqs.71,72
	K <& \frac{1}{2-\frac{\theta}{\sqrt{k}}}\bigg(\frac{1}{\underline{\mu}_{\mathbf{D}}}+1\bigg), \label{theo8mainresponse31} \\
	k d <& \frac{1}{2-\frac{\theta}{\sqrt{k}}}\bigg(\frac{1}{\mu_{\mathbf{D}}}+d-\frac{d\nu_{\mathbf{D}}}{\mu_{\mathbf{D}}}+\frac{\nu_{\mathbf{D}}}{\mu_{\mathbf{D}}}\bigg), \label{theo8mainresponse32}
\end{align}
where $K=kd=k\times 1$ implies the total sparsity in (\ref{theo8mainresponse31}). Since $\frac{1}{2-\frac{\theta}{\sqrt{k}}}>\frac{1}{2}$, the upper bounds in (\ref{theo8mainresponse31}) and (\ref{theo8mainresponse32}) provide an intuitive improvement compared to those in (\ref{conven1}) and (\ref{conven2}). This indicates that our derived reconstructible sparsity can be better than those in (\ref{conven1}) and (\ref{conven2}) even if the most restricted constraint is supposed in (\ref{theo8mainresponse1}), which completes our demonstration.}
\end{rmk4}

\begin{rmk11}\label{rmk5rmk5} % Rem.5
\emph{By using the bound of (\ref{gersthey1}) and based on the proof in Appendix \ref{proofoftheo8}, we obtain the following conditions for reconstructible sparsities of the least and most restricted:
\begin{align} % eqs.73,74
	& k<\Bigg(-\frac{1}{2}+\sqrt{\frac{5}{4}-\frac{(\prod_{t=1}^{n}d_t-1)\tau_{\mathbf{\Upsilon}}^n-1}{\varpi_{\mathbf{\Upsilon}}^n\prod_{t=1}^{n}d_t}}\Bigg)^2, \nonumber \\
	& k<\frac{\Big(\frac{1-(\prod_{t=1}^{n}d_t-1)\tau_{\mathbf{\Upsilon}}^n}{\varpi_{\mathbf{\Upsilon}}^n\prod_{t=1}^{n}d_t}\Big)^2+1+\frac{2}{\varpi_{\mathbf{\Upsilon}}^n\prod_{t=1}^{n}d_t}}{2+\frac{2}{\varpi_{\mathbf{\Upsilon}}^n\prod_{t=1}^{n}d_t}-\theta},  \nonumber
\end{align}
where $\theta$ is similarly defined in} \emph{Remark~\ref{remarkremark4}} \emph{which satisfies $\theta>0$.}
\end{rmk11}

In the following theorem, we study the rate of convergence of T-GBOMP in the noiseless scenario, which provides useful insights, in particular, revealing that the residual power of T-GBOMP decreases exponentially as the number of iterations increases.

\begin{theoremmaindescription5}\label{theo5} % T3
Given a matrix set $\mathbf{\Upsilon}=\{\mathbf{D}_i\in\mathbb{C}^{M_i\times s_i d_i} , 1\le i\le n\}$ $(n\geq1)$, if the MIP satisfies $\underline{W}_{\mathbf{\Upsilon},\prod_{t=1}^{n}k_n+s k}>0$, then the residual of T-GBOMP satisfies
\begin{align}
	\big\|\mathcal{R}^{l+1}\big\|^2_F \leq \Bigg(1-\frac{s\underline{W}_{\mathbf{\Upsilon},\prod_{t=1}^{n}k_n+sl}}{k\overline{W}^{\frac{1}{2}}_{\mathbf{\Upsilon},\prod_{t=1}^{n}k_n+sl}\overline{W}_{\mathbf{\Upsilon},s}}\Bigg)^{l+1}\big\|\mathcal{Y}\big\|^2_F .\nonumber
\end{align}
\end{theoremmaindescription5}

\begin{IEEEproof}
See Appendix~\ref{proofoftheo5}.
\end{IEEEproof}

\subsection{Noisy Recovery Conditions}\label{S4.2}

In practice, the tensor measurements are corrupted by noise. This subsection presents several important results guaranteeing the reliable recovery of the T-GBOMP algorithm with the help of the defined MIP concepts. As illustrated in Algorithm~\ref{alg:T-GBOMP}, T-GBOMP stops the iteration procedure when $l= k$ or $\|\mathcal{R}^l\|_2\leq\epsilon$. The following theorem reveals an upper bound on the error between the original and the estimated tensors when T-GBOMP is terminated.

\begin{theoremmaindescription1}\label{theo1} % T4
Given a matrix set $\mathbf{\Upsilon}=\{\mathbf{D}_i\in\mathbb{C}^{M_i\times s_i d_i}, 1\le i\le n\}$ $(n\geq1)$, if the iterative procedure of T-GBOMP satisfies $\|\mathcal{R}^{l^*}\|_F\leq\epsilon$, $\underline{W}_{\mathbf{\Upsilon},\prod_{t=1}^nk_t+k}>0$ and $\underline{W}_{\mathbf{\Upsilon},\prod_{t=1}^nk_t+sl^*}>0$ for $l^*\leq k$, then the following inequality with respect to $\hat{\mathcal{X}}$ holds:
\begin{align}\label{theo1main} % eq.76
	&\|\mathcal{X}-\hat{\mathcal{X}}\|_F \leq \frac{2\overline{W}^{\frac{1}{2}}_{\mathbf{\Upsilon},\prod_{t=1}^nk_t+k}\epsilon }{\underline{W}^{\frac{1}{2}}_{\mathbf{\Upsilon},\prod_{t=1}^nk_t+k}\underline{W}^{\frac{1}{2}}_{\mathbf{\Upsilon},\prod_{t=1}^nk_t+sl^*}}\nonumber\\
	&\qquad+\frac{ 2\Big(\overline{W}^{\frac{1}{2}}_{\mathbf{\Upsilon},\prod_{t=1}^nk_t+k}+\underline{W}^{\frac{1}{2}}_{\mathbf{\Upsilon},\prod_{t=1}^nk_t+sl^*}\Big)\|\mathcal{N}\|_F}{\underline{W}^{\frac{1}{2}}_{\mathbf{\Upsilon},\prod_{t=1}^nk_t+k}\underline{W}^{\frac{1}{2}}_{\mathbf{\Upsilon},\prod_{t=1}^nk_t+sl^*}},
\end{align}
where $\overline{W}_{\mathbf{\Upsilon},\prod_{t=1}^nk_t+k}$, $\underline{W}_{\mathbf{\Upsilon},\prod_{t=1}^nk_t+sl^*}$ and $\underline{W}_{\mathbf{\Upsilon},\prod_{t=1}^nk_t+k}$ can be obtained by \textbf{Corollary~\ref{Corollary6}}.
\end{theoremmaindescription1}

\begin{IEEEproof}
See Appendix \ref{proofoftheo1}.
\end{IEEEproof}

Even if \textbf{Theorem \ref{theo1}} is illustrated mainly based on the defined MIP framework, it seems to be complex. We now present intuitive explanations of \textbf{Theorem~\ref{theo1}} for adding the understanding of the theoretical reconstruction error. To begin with, we provide \textbf{Proposition~\ref{prop1}} that describes the lower and upper bounds of the multiplication of the shadow block sparsity.

\begin{proposition1}\label{prop1} % Prop.2
Given a $k$ block-sparse tensor with $n$ modes, its shadow block sparsity $\{k_i\}$ $(i\in\{1,2,\cdots,n\})$ satisfies
\begin{align} % eq.77
	k\leq k_1\times k_2\times\cdots\times k_n \leq k^n, \label{shadowsparsityminlow}
\end{align}
\end{proposition1}

The proof of \textbf{Proposition \ref{prop1}} is omitted, since it can be derived by considering the spatial geometric characteristic of the $k$ nonzero tensor blocks. The lower bound in (\ref{shadowsparsityminlow}) can be achieved by assuming that the $k$ nonzero tensor blocks are arranged regularly along one of the tensor modes, e.g., the $t$th $(t\in\{1,2,\cdots,n\})$ tensor mode, resulting in $k_t=k$, while the remaining shadow block sparsity is equal to 1. On the other hand, if the $k$ nonzero tensor blocks are arranged unobstructed throughout the entire tensor, the upper bound in (\ref{shadowsparsityminlow}) can be obtained with $k_i=k$ $(i\in\{1,2,\cdots,n\})$.

Based on \textbf{Proposition \ref{prop1}} and the aforementioned discussions, the following remark presents results investigating some particular tensor structures for \textbf{Theorem \ref{theo1}}, which offer intuitive relationships between reconstruction error and MIP.

\begin{rmk7}\label{rmk7rmk7} % Rem.6
\emph{Considering that the tensor structure achieves the lower bound in \textbf{Proposition~\ref{prop1}} with $d_1=d_2=\cdots=d_n=d$, $l^*$ satisfies $\frac{k}{s}\leq l^*\leq k$ and $\epsilon=\|\mathcal{N}\|_F$, then (\ref{theo1main}) changes into
\begin{align} % eqs.78,79
	&\|\mathcal{X}-\hat{\mathcal{X}}\|_F \nonumber\\
	&\leq\frac{\Big(4\overline{W}^{\frac{1}{2}}_{\mathbf{\Upsilon},2k(k+1)^{n-1}}+2\underline{W}^{\frac{1}{2}}_{\mathbf{\Upsilon},2k(k+1)^{n-1}}\Big)\|\mathcal{N}\|_F}{\underline{W}^{\frac{1}{2}}_{\mathbf{\Upsilon},2k(k+1)^{n-1}}\underline{W}^{\frac{1}{2}}_{\mathbf{\Upsilon},(s+1)k(sk+1)^{n-1}}} \label{remark51} \\
	&\leq\frac{2\sqrt{10}\|\mathcal{N}\|_F}{1-\sum_{t=0}^{n-1}\underline{C}^{d^n}_t\tau_{\mathbf{\Upsilon},t}^{n-t}-\sum_{t=0}^{n-1}C^{(s+1)k(sk+1)^{n-1}}_t\varpi_{\mathbf{\Upsilon},t}^nd^n}, \label{remark52}
\end{align}
where (\ref{remark51}) is because $\overline{W}_{\mathbf{\Upsilon},x}$ and $\underline{W}_{\mathbf{\Upsilon},x}$ are monotonically increasing and decreasing functions of the variable $x$, respectively, and (\ref{remark52}) follows from $4\overline{W}^{\frac{1}{2}}_{\mathbf{\Upsilon},2k(k+1)^{n-1}}+2\underline{W}^{\frac{1}{2}}_{\mathbf{\Upsilon},2k(k+1)^{n-1}}\leq2\sqrt{10}$ and $2k(k+1)^{n-1}\leq(s+1)k(sk+1)^{n-1}$. Clearly, the upper bound of the reconstruction error $\|\mathcal{X}-\hat{\mathcal{X}}\|_F$ decreases with the reductions in block sparsity, mutual block coherence and mutual sub-coherence, indicating an improved recovery performance. Compared with the existing results mainly based on the NP-hard RIP metric, e.g., \cite{Blumensath2009} and \cite{yDu2018}, the derived bound in (\ref{remark52}) provides a computable theoretical reconstruction error, making it more practically applicable.}

\emph{Let us come to the tensor structure achieving the upper bound in \textbf{Proposition \ref{prop1}}, which seems to offer the most restricted results. Similar to the assumption and derivation of (\ref{remark52}), the reconstruction error in \textbf{Theorem \ref{theo1}} can be further bounded by
\begin{align} % eq.80
	&\|\mathcal{X}-\hat{\mathcal{X}}\|_F \nonumber\\
	&\leq\frac{\Big(4\overline{W}^{\frac{1}{2}}_{\mathbf{\Upsilon},2^nk^n}+2\underline{W}^{\frac{1}{2}}_{\mathbf{\Upsilon},2^nk^n}\Big)\|\mathcal{N}\|_F}{\underline{W}^{\frac{1}{2}}_{\mathbf{\Upsilon},2^nk^n}\underline{W}^{\frac{1}{2}}_{\mathbf{\Upsilon},(s+1)^nk^n}}  \nonumber\\ &
	\leq\frac{2\sqrt{10}\|\mathcal{N}\|_F}{1-\sum_{t=0}^{n-1}\underline{C}^{d^n}_t\tau_{\mathbf{\Upsilon},t}^{n-t}-\sum_{t=0}^{n-1}C^{(s+1)^nk^n}_t\varpi_{\mathbf{\Upsilon},t}^nd^n} . \nonumber
\end{align}
Define the following two functions:
\begin{align} % eqs.81,82
	&f(\varpi_{\mathbf{\Upsilon}},\tau_{\mathbf{\Upsilon}})\nonumber\\
	&=\frac{2\sqrt{10}\|\mathcal{N}\|_F}{1-\sum_{t=0}^{n-1}\underline{C}^{d^n}_t\tau_{\mathbf{\Upsilon},t}^{n-t}-\sum_{t=0}^{n-1}C^{(s+1)k(sk+1)^{n-1}}_t\varpi_{\mathbf{\Upsilon},t}^nd^n}, \nonumber\\
	&g(\varpi_{\mathbf{\Upsilon}},\tau_{\mathbf{\Upsilon}})\nonumber\\
	&=\frac{2\sqrt{10}\|\mathcal{N}\|_F}{1-\sum_{t=0}^{n-1}\underline{C}^{d^n}_t\tau_{\mathbf{\Upsilon},t}^{n-t}-\sum_{t=0}^{n-1}C^{(s+1)^nk^n}_t\varpi_{\mathbf{\Upsilon},t}^nd^n}. \nonumber
\end{align}
Then, we obtain
\begin{align} % eq.83
	&\frac{g(\varpi_{\mathbf{\Upsilon}},\tau_{\mathbf{\Upsilon}})}{f(\varpi_{\mathbf{\Upsilon}},\tau_{\mathbf{\Upsilon}})} \nonumber\\
	& = 1+\frac{\sum_{t=0}^{n-1}(C^{(s+1)k(sk+1)^{n-1}}_t-C^{(s+1)^nk^n}_t)\varpi_{\mathbf{\Upsilon},t}^nd^n}{1-\sum_{t=0}^{n-1}\underline{C}^{d^n}_t\tau_{\mathbf{\Upsilon},t}^{n-t}-\sum_{t=0}^{n-1}C^{(s+1)^nk^n}_t\varpi_{\mathbf{\Upsilon},t}^nd^n} \nonumber\\
	&= 1 + \mathbf{\Delta}^*, \label{remark58}
\end{align}
where $\mathbf{\Delta}^*=\frac{\sum_{t=0}^{n-1}(C^{(s+1)k(sk+1)^{n-1}}_t-C^{(s+1)^nk^n}_t)\varpi_{\mathbf{\Upsilon},t}^nd^n}{1-\sum_{t=0}^{n-1}\underline{C}^{d^n}_t\tau_{\mathbf{\Upsilon},t}^{n-t}-\sum_{t=0}^{n-1}C^{(s+1)^nk^n}_t\varpi_{\mathbf{\Upsilon},t}^nd^n}$. For (\ref{remark58}), we have
\begin{subequations}
\begin{numcases}{}	% eqs.84a,84b,84.c	
	\frac{\partial\mathbf{\Delta}^*}{\partial \varpi_{\Upsilon}}\geq 0,\label{remark512a}\\
	\frac{\partial\mathbf{\Delta}^*}{\partial \tau_{\Upsilon}}\geq 0,\label{remark512b}\\
	\frac{\partial\mathbf{\Delta}^*}{\partial k} \leq 0.\label{remark512c}
\end{numcases}
\end{subequations}
These results provide intuitive theoretical analyses on the question how can the general block-sparse tensor structure approaches the tensor structure that achieves the lower bound in (\ref{shadowsparsityminlow}), with respect to several key parameters in block-sparse tensor recovery. Note that in order to achieve the approximation of the tensor structure mentioned above, the difference $\mathbf{\Delta}^*$ should be sufficiently small. Specially, the partial derivatives derived in (\ref{remark512a})-(\ref{remark512b}) indicate that small coherence in terms of mutual block coherence and mutual sub-coherence can contribute to the desirable block-sparse tensor structure, i.e., achieving the lower bound in (\ref{shadowsparsityminlow}). The result (\ref{remark512c}) indicates that when the block sparsity $k$ is sufficiently small, the reconstruction error can be close to that of the satisfactory tensor structure mentioned above. Furthermore, when the total sparsity $K=k d$ is fixed, we have $\frac{\partial\mathbf{\Delta}^*}{\partial d} \leq 0$. 
This unveils an important conclusion related to the block structure characteristic of tensor recovery, i.e., the larger the block length, the closer the reconstruction error compared to that of the desirable tensor structure. It indicates that a stronger block structure can result in better recovery performance.}
\end{rmk7}

In the noisy scenario, $\epsilon$ should be set appropriately to avoid late or early termination of the algorithm \cite{jwang2017mols}. It is worth noting that early cessation leads to an insufficient recovery of the tensor, while  stopping too late results in the tensor being disturbed by more noise, i.e., over-fitting. Note that in \textbf{Theorem~\ref{theo1}}, we consider the case where $\|\mathcal{R}^{l^*}\|_F\leq\epsilon$ for $l^*\leq k$. As given in Algorithm \ref{alg:T-GBOMP}, T-GBOMP is forced to iterate for $k$ iterations. However, the scenario where $\|\mathcal{R}^{l^*}\|_F\leq\epsilon$ is not met when $l^*\leq k$ naturally occurs since $\epsilon$ is a preset parameter. The following theorem considers this scenario, and provides a recovery condition demonstrating that the algorithm can perform reliable recovery in at most $k$ iterations.

\begin{theoremmaindescription2}\label{theo2} % T5
Given a matrix set $\mathbf{\Upsilon}=\{\mathbf{D}_i\in\mathbb{C}^{M_i\times s_i d_i}, 1\le i\le n\}$ $(n\geq1)$, suppose that $\|\mathcal{R}^{l^*}\|_F\leq\epsilon$ is not met for $l^*\leq k$. If $\underline{W}_{\mathbf{\Upsilon},ks}>0$, $\underline{W}_{\mathbf{\Upsilon},k^n}>0$, and the ${\rm SNR}$ satisfies
\begin{subequations}
	\begin{numcases}{}	% eqs.86a,86b		
		{\rm SNR} > \Big(\sqrt{k}\overline{W}^{\frac{1}{2}}_{\mathbf{\Upsilon},s}\overline{W}^{\frac{1}{2}}_{\mathbf{\Upsilon},k}+\sqrt{ks}\overline{W}^{\frac{1}{2}}_{\mathbf{\Upsilon},ks}\overline{W}^{\frac{1}{2}}_{\mathbf{\Upsilon},k}\Big)^2 \nonumber \\
		\qquad\quad \times \Bigg(\bigg(\frac{\sqrt{s}}{\sqrt{k}}\underline{W}^{\frac{1}{2}}_{\mathbf{\Upsilon},ks}\underline{W}^{\frac{1}{2}}_{\mathbf{\Upsilon},k}-k\varpi_{\mathbf{\Upsilon}}d_1\nonumber\\
	\qquad\qquad\;	-\frac{(ks\varpi_{\mathbf{\Upsilon}}d_1)^2}{\underline{W}_{\mathbf{\Upsilon},ks}}
			\bigg)\, {\rm MAR}_* \Bigg)^{-2}\! ,\qquad n=1, \label{theo2maina} \\
	  {\rm SNR} > \Big(\sqrt{k}\overline{W}^{\frac{1}{2}}_{\mathbf{\Upsilon},s}\overline{W}^{\frac{1}{2}}_{\mathbf{\Upsilon},\prod_{t=1}^{n}k_t}\nonumber\\
	  \qquad\qquad+\sqrt{ks}\overline{W}^{\frac{1}{2}}_{\mathbf{\Upsilon},ks}\overline{W}^{\frac{1}{2}}_{\mathbf{\Upsilon},\prod_{t=1}^{n}k_t}\Big)^2 \nonumber \\
		\qquad\quad \times \Bigg(\bigg(\frac{\sqrt{s}}{\sqrt{k}}\underline{W}^{\frac{1}{2}}_{\mathbf{\Upsilon},ks}\underline{W}^{\frac{1}{2}}_{\mathbf{\Upsilon},k^n}\! -k^n\varpi_{\mathbf{\Upsilon},n-1}^n\prod_{t=1}^{n}d_t\nonumber\\
		 -\frac{k^{n+1}s\varpi_{\mathbf{\Upsilon},n-1}^{2n} (\prod_{t=1}^{n}d_t)^2}{\underline{W}_{\mathbf{\Upsilon},ks}}
			\bigg)\, {\rm MAR}_*\Bigg)^{-2}\! , \quad
			 n>1 , \label{theo2mainb}
	\end{numcases}
\end{subequations}
then T-GBOMP selects all the support indices in at most $k$ iterations.
\end{theoremmaindescription2}

\begin{IEEEproof}
See Appendix \ref{proofoftheo2}.
\end{IEEEproof}

\textbf{Theorem \ref{theo2}} may not be intuitive in revealing the connection between various MIP metrics and the necessary SNR level for reliable recovery. To this end, the following remark is provided for a detailed descriptions.

\begin{rmk8}\label{rmkrmk8} % Rem.7
\emph{Based on (\ref{gersthey22}) in \textbf{Lemma \ref{lemma4}}, (\ref{theo2maina}) and (\ref{theo2mainb}) in \textbf{Theorem \ref{theo2}} can be integrated into the following inequality:
\begin{align} % eq.87
	\sqrt{{\rm SNR}} 
	&> \Bigg(\frac{1}{\sqrt{s}}\overline{W}^{\frac{1}{2}}_{\mathbf{\Upsilon},s}+\overline{W}^{\frac{1}{2}}_{\mathbf{\Upsilon},ks}\Bigg) \sqrt{k}\overline{W}^{\frac{1}{2}}_{\mathbf{\Upsilon},\prod_{t=1}^{n}k_t} \nonumber \\
	& \times \Bigg(\bigg(\frac{1}{\sqrt{k}}\underline{W}^{\frac{1}{2}}_{\mathbf{\Upsilon},ks}\underline{W}^{\frac{1}{2}}_{\mathbf{\Upsilon},k^n} 
	-\frac{s\varpi_{\mathbf{\Upsilon},n-1}^{2n}k^{\frac{n}{2}+1}\Big(\prod\limits_{t=1}^{n}d_t\Big)^2}{\underline{W}_{\mathbf{\Upsilon},ks}}  \nonumber\\
	&\hspace{2.5em}\enspace-\sqrt{k^n}\varpi_{\mathbf{\Upsilon},n-1}^n\prod_{t=1}^{n}d_t\bigg)\, {\rm MAR}_*  \Bigg)^{-1}. \label{snrminremark6}
\end{align}
Then, letting $d_1=d_2=\cdots=d_n=d$, (\ref{snrminremark6}) changes into 
\begin{align} % eq.88
	&\sqrt{{\rm SNR}} \nonumber\\
	&> \Bigg(\frac{1}{\sqrt{s}}\sqrt{1+\sum_{l=0}^{n-1}\underline{C}^{d^n}_l\tau_{\mathbf{\Upsilon},l}^{n-l}+\sum_{l=0}^{n-1}C^{s}_l\varpi_{\mathbf{\Upsilon},l}^nd^n}\nonumber\\
	&\qquad+\sqrt{1+\sum_{l=0}^{n-1}\underline{C}^{d^n}_l\tau_{\mathbf{\Upsilon},l}^{n-l}+\sum_{l=0}^{n-1}C^{ks}_l\varpi_{\mathbf{\Upsilon},l}^nd^n}\Bigg) \nonumber\\
	&\quad\times\sqrt{k}\sqrt{1+\sum_{l=0}^{n-1}\underline{C}^{d^n}_l\tau_{\mathbf{\Upsilon},l}^{n-l}+\sum_{l=0}^{n-1}C^{\prod_{t=1}^{n}k_t}_l\varpi_{\mathbf{\Upsilon},l}^nd^n}\nonumber\\
	&\quad\times  \Bigg(\bigg(\frac{1}{\sqrt{k}}\Big(1-\sum_{l=0}^{n-1}\underline{C}^{d^n}_l\tau_{\mathbf{\Upsilon},l}^{n-l}-\sum_{l=0}^{n-1}C^{ks}_l\varpi_{\mathbf{\Upsilon},l}^nd^n\Big)^{\frac{1}{2}}\nonumber\\
	&\qquad\times\Big(1-\sum_{l=0}^{n-1}\underline{C}^{d^n}_l\tau_{\mathbf{\Upsilon},l}^{n-l}-\sum_{l=0}^{n-1}C^{k^n}_l\varpi_{\mathbf{\Upsilon},l}^nd^n\Big)^{\frac{1}{2}} \nonumber\\
	&\quad-\frac{s\varpi_{\mathbf{\Upsilon},n-1}^{2n}k^{\frac{n}{2}+1}d^{2n}}{1-\sum_{l=0}^{n-1}\underline{C}^{d^n}_l\tau_{\mathbf{\Upsilon},l}^{n-l}-\sum_{l=0}^{n-1}C^{ks}_l\varpi_{\mathbf{\Upsilon},l}^nd^n}  \nonumber\\
	&\quad-k^{\frac{n}{2}}\varpi_{\mathbf{\Upsilon},n-1}^nd^n\bigg)\, {\rm MAR}_*  \Bigg)^{-1}. \label{snrminremark62}
\end{align}
Since the prerequisites that $\underline{W}_{\mathbf{\Upsilon},ks}>0$ and  $\underline{W}_{\mathbf{\Upsilon},k^n}>0$ are given in \textbf{Theorem \ref{theo2}}, we have 
\begin{align} % eqs.89,90
	&	 \sum_{l=0}^{n-1}\underline{C}^{d^n}_l\tau_{\mathbf{\Upsilon},l}^{n-l}+\sum_{l=0}^{n-1}C^{ks}_l\varpi_{\mathbf{\Upsilon},l}^nd^n<1,\label{precodition1}\\
	& \sum_{l=0}^{n-1}\underline{C}^{d^n}_l\tau_{\mathbf{\Upsilon},l}^{n-l}+\sum_{l=0}^{n-1}C^{k^n}_l\varpi_{\mathbf{\Upsilon},l}^nd^n<1.\label{precodition2}
\end{align} 
Furthermore, (\ref{precodition1}) and (\ref{precodition2}) indicate that
\begin{align} % eqs.91,92
	&	 \sum_{l=0}^{n-1}\underline{C}^{d^n}_l\tau_{\mathbf{\Upsilon},l}^{n-l}+\sum_{l=0}^{n-1}C^{s}_l\varpi_{\mathbf{\Upsilon},l}^nd^n<1,\label{precodition3}\\
	& \sum_{l=0}^{n-1}\underline{C}^{d^n}_l\tau_{\mathbf{\Upsilon},l}^{n-l}+\sum_{l=0}^{n-1}C^{\prod_{t=1}^{n}k_t}_l\varpi_{\mathbf{\Upsilon},l}^nd^n<1,\label{precodition4}
\end{align} 
respectively.
Thus, as for the right-hand side in (\ref{snrminremark62}), (\ref{precodition1})-(\ref{precodition4}) lead to the following limits:
\begin{align} % eqs.93-98
	&\lim_{n\to\infty}	\sum_{l=0}^{n-1}\underline{C}^{d^n}_l\tau_{\mathbf{\Upsilon},l}^{n-l}+\sum_{l=0}^{n-1}C^{ks}_l\varpi_{\mathbf{\Upsilon},l}^nd^n=0, \label{eqLim1} \\
	&\lim_{n\to\infty}\sum_{l=0}^{n-1}\underline{C}^{d^n}_l\tau_{\mathbf{\Upsilon},l}^{n-l}+\sum_{l=0}^{n-1}C^{k^n}_l\varpi_{\mathbf{\Upsilon},l}^nd^n=0, \label{eqLim2} \\
	&\lim_{n\to\infty}\sum_{l=0}^{n-1}\underline{C}^{d^n}_l\tau_{\mathbf{\Upsilon},l}^{n-l}+\sum_{l=0}^{n-1}C^{s}_l\varpi_{\mathbf{\Upsilon},l}^nd^n=0, \label{eqLim3} \\
	&\lim_{n\to\infty}\sum_{l=0}^{n-1}\underline{C}^{d^n}_l\tau_{\mathbf{\Upsilon},l}^{n-l}+\sum_{l=0}^{n-1}C^{\prod_{t=1}^{n}k_t}_l\varpi_{\mathbf{\Upsilon},l}^nd^n=0, \label{eqLim4} \\
	&\lim_{n\to\infty}k^{\frac{n}{2}+1}\varpi_{\mathbf{\Upsilon},n-1}^{2n}d^{2n}=0, \label{eqLim5} \\
	&\lim_{n\to\infty}k^{\frac{n}{2}}\varpi_{\mathbf{\Upsilon},n-1}^nd^n=0.\label{limits}
\end{align}
The condition of SNR presented in (\ref{snrminremark62}) can be reformulated intuitively based on the aforementioned limits, i.e.,
\begin{align}\label{snrlast} % eq.99
	&\lim_{n\to\infty}	\sqrt{\rm SNR}>\frac{\big(\frac{1}{\sqrt{s}}+1\big)k}{{\rm MAR}_*}.
\end{align}
Denote the right-hand side of (\ref{snrlast}) as a function $h(k,s,{\rm MAR}_*)$. The observation that $\frac{\partial h(k,s,{\rm MAR}_*)}{\partial s}<0$ indicates that the larger the selection parameter $s$, the lower the SNR required for reliable recovery. This indicates that if it is possible to select more atoms in each iteration, then within a maximum of $k$ iterations, the likelihood of the algorithm selecting all correct atoms increases, so as to achieve the desired performance with lower SNR. Meanwhile, the observation that $\frac{\partial h(k,s,{\rm MAR}_*)}{\partial k}>0$ reveals that a larger block sparsity $k$ requires a more stringent SNR for reliable recovery. This result is consistent with the upper bound of reconstruction error provided in \textbf{Theorem \ref{theo1}} which indicate that a lower reconstruction error may require a lower block sparsity level. Moreover, as ${\rm MAR}_*$ is defined in \textbf{Definition \ref{defofmar}}, (\ref{snrlast}) can be adjusted into 
\begin{align}
	&\lim_{n\to\infty}	\sqrt{\rm SNR}>\frac{\big(\frac{1}{\sqrt{s}}+1\big)\sqrt{k}\big\|\mathcal{X}\big\|_F}{\min\limits_{(i_1,\cdots,i_n)\in\mathbf{\Xi}} \big\|\mathcal{X}_{[i_1,\cdots,i_n]}\big\|_F}.\nonumber
\end{align}
It can be seen that $k$ and $s$ are on the same scale that contributes to $h(k,s,{\rm MAR}_*)$, implying that an increase in the selection parameter $s$ can effectively counteract the performance loss resulting from the increased block sparsity. Since $\frac{\partial h(k,s,{\rm MAR}_*)}{\partial {\rm MAR}_*}\! <\! 0$, a lower SNR bound is acquired as the support tensor blocks becomes more powerful. The results related to the limits in (\ref{eqLim1})-(\ref{limits}) indicate a monotonic decreasing property of the bound in the right-hand side of (\ref{snrminremark62}) with respect to the number of tensor modes $n$. Therefore, this analysis unveils that a high-dimensional tensor recovery can provide strong robustness.}
\end{rmk8}

The following theorem is an extension of \textbf{Theorem \ref{theo1}}, which considers the scenario where the SNR satisfies the conditions in  \textbf{Theorem \ref{theo2}}, and $\|\mathcal{R}^{l^*}\|_F\leq\epsilon$ is not met for $l^*\leq k$.

\begin{theoremmaindescription3}\label{theo3} % T6
Given a matrix set $\mathbf{\Upsilon}=\{\mathbf{D}_i\in\mathbb{C}^{M_i\times s_i d_i}, 1\le i\le n\}$ $(n\geq1)$, suppose that $\|\mathcal{R}^{l^*}\|_F\leq\epsilon$ is not met for $l^*\leq k$. If $\underline{W}_{\mathbf{\Upsilon},ks}>0$, $\underline{W}_{\mathbf{\Upsilon},\prod_{t=1}^{n}k_t+k}>0$, and the ${\rm SNR}$ satisfies (\ref{theo2maina}) and (\ref{theo2mainb}), then the output $\hat{\mathcal{X}}$ satisfies
\begin{subequations}
	\begin{numcases}{}	% eqs.101a,101b		
		\big\|\hat{\mathcal{X}}-\mathcal{X}\big\|_F
			\leq\frac{\|\mathcal{N}\|_F}{\underline{W}^{\frac{1}{2}}_{\mathbf{\Upsilon},k}}, s=1, \label{theo3maina} \\
	  \big\|\hat{\mathcal{X}}-\mathcal{X}\big\|_F \leq \Bigg(1+\frac{\overline{W}^{\frac{1}{2}}_{\mathbf{\Upsilon},\prod_{t=1}^{n}k_t+k}}{\underline{W}^{\frac{1}{2}}_{\mathbf{\Upsilon},ks}}\Bigg)\nonumber\\
	  \qquad\qquad\qquad\times\frac{2\|\mathcal{N}\|_F}{\underline{W}^{\frac{1}{2}}_{\mathbf{\Upsilon},\prod_{t=1}^{n}k_t+k}},   s>1. \label{theo3mainb}
  \end{numcases}
\end{subequations}
\end{theoremmaindescription3}

\begin{IEEEproof}
See Appendix \ref{proofoftheo3}.
\end{IEEEproof}

 We now present an insight to the block orthogonal tensor recovery model, where the measurement matrix satisfying block orthogonal has been proved to offer more assured theoretical guarantees.

\begin{rmk9}\label{rmk9rmk9} % Rem.8
\emph{Consider that the measurement matrices satisfy block orthogonality among the measurement matrix set $\mathbf{\Upsilon}$, i.e., $\nu_{\mathbf{D}_t}=0$ $(t\in\{1,2,\cdots,n\})$, and thus $\tau_{\mathbf{\Upsilon},l}=0$ $(l\in\{0,1,\cdots,n\})$. Let $s=1$. Then, the condition in \textbf{Theorem~\ref{theo3}} becomes
\begin{align}	% eq.102
	&\big\|\hat{\mathcal{X}}-\mathcal{X}\big\|_F
	\leq\frac{\|\mathcal{N}\|_F}{\sqrt{1  -\sum_{l=0}^{n-1}C^{k}_l\varpi_{\mathbf{\Upsilon},l}^n \prod\limits_{t=1}^{n}d_t}}. \label{remark71} 			 
\end{align}
By defining the right-hand sides of (\ref{theo3maina}) and (\ref{remark71}) as functions $p(\varpi_{\mathbf{\Upsilon}},\tau_{\mathbf{\Upsilon}})$ and $q(\varpi_{\mathbf{\Upsilon}},\tau_{\mathbf{\Upsilon}})$, respectively, we have
\begin{align}	% eq.103
	&\frac{q(\varpi_{\mathbf{\Upsilon}},\tau_{\mathbf{\Upsilon}})}{p(\varpi_{\mathbf{\Upsilon}},\tau_{\mathbf{\Upsilon}})}\nonumber\\
	& = \Bigg(\frac{1  -\sum_{l=0}^{n-1}C^{k}_l\varpi_{\mathbf{\Upsilon},l}^n \prod_{t=1}^{n}d_t}{1  -\sum_{l=0}^{n-1}\underline{C}^{\prod_{t=1}^nd_t}_l\tau_{\mathbf{\Upsilon},l}^{n-l}-\sum_{l=0}^{n-1}C^{k}_l\varpi_{\mathbf{\Upsilon},l}^n\prod_{t=1}^nd_t}\Bigg)^{\frac{1}{2}}		\nonumber\\
	& = \Bigg(1\!\!+\!\!\frac{\sum_{l=0}^{n-1}\underline{C}^{\prod_{t=1}^nd_t}_l\tau_{\mathbf{\Upsilon},l}^{n-l}}{1  \!\!-\!\!\sum_{l=0}^{n-1}\underline{C}^{\prod_{t=1}^nd_t}_l\tau_{\mathbf{\Upsilon},l}^{n-l}\!\!-\!\!\sum_{l=0}^{n-1}C^{k}_l\varpi_{\mathbf{\Upsilon},l}^n\prod_{t=1}^nd_t}\Bigg)^{\frac{1}{2}}  \nonumber\\ &
 = (1+\mathbf{\underline{\Delta}})^{\frac{1}{2}},\nonumber
\end{align}
where $\mathbf{\underline{\Delta}}\! =\! \frac{\sum_{l=0}^{n-1}\underline{C}^{\prod_{t=1}^nd_t}_l\tau_{\mathbf{\Upsilon},l}^{n-l}}{1  -\sum_{l=0}^{n-1}\underline{C}^{\prod_{t=1}^nd_t}_l\tau_{\mathbf{\Upsilon},l}^{n-l}-\sum_{l=0}^{n-1}C^{k}_l\varpi_{\mathbf{\Upsilon},l}^n\prod_{t=1}^nd_t}$. In general, as the block lengths increase, the dimensions of the measurement matrix blocks are increased correspondingly, leading to the increases of $\varpi_{\mathbf{\Upsilon},l}^n \prod\limits_{t=1}^{n}d_t$ and $\tau^{n-l}_{\mathbf{\Upsilon},l}$. Then, we obtain that the $\mathbf{\underline{\Delta}}$ is increased with the increase of block lengths. It reveals that the reconstruction error difference between the block orthogonal tensor recovery and non-orthogonal tensor recovery is further exacerbated as the block structure property becomes stronger. This necessitates the design of measurement matrix that obeys block orthogonality for better recovery performance guarantees, which can be achieved through methodologies like Schmidt orthogonalization \cite{OLS1991}.}
\end{rmk9}

\textbf{Theorems \ref{theo1}} to \textbf{\ref{theo3}} present  reliable recovery conditions of T-GBOMP by considering the level of residual power within $k$ iterations as a prerequisite. Now we naturally turn to the study of bounding the residual power. To begin with, we first provide the following lemma that characterizes the relationship between the residuals in different numbers of iterations.

\begin{lemma1}\label{lemma1} % L3
Given a matrix set $\mathbf{\Upsilon}=\{\mathbf{D}_i\in\mathbb{C}^{M_i\times s_id_i}, 1\le i\le n\}$ $(n\geq1)$, if $\underline{W}_{\mathbf{\Upsilon},|\mathbf{\Lambda}_{\varrho}^l\cup\mathbf{\Xi}^{c+\Delta c-1}|}>0$, $c\geq l$ and $\Delta c>0$ for any integer $c$, and $\varrho\in\big\{1,\cdots,\max\big\{0,\big\lceil\log_2\frac{L}{s}\big\rceil\big\}+1\big\}$, the residual $\mathcal{R}^{c+\Delta c}$ in the $(c+\Delta c)$ iteration satisfies
\begin{align} % eq.104
	& \big\|\mathcal{R}^{c+\Delta c}\big\|^2_F \nonumber\\
	&- \big\|\mathbf{D}_{n_{\mathbf{\Lambda}_n^l\backslash\mathbf{\Lambda}^l_{n_{{\varrho}}}}}\otimes\cdots\otimes\mathbf{D}_{1_{\mathbf{\Lambda}_1^l\backslash\mathbf{\Lambda}^l_{1_{\varrho}}}} {\rm vec}\big(\mathcal{X}_{\mathbf{\Lambda}^l\backslash\mathbf{\Lambda}^l_{\varrho}}\big)\!\!+\!\!{\rm vec}(\mathcal{N})\big\|^2_F \nonumber \\
	&  \leq G_{\mathbf{\Upsilon},\varrho,c,\Delta c}
		\bigg(\big\|{\rm vec}(\mathcal{R}^{c})\big\|^2_F \nonumber\\
	&\quad	\!\!-\!\! \Big\|\mathbf{D}_{n_{\mathbf{\Lambda}_n^l\backslash\mathbf{\Lambda}^l_{n_{{\varrho}}}}}\otimes\cdots\otimes\mathbf{D}_{1_{\mathbf{\Lambda}_1^l\backslash\mathbf{\Lambda}^l_{1_{\varrho}}}} {\rm vec}\big(\mathcal{X}_{\mathbf{\Lambda}^l\backslash\mathbf{\Lambda}^l_{\varrho}}\big) 
	\!\!+\!\! {\rm vec}(\mathcal{N})\Big\|^2_F\bigg),\nonumber
\end{align}
where
\begin{align} % eq.105
	G_{\mathbf{\Upsilon},\varrho,c,\Delta c}=\exp\Bigg(-\frac{\Delta c\underline{W}_{\mathbf{\Upsilon},|\mathbf{\Lambda}_{\varrho}^l\cup\mathbf{\Xi}^{c+\Delta c-1}|}}{\Big\lceil\frac{|\mathbf{\Lambda}^l_{\varrho}|}{s}\Big\rceil\overline{W}_{\mathbf{\Upsilon},s}}\Bigg). \nonumber
\end{align}
\end{lemma1}

\begin{IEEEproof}
See Appendix \ref{proofoflemma1}.
\end{IEEEproof}

 In the following remark, we present the results related to the Frobenius norm of residual tensors in different iterations, which is regarded as the true residual error.

\begin{rmk10}\label{Rem-new9} % Rem.9
\emph{For the $l$th $(0< l< k)$ iteration, denote the tensor index sets selected in the $(l+1)$th iterations as $\mathbf{\Xi}^{l+1}$. Following the proof in Appendix \ref{proofoftheo5}, we have
\begin{align} % eq.106
	\|\mathcal{R}^l - \mathcal{R}^{l+1}\|^2_F &= \|\mathbf{P}_{\ddot{\mathbf{D}}_{\mathbf{\Xi}^{l+1}}}{\rm vec}(\mathcal{R}^l)\|^2_F % \nonumber \\ &
	\leq\|{\rm vec}(\mathcal{R}^l)\|^2_F  \nonumber\\ &
	\leq\Bigg(1-\frac{s\underline{W}_{\mathbf{\Upsilon},\prod_{t=1}^{n}k_n+s(l-1)}}{k\overline{W}^{\frac{1}{2}}_{\mathbf{\Upsilon},\prod_{t=1}^{n}k_n+s(l-1)}\overline{W}_{\mathbf{\Upsilon},s}}\Bigg)^{l}\big\|\mathcal{Y}\big\|^2_F. \nonumber
\end{align}
For a more intuitive representation, let $d_1=d_2=\cdots=d_n=d$, and we have
\begin{align} % eq.107
	&\|\mathcal{R}^l - \mathcal{R}^{l+1}\|^2_F 
		\leq\Bigg(1-\nonumber\\
		&\frac{s(1  -\sum_{l=0}^{n-1}\underline{C}^{d^n}_l\tau_{\mathbf{\Upsilon},l}^{n-l}-\sum_{l=0}^{n-1}C^{\prod_{t=1}^{n}k_n+s(l-1)}_l\varpi_{\mathbf{\Upsilon},l}^nd^n)}{k(1  +\sum_{l=0}^{n-1}\underline{C}^{d^n}_l\tau_{\mathbf{\Upsilon},l}^{n-l}+\sum_{l=0}^{n-1}C^{\prod_{t=1}^{n}k_n+s(l-1)}_l\varpi_{\mathbf{\Upsilon},l}^nd^n)^{\frac{1}{2}}}\nonumber\\
	&\times\frac{1}{1  +\sum_{l=0}^{n-1}\underline{C}^{d^n}_l\tau_{\mathbf{\Upsilon},l}^{n-l}+\sum_{l=0}^{n-1}C^{s}_l\varpi_{\mathbf{\Upsilon},l}^nd^n}\Bigg)^{l}\big\|\mathcal{Y}\big\|^2_F.
		\nonumber
\end{align}
It can be observed that as the number of iteration $l$ increases, the upper bound of the residual tensor error gradually approaches 0, which indicates that the significant tensor blocks are being subtracted from the measurement tensor. It can be roughly assumed that all the correct tensor blocks have been selected if $\|\mathcal{R}^l - \mathcal{R}^{l+1}\|^2_F $ tends very close to 0. Similar to the analysis in \textbf{Remark \ref{rmkrmk8}}, we obtain
\begin{align} % eq.108
	\lim_{n\to\infty}\|\mathcal{R}^l - \mathcal{R}^{l+1}\|^2_F 
		\leq\Big(1-\frac{s}{k}\Big)^{l}\big\|\mathcal{Y}\big\|^2_F.
		\nonumber
\end{align}
This result indicates that as $s$ increases, the significant tensor blocks are more likely to be selected correctly, resulting in a faster convergence speed that equalizes $\mathcal{R}^l$ and $\mathcal{R}^{l+1}$. A special case arises when $s=k$, wherein  we observe $\lim_{n\to\infty}\|\mathcal{R}^l - \mathcal{R}^{l+1}\|^2_F=0$, which indicates that all the correct tensor blocks are selected during the first iteration. Moreover, it is worth mentioning that the convergence speed increases with the increase of the tensor mode $n$ when the other parameters are fixed. This also unveils that the derived bound of the residual error becomes smaller as the tensor mode $n$ increases. That is, the high-dimensional tensor recovery provides tight theoretical guarantees.}
\end{rmk10}

Based on \textbf{Lemma~\ref{lemma1}}, we develop the following theorem, which indicates that after a specified number of iterations related to the number of remaining supports to be selected, the residual is upper bounded by a quantity with respect to the MIP and noise power.

\begin{theoremmaindescription6}\label{theo6} % T7
Given a matrix set $\mathbf{\Upsilon}=\{\mathbf{D}_i\in\mathbb{C}^{M_i\times s_i d_i}, 1\le i\le n\}$ $(n\geq1)$, let $\mathbf{\Lambda}^l$ be the set consisting of the remaining support block tensors after $l$ iterations with $|\mathbf{\Lambda}^l|=\theta$, and $\alpha$ be an integer such that $\alpha={\rm inf}\bigg\{x\in\Big\{1,\cdots,\max\Big\{0,\Big\lceil\log\frac{e|\mathbf{\Lambda}^l|}{s}\Big\rceil\Big\}+1\Big\},\frac{\exp(x-2) s (e-1)}{\exp(x-1) -1+(e-1)x}\bigg\}$. If the MIP of $\mathbf{\Upsilon}$ satisfies $\underline{W}_{\mathbf{\Upsilon},k}>0$, $\underline{W}_{\mathbf{\Upsilon},sl+s+\theta}>0$ and $\underline{W}_{\mathbf{\Upsilon},sl+s\theta+\theta}>0$, then we have
\begin{align} % eq.109
	\big\|\mathcal{R}^{l+|\mathbf{\Lambda}^l|}\big\|_F \leq \xi \|\mathcal{N}\|_F , \nonumber
\end{align}
where
\begin{align} % eqs.110-112
	  & \xi = \Big(\omega(1+\gamma)\overline{W}_{\mathbf{\Upsilon},\theta^n}+\frac{\underline{W}_{\mathbf{\Upsilon},sl+s+\theta}}{\overline{W}_{\mathbf{\Upsilon},s}}(1+\gamma)\overline{W}_{\mathbf{\Upsilon},\theta^n}\eta^{-1}\Big)^{\frac{1}{2}}  \nonumber\\
	  &\hspace{2em}\times\big(1+\big(\frac{1}{1-\beta}(1+\gamma^{-1})\big)^{\frac{1}{2}}\big)\nonumber\\
	  &\hspace{2em}\times\underline{W}^{-\frac{1}{2}}_{\mathbf{\Upsilon},k}\Big(1-\nonumber\\
	  &\hspace{2em}\Big(\frac{\omega(1\!+\!\gamma)\overline{W}_{\mathbf{\Upsilon},\theta^n}\!+\!\frac{\underline{W}_{\mathbf{\Upsilon},sl+s+\theta}}{\overline{W}_{\mathbf{\Upsilon},s}}(1\!+\!\gamma)\overline{W}_{\mathbf{\Upsilon},\theta^n}\eta^{-1}}{\underline{W}_{\mathbf{\Upsilon},k}}\Big)^{\frac{1}{2}}\Big)^{-1}\nonumber\\
	  &\hspace{1.5em}+\frac{(1+\gamma^{-1})^{\frac{1}{2}}}{(1-\beta)^{\frac{1}{2}}}, \nonumber \\
  & \omega = \max\Big\{\frac{1}{\eta(1-\eta\beta)},1-\frac{\underline{W}_{\mathbf{\Upsilon},sl+s+\theta}}{\overline{W}_{\mathbf{\Upsilon},s}}\Big\}, \nonumber \\
	& \beta = \exp\Big(-\frac{\alpha\underline{W}_{\mathbf{\Upsilon},sl+s\theta+\theta}}{\overline{W}_{\mathbf{\Upsilon},s}}\Big), \nonumber
\end{align}
with $\eta\beta<1$ and $\gamma>0$.
\end{theoremmaindescription6}

\begin{IEEEproof}
See Appendix \ref{proofoftheo6}.
\end{IEEEproof}

It can be observed that if T-GBOMP has already operated $l$ iterations, then the algorithm iterates at most $|\mathbf{\Lambda}^l|$ additional iterations to guarantee that the residual falls below $\xi\|\mathcal{N}\|_F$ with the MIP satisfying the conditions in \textbf{Theorem~\ref{theo6}}. Note that when $l=0$, we have $|\mathbf{\Lambda}^0|=k$. Based on these discussions, we present the following corollary.

\begin{Corollary7}\label{Corollary7} % C5
Given a matrix set $\mathbf{\Upsilon}=\{\mathbf{D}_i\in\mathbb{C}^{M_i\times s_i d_i}, 1\le i\le n\}$ $(n\geq1)$, let $\alpha$ be an integer such that $\alpha={\rm inf}\Big\{x\in\Big\{1,\cdots,\max\Big\{0,\Big\lceil\log\frac{e k}{s}\Big\rceil\Big\}+1\Big\},\frac{\exp(x-2)s(e-1)}{\exp(x-1)-1+(e-1)x}\Big\}$. If the MIP of $\mathbf{\Upsilon}$ satisfies $\underline{W}_{\mathbf{\Upsilon},s k+k}>0$, then we have
\begin{align} % eq.113
  \|\mathcal{R}^{k}\|_F\leq\xi^*\|\mathcal{N}\|_F, \nonumber
\end{align}
where
\begin{align} % eqs.114-116
	  & \xi^* = \Big(\frac{1}{\eta(1-\frac{1}{e})}(1+\gamma)\overline{W}_{\mathbf{\Upsilon},k^n}\nonumber\\
	  &\hspace{3em}+\frac{\underline{W}_{\mathbf{\Upsilon},s+k}}{\overline{W}_{\mathbf{\Upsilon},s}}(1+\gamma)\overline{W}_{\mathbf{\Upsilon},k^n}\eta^{-1}\Big)^{\frac{1}{2}}\nonumber\\
	  &\hspace{2em}\times\big(1+\big(\frac{1}{1-\beta}(1+\gamma^{-1})\big)^{\frac{1}{2}}\big)\nonumber\\
	  &\hspace{2em}\times\underline{W}^{-\frac{1}{2}}_{\mathbf{\Upsilon},k}\Big(1-\nonumber\\
	  &\Big(\frac{\frac{1}{\eta(1-\frac{1}{e})}(1+\gamma)\overline{W}_{\mathbf{\Upsilon},k^n}+\frac{\underline{W}_{\mathbf{\Upsilon},s+k}}{\overline{W}_{\mathbf{\Upsilon},s}}(1+\gamma)\overline{W}_{\mathbf{\Upsilon},k^n}\eta^{-1}}{\underline{W}_{\mathbf{\Upsilon},k}}\Big)^{\frac{1}{2}}\Big)^{-1}\nonumber\\
	  &\hspace{2em}+\frac{(1+\gamma^{-1})^{\frac{1}{2}}}{(1-\beta)^{\frac{1}{2}}}, \label{xixingixng} \\
	& \eta = \frac{1}{e} \exp\Big(\frac{\alpha\underline{W}_{\mathbf{\Upsilon},sk+k}}{\overline{W}_{\mathbf{\Upsilon},s}}\Big), \enspace
	\beta = \exp\Big(-\frac{\alpha\underline{W}_{\mathbf{\Upsilon},sk+k}}{\overline{W}_{\mathbf{\Upsilon},s}}\Big), \nonumber
\end{align}
with $\eta\beta<1$ and $\gamma>0$.
\end{Corollary7}

Note that in \textbf{Corollary \ref{Corollary7}}, $\eta\beta=\frac{1}{e}\leq1$. Thus,
\begin{align} % eq.117
	\frac{1}{\eta(1-\eta\beta)} &= \frac{1}{\frac{1}{e} \exp\Big(\frac{\alpha\underline{W}_{\mathbf{\Upsilon},sk+k}}{\overline{W}_{\mathbf{\Upsilon},s}}\Big)(1-\frac{1}{e})} \nonumber\\
	  &\geq 1 - \frac{\underline{W}_{\mathbf{\Upsilon},sl+s+\theta}}{\overline{W}_{\mathbf{\Upsilon},s}},\nonumber
\end{align}
and we have $\omega=\frac{1}{\eta\big(1-\frac{1}{e}\big)}$.

\begin{rmk5} % Rem.10
\emph{The existing result given in \cite{jWang2016tsp} for generalized OMP indicates that when the RIP is constrained to a derived degree, the number of iterations should be $\max\big\{K,\frac{\lceil8K\rceil}{s}\big\}$ for the residual to be less than a product of the noise power and a constant related to the RIP indicator, where $K=k d$ is the total sparsity. However, since the calculation of the RIP is NP-hard, the result of \cite{jWang2016tsp} is not verifiable.
By contrast, the MIP is computable, and our result in \textbf{Corollary \ref{Corollary7}} reveals that when the number of iterations is equal to $k$, the residual can be smaller than $\xi^*\|\mathcal{N}\|_F$ with $\underline{W}_{\mathbf{\Upsilon},sk+k}>0$. 
Intuitively, due to the consideration of the block structure, we have $k<K$ for $d>1$. The T-GBOMP exhibits significant superiority  in terms of  lower iterative complexity over that of the generalized OMP.  In particular, setting $s=1$ and $d=1$, we obtain that the number of iterations for the generalized OMP derived in \cite{jWang2016tsp} is $8K$ and the parallel result in \textbf{Corollary \ref{Corollary7}} is $K$. 
We present two explanations for this phenomenon. The first is that the MIP condition $\underline{W}_{\mathbf{\Upsilon},sk+k}>0$ in \textbf{Corollary \ref{Corollary7}} is much more restrictive than the RIP constraint in \cite{jWang2016tsp}, and hence the number of iterations in \textbf{Corollary \ref{Corollary7}} is much smaller than that in \cite{jWang2016tsp}. Since computing RIP is NP-hard, this viewpoint cannot be verified. The second is that the result given in \textbf{Corollary \ref{Corollary7}} is indeed better than the one given in \cite{jWang2016tsp}, which provides a lower number of iterations for reliable recovery, resulting in less consumption of computing resources. In any case, since MIP is computationally friendly, \textbf{Corollary \ref{Corollary7}} is useful in presenting a verifiable prerequisite for determining whether the signal components in the residual are fully extracted.}
\end{rmk5}

Similar to \textbf{Theorem \ref{theo3}}, the following theorem demonstrates that the estimated error is upper bounded by a more relaxed quantity with respect to MIP and noise power under a less restrictive MIP condition.

\begin{theoremmaindescription7}\label{theo7} % T8
Given a matrix set $\mathbf{\Upsilon}=\{\mathbf{D}_i\in\mathbb{C}^{M_i\times s_i d_i}, 1\le i\le n\}$ $(n\geq1)$, let $\alpha$ be an integer such that $\alpha={\rm inf}\Big\{x\in\Big\{1,\cdots,\max\Big\{0,\Big\lceil\log\frac{ek}{s}\Big\rceil\Big\}+1\Big\},\frac{\exp(x-2)s(e-1)}{\exp(x-1)-1+(e-1)x}\Big\}$. If the MIP of $\mathbf{\Upsilon}$ satisfies that $\underline{W}_{\mathbf{\Upsilon},sk+k}>0$, then we have
\begin{align} % eq.118
	&\big\|\hat{\mathcal{X}}-\mathcal{X}\big\|_F \nonumber\\
	&\leq \frac{2\Big(\underline{W}^{-\frac{1}{2}}_{\mathbf{\Upsilon},\prod_{t=1}^{n}k_t+ks}(\xi^*+1)\overline{W}^{\frac{1}{2}}_{\mathbf{\Upsilon},\prod_{t=1}^{n}k_t+k}+1\Big)}{\underline{W}^{\frac{1}{2}}_{\mathbf{\Upsilon},\prod_{t=1}^{n}k_t+k}}\|\mathcal{N}\|_F, \label{theo72}
\end{align}
where $\xi^*$ is given in (\ref{xixingixng}).
\end{theoremmaindescription7}

\begin{IEEEproof}
See Appendix \ref{proofoftheo7}.
\end{IEEEproof}

\begin{rmk3}\label{remL11} % Rem.11
\emph{From \textbf{Theorems \ref{theo3}} and \textbf{\ref{theo7}}, when $\|\mathcal{N}\|_F=0$, we have $\hat{\mathcal{X}}=\mathcal{X}$, indicating that the block-sparse tensor can be recovered exactly in the noiseless scenario. The variables on the right-hand sides of equations (\ref{theo3maina}), (\ref{theo3mainb}) and (\ref{theo72}) are the computable MIP, which not only presents strong theoretical interpretability for the feasibility of the block-sparse tensor recovery with low error reconstruction, but also provides a basis for making decisions regarding reliable performance guarantees before operation of the algorithm.}
\end{rmk3}

\section{Simulation Results}\label{S5}

We empirically investigate the recovery performance of T-GBOMP in reconstructing block-sparse tensors under both noiseless and noisy conditions. It is important to note that we not only compare the recovery performance using various metrics but also demonstrate the resilience of different algorithms to suboptimal coherence of measurement matrices. We adopt similar testing metrics as those used in \cite{Kim2020tit, jwang2017mols}. In particular, the exact recovery ratio (ERR) of block-sparse tensor reconstruction is examined as a function of block sparsity and the number of measurements in the noise-free scenario. In the noisy scenario, false alarm/miss-detection ratio and the normalized mean square error (NMSE) are employed to assess the recovery performance of various approaches. To explore the robustness against unsatisfactory MIP, an effortless method is to use the hybrid dictionary as presented in \cite{Soussen2013tit,liyang2023tvt}, where the conventional matrix coherence is close to 1. This measurement matrix substantially diminishes the performance of the compared algorithms, with the extent of performance degradation varying across different algorithms, thereby enabling a more comprehensive insight into the merits and demerits of each algorithm. In a similar fashion, we also produce measurement matrices with conventional matrix coherence nearly 1, but by generating small-dimensional Gaussian matrices. The number of tensor modes is fixed to $3$. The measurement matrices satisfy $\prod_{i=1}^3M_i\! =\! 144$ or $\prod_{i=1}^3M_i\! =\! 216$, and $\prod_{i=1}^3N_i\! =\! 512$ or $\prod_{i=1}^3N_i\! =\! 1024$ with each element drawn independently and identically from a Gaussian distribution. When $\prod_{i=1}^3M_i\! =\! 216$ and $M_1\! =\! M_2\! =\! M_3$, then we have $M_1\! =\! M_2\! =\! M_3\! =\! 6$, which are small-dimensional measurement matrices whose conventional matrix coherence is about 0.75. Meanwhile, measurement matrices with low matrix coherence are also generated for performance comparison, and they have a coherence value of approximately 0.23. Moreover, these matrices are created as block orthogonal matrices with $\tau_{\mathbf{\Upsilon}}\! =\! 0$ and are normalized to have unit column norm. For each realization of a block-sparse tensor, its block support is chosen uniformly at random, and nonzero elements are 1) drawn independently from a standard Gaussian distribution, or 2) drawn independently from the set $\{\pm1\}$. These two types of tensors are referred to as the block-sparse Gaussian tensor and the block-sparse 2-ary pulse amplitude modulation (2-PAM) tensor, respectively \cite{jwang2017mols}. In our experiments, we perform 1,000 independent trials for each point of the approaches, and consider the following recovery algorithms: 1)~BOMP \cite{Eldar2010}; 2)~BOLS \cite{liyang2022}; 3)~T-OMP \cite{AraUjo2019}; 4)~T-GOMP ($s\! =\! 2$); 5)~T-GOMP ($s\! =\! 3$); 6)~T-BOMP; 7)~T-GBOMP ($s\! =\! 2$); and 8)~T-GBOMP ($s\! =\! 3$).

\begin{figure*}[!t]
\vspace*{-1mm}
\centering
	\subfigure[$\prod_{i=1}^3M_i=144$, $\prod_{i=1}^3N_i=512$, Gaussian tensors]{\label{fig3a}
		\includegraphics[width=0.47\linewidth]{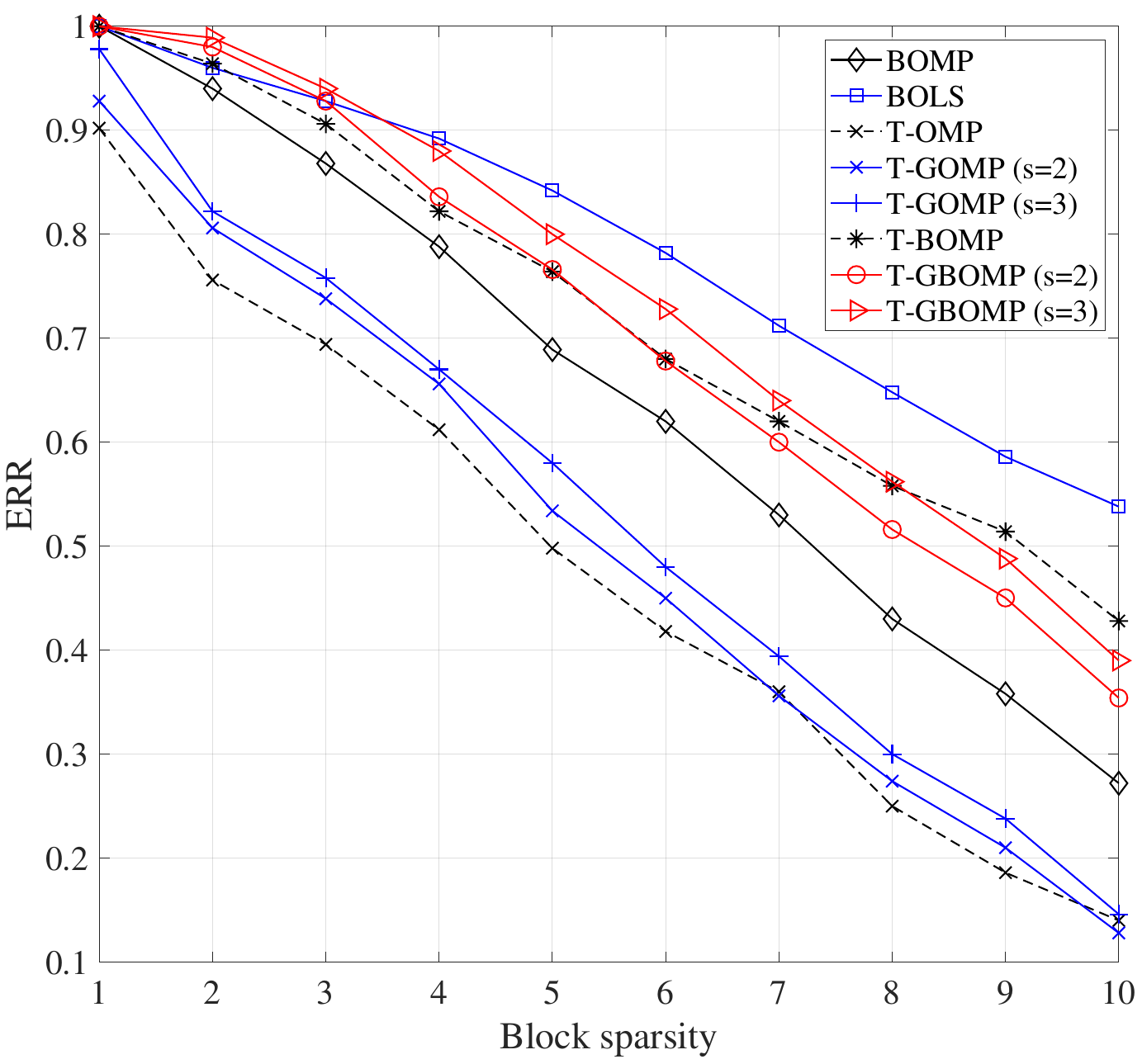}}
	\hspace{0.01\linewidth}
	\subfigure[$\prod_{i=1}^3M_i=144$, $\prod_{i=1}^3N_i=512$, 2-PAM tensors]{\label{fig3b}
		\includegraphics[width=0.47\linewidth]{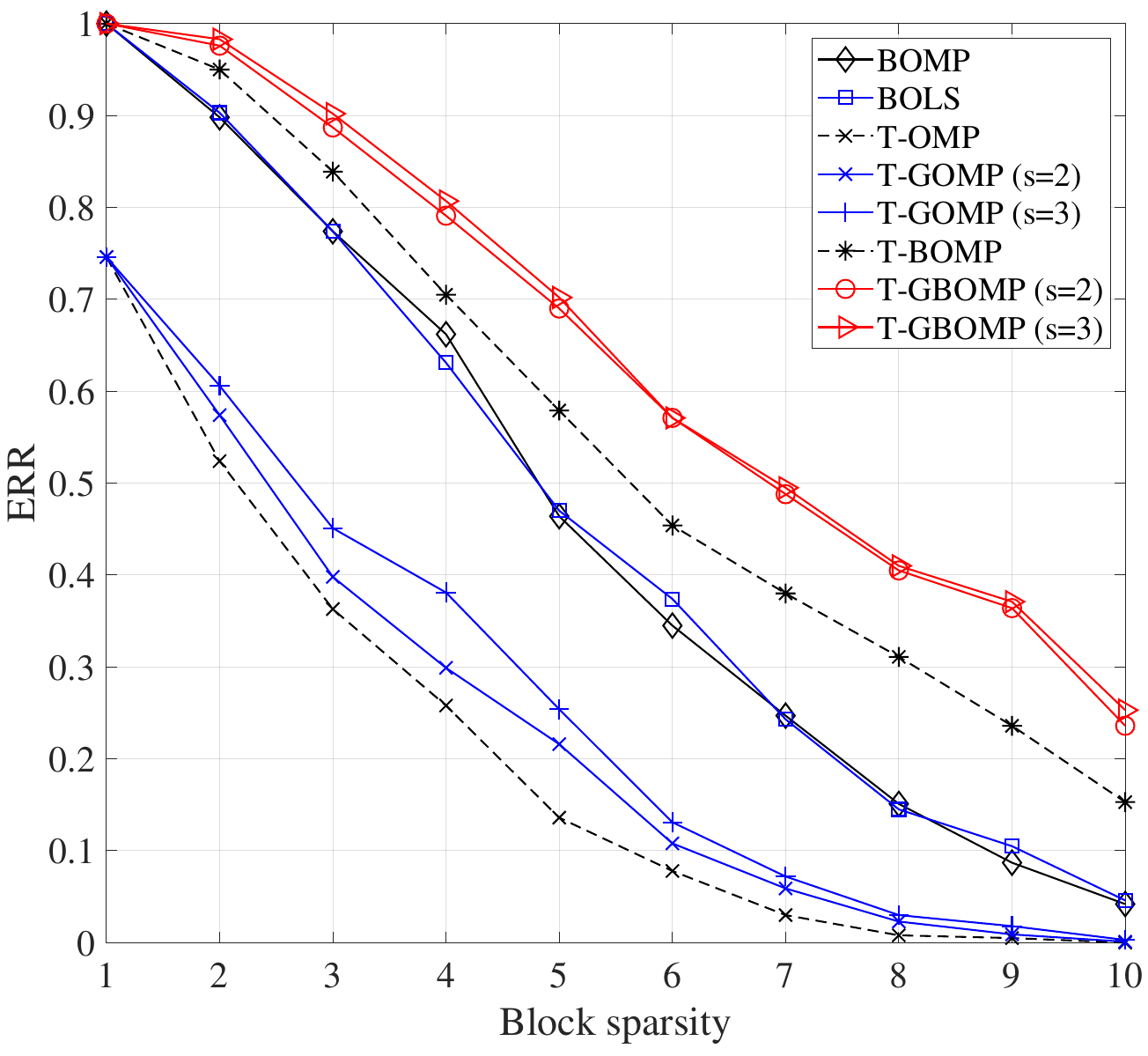}}
	\hspace{0.01\linewidth}
	\subfigure[$\prod_{i=1}^3M_i=216$, $\prod_{i=1}^3N_i=512$, Gaussian tensors]{\label{fig3c}
		\includegraphics[width=0.47\linewidth]{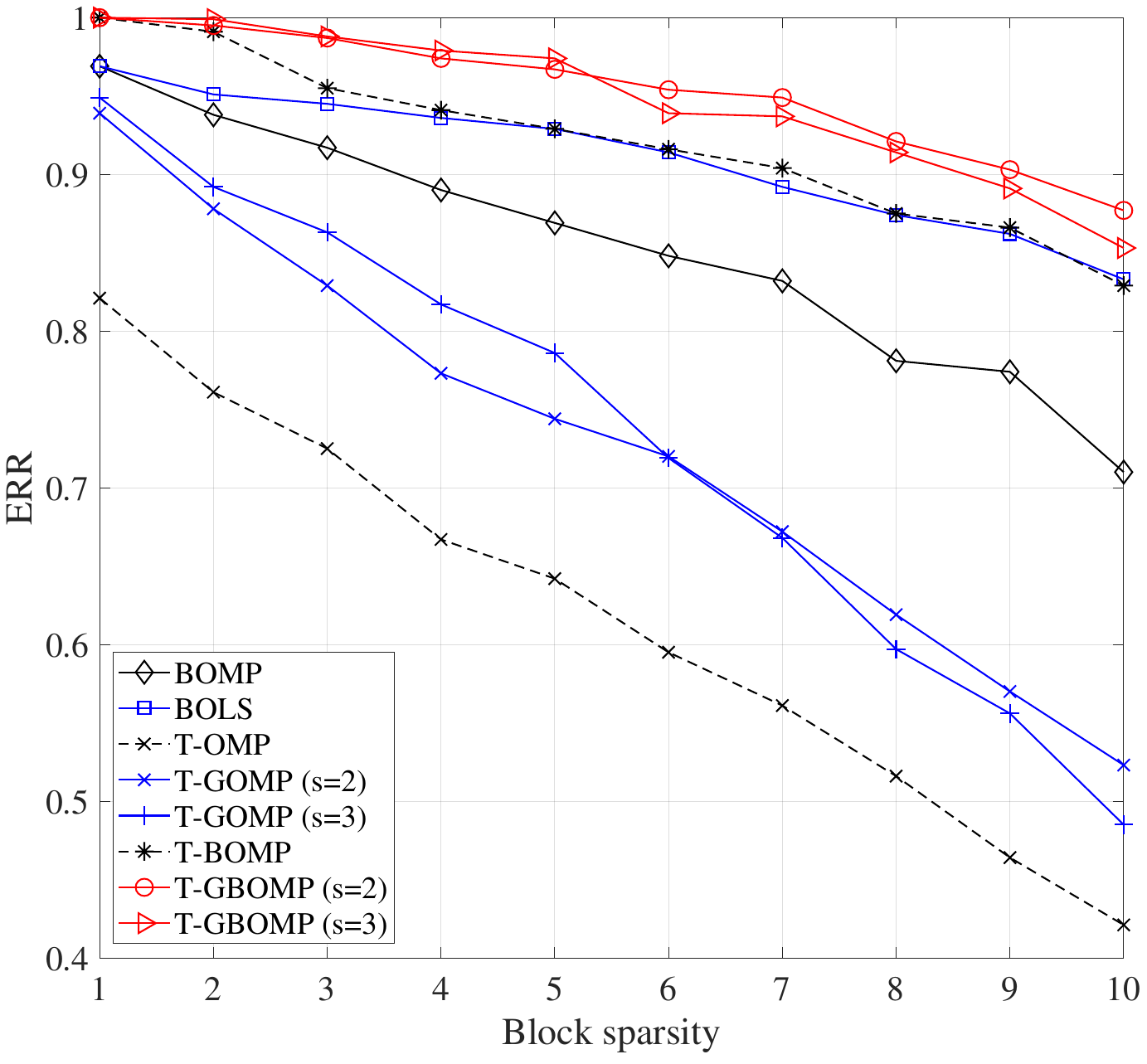}}
		\hspace{0.01\linewidth}
	\subfigure[$\prod_{i=1}^3M_i=216$, $\prod_{i=1}^3N_i=512$, 2-PAM tensors]{\label{fig3d}
		\includegraphics[width=0.47\linewidth]{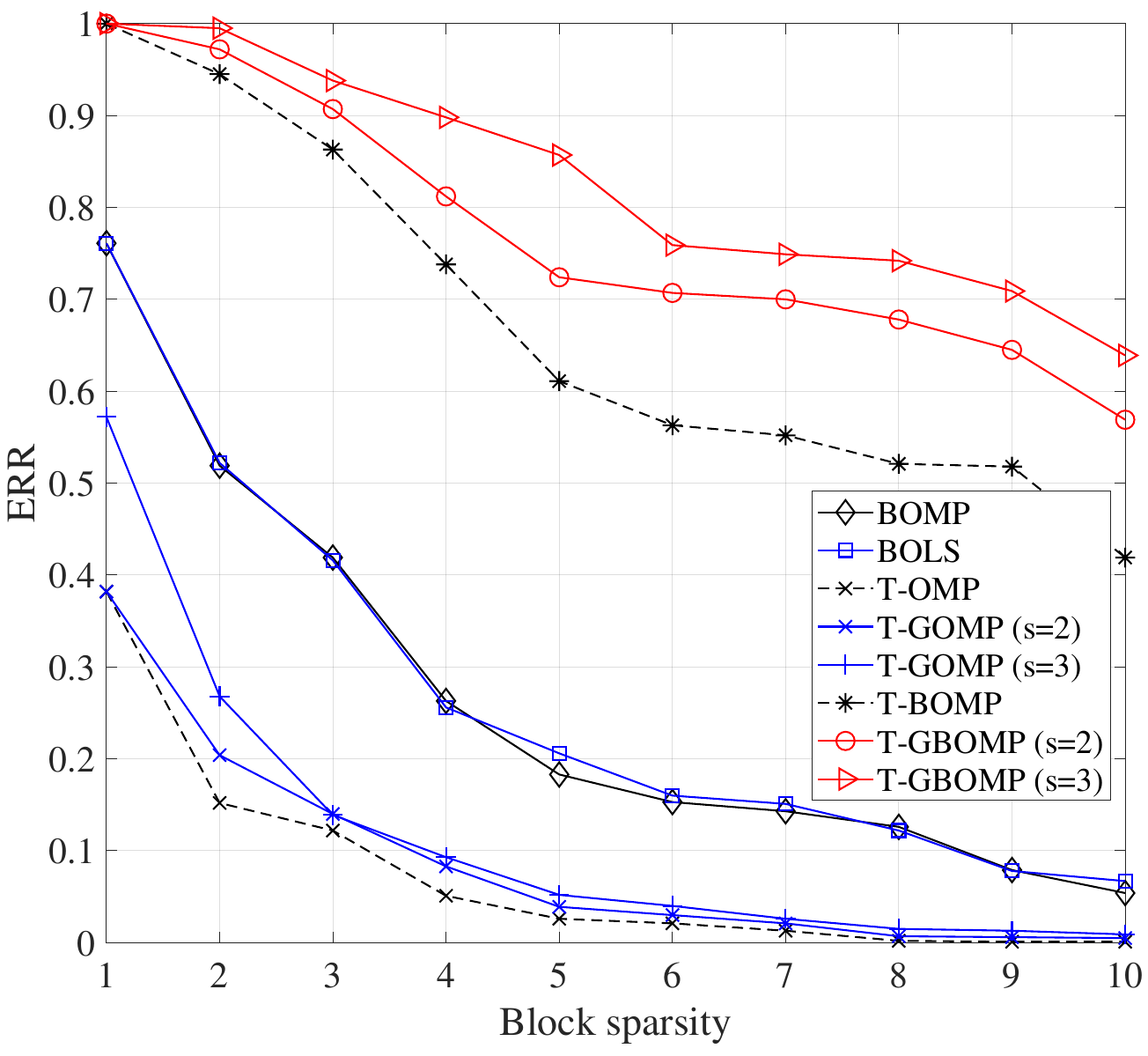}}
\vspace*{-1mm}
\caption{ERR for recovering block-sparse tensors as a function of the block sparsity.}
\vspace*{2mm}
\label{ERRblocksparsityESULTS} % Fig.3
\end{figure*}

\subsection{The Noiseless Scenario}\label{S5.1}

In the noiseless case, the ERR is used as a metric to evaluate the performance of various algorithms, which is defined as \cite{Kim2020tit}
\begin{align}
	\text{ERR} = \frac{\text{number of exact reconstructions}}{\text{number of total trials}}.\nonumber
\end{align}

In Fig.~\ref{ERRblocksparsityESULTS}, we depict the ERR as a function of the block sparsity $k$. The block lengths are fixed at $d_1\! =\! 2$, $d_2\! =\! 2$ and $d_3\! =\! 1$, resulting in a total sparsity of $K\! =\! d_1d_2d_3 k\! =\! 4 k$. The conventional matrix coherence of the measurement matrices is around 0.75, which is unsatisfactory for recovery and can significantly degrade the performance of the algorithms. The mutual block coherence, as defined in \textbf{Definition \ref{crossblockcoherece}}, is about 0.6. It is evident that due to the unsatisfactory MIP of the measurement matrices, none of the compared approaches can consistently achieve exact recovery. However, it can be seen that T-GBOMP ($s\! =\! 2$) and T-GBOMP ($s\! =\! 3$) provide better assurance of exact recovery, outperforming other algorithms when the block sparsity is less than or equal to 2, wherein $k\! =\! 1$ is the only possible point for the algorithms to achieve exact recovery, i.e., $\text{ERR}\! =\! 1$. As can be seen from Fig.~\ref{fig3a}, T-GBOMP ($s\! =\! 2$) and T-GBOMP ($s\! =\! 3$) perform worse than BOLS in recovering the block-sparse Gaussian tensors with $\prod_{i=1}^3M_i\! =\! 144$ and $\prod_{i=1}^3N_i\! =\! 512$ when a block sparsity is greater than 3, which indicates that the OLS-framework may be highly competitive in providing reliable recovery in the scenarios with unsatisfactory MIP and high block sparsity. For 2-PAM tensors, Figs.~\ref{fig3b} and \ref{fig3d} show that T-BOMP, T-GBOMP ($s\! =\! 2$) and T-GBOMP ($s\! =\! 3$) exhibit significant performance improvements over the other algorithms. This suggests that tensor structure information is beneficial for reliable recovery, as T-BOMP outperforms BOMP and BOLS. Moreover, exploiting block sparsity yields more assured results, since T-GBOMP ($s\! =\! 2$) and T-GBOMP ($s\! =\! 3$) outperform T-GOMP ($s\! =\! 2$) and T-GOMP ($s\! =\! 3$).
As T-GBOMP performs better than T-BOMP, and T-BOMP performs better than BOMP, it becomes evident that the larger the tensor mode, the better the recovery performance of the algorithms. This aligns with the statement in Section~\ref{S4}, indicating that a larger tensor mode can offer more reliable recovery performance. Furthermore, in general, T-GBOMP ($s\! =\! 3$) outperforms T-GBOMP ($s\! =\!2$), indicating that a larger selection parameter can indeed enhance the accuracy of recovery by increasing the likelihood of selecting all the correct atoms.
We now verify the analytic results related to reconstructible sparsity derived in Subsection~\ref{S4.1}. As illustrated in \emph{Remark~\ref{rmk5rmk5}}, based on the parameter settings in Fig.~\ref{ERRblocksparsityESULTS}, the most restricted reconstructible sparsity can be calculated by $\Big(-\frac{1}{2}+\sqrt{\frac{5}{4}-\frac{-1}{0.6^3\times4}}\Big)^2\! \approx\! 1$, which is consistent with the results given in Fig.~\ref{ERRblocksparsityESULTS}. Observe that T-GBOMP ($s\! =\! 2$) and T-GBOMP ($s\! =\! 3$) can achieve 100\% ERR when the block sparsity is less than or equal to 1.

\begin{figure*}[!b]
%\vspace*{-4mm}
\centering
	\subfigure[$k=2$, $\prod_{i=1}^3d_i=8$, $\prod_{i=1}^3N_i=1024$]{\label{ERRMRESULTSA}
		\includegraphics[width=0.47\linewidth]{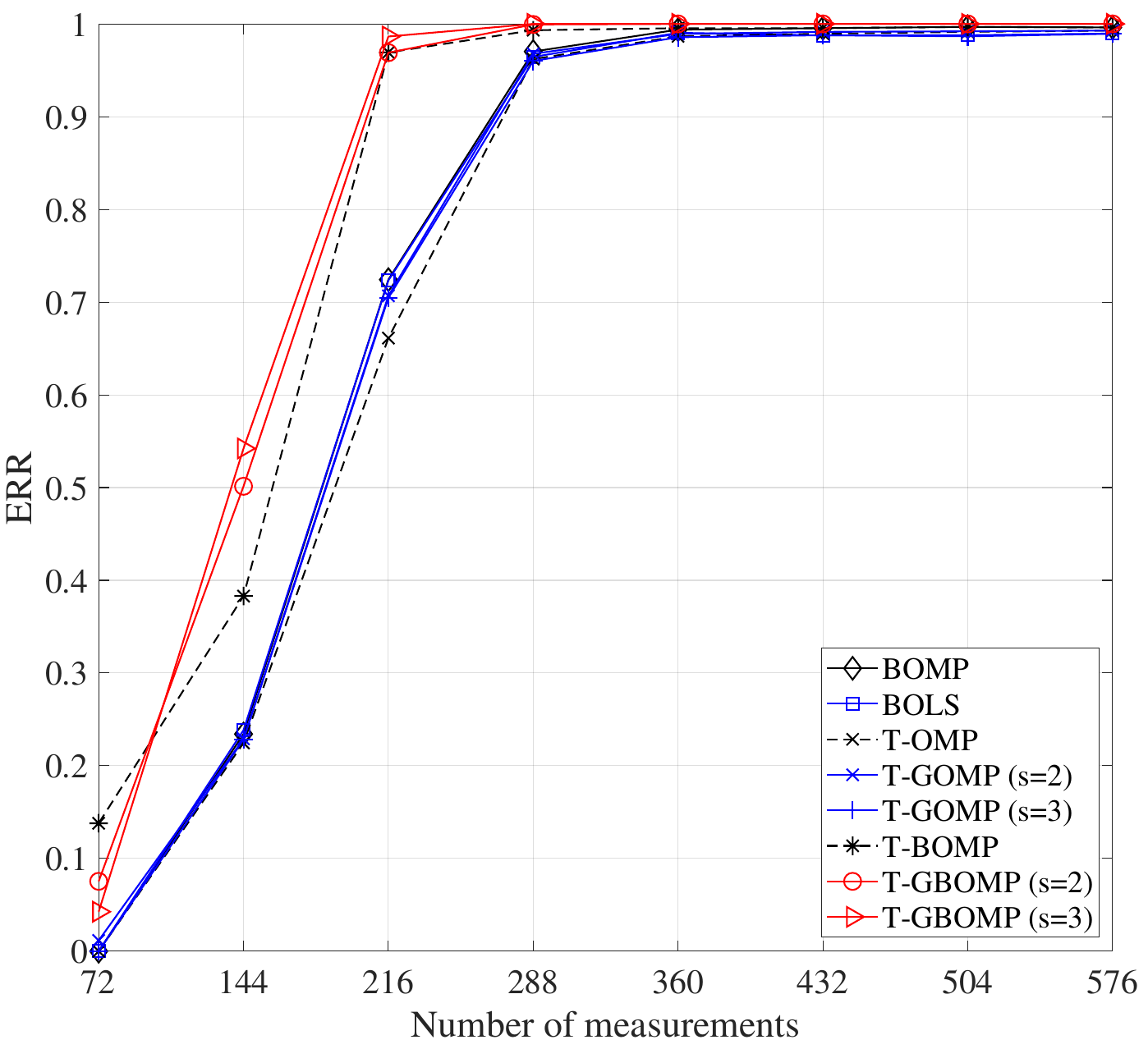}}
	\hspace{0.01\linewidth}
	\subfigure[$k=1$, $\prod_{i=1}^3d_i=16$, $\prod_{i=1}^3N_i=1024$]{\label{ERRMRESULTSB}
		\includegraphics[width=0.47\linewidth]{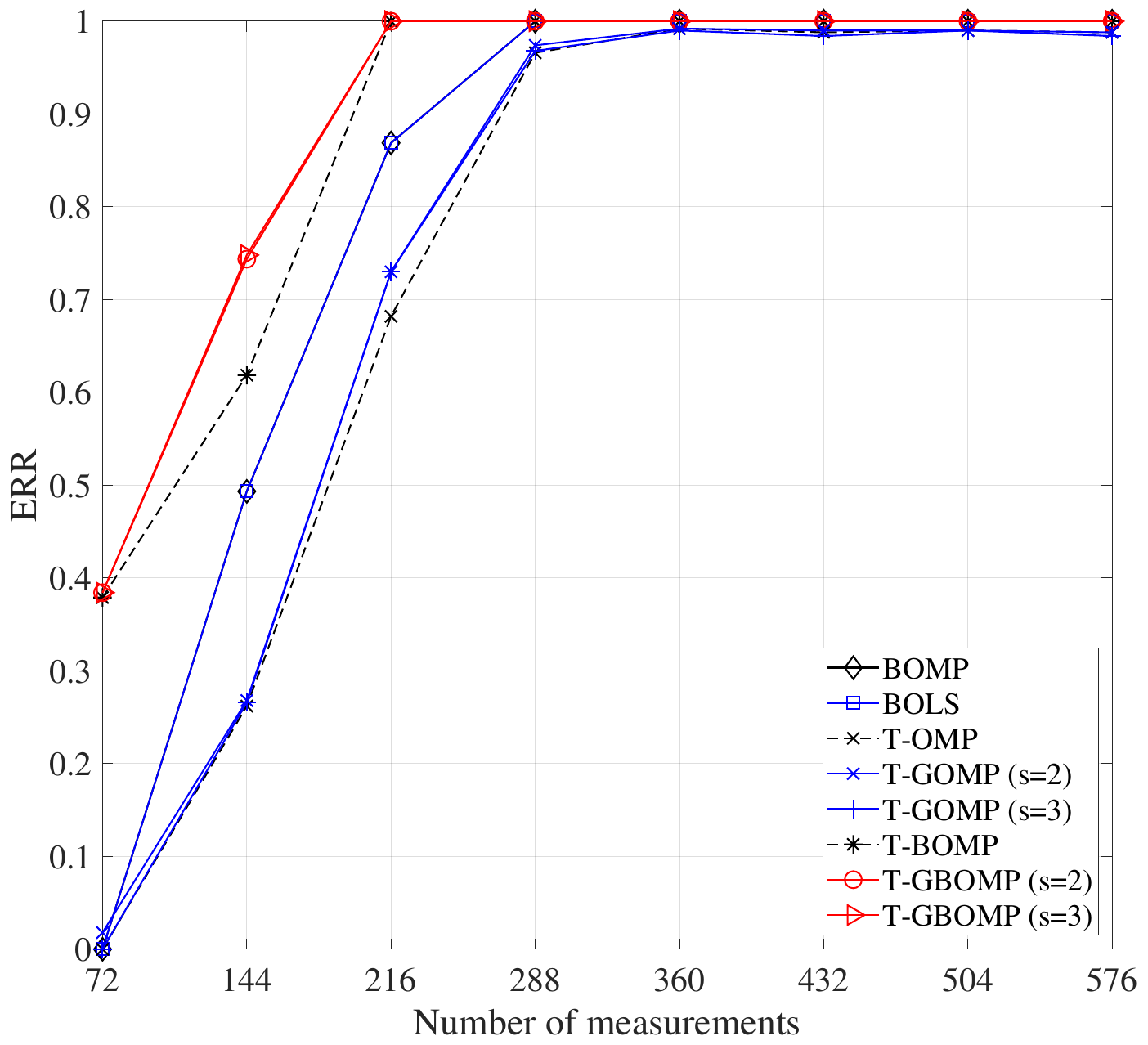}}
\vspace*{-1mm}
\caption{ERR for recovering block-sparse 2-PAM tensors as a function of the number of measurements.}
\label{ERRMRESULTS} % Fig.4
\end{figure*}

In Fig.~\ref{ERRMRESULTS}, where measurement matrices are generated to have low matrix coherence, the ERR is plotted as a function of the number of measurements. It can be observed that T-GBOMP ($s\! =\! 2$) and T-GBOMP ($s\! =\! 3$) outperform the other compared methods, with T-GBOMP ($s\!=\! 3$) achieving the best ERR when the number of measurements exceeds 144. When the total sparsity is fixed, i.e., $K\! =\! k\prod_{i=1}^3d_i\! =\! 16$, the recovery performance of T-GBOMP ($s\! =\! 2$) and T-GBOMP ($s\! =\! 3$) are improved as the block length increases. This indicates that a stronger block-structure characteristic leads to a more assured recoverability of the algorithms. It is worth noting that the block-structure characteristic is crucial for the reliability of recovery. For instance, in Fig.~\ref{ERRMRESULTSA}, for T-GBOMP ($s\! =\! 3$) to attain the 100\% ERR, the number of measurements is larger than 288. In contrast, as can be seen from Fig.~\ref{ERRMRESULTSB}, T-GBOMP ($s\! =\! 3$) can provide 100\% ERR performance when the number of measurements exceeds 216.
Furthermore, since the total sparsity remains constant, the corresponding ERRs of T-GOMP ($s\! =\! 2$) and T-GOMP ($s\! =\! 3$) are similar given a same number of measurements, as can be seen from Figs.~\ref{ERRMRESULTSA} and \ref{ERRMRESULTSB}. This finding aligns with the expectation that these algorithms do not leverage the block structure of the tensors being recovered, and therefore are not significantly influenced by changes in the block structure given a fixed total sparsity.

\subsection{The Noisy Scenario}\label{S5.2}

The false alarm/miss-detection and NMSE are used as metrics to evaluate recovery performance of different approaches. The false alarm ratio is expressed as 
\begin{align}
	\text{False alarm ratio}=\frac{|\hat{\mathbf{\Xi}}\setminus\mathbf{\Xi}|}{k\prod_{i=1}^3d_i},\nonumber
\end{align}
which is equivalent to the miss-detection ratio since $|\hat{\mathbf{\Xi}}|=|\mathbf{\Xi}|=k\prod_{i=1}^3d_i$, where $\hat{\mathbf{\Xi}}$ is the estimated set consisting of the indices corresponding to nonzero tensor elements, and $\mathbf{\Xi}$ denotes the correct index set. The NMSE is expressed as 
\begin{align}
	\text{NMSE} = \frac{\|\hat{\mathcal{X}}-\mathcal{X}\|^2_F}{\|\mathcal{X}\|^2_F},\nonumber
\end{align}
where $\hat{\mathcal{X}}$ represents the estimated block-sparse tensor, and $\mathcal{X}$ denotes the true block-sparse tensor.

\begin{figure*}[!tp]
\centering
	\subfigure[$k=2$, $\prod_{i=1}^{3}M_i=144$]{\label{falsealarmnsra}
		\includegraphics[width=0.47\linewidth]{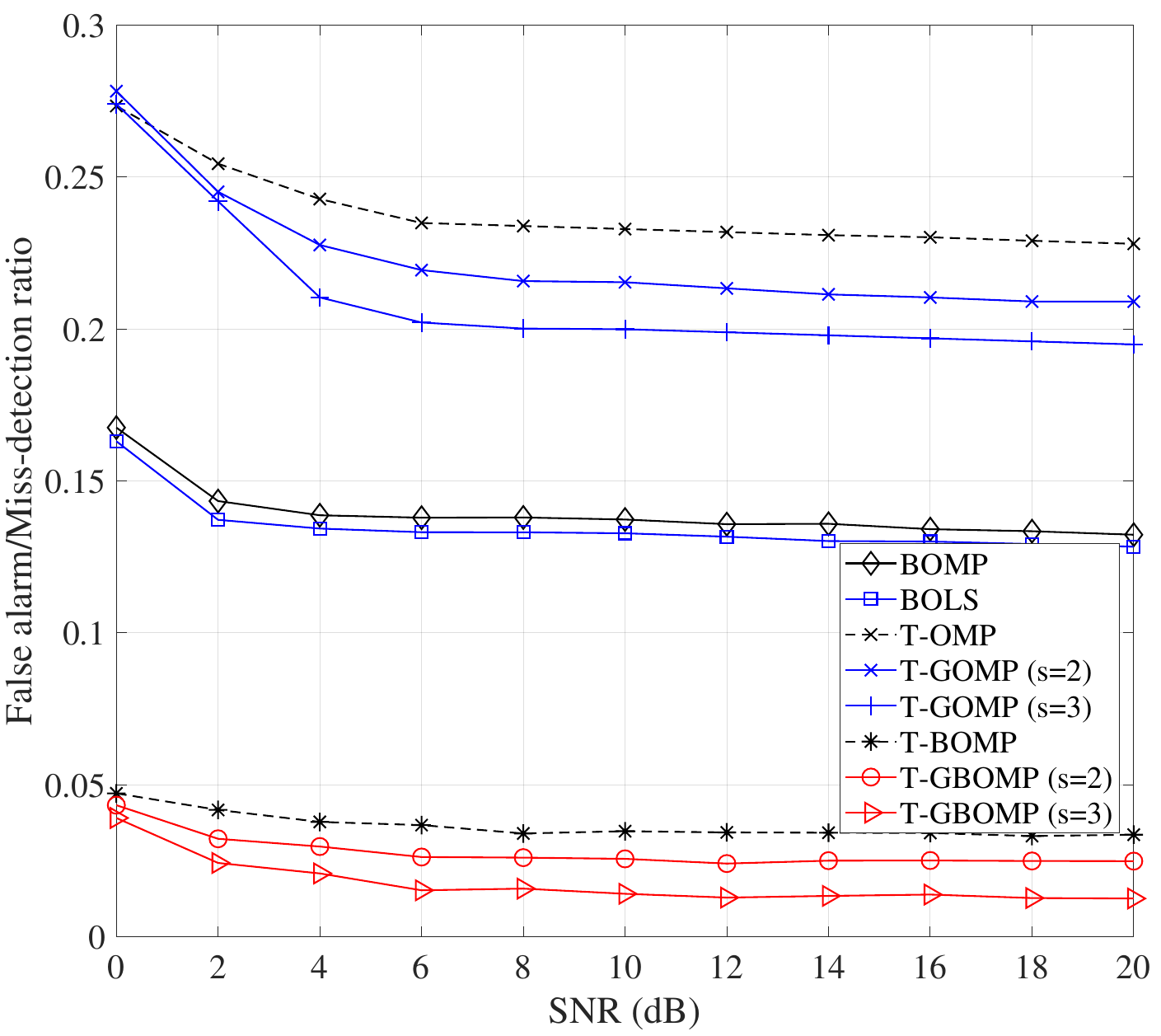}}
	\hspace{0.01\linewidth}
	\subfigure[$k=2$, $\prod_{i=1}^{3}M_i=216$]{\label{falsealarmnsrb}
		\includegraphics[width=0.47\linewidth]{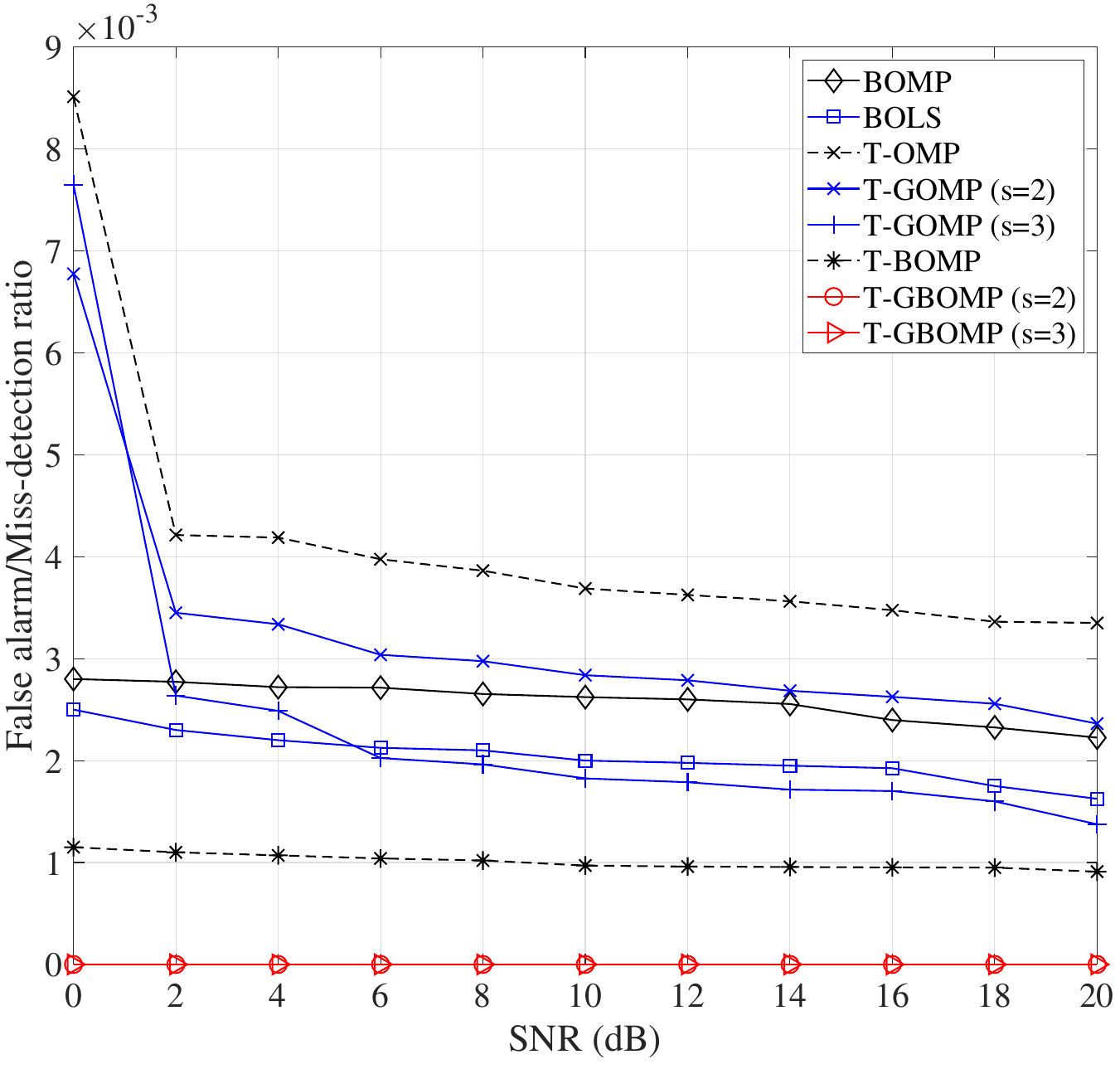}}
\vspace*{-1mm}
\caption{False alarm/Miss-detection ratio for recovering block-sparse 2-PAM tensors as a function of SNR.}
\label{falsealarmnsr} % Fig.5
\vspace*{2mm}
\end{figure*}

As illustrated in Fig.~\ref{falsealarmnsr}, T-GBOMP ($s\! =\! 3$) achieves the best false alarm/miss-detection ratio among all the compared algorithms. This indicates that a larger selection parameter $s$ can provide more assured performance in identifying the correct tensor supports. When the compression ratio becomes larger but still remains below $\frac{1}{2}$ as depicted in Fig.~\ref{falsealarmnsrb}, T-GBOMP ($s\! =\! 2$) and T-GBOMP ($s\! =\! 3$) achieve exceptionally good detection performance, surpassing the other algorithms, with the false alarm/miss-detection ratios of T-GBOMP ($s\! =\! 2$) and T-GBOMP ($s\! =\! 3$) reaching 0 across all compared SNR values. In general, algorithms that exploit block sparsity outperform those that do not utilize block sparsity, indicating that the block-structure characteristic is indeed a beneficial factor for reliable recovery. Furthermore, it can be observed that the methods using block-sparse tensor structure, i.e., T-BOMP, T-GBOMP ($s\! =\! 2$) and T-GBOMP ($s\, =\, 3$), provide better false alarm/miss-detection ratios than BOMP and BOLS, which only utilize block sparsity without employing tensor structure. This is due to the vectorization in conventional non-tensor algorithms which causes the loss of certain high-dimensional feature information in sparse recovery.

\begin{figure*}[!tp]
\centering
	\subfigure[$k=2$, $\prod_{i=1}^3d_i=4$]{\label{nmsesnra}
		\includegraphics[width=0.47\linewidth]{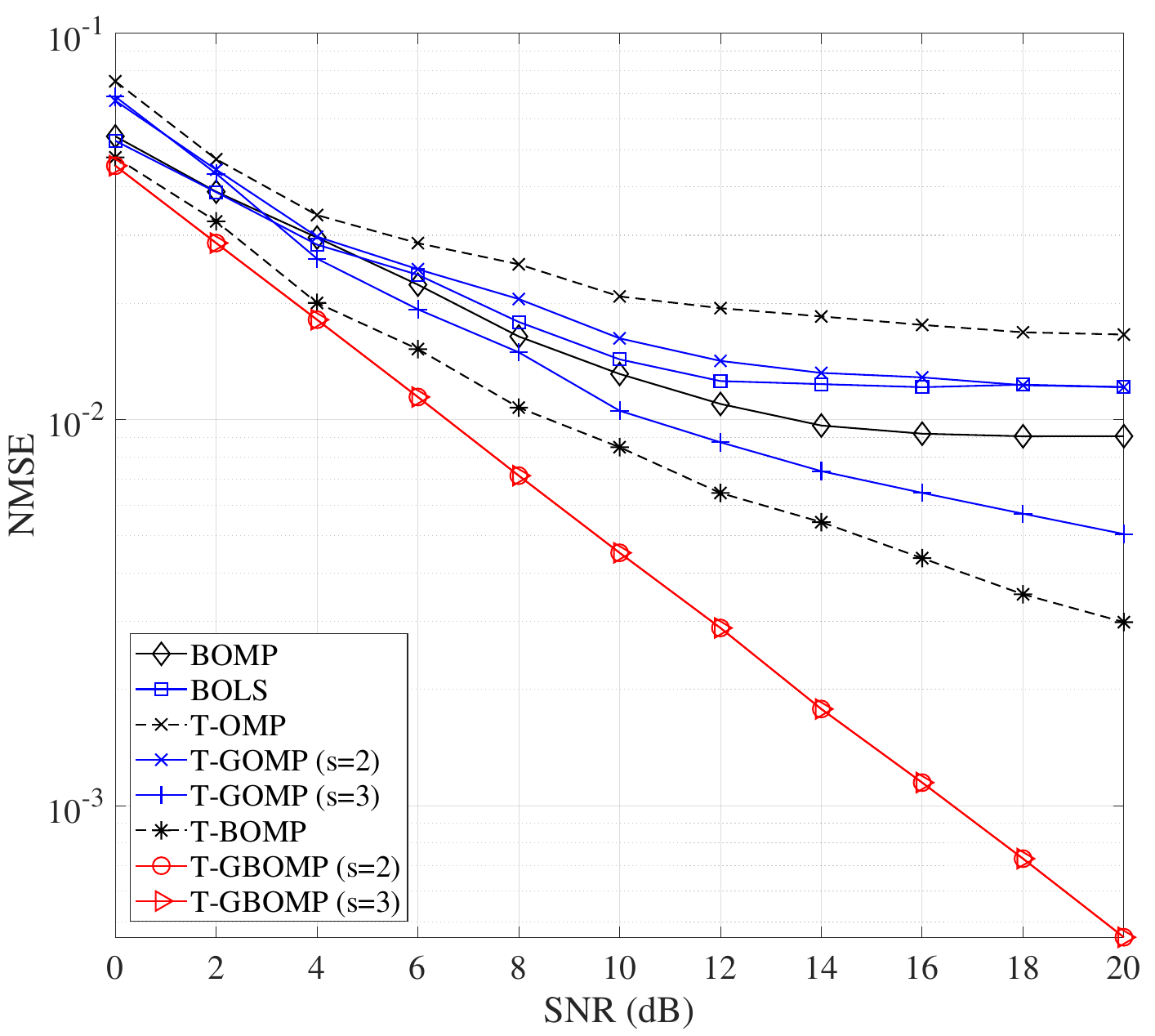}}
	\hspace{0.01\linewidth}
	\subfigure[$k=4$, $\prod_{i=1}^3d_i=4$]{\label{nmsesnrb}
		\includegraphics[width=0.47\linewidth]{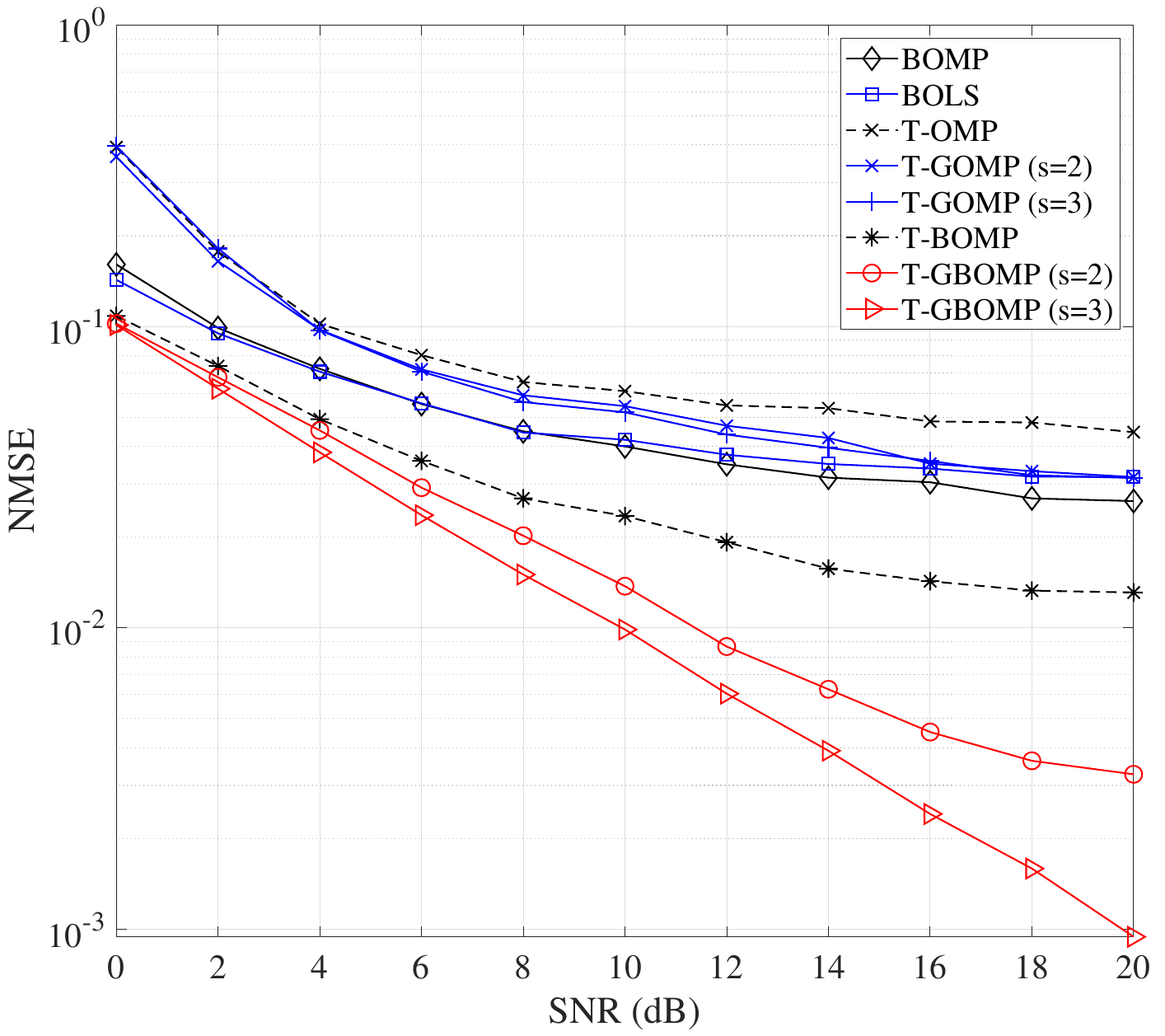}}
\vspace*{-1mm}
\caption{NMSE for recovering block-sparse 2-PAM tensors as a function of SNR.}
\label{nmsesnr} % Fig.6
\vspace*{2mm}
\end{figure*}

As can be observed from Fig.~\ref{nmsesnr}, where measurement matrices are generated with low matrix coherence, T-GBOMP ($s\! =\! 3$) generally achieves the lowest NMSE among all the compared methods. The results of Fig.~\ref{nmsesnr} show that as the block sparsity increases, the performance of T-BOMP, T-GBOMP ($s\! =\! 2$) and T-GBOMP ($s\! =\! 3$) deteriorate, but they still outperform the other approaches significantly. We can now discuss the theoretical results derived in Subsection~\ref{S4.2}. Firstly, regarding the lower bound of SNR required for reliable recovery, we consider the intuitive result presented in \emph{Remark~\ref{rmkrmk8}}, taking into account the parameter settings of T-GBOMP ($s\! =\! 3$) in Fig.~\ref{nmsesnra}. Since block-sparse 2-PAM tensors are generated in the simulation of Fig.~\ref{nmsesnra}, the MAR is equal to 1 as defined in \textbf{Definition~\ref{defofmar}}. The lower bound of SNR required for reliable recovery in \emph{Remark~\ref{rmkrmk8}} is thus equal to $10\log_{10}\big(2\big(\frac{1}{\sqrt{3}}+1\big)\big)^2\approx10$\,dB. This holds true, as the NMSE is relatively low, and remains under $10^{-2}$ when the SNR exceeds 10\,dB in Fig.~\ref{nmsesnra}. It is worth mentioning that the result given in \emph{Remark~\ref{rmkrmk8}} is less restrictive than those in \textbf{Theorem~\ref{theo2}} due to the asymptotic analysis, which indicates that the correctness of the result in \emph{Remark~\ref{rmkrmk8}} also demonstrates the truth in \textbf{Theorem~\ref{theo2}}. Secondly, regarding the reconstruction error which can be reflected by the NMSE, we consider the results presented in \textbf{Theorem~\ref{theo3}}. Similarly, we assume a less restricted scenario for calculating the reconstruction error bound, where each element in the measurement matrix of Fig.~\ref{nmsesnra} follows an independent and identically distributed Gaussian distribution with zero mean and variance $\frac{1}{\prod_{i=1}^3M_i}$. In Fig.~\ref{nmsesnra}, when $\text{SNR}\! =\! 10$\,dB and $K\! =\! k\prod_{i=1}^3d_i\! =\! 8$, we have $\mathbb{E}(\|\mathcal{N}\|_F)\! =\! \sqrt{K\times10^{-\frac{10}{10}}}\! =\! \frac{2\sqrt{5}}{5}$. Therefore, according to \textbf{Theorem~\ref{theo3}}, we have $\text{NMSE}\! =\! \frac{\|\hat{\mathcal{X}}-\mathcal{X}\|^2_F}{\|\mathcal{X}\|^2_F}\! \leq\! \frac{(4\|\mathcal{N}\|_F)^2}{K^2}\! =\! \frac{16(\frac{2\sqrt{5}}{5})^2}{64}\! =\! \frac{1}{5}$. This result is consistent with the findings in Fig.~\ref{nmsesnra}, indicating that the other theoretical results derived regarding the reconstruction error are also valid. Note that the analytical results presented are more pessimistic than the numerical results, since our analyses based on the MIP framework essentially consider worst-case scenarios.

\section{Conclusions}\label{S6}

In this paper, we have defined new MIP concepts measuring the uncertainty relations in a matrix set, formulated block-sparse tensor recovery modeling, and elaborated T-GBOMP as a generalized algorithmic basis for theoretical derivation. Then, we have provided in-depth theoretical analyses for T-GBOMP, in terms of exact recovery conditions in the noiseless case and reliable reconstruction conditions in noisy scenarios. Specifically, we have presented an exact recovery condition and closed-form sufficient conditions that ensure T-GBOMP can recover the block-sparse tensor exactly if the block sparsity is smaller than a derived upper bound. We have further derived error bounds and the residual convergence level of T-GBOMP after a specified number of iterations, in the presence of noise. Moreover, the SNR required for reliable recovery with respect to the newly defined version of MIP has also been developed. We have further highlighted that the T-GBOMP includes the existing greedy algorithms as special cases and it can be extended to other tensor variants, indicating that the theoretical results developed in this paper are quite broad.

\appendix

\subsection{Proof of \textbf{Lemma~\ref{lemma5}}}\label{proofoflemma5} % Ap-A

\begin{IEEEproof}	
Use elementary column transformation to form $\mathbf{T}\in\mathbb{C}^{\prod_{t=1}^nM_t\times s_td_t}$ as a concatenation of $\prod_{t=1}^ns_t$ column-block submatrices,  of sizes $\prod_{t=1}^nM_t\times\prod_{t=1}^nd_t$, where the $c$th $(c\in\{1,\cdots,\prod_{t=1}^ns_t\})$ block of the transformed matrix is $\mathbf{D}_{1_{[i_1]}}\otimes\cdots\otimes\mathbf{D}_{n_{[i_n]}}$, and $i_1\in\{1,\cdots,s_1\},\cdots,i_n\in\{1,\cdots,s_n\}$. We denote this new transformed matrix as $\mathbf{Q}$, and
the eigenvalues of $\mathbf{Q}^{\rm H}\mathbf{Q}$ are equal to those of $\mathbf{T}^{\rm H}\mathbf{T}$. To lower bound $\lambda_{\min}$, we prove that the matrix $\mathbf{Q}^{\rm H}\mathbf{Q}-\lambda \mathbf{I}$ is nonsingular under the condition $\lambda < 1-(d-1)\nu_D-(s-1)d\mu_D$ when $(d-1)\nu_D+(s-1)d\mu_D < 1$. This is equivalent to proving that for any nonzero vector $\mathbf{f}=\big[\mathbf{f}_{[1]}^{\rm T},\mathbf{f}_{[2]}^{\rm T},\cdots,\mathbf{f}^{\rm T}_{[s]}\big]^{\rm T}\in\mathbb{C}^{\prod_{t=1}^n s_t\prod_{t=1}^n d_t}$, $(\mathbf{Q}^{\rm H}\mathbf{Q}-\lambda \mathbf{I})\mathbf{f}\neq \mathbf{0}$. Without loss of generality, we assume $\|\mathbf{f}_{[1]}\|_2\geq\|\mathbf{f}_{[2]}\|_2\geq \cdots\geq \|\mathbf{f}_{[s]}\|_2$. The $\ell_2$-norm of the first block of $(\mathbf{Q}^{\rm H}\mathbf{Q}-\lambda \mathbf{I})\mathbf{f}$ satisfies
\begin{align}\label{thefirstblock1} % eq.122
	&\Big\|\big[(\mathbf{Q}^{\rm H}\mathbf{Q}-\lambda \mathbf{I})\mathbf{f}\big]_{[1]}\Big\|_2\nonumber\\
	  & = \bigg\|\big(\mathbf{Q}_{[1]}^{\rm H}\mathbf{Q}_{[1]}-\lambda \mathbf{I}\big)\mathbf{f}_{[1]} + \sum_{i=2}^{\prod_{t=1}^ns_t} \mathbf{Q}_{[1]}^{\rm H}\mathbf{Q}_{[i]}\mathbf{f}_{[i]}\bigg\|_2 \nonumber \\
		& \geq \big\|\mathbf{Q}_{[1]}^{\rm H}\mathbf{Q}_{[1]}\mathbf{f}_{[1]}\big\|_2 - \lambda \| \mathbf{f}_{[1]}\|_2
		- \bigg\|\sum_{i=2}^{\prod_{t=1}^ns_t} \mathbf{Q}_{[1]}^{\rm H}\mathbf{Q}_{[i]}\mathbf{f}_{[i]}\bigg\|_2\nonumber\\
		& \geq \big\|\mathbf{Q}_{[1]}^{\rm H}\mathbf{Q}_{[1]}\mathbf{f}_{[1]}\big\|_2 - \lambda \| \mathbf{f}_{[1]}\|_2\nonumber\\
		&\hspace{1em}- \sum_{t=0}^{n-1}C_t\varpi_{\mathbf{\Upsilon},t}^n\prod^{n}_{i=1}d_i  \max\limits_{l\in\{C_t\}}\|\mathbf{f}_{[l]}\|_2,
\end{align}
where the set $\{C_t\}$ consists of the indices corresponding to $C_t$ as defined in \textbf{Proposition \ref{prop2}}.
Denote $\mathbf{Q}_{[1]}^{\rm H}\mathbf{Q}_{[1]}=\mathbf{I}+\mathbf{A}$, where $\forall i$, $\mathbf{A}_{(i,i)}=0$. Based on Ger\v{s}gorin's disc theorem \cite{matrixanalysis}
\begin{align}
	\|\mathbf{A}\|_2\leq \sum_{t=0}^{n-1}\underline{C}_t\tau_{\mathbf{\Upsilon},t}^{n-t}.\nonumber
\end{align}
Thus,
\begin{align}\label{thefirstblock3} % eq.124
	\big\|\mathbf{Q}_{[1]}^{\rm H}\mathbf{Q}_{[1]}\big\|_2 &\geq \|\mathbf{I}\|_2-\|\mathbf{A}\|_2
	\geq 1 -  \sum_{t=0}^{n-1}\underline{C}_t\tau_{\mathbf{\Upsilon},t}^{n-t} .
\end{align}
Combining (\ref{thefirstblock1}) and (\ref{thefirstblock3}) yields
\begin{align}
	&\Big\|\big[(\mathbf{Q}^{\rm H}\mathbf{Q}-\lambda \mathbf{I})\mathbf{f}\big]_{[1]}\Big\|_2\nonumber\\
		& \geq \Bigg( 1 - \sum_{t=0}^{n-1}\underline{C}_t\tau_{\mathbf{\Upsilon},t}^{n-t} - \lambda \Bigg)\|\mathbf{f}_{[1]}\|_2\nonumber\\
		&\hspace{1em}- \sum_{t=0}^{n-1}C_t\varpi_{\mathbf{\Upsilon},t}^n\prod^{n}_{i=1}d_i  \max\limits_{l\in\{C_t\}}\|\mathbf{f}_{[l]}\|_2  \nonumber \\
	& > \sum_{t=0}^{n-1}C_t\varpi_{\mathbf{\Upsilon},t}^n\prod^{n}_{i=1}d_i\|\mathbf{f}_{[1]}\|_2\nonumber\\
	&\hspace{1em}	-\sum_{t=0}^{n-1}C_t\varpi_{\mathbf{\Upsilon},t}^n\prod^{n}_{i=1}d_i  \max\limits_{l\in\{C_t\}}\|\mathbf{f}_{[l]}\|_2 \nonumber\\
	& = \sum_{t=0}^{n-1}C_t\varpi_{\mathbf{\Upsilon},t}^n\prod^{n}_{i=1}d_i	\Big(\|\mathbf{f}_{[1]}\|_2-\max\limits_{l\in\{C_t\}}\|\mathbf{f}_{[l]}\|_2\Big)
	\geq 0 .\nonumber
\end{align}
This reveals that $\big(\mathbf{Q}^{\rm H}\mathbf{Q}-\lambda \mathbf{I}\big)\mathbf{f}\neq \mathbf{0}$ and $\lambda_{\min}\geq 1-\sum_{t=0}^{n-1}\underline{C}_t\tau_{\mathbf{\Upsilon},t}^{n-t}-\sum_{t=0}^{n-1}C_t\varpi_{\mathbf{\Upsilon},t}^n\prod^{n}_{i=1}d_i$ holds.
	
The inequality $\lambda_{\max}\leq1+\sum_{t=0}^{n-1}\underline{C}_t\tau_{\mathbf{\Upsilon},t}^{n-t}+\sum_{t=0}^{n-1}C_t\varpi_{\mathbf{\Upsilon},t}^n\prod^{n}_{i=1}d_i$ can be proved similarly.
\end{IEEEproof}

\subsection{Proof of \textbf{Lemma~\ref{lemma4}}}\label{proofoflemmager} % Ap-B

\begin{IEEEproof}
Based on matrix norm property, we have
\begin{align} % eqs.126-129
	\|\mathbf{D}\|_2&=\sqrt{\|\mathbf{D}^{\rm H}\mathbf{D}\|_2}
	=\sqrt{\rho(\mathbf{D}^{\rm H}\mathbf{D})}\nonumber\\
	&\leq\sqrt{\min\{\rho_r(\mathbf{D}^{\rm H}\mathbf{D}),\rho_c(\mathbf{D}^{\rm H}\mathbf{D})\}}\label{proflemma23}\\
	&\leq\sqrt{\min\{\rho_r(\mathbf{D}^{\rm H})\rho_r(\mathbf{D}),\rho_c(\mathbf{D}^{\rm H})\rho_c(\mathbf{D})\}}\label{proflemma24}\\
	&=\sqrt{\min\{\rho_c(\mathbf{D})\rho_r(\mathbf{D}),\rho_r(\mathbf{D})\rho_c(\mathbf{D})\}}\label{proflemma25}\\
	&=\sqrt{\rho_c(\mathbf{D})\rho_r(\mathbf{D})},\nonumber
\end{align}
where $\rho(\cdot)$ denotes the spectral radius on its objective, (\ref{proflemma23}) is because $\rho_r(\mathbf{D})$ and $\rho_c(\mathbf{D})$ are matrix norms as discussed in \cite{Eldar2010}, and the spectral radius of a square matrix is less than or equal to any other matrix norm, (\ref{proflemma24}) follows from the sub-multiplicativity of matrix norm, and (\ref{proflemma25}) is because $\rho_r(\mathbf{D}^{\rm H})=\rho_c(\mathbf{D})$ and $\rho_c(\mathbf{D}^{\rm H})=\rho_r(\mathbf{D})$ \cite{Eldar2010}. 
This completes the proof.
\end{IEEEproof}

\subsection{Proof of \textbf{Corollary~\ref{Corollary4}}}\label{proofofCorollary4} % Ap-C

\begin{IEEEproof}	
Observe that
\begin{align}% eqs.130,131
	\big\| \mathbf{T}^{\rm H}\mathbf{T}_{*}\big\|_2 & \leq \min\big\{\rho_r(\mathbf{T}^{\rm H}\mathbf{T}_{*}),\rho_c(\mathbf{T}^{\rm H}\mathbf{T}_{*})\big\} \label{proofofCorollary41} \\
	& \leq\min \bigg\{\varpi_{\mathbf{\Upsilon}}^n\prod_{t=1}^{n}l_td_t,\varpi_{\mathbf{\Upsilon}}^n\prod_{t=1}^{n}l^*_td_t\bigg\},\label{proofofCorollary42}
\end{align}
where (\ref{proofofCorollary41}) is from \textbf{Lemma~\ref{lemma4}}, and (\ref{proofofCorollary42}) is from the definition of mutual block coherence.
On the other hand,
\begin{align} % eqs.132,133
	&\big\|\mathbf{T}^{\rm H}\mathbf{T}_{*}\big\|_2 \nonumber\\
	&\geq  \sigma_{\min}(\mathbf{T})\sigma_{\min}(\mathbf{T}_{*}) \nonumber \\
	&\geq  \Bigg( 1 - \bigg(\prod^{n}_{t=1}d_t-1\bigg)\tau_{\mathbf{\Upsilon}}^n - \bigg(\prod_{t=1}^nl_t-1\bigg)\varpi_{\mathbf{\Upsilon}}^n\prod^{n}_{t=1}d_t \Bigg)^{\frac{1}{2}} \nonumber \\
	&\hspace{1em} \times \Bigg( 1 - \bigg(\prod^{n}_{t=1}d_t-1\bigg)\tau_{\mathbf{\Upsilon}}^n - \bigg(\prod_{t=1}^nl^*_t-1\bigg)\varpi_{\mathbf{\Upsilon}}^n\prod^{n}_{t=1}d_t \Bigg)^{\frac{1}{2}} \label{coro1last} \\
	&= \underline{W}_{\mathbf{\Upsilon},\prod_{t=1}^nl_t}^{\frac{1}{2}}\underline{W}_{\mathbf{\Upsilon},\prod_{t=1}^nl^*_t}^{\frac{1}{2}},\nonumber
\end{align}
where (\ref{coro1last}) is derived based on \textbf{Lemma~\ref{lemma4}}, and $\sigma_{\min}(\cdot)$ represents the minimum singular value  of its target. This completes the proof.
\end{IEEEproof}

\subsection{Proof of \textbf{Corollary \ref{Corollary6}}}\label{proofofCorollary6} % Ap-D

\begin{IEEEproof}
Observe that
\begin{align} % eq.134
	&\|\mathcal{X}\times_1\mathbf{D}_1\times_2\cdots\times_n\mathbf{D}_n\|_F \nonumber\\
	&= \|\mathcal{X}_{\mathbf{\Xi}}\times_1\overline{\mathbf{D}}_1\times_2\cdots\times_n\overline{\mathbf{D}}_n\|_F \nonumber \\ &
	= \|(\overline{\mathbf{D}}_n\otimes\overline{\mathbf{D}}_{n-1}\otimes\cdots\otimes\overline{\mathbf{D}}_1){\rm vec}(\mathcal{X}_{\mathbf{\Xi}})\|_F \nonumber \\
	&\leq  \|\overline{\mathbf{D}}_n\otimes\overline{\mathbf{D}}_{n-1}\otimes\cdots\otimes\overline{\mathbf{D}}_1\|_2\|{\rm vec}(\mathcal{X}_{\mathbf{\Xi}})\|_F  \nonumber \\
  &\leq  \overline{W}_{\mathbf{\Upsilon},\prod_{t=1}^nk_t}^{\frac{1}{2}}\|\mathcal{X}\|_F, \label{lemma1mapping2}
\end{align}
where $\overline{\mathbf{D}}_i$ $(i\in[1,2,\cdots,n])$ are the column-block submatrices of $\mathbf{D}_i$ whose column indices correspond to the shadow block sparsity of $\mathcal{X}$, ${\rm vec}(\mathcal{X})\in\mathbb{C}^{k_1d_1k_2d_2\cdots k_nd_n}$ is the vectorized $\mathcal{X}_{\mathbf{\Xi}}$, and (\ref{lemma1mapping2}) is based on \textbf{Lemma \ref{lemma5}} and due to the fact that $\|{\rm vec}(\mathcal{X}_{\mathbf{\Xi}})\|_F=\|\mathcal{X}\|_F$. On the other hand,
\begin{align} % eq.135
	&\|\mathcal{X}\times_1\mathbf{D}_1\times_2\cdots\times_n\mathbf{D}_n\|_F \nonumber\\
	&\geq 
	  \sigma_{\min}\big(\overline{\mathbf{D}}_n\otimes\overline{\mathbf{D}}_{n-1}\otimes\cdots\otimes\overline{\mathbf{D}}_1\big)\|{\rm vec}(\mathcal{X}_{\mathbf{\Xi}})\|_F \nonumber \\ 
	  &
	\geq \underline{W}_{\mathbf{\Upsilon},\prod_{t=1}^nk_t}^{\frac{1}{2}}\|\mathcal{X}\|_F, \label{lemma3mapping2}
\end{align}
where $\sigma_{\min}(\cdot)$ denotes the smallest singular value of its argument, and (\ref{lemma3mapping2}) is based on \textbf{Lemma \ref{lemma5}}.

The proofs of (\ref{Corollary6main2}) and (\ref{proposition3MAIN}) are similar, and the proof of (\ref{proposition4MAIN}) is essentially an extension of those of the \cite[Proposition 3.1]{cosamp2009} and \cite[Lemma 6]{jwang2017mols}. This completes the proof.
\end{IEEEproof}

\subsection{Proof of \textbf{Theorem \ref{theo4}}}\label{proofoftheo4} % Ap-E

\begin{IEEEproof}
Note that
\begin{align} % eq.136
	&\big\|\mathcal{R}^l\times_1\mathbf{D}^{\rm H}_{1_{[i^*_1]}}\times_2\cdots\times_n\mathbf{D}^{\rm H}_{n_{[i^*_n]}}\big\|_F\nonumber\\
	& = \max\limits_{(i_1,i_2,\cdots,i_n)\in\mathbf{\Xi}} \big\|\mathcal{R}^l\times_1\mathbf{D}^{\rm H}_{1_{[i_1]}}\times_2\cdots\times_n\mathbf{D}^{\rm H}_{n_{[i_n]}}\big\|_F. \label{theo41}
\end{align}
Letting $q_s$ be the $s$th largest element in $\mathbf{\Psi}$, we have
\begin{align} % eq.137
	q_s
	 \leq \frac{1}{\sqrt{s}}
		\sqrt{\sum_{(i_1,\cdots,i_n)\in\mathbf{\Psi}}\big\|\mathcal{R}^l\times_1\mathbf{D}^{\rm H}_{1_{[i_1]}}\times_2\cdots\times_n\mathbf{D}^{\rm H}_{n_{[i_n]}}\big\|_F^2}. \label{theo42}
\end{align}
If the bound in (\ref{theo42}) is less than (\ref{theo41}), i.e.,
\begin{align} % eq.138
	\frac{\sqrt{\sum_{(i_1,\cdots,i_n)\in\mathbf{\Psi}} \big\|\mathcal{R}^l\times_1\mathbf{D}^{\rm H}_{1_{[i_1]}}\times_2\cdots\times_n\mathbf{D}^{\rm H}_{n_{[i_n]}}\big\|_F^2}}{\sqrt{s} \big\|\mathcal{R}^l\times_1\mathbf{D}^{\rm H}_{1_{[i^*_1]}}\times_2\cdots\times_n\mathbf{D}^{\rm H}_{n_{[i^*_n]}} \big\|_F}<1, \label{xiaoyu1}
\end{align}
then in the next iteration at least $c$ correct indices are selected. Since $\ddot{\mathbf{D}}_{\mathbf{\Xi}}\ddot{\mathbf{D}}_{\mathbf{\Xi}}^{\dagger}$ is the orthogonal projector onto ${\rm span}(\ddot{\mathbf{D}}_{\mathbf{\Xi}})$ \cite{liyang2022}, we have $\ddot{\mathbf{D}}_{\mathbf{\Xi}}\ddot{\mathbf{D}}_{\mathbf{\Xi}}^{\dagger} {\rm vec}(\mathcal{R}^l)={\rm vec}(\mathcal{R}^l)$. Thus, the left-hand side of (\ref{xiaoyu1}) becomes
\begin{align} % eqs.139-141
	& \frac{\sqrt{\sum\limits_{(i_1,\cdots,i_n)\in\mathbf{\Psi}}\big\|\big(\mathbf{D}^{\rm H}_{n_{[i_n]}}\otimes\cdots\otimes\mathbf{D}^{\rm H}_{1_{[i_1]}}\big) \ddot{\mathbf{D}}_{\mathbf{\Xi}}\ddot{\mathbf{D}}_{\mathbf{\Xi}}^{\dagger}\, {\rm vec}(\mathcal{R}^l)\big\|_F^2}}{\sqrt{s} \big\|\big(\mathbf{D}^{\rm H}_{n_{[i^*_n]}}\otimes\cdots\otimes\mathbf{D}^{\rm H}_{1_{[i^*_1]}}\big){\rm vec}(\mathcal{R}^l)\big\|_F} \nonumber \\
	&  =\!\! \frac{\sqrt{\sum\limits_{(i_1,\cdots,i_n)\in\mathbf{\Psi}}\!\!\big\|\!\big(\mathbf{D}^{\rm H}_{n_{[i_n]}}\!\!\otimes\!\cdots\!\otimes\!\mathbf{D}^{\rm H}_{1_{[i_1]}}\big) \!\big(\ddot{\mathbf{D}}_{\mathbf{\Xi}}^{\dagger}\big)^{\rm H}\ddot{\mathbf{D}}^{\rm H}_{\mathbf{\Xi}}\, {\rm vec}(\mathcal{R}^l)\!\big\|_F^2}}{\sqrt{s} \big\|\big(\mathbf{D}^{\rm H}_{n_{[i^*_n]}}\otimes\cdots\otimes\mathbf{D}^{\rm H}_{1_{[i^*_1]}}\big) {\rm vec}(\mathcal{R}^l)\big\|_F} \label{exact1} \\
	&  \leq\frac{1}{\sqrt{s}} \big\|\ddot{\mathbf{D}}_{\mathbf{\Psi}}^{\rm H}(\ddot{\mathbf{D}}^{\dagger}_{\mathbf{\Xi}})^{\rm H}\big\|_2 \frac{\big\|\ddot{\mathbf{D}}^{\rm H}_{\mathbf{\Xi}} {\rm vec}(\mathcal{R}^l)\big\|_F}{\big\|\big(\mathbf{D}^{\rm H}_{n_{[i^*_n]}}\otimes\cdots\otimes\mathbf{D}^{\rm H}_{1_{[i^*_1]}}\big) {\rm vec}(\mathcal{R}^l)\big\|_F} \label{exact2} \\
	&  =\frac{Z_{\mathcal{R}^l}}{\sqrt{s}} \big\|\ddot{\mathbf{D}}^{\dagger}_{\mathbf{\Xi}}\ddot{\mathbf{D}}_{\mathbf{\Psi}}\big\|_2, \label{exact3}
\end{align}
where (\ref{exact1}) is because $\ddot{\mathbf{D}}_{\mathbf{\Xi}}\ddot{\mathbf{D}}_{\mathbf{\Xi}}^{\dagger}$ is Hermitian matrix, (\ref{exact2}) is from the sub-multiplicativity of matrix norm, and (\ref{exact3}) is due to the definition of the function $Z_{\mathcal{R}^l}$. Then, following the similar proofs in \cite{liyang2022, Eldar2010}, we can prove that T-GBOMP selects a new tensor block participating in the unique solution of $\mathcal{Y}=\mathcal{X}\times_1\mathbf{D}_1\times_2\mathbf{D}_2\times_3\cdots\times_n\mathbf{D}_n$. This completes the proof.
\end{IEEEproof}

\subsection{Proof of \textbf{Theorem \ref{theo8}}}\label{proofoftheo8} % Ap-F

\begin{IEEEproof}
Observe that
\begin{align} % eqs.142-1444
	&\frac{Z_{\mathcal{R}^l}}{\sqrt{s}}\big\|\ddot{\mathbf{D}}^{\dagger}_{\mathbf{\Xi}}\ddot{\mathbf{D}}_{\mathbf{\Psi}}\big\|_2 \nonumber\\
	&= \frac{Z_{\mathcal{R}^l}}{\sqrt{s}}\Big\|\big(\ddot{\mathbf{D}}_{\mathbf{\Xi}}^{\rm H}\ddot{\mathbf{D}}_{\mathbf{\Xi}}\big)^{-1}\ddot{\mathbf{D}}_{\mathbf{\Xi}}^{\rm H}\ddot{\mathbf{D}}_{\mathbf{\Psi}}\Big\|_2 \nonumber \\
	& \leq \frac{Z_{\mathcal{R}^l}}{\sqrt{s}} \Big\|\big(\ddot{\mathbf{D}}_{\mathbf{\Xi}}^{\rm H}\ddot{\mathbf{D}}_{\mathbf{\Xi}}\big)^{-1}\Big\|_2 \big\| \ddot{\mathbf{D}}_{\mathbf{\Xi}}^{\rm H} \ddot{\mathbf{D}}_{\mathbf{\Psi}}\big\|_2 \label{observe2} \\
	& \leq \frac{Z_{\mathcal{R}^l}}{\sqrt{s}} \Big\|\big(\ddot{\mathbf{D}}_{\mathbf{\Xi}}^{\rm H}\ddot{\mathbf{D}}_{\mathbf{\Xi}}\big)^{-1}\Big\|_2\times\max\Big\{\varpi_{\Upsilon}^nk\prod_{t=1}^{n}d_t,\varpi_{\Upsilon}^ns\prod_{t=1}^{n}d_t\Big\} \label{observe3} \\
	& \leq \frac{Z_{\mathcal{R}^l}}{\sqrt{s}}\Big\|\big(\ddot{\mathbf{D}}_{\mathbf{\Xi}}^{\rm H}\ddot{\mathbf{D}}_{\mathbf{\Xi}}\big)^{-1}\Big\|_2\, \varpi_{\Upsilon}^nk\prod_{t=1}^{n}d_t, \label{observe4}
\end{align}
where (\ref{observe2}) is from the sub-multiplicativity, (\ref{observe3}) is based on \textbf{Corollary~\ref{Corollary6}}, and (\ref{observe4}) is because $s\leq k$.
	
It remains to derive an upper bound on $\big\|\big(\ddot{\mathbf{D}}_{\mathbf{\Xi}}^{\rm H}\ddot{\mathbf{D}}_{\mathbf{\Xi}}\big)^{-1}\big\|_2$. Let $\ddot{\mathbf{D}}_{\mathbf{\Xi}}^{\rm H}\ddot{\mathbf{D}}_{\mathbf{\Xi}}=\mathbf{I}+\mathbf{A}$, where $\mathbf{A}\in\mathbb{C}^{k\prod_{t=1}^nd_t\times k\prod_{t=1}^nd_t}$ is a symmetric matrix with $\mathbf{A}_{(i,i)}=0$, $\forall i$. Thus, based on \textbf{Corollary \ref{Corollary1}}, we have
\begin{align} % eq.145
	\|\mathbf{A}\|_2\leq \bigg(\prod_{t=1}^nd_t-1\bigg)\tau_{\mathbf{\Upsilon}}^n+(k-1)\varpi_{\mathbf{\Upsilon}}^n\prod_{t=1}^{n}d_t.\label{auuper}
	\end{align}
The assumption now implies that $\Big(\prod_{t=1}^nd_t-1\Big)\tau_{\mathbf{\Upsilon}}^n+(k-1)\varpi_{\mathbf{\Upsilon}}^n\prod_{t=1}^{n}d_t<1$. Based on \cite[Corollary 5.6.16]{matrixanalysis}, we obtain
\begin{align} % eqs.146-149
	&\big\|\big(\ddot{\mathbf{D}}_{\mathbf{\Xi}}^{\rm H}\ddot{\mathbf{D}}_{\mathbf{\Xi}}\big)^{-1}\big\|_2 = \big\|(\mathbf{I}+\mathbf{A})^{-1}\big\|_2 =
	\bigg\|\sum_{t=0}^{\infty}(-\mathbf{A})^t\bigg\|_2 \label{th8prof1} \\
	& \leq \sum_{t=0}^{\infty}\|\mathbf{A}\|_2^t \label{th8prof3} \\
	& = \frac{1}{1-\|\mathbf{A}\|_2} \label{th8prof4} \\
	& \leq \frac{1}{1-\Big(\prod_{t=1}^nd_t-1\Big)\tau_{\mathbf{\Upsilon}}^n-(k-1)\varpi_{\mathbf{\Upsilon}}^n\prod_{t=1}^{n}d_t},\label{th8prof5}
\end{align}
where (\ref{th8prof3}) is from the triangle inequality, (\ref{th8prof4}) is the sum of proportional sequence, and (\ref{th8prof5}) follows by using (\ref{auuper}).

Combining (\ref{th8prof5}) and (\ref{observe4}) yields
\begin{align} % eqs.150,151
  &\frac{Z_{\mathcal{R}^l}}{\sqrt{s}}\big\|\ddot{\mathbf{D}}^{\dagger}_{\mathbf{\Xi}}\ddot{\mathbf{D}}_{\mathbf{\Psi}}\big\|_2 \nonumber\\
  &\leq \frac{Z_{\mathcal{R}^l}}{\sqrt{s}}\, \frac{\varpi_{\Upsilon}^nk\prod_{t=1}^{n}d_t}{1-\Big(\prod_{t=1}^nd_t-1\Big)\tau_{\mathbf{\Upsilon}}^n-(k-1)\varpi_{\mathbf{\Upsilon}}^n\prod_{t=1}^{n}d_t} \label{theo8last} \\
  &< 1. \label{theo8last1}
\end{align}
Then, (\ref{theo8main}) follows from (\ref{theo8last}) and (\ref{theo8last1}), and this completes the proof.
\end{IEEEproof}

\subsection{Proof of \textbf{Corollary \ref{Corollary8}}}\label{proofofcoro8} % Ap-G

\begin{IEEEproof}
Letting $Z_{\mathcal{R}^l}=\sqrt{k}$, we obtain the following cubic inequality with respect to $k$:
\begin{align} % eq.152
	\big(k^{\frac{1}{2}}\big)^3+\sqrt{s}\big(k^{\frac{1}{2}}\big)^2+\delta<0, \label{sancifangcheng}
\end{align}
where $\delta$ is given by (\ref{eqC4delta}) in \textbf{Corollary~\ref{Corollary8}}.
Setting $k^{\frac{1}{2}}=x-\frac{\sqrt{s}}{3}$, (\ref{sancifangcheng}) changes into
\begin{align} % 153
	x^3+Px+Q < 0. \label{diersanci}
\end{align}
The real-domain solution of (\ref{diersanci}) based on the Cardano formula is
\begin{align} % eq.154
	x < \sqrt[3]{-\frac{Q}{2}+\sqrt{\Delta}}+\sqrt[3]{-\frac{Q}{2}-\sqrt{\Delta}}. \label{Cardanosanci}
\end{align}
(\ref{Corollary8main}) is the consequence of combining (\ref{Cardanosanci}) and $k^{\frac{1}{2}}=x-\frac{\sqrt{s}}{3}$.
\end{IEEEproof}

\subsection{Proof of \textbf{Theorem~\ref{theo5}}}\label{proofoftheo5} % Ap-H

\begin{IEEEproof}	
Note that
\begin{align} % eqs.155-160
	&\|\mathcal{R}^l - \mathcal{R}^{l+1}\|_F\nonumber\\
	 &= \big\|\mathbf{P}^{\bot}_{\ddot{\mathbf{D}}_{\mathbf{\Xi}^{l}}}{\rm vec}(\mathcal{Y}) - \mathbf{P}^{\bot}_{\ddot{\mathbf{D}}_{\mathbf{\Xi}^{l+1}}}{\rm vec}(\mathcal{Y})\big\|_F \label{proofoftheo51} \\
	&= \Big\|\big(\mathbf{I}-\mathbf{P}_{\ddot{\mathbf{D}}_{\mathbf{\Xi}^{l}}}{\rm vec}(\mathcal{Y})\big) - \big(\mathbf{I} - \mathbf{P}_{\ddot{\mathbf{D}}_{\mathbf{\Xi}^{l+1}}}\big) {\rm vec}(\mathcal{Y})\Big\|_F \label{proofoftheo52} \\
  &= \big\|\mathbf{P}_{\ddot{\mathbf{D}}_{\mathbf{\Xi}^{l+1}}}{\rm vec}(\mathcal{Y}) - \mathbf{P}_{\ddot{\mathbf{D}}_{\mathbf{\Xi}^{l+1}}}\mathbf{P}_{\ddot{\mathbf{D}}_{\mathbf{\Xi}^{l}}}{\rm vec}(\mathcal{Y})\big\|_F \label{proofoftheo53} \\
	&= \|\mathbf{P}_{\ddot{\mathbf{D}}_{\mathbf{\Xi}^{l+1}}}\big(\mathbf{I}-\mathbf{P}_{\ddot{\mathbf{D}}_{\mathbf{\Xi}^{l}}}\big) {\rm vec}(\mathcal{Y})\|_F 
	 = \|\mathbf{P}_{\ddot{\mathbf{D}}_{\mathbf{\Xi}^{l+1}}}{\rm vec}(\mathcal{R}^l)\|_F \nonumber \\
	&\geq \big\|\mathbf{P}_{\ddot{\mathbf{D}}_{\mathbf{\Theta}^{l+1}}}{\rm vec}(\mathcal{R}^l)\big\|_F \label{proofoftheo56} \\
	&= \Big\|\big(\mathbf{P}_{\ddot{\mathbf{D}}_{\mathbf{\Theta}^{l+1}}}\big)^{\rm H} {\rm vec}(\mathcal{R}^l)\Big\|_F \label{proofoftheo57} \\
	&= \Big\|\big(\ddot{\mathbf{D}}_{\mathbf{\Theta}^{l+1}}\ddot{\mathbf{D}}_{\mathbf{\Theta}^{l+1}}^{\dagger}\big)^{\rm H} {\rm vec}\big(\mathcal{R}^l\big)\big\|_F\nonumber\\
	 & = \big\|(\ddot{\mathbf{D}}_{\mathbf{\Theta}^{l+1}}^{\dagger})^{\rm H}\ddot{\mathbf{D}}_{\mathbf{\Theta}^{l+1}}^{\rm H}{\rm vec}(\mathcal{R}^l)\big\|_F \nonumber \\
	&\geq \frac{\big\|\ddot{\mathbf{D}}_{\mathbf{\Theta}^{l+1}}^{\rm H}{\rm vec}(\mathcal{R}^l)\big\|_F}{\overline{W}^{\frac{1}{2}}_{\mathbf{\Upsilon},s}} \label{proofoftheo510} ,
\end{align}
where $\mathbf{\Theta}^{l+1}$ is the selected index set in the $(l+1)$th iteration as given in Algorithm \ref{alg:T-GBOMP}, (\ref{proofoftheo51}) and (\ref{proofoftheo52}) are due to the definitions of the residual matrix and orthogonal projection matrix, respectively, (\ref{proofoftheo53}) and (\ref{proofoftheo56}) are because ${\rm span}\big(\ddot{\mathbf{D}}_{\mathbf{\Xi}^{l}}\big)\subseteq{\rm span}\big(\ddot{\mathbf{D}}_{\mathbf{\Xi}^{l+1}}\big)$ and ${\rm span}\big(\ddot{\mathbf{D}}_{\mathbf{\Theta}^{l+1}}\big)\subseteq{\rm span}\big(\ddot{\mathbf{D}}_{\mathbf{\Xi}^{l+1}}\big)$, respectively, (\ref{proofoftheo57}) is from the property of projection matrix, and (\ref{proofoftheo510}) is based on \textbf{Corollary~\ref{Corollary6}}.
	
It remains to develop the lower bound on $\big\|\ddot{\mathbf{D}}_{\mathbf{\Theta}^{l+1}}^{\rm H}{\rm vec}(\mathcal{R})\big\|_F$. Let $\mathbf{\Theta}^*=\arg \max\limits_{\mathbf{\Theta}:|\mathbf{\Theta}|=s} \sum\limits_{(i_1,i_2,\cdots,i_n)\in\mathbf{\Theta}} \big\|\mathcal{R}^l\times_1\mathbf{D}^H_{1_{[i_1]}}\times_2\cdots\times_n\mathbf{D}^H_{n_{[i_n]}}\big\|^2_F$. Then, we have
\begin{align} % eq.161
	&\big\|\ddot{\mathbf{D}}_{\mathbf{\Theta}^{l+1}}^{\rm H}{\rm vec}(\mathcal{R})\big\|^2_F \nonumber\\
	&=
	  \sum_{(i_1,i_2,\cdots,i_n)\in\mathbf{\Theta}^{l+1}}
		\big\|\mathcal{R}^l\times_1\mathbf{D}^{\rm H}_{1_{[i_1]}}\times_2\cdots\times_n\mathbf{D}^{\rm H}_{n_{[i_n]}}\big\|^2_F \nonumber \\
	&= \max\limits_{\mathbf{\Theta}:|\mathbf{\Theta}|=s} \sum_{(i_1,i_2,\cdots,i_n)\in\mathbf{\Theta}}
		\big\|\mathcal{R}^l\times_1\mathbf{D}^{\rm H}_{1_{[i_1]}}\times_2\cdots\times_n\mathbf{D}^{\rm H}_{n_{[i_n]}}\big\|^2_F \nonumber \\
	&\geq \sum_{(i_1,i_2,\cdots,i_n)\in\mathbf{\Theta}^*}
		\big\|\mathcal{R}^l\times_1\mathbf{D}^{\rm H}_{1_{[i_1]}}\times_2\cdots\times_n\mathbf{D}^{\rm H}_{n_{[i_n]}}\big\|^2_F \nonumber \\
	&= \max\limits_{\mathbf{\Theta}:|\mathbf{\Theta}|=s} \sum_{(i_1,i_2,\cdots,i_n)\in\mathbf{\Theta}}
		\big\|\mathcal{R}^l\times_1\mathbf{D}^{\rm H}_{1_{[i_1]}}\times_2\cdots\times_n\mathbf{D}^{\rm H}_{n_{[i_n]}}\big\|^2_F . \label{proofoftheo521}
\end{align}
Combining (\ref{proofoftheo510}) and (\ref{proofoftheo521}), we have
\begin{align} % eq.162
	&\big\|\mathcal{R}^l\big\|^2_F - \big\|\mathcal{R}^{l+1}\big\|^2_F \nonumber\\
	&= \big\|\mathcal{R}^l - \mathcal{R}^{l+1}\big\|^2_F \nonumber \\
	&\geq \frac{\max\limits_{\mathbf{\Theta}:|\mathbf{\Theta}|=s} \sum\limits_{(i_1,i_2,\cdots,i_n)\in\mathbf{\Theta}}
		\big\|\mathcal{R}^l\times_1\mathbf{D}^{\rm H}_{1_{[i_1]}}\times_2\cdots\times_n\mathbf{D}^{\rm H}_{n_{[i_n]}}\big\|^2_F}{\overline{W}_{\mathbf{\Upsilon},s}} . \label{proofoftheo522}
\end{align}
Moreover, since $s\leq k$,
\begin{align} % eqs.163-168
	& \max\limits_{\mathbf{\Theta}:|\mathbf{\Theta}|=s} \sum_{(i_1,i_2,\cdots,i_n)\in\mathbf{\Theta}}
		\big\|\mathcal{R}^l\times_1\mathbf{D}^{\rm H}_{1_{[i_1]}}\times_2\cdots\times_n\mathbf{D}^{\rm H}_{n_{[i_n]}}\big\|^2_F \nonumber \\
	&\geq \frac{s}{k} \sum_{(i_1,\cdots,i_n)\in\mathbf{\Xi}}
		\big\|\mathcal{R}^l\times_1\mathbf{D}^{\rm H}_{1_{[i_1]}}\times_2\cdots\times_n\mathbf{D}^{\rm H}_{n_{[i_n]}}\big\|^2_F \nonumber \\
	&\geq \frac{s}{k} \sum_{(i_1,\cdots,i_n)\in\mathbf{\Xi}\cup\mathbf{\Xi}^l}
		\big\|\mathcal{R}^l\times_1\mathbf{D}^{\rm H}_{1_{[i_1]}}\times_2\cdots\times_n\mathbf{D}^{\rm H}_{n_{[i_n]}}\big\|^2_F \label{proofoftheo552} \\
	&= \frac{s}{k} \Big\|\ddot{\mathbf{D}}_{\mathbf{\Xi}\cup\mathbf{\Xi}^l}^{\rm H}\Big({\rm vec}(\mathcal{Y}) - \big(\mathbf{D}_{n_{\mathbf{\Xi}^l_n}}\otimes\cdots\otimes\mathbf{D}_{1_{\mathbf{\Xi}^l_1}}\big)\, {\rm vec}(\mathcal{X}_{\mathbf{\Xi}^l})\Big)\Big\|_F^2 \label{proofoftheo553} \\
	&= \frac{s}{k}\Big\| \ddot{\mathbf{D}}_{\mathbf{\Xi}\cup\mathbf{\Xi}^l}^{\rm H}\Big( \big(\mathbf{D}_{n_{\mathbf{\Xi}_n}}\otimes\cdots\otimes\mathbf{D}_{1_{\mathbf{\Xi}_1}}\big) {\rm vec}(\mathcal{X}) \nonumber\\
	&\hspace{6em}- \big(\mathbf{D}_{n_{\mathbf{\Xi}^l_n}}\otimes\cdots\otimes\mathbf{D}_{1_{\mathbf{\Xi}^l_1}} \big) {\rm vec}(\mathcal{X}_{\mathbf{\Xi}^l}) \Big)\Big\|_F^2 \nonumber \\
	&= \frac{s}{k}\Big\|\ddot{\mathbf{D}}_{\mathbf{\Xi}\cup\mathbf{\Xi}^l}^{\rm H}\ddot{\mathbf{D}}_{\mathbf{\Xi}\cup\mathbf{\Xi}^l} \big({\rm vec}(\mathcal{X}_{\mathbf{\Xi}})-{\rm vec}(\mathcal{X}_{\mathbf{\Xi}^l})\big)\Big\|_F^2 \label{proofoftheo555} \\
	&\geq \frac{s}{k}\, \underline{W}_{\mathbf{\Upsilon},\prod_{t=1}^{n}k_n+sl} \big\|{\rm vec}(\mathcal{X}_{\mathbf{\Xi}})-{\rm vec}(\mathcal{X}_{\mathbf{\Xi}^l})\big\|^2_F \label{proofoftheo556} \\
	&\geq \frac{s\underline{W}_{\mathbf{\Upsilon},\prod_{t=1}^{n}k_n+sl}}{k\overline{W}^{\frac{1}{2}}_{\mathbf{\Upsilon},\prod_{t=1}^{n}k_n+sl}} \big\|\ddot{\mathbf{D}}_{\mathbf{\Xi}\cup\mathbf{\Xi}^l} \big({\rm vec}(\mathcal{X}_{\mathbf{\Xi}})-{\rm vec}(\mathcal{X}_{\mathbf{\Xi}^l})\big)\big\|^2_F \label{proofoftheo557} \\
	&= \frac{s\underline{W}_{\mathbf{\Upsilon},\prod_{t=1}^{n}k_n+sl}}{k\overline{W}^{\frac{1}{2}}_{\mathbf{\Upsilon},\prod_{t=1}^{n}k_n+sl}} \big\|\mathcal{R}^l\big\|^2_F , \label{proofoftheo558}
\end{align}
where (\ref{proofoftheo552}) is because $\sum_{(i_1,\cdots,i_n)\in\mathbf{\Xi}^l}
	\big\|\mathcal{R}^l\times_1\mathbf{D}^{\rm H}_{1_{[i_1]}}\times_2\cdots\times_n\mathbf{D}^{\rm H}_{n_{[i_n]}}\big\|^2_F=0$, (\ref{proofoftheo553}) is due to (\ref{tensorCSmodel}), (\ref{proofoftheo555}) is because $\big(\mathbf{D}_{n_{\mathbf{\Xi}_n}}\otimes\cdots\otimes\mathbf{D}_{1_{\mathbf{\Xi}_1}}\big)\, {\rm vec}(\mathcal{X}_{\mathbf{\Xi}})=\big(\mathbf{D}_{n_{\mathbf{\Xi}_n\cup\mathbf{\Xi}^l_n}}\otimes\cdots\otimes\mathbf{D}_{1_{\mathbf{\Xi}_1\cup\mathbf{\Xi}^l_1}}\big)\, {\rm vec}(\mathcal{X}_{\mathbf{\Xi}\cup\mathbf{\Xi}^l})$, (\ref{proofoftheo556}) and (\ref{proofoftheo557}) are from \textbf{Corollary~\ref{Corollary6}}.
	
Combining (\ref{proofoftheo522}) and (\ref{proofoftheo558}) yields
\begin{align}
	\big\|\mathcal{R}^l\big\|^2_F - \big\|\mathcal{R}^{l+1}\big\|^2_F \geq \frac{s\underline{W}_{\mathbf{\Upsilon},\prod_{t=1}^{n}k_n+sl}}{k\overline{W}^{\frac{1}{2}}_{\mathbf{\Upsilon},\prod_{t=1}^{n}k_n+sl}\overline{W}_{\mathbf{\Upsilon},s}}\big\|\mathcal{R}^l\big\|^2_F .\nonumber
\end{align}
Thus, we have
\begin{align} % eq. 170
	\big\|\mathcal{R}^{l+1}\big\|^2_F &\leq \Bigg(1-\frac{s\underline{W}_{\mathbf{\Upsilon},\prod_{t=1}^{n}k_n+sl}}{k\overline{W}^{\frac{1}{2}}_{\mathbf{\Upsilon},\prod_{t=1}^{n}k_n+sl}\overline{W}_{\mathbf{\Upsilon},s}}\Bigg)\big\|\mathcal{R}^l\big\|^2_F \nonumber \\
	&\leq \prod_{j=0}^l\Bigg(1-\frac{s\underline{W}_{\mathbf{\Upsilon},\prod_{t=1}^{n}k_n+sj}}{k\overline{W}^{\frac{1}{2}}_{\mathbf{\Upsilon},\prod_{t=1}^{n}k_n+sj}\overline{W}_{\mathbf{\Upsilon},s}}\Bigg) \big\|\mathcal{R}^0\big\|^2_F \nonumber \\
	&\leq \Bigg(1-\frac{s\underline{W}_{\mathbf{\Upsilon},\prod_{t=1}^{n}k_n+sl}}{k\overline{W}^{\frac{1}{2}}_{\mathbf{\Upsilon},\prod_{t=1}^{n}k_n+sl}\overline{W}_{\mathbf{\Upsilon},s}}\Bigg)^{l+1}\big\|\mathcal{Y}\big\|^2_F .\nonumber
\end{align}
This completes the proof.
\end{IEEEproof}

\subsection{Proof of \textbf{Theorem \ref{theo1}}}\label{proofoftheo1} % Ap-I

\begin{IEEEproof}
Denoting $\mathcal{Q}^{l^*}$ as the best $k$ block-sparse tensor approximation of $\mathcal{X}^{l^*}$, and $k_n$ as the shadow block sparsity in the $n$th mode, we have
\begin{align} % eqs.171-175
	&\big\|\mathcal{Q}^{l^*}-\mathcal{X}\big\|_F \nonumber\\
	&= \big\|\mathcal{Q}^{l^*}-\mathcal{X}^{l^*}+\mathcal{X}^{l^*}-\mathcal{X}\big\|_F\nonumber\\
	&	\leq \big\|\mathcal{Q}^{l^*}-\mathcal{X}^{l^*}\big\|_F + \big\|\mathcal{X}^{l^*}-\mathcal{X}\big\|_F \label{proofthe11} \\
	&\leq 2\big\|\mathcal{X}^{l^*}-\mathcal{X}\big\|_F \label{proofthe12} \\
	&\leq \frac{2\big\|(\mathcal{X}^{l^*}-\mathcal{X})\times_1\mathbf{D}_1\times_2\cdots\times_n\mathbf{D}_n\big\|_F}{\underline{W}_{\mathbf{\Upsilon},\prod_{t=1}^{n}k_t+sl^*}^{\frac{1}{2}}} \label{proofthe13} \\
	&\leq \frac{2\big(\|\mathcal{R}^{l^*}\|_F+\|\mathcal{N}\|_F\big)}{\underline{W}_{\mathbf{\Upsilon},\prod_{t=1}^{n}k_t+sl^*}^{\frac{1}{2}}} \label{proofthe14} \\
	&\leq \frac{2\big(\epsilon+\|\mathcal{N}\|_F\big)}{\underline{W}_{\mathbf{\Upsilon},\prod_{t=1}^{n}k_t+sl^*}^{\frac{1}{2}}} , \label{proofthe15}
\end{align}
where (\ref{proofthe11}) is due to the triangle inequality, (\ref{proofthe12}) is because $\mathcal{Q}^{l^*}$ is the best $k$ block-sparse tensor approximation to $\mathcal{X}^{l^*}$, (\ref{proofthe13}) is from \textbf{Corollary \ref{Corollary6}} and the fact that each shadow block sparsity of the tensor $\mathcal{X}^{l^*}-\mathcal{X}$ can increase by up to $sl^{*}$, and (\ref{proofthe14}) is because $\mathcal{R}^{l^*}=\mathcal{Y}-\mathcal{X}^{l^*}\times_1\mathbf{D}_1\times_2\cdots\times_n\mathbf{D}_n=(\mathcal{X}^{l^*}-\mathcal{X})\times_1\mathbf{D}_1\times_2\cdots\times_n\mathbf{D}_n+\mathcal{N}$.

Further,
\begin{align} % eqs.176-180
	&\|\mathcal{Q}^{l^*}-\mathcal{X}\|_F\nonumber\\
	 &\geq \frac{\|(\mathcal{Q}^{l^*}-\mathcal{X})\times_1\mathbf{D}_1\times_2\cdots\times_n\mathbf{D}_n\|_F}{\overline{W}^{\frac{1}{2}}_{\mathbf{\Upsilon},\prod_{t=1}^{n}k_t+k}} \label{the1lower1} \\
	&= \frac{\|\mathcal{Q}^{l^*}\times_1\mathbf{D}_1\times_2\cdots\times_n\mathbf{D}_n-\mathcal{Y}+\mathcal{N}\|_F}{\overline{W}^{\frac{1}{2}}_{\mathbf{\Upsilon},\prod_{t=1}^{n}k_t+k}} \nonumber \\
	&\geq \frac{\|\mathcal{Q}^{l^*}\times_1\mathbf{D}_1\times_2\cdots\times_n\mathbf{D}_n-\mathcal{Y}\|_F-\|\mathcal{N}\|_F}{\overline{W}^{\frac{1}{2}}_{\mathbf{\Upsilon},\prod_{t=1}^{n}k_t+k}} \label{the1lower2} \\
	&\geq \frac{\|\hat{\mathcal{X}}\times_1\mathbf{D}_1\times_2\cdots\times_n\mathbf{D}_n-\mathcal{Y}\|_F-\|\mathcal{N}\|_F}{\overline{W}^{\frac{1}{2}}_{\mathbf{\Upsilon},\prod_{t=1}^{n}k_t+k}} \label{the1lower3} \\
	&= \frac{\|(\hat{\mathcal{X}}-\mathcal{X})\times_1\mathbf{D}_1\times_2\cdots\times_n\mathbf{D}_n-\mathcal{N}\|_F-\|\mathcal{N}\|_F}{\overline{W}^{\frac{1}{2}}_{\mathbf{\Upsilon},\prod_{t=1}^{n}k_t+k}} \nonumber \\
	&\geq \frac{\|(\hat{\mathcal{X}}-\mathcal{X})\times_1\mathbf{D}_1\times_2\cdots\times_n\mathbf{D}_n\|_F-2\|\mathcal{N}\|_F}{\overline{W}^{\frac{1}{2}}_{\mathbf{\Upsilon},\prod_{t=1}^{n}k_t+k}} \label{the1lower4} \\
	&\geq \frac{\underline{W}^{\frac{1}{2}}_{\mathbf{\Upsilon},\prod_{t=1}^{n}k_t+k}\|(\hat{\mathcal{X}}-\mathcal{X})\|_F-2\|\mathcal{N}\|_F}{\overline{W}^{\frac{1}{2}}_{\mathbf{\Upsilon},\prod_{t=1}^{n}k_t+k}} , \label{the1lower5}
\end{align}
where (\ref{the1lower1}) and (\ref{the1lower5}) are derived by \textbf{Corollary \ref{Corollary6}} and due to the fact that the $n$th mode shadow block sparsity of $\mathcal{Q}^{l^*}-\mathcal{X}$ is up to $k_n+k$, (\ref{the1lower2}) and (\ref{the1lower4}) are based on the triangle inequality, and (\ref{the1lower3}) is because $\mathcal{Q}^{l^*}$ is supported on $\mathbf{\Xi}^{l^*}$ and $\hat{\mathcal{X}}=\arg \min\limits_{\mathcal{X}:{\rm supp}(\mathcal{X})=\mathbf{\Xi}^{l^*}}\big\|{\rm vec}(\mathcal{Y})-\sum_{(i_1,i_2,\cdots,i_n)\in\mathbf{\Xi}^{l^*}}(\mathbf{D}_{n_{[i_n]}}\otimes\cdots\otimes\mathbf{D}_{1_{[i_1]}}){\rm vec}(\mathcal{X})_{[i_1,\cdots,i_n]}\big\|_F$. Finally, the proof is completed by combining (\ref{proofthe15}) and (\ref{the1lower5}).
\end{IEEEproof}

\subsection{Proof of \textbf{Theorem \ref{theo2}}}\label{proofoftheo2} % Ap-J

\begin{IEEEproof}
The proof is based on induction of selecting at least one correct tensor block for each iteration. Note that this proof technique is similar in spirit to those in \cite{jwang2017mols,Kim2020tit}.
	
\emph{a)}~First, we prove that T-GBOMP succeeds at choosing the correct index in the first iteration. According to Algorithm \ref{alg:T-GBOMP}, T-GBOMP chooses the index set
\begin{align} % eq.181
	&\mathbf{\Xi}^{1} \nonumber\\
	&= \arg\max\limits_{\mathbf{\Theta}:|\mathbf{\Theta}|=s} \sum_{(i_1,\cdots,i_n)\in\mathbf{\Theta}}
		\big\|\mathcal{R}^0\times_1\mathbf{D}^{\rm H}_{1_{[i_1]}}\times_2\cdots\times_n\mathbf{D}^{\rm H}_{n_{[i_n]}}\big\|_F	\nonumber \\
	&= \arg\max\limits_{\mathbf{\Theta}:|\mathbf{\Theta}|=s} \sum_{(i_1,\cdots,i_n)\in\mathbf{\Theta}}
		\big\|\mathcal{Y}\times_1\mathbf{D}^{\rm H}_{1_{[i_1]}}\times_2\cdots\times_n\mathbf{D}^{\rm H}_{n_{[i_n]}}\big\|_F, \nonumber
\end{align}
where
\begin{align}
	&\mathbf{\Theta}=\{(\mathbf{\Theta}_{1_1},\mathbf{\Theta}_{2_1},\cdots,\mathbf{\Theta}_{n_1}),(\mathbf{\Theta}_{1_2},\mathbf{\Theta}_{2_2},\cdots,\mathbf{\Theta}_{n_2}),\nonumber\\
	&\hspace{2.7em}\cdots,(\mathbf{\Theta}_{1_k},\mathbf{\Theta}_{2_k},\cdots,\mathbf{\Theta}_{n_k})\}.\nonumber
\end{align} 

Since $s\leq k$, we have
\begin{align} % eqs.182-185
	& \sum_{(i_1,\cdots,i_n)\in\mathbf{\Xi}^1}
	  \big\|\mathcal{Y}\times_1\mathbf{D}^{\rm H}_{1_{[i_1]}}\times_2\cdots\times_n\mathbf{D}^{\rm H}_{n_{[i_n]}}\big\|_F \nonumber \\
	&  \geq \sqrt{\frac{s}{k}} \sum_{(i_1,\cdots,i_n)\in\mathbf{\Xi}}
	  \big\|\mathcal{Y}\times_1\mathbf{D}^{\rm H}_{1_{[i_1]}}\times_2\cdots\times_n\mathbf{D}^{\rm H}_{n_{[i_n]}}\big\|_F \label{prooftheo2main21} \\
	&  = \sqrt{\frac{s}{k}} \sum_{(i_1,\cdots,i_n)\in\mathbf{\Xi}} \big\|(\mathcal{X}\times_1\mathbf{D}_1\times_2\cdots\times_n\mathbf{D}_n+\mathcal{N})\nonumber\\
	&\hspace{9em}\times_1\mathbf{D}^{\rm H}_{1_{[i_1]}}\times_2\cdots\times_n\mathbf{D}^{\rm H}_{n_{[i_n]}}\big\|_F \label{prooftheo2main22} \\
	&  \geq \sqrt{\frac{s}{k}} \sum_{(i_1,\cdots,i_n)\in\mathbf{\Xi}} \Big(\big\|(\mathcal{X}\times_1\mathbf{D}_1\times_2\cdots\times_n\mathbf{D}_n)\nonumber\\
	&\hspace{9em}\times_1\mathbf{D}^{\rm H}_{1_{[i_1]}}\times_2\cdots\times_n\mathbf{D}^{\rm H}_{n_{[i_n]}}\big\|_F \nonumber \\
	& \hspace{8em}- \big\|\mathcal{N}\times_1\mathbf{D}^{\rm H}_{1_{[i_1]}}\times_2\cdots\times_n\mathbf{D}^{\rm H}_{n_{[i_n]}}\big\|_F\Big) \label{prooftheo2main23} \\
	& = \sqrt{\frac{s}{k}} \sum_{(i_1,\cdots,i_n)\in\mathbf{\Xi}} \Big(\big\|\mathcal{X}_{\mathbf{\Xi}}\times_1 \big(\mathbf{D}^{\rm H}_{1_{[i_1]}}\mathbf{D}_{1_{\mathbf{\Xi}_1}}\big)\nonumber\\
	&\hspace{9em}\times_2\cdots \times_n \big(\mathbf{D}^{\rm H}_{n_{[i_n]}}\mathbf{D}_{n_{\mathbf{\Xi}_n}}\big)\big\|_F \nonumber \\
	& \hspace{8em} -\big\|\mathcal{N}\times_1\mathbf{D}^{\rm H}_{1_{[i_1]}}\times_2\cdots\times_n\mathbf{D}^{\rm H}_{n_{[i_n]}}\big\|_F\Big) \nonumber \\
	& \quad \geq \sqrt{sk} \Big(\underline{W}^{\frac{1}{2}}_{\mathbf{\Upsilon},\prod_{t=1}^{n}k_t}\underline{W}^{\frac{1}{2}}_{\mathbf{\Upsilon},1}\|\mathcal{X}\|_F-	 \overline{W}^{\frac{1}{2}}_{\mathbf{\Upsilon},1}\|\mathcal{N}\|_F\Big), \label{prooftheo2main24}
\end{align}
where $\mathbf{D}_{1_{\mathbf{\Xi}_1}}\in\mathbb{C}^{M_1\times sd}, \mathbf{D}_{2_{\mathbf{\Xi}_2}}\in\mathbb{C}^{M_2\times sd},\cdots,\mathbf{D}_{n_{\mathbf{\Xi}_n}}\in\mathbb{C}^{M_n\times sd}$ indexed by $\mathbf{\Xi}_1,\mathbf{\Xi}_2,\cdots,\mathbf{\Xi}_n$ are the column-block submatrices of $\mathbf{D}_1,\mathbf{D}_2,\cdots,\mathbf{D}_n$, respectively, (\ref{prooftheo2main21}) is due to the fact that the average of the correlation between $s$ selected blocks and the residual in the first iteration is the largest and should be no less than that of any other subset with cardinality being no less than $s$, (\ref{prooftheo2main22}) is from (\ref{tensorCSmodel}), (\ref{prooftheo2main23}) follows from the triangle inequality, and (\ref{prooftheo2main24}) is derived based on \textbf{Corollary~\ref{Corollary6}}.

Consider the case in which no correct index is selected at the first iteration, i.e., there exists at least one tensor index set satisfying that $\mathbf{\Xi}_i\cap\mathbf{\Xi}^1_i=\mathbf{\emptyset}$ $(i\in\{1,2,\cdots,n\})$; we have
\begin{align} % eq.186
	& \sum_{(i_1,\cdots,i_n)\in\mathbf{\Xi}^1}
	  \big\|\mathcal{Y}\times_1\mathbf{D}^{\rm H}_{1_{[i_1]}}\times_2\cdots\times_n\mathbf{D}^{\rm H}_{n_{[i_n]}}\big\|_F \nonumber \\
	&  = \sum_{(i_1,\cdots,i_n)\in\mathbf{\Xi}^1} \big\|\big(\mathcal{X}_{\mathbf{\Xi}}\times_1\mathbf{D}_{1_{\mathbf{\Xi}_1}}\times_2\cdots\times_n\mathbf{D}_{n_{\mathbf{\Xi}_n}}+\mathcal{N}\big)\nonumber\\
	&\hspace{7em}\times_1\mathbf{D}^{\rm H}_{1_{[i_1]}}\times_2\cdots\times_n\mathbf{D}^{\rm H}_{n_{[i_n]}}\big\|_F \nonumber \\
	&  \leq \sum_{(i_1,\cdots,i_n)\in\mathbf{\Xi}^1}\Big(\big\|\big(\mathcal{X}_{\mathbf{\Xi}}\times_1\mathbf{D}_{1_{\mathbf{\Xi}_1}}\times_2\cdots\times_n\mathbf{D}_{n_{\mathbf{\Xi}_n}}\big)\nonumber\\
	&\hspace{7.5em}\times_1\mathbf{D}^{\rm H}_{1_{[i_1]}}\times_2\cdots\times_n\mathbf{D}^{\rm H}_{n_{[i_n]}}\big\|_F \nonumber \\
	& \hspace{7em} + \big\|\mathcal{N}\times_1\mathbf{D}^{\rm H}_{1_{[i_1]}}\times_2\cdots\times_n\mathbf{D}^{\rm H}_{n_{[i_n]}}\big\|_F\Big) \nonumber \\
	&  = \sum_{(i_1,\cdots,i_n)\in\mathbf{\Xi}^1}\Big(\big\|\mathcal{X}_{\mathbf{\Xi}}\times_1\big(\mathbf{D}^{\rm H}_{1_{[i_1]}}\mathbf{D}_{1_{\mathbf{\Xi}_1}}\big)\nonumber\\
	&\hspace{7.5em}\times_2\cdots\times_n\big(\mathbf{D}^{\rm H}_{n_{[i_n]}}\mathbf{D}_{n_{\mathbf{\Xi}_n}}\big)\big\|_F \nonumber \\
	& \hspace{7em} + \big\|\mathcal{N}\times_1\mathbf{D}^{\rm H}_{1_{[i_1]}}\times_2\cdots\times_n\mathbf{D}^{\rm H}_{n_{[i_n]}}\big\|_F\Big) \nonumber \\
	& \leq s \bigg(\varpi^n_{\mathbf{\Upsilon},n-1}\prod_{t=1}^nk_td_t\bigg)\|\mathcal{X}\|_F+s\overline{W}^{\frac{1}{2}}_{\mathbf{\Upsilon},1}\|\mathcal{N}\|_F \label{ApJT5-1} ,
\end{align}
where $s$ is the number of selected indices during the first iteration of T-GBOMP, $\sum_{t=0}^{n-1}\underline{C}_t=\prod_{i=1}^nd_i-1$, $\overline{W}_{\mathbf{\Upsilon},1}$ can be obtained by \textbf{Corollary~\ref{Corollary6}}, and (\ref{ApJT5-1}) is derived from the scenario where only one tensor index set is selected incorrectly, which could present the strictest upper bound in (\ref{ApJT5-1}). 

Then, if
\begin{align} % eq.187
	&\sqrt{s k}\Big(\underline{W}^{\frac{1}{2}}_{\mathbf{\Upsilon},\prod_{t=1}^{n}k_t}\underline{W}^{\frac{1}{2}}_{\mathbf{\Upsilon},1}\|\mathcal{X}\|_F-	 \overline{W}^{\frac{1}{2}}_{\mathbf{\Upsilon},1}\|\mathcal{N}\|_F\Big)\nonumber\\
	&\qquad> s\bigg(\varpi^n_{\mathbf{\Upsilon},n-1}\prod_{t=1}^nk_td_t\bigg)\|\mathcal{X}\|_F+s\overline{W}^{\frac{1}{2}}_{\mathbf{\Upsilon},1}\|\mathcal{N}\|_F,\label{contradicted}
\end{align}
(\ref{prooftheo2main24}) is contradicted.

Reorganizing (\ref{contradicted}), we have
\begin{align} % eq.188
	\frac{\|\mathcal{X}\|_F}{\|\mathcal{N}\|_F} > \frac{\big(s+\sqrt{s k}\big)\overline{W}^{\frac{1}{2}}_{\mathbf{\Upsilon},1}}{\sqrt{sk}\underline{W}^{\frac{1}{2}}_{\mathbf{\Upsilon},\prod_{t=1}^{n}k_t}\underline{W}^{\frac{1}{2}}_{\mathbf{\Upsilon},1}-s\Big(\varpi^n_{\mathbf{\Upsilon},n-1}\prod_{t=1}^nk_td_t\Big)}. \label{simpcontradicted}
\end{align}
On the other hand,
\begin{align} % eqs.189,190
	\frac{\|\mathcal{X}\|_F}{\|\mathcal{N}\|_F}&\geq \frac{\|\mathcal{X}\times_1\mathbf{D}_1\times_2\cdots\times_n\mathbf{D}_n\|_F}{\overline{W}^{\frac{1}{2}}_{\mathbf{\Upsilon},\prod_{t=1}^{n}k_t}\|\mathcal{N}\|_F}\label{simpcontradicted1}\\
	&= \frac{\sqrt{{\rm SNR}}}{\overline{W}^{\frac{1}{2}}_{\mathbf{\Upsilon},\prod_{t=1}^{n}k_t}}, \label{simpcontradicted2}
\end{align}
where (\ref{simpcontradicted1}) is from \textbf{Corollary \ref{Corollary6}} and (\ref{simpcontradicted2}) is due to the definition of SNR.

Therefore, combining (\ref{simpcontradicted}) and (\ref{simpcontradicted2}), we obtain that if
\begin{align} % eq.191
	\sqrt{{\rm SNR}} > \frac{\big(\frac{\sqrt{k}}{\sqrt{s}}+1\big)\overline{W}^{\frac{1}{2}}_{\mathbf{\Upsilon},1}\overline{W}^{\frac{1}{2}}_{\mathbf{\Upsilon},\prod_{t=1}^{n}k_t}}{\frac{\sqrt{k}}{\sqrt{s}}\underline{W}^{\frac{1}{2}}_{\mathbf{\Upsilon},\prod_{t=1}^{n}k_t}\underline{W}^{\frac{1}{2}}_{\mathbf{\Upsilon},1}-\varpi^n_{\mathbf{\Upsilon},n-1}\prod_{t=1}^nk_td_t},\label{simpcontradictedcondition}
\end{align}
then at least one correct tensor block is selected at T-GBOMP's first iteration.

\emph{b)}~Second, supposing that T-GBOMP selects at least one correct tensor block at each of the previous $l$ $(1\leq l<k)$ iterations, we prove that the algorithm selects at least one correct tensor block at the $(l+1)$-th iteration. Denote $p_1$ as the largest element in the set $\big\{\big\|\mathcal{R}^l\times_1\mathbf{D}^{\rm H}_{1_{[i_1]}}\times_2\cdots\times_n\mathbf{D}^{\rm H}_{n_{[i_n]}}\big\|_F\big\}_{(i_1,\cdots,i_n)\in\mathbf{\Xi}\backslash\mathbf{\Xi}^l}$. The following inequality holds:
\begin{align} % eqs.192-196
	&p_1\nonumber\\
	 &\geq \frac{1}{\sqrt{|\mathbf{\Xi}\backslash\mathbf{\Xi}^l|}}
	\Bigg(\sum_{(i_1,i_2,\cdots,i_n)\in\mathbf{\Xi}\backslash\mathbf{\Xi}^l} \big\|\mathcal{R}^l\times_1\mathbf{D}^{\rm H}_{1_{[i_1]}}\nonumber\\
	&\hspace{11em}\times_2\cdots\times_n\mathbf{D}^{\rm H}_{n_{[i_n]}}\big\|^2_F\Bigg)^{\frac{1}{2}} \label{cauchy1} \\
	& \geq \frac{1}{\sqrt{k-c}}\Big(\big\|\ddot{\mathbf{D}}_{\mathbf{\Xi}\backslash\mathbf{\Xi}^l} \mathbf{P}^{\bot}_{\ddot{\mathbf{D}}_{\mathbf{\Xi}^l}}{\rm vec}(\mathcal{Y})\big\|_F\Big) \nonumber \\
	&= \frac{1}{\sqrt{k\! -\! c}} \Big(\big\|\ddot{\mathbf{D}}_{\mathbf{\Xi}\backslash\mathbf{\Xi}^l} \mathbf{P}^{\bot}_{\ddot{\mathbf{D}}_{\mathbf{\Xi}^l}}\big(\mathbf{D}_{n_{\mathbf{\Xi}_n\backslash\mathbf{\Xi}_n^l}}\otimes\cdots\otimes\mathbf{D}_{1_{\mathbf{\Xi}_1\backslash\mathbf{\Xi}_1^l}}\big)\nonumber\\
	&\hspace{5em}\times {\rm vec}\big(\mathcal{X}_{\mathbf{\Xi}\backslash\mathbf{\Xi}^l}\big) + \ddot{\mathbf{D}}_{\mathbf{\Xi}\backslash\mathbf{\Xi}^l} \mathbf{P}^{\bot}_{\ddot{\mathbf{D}}_{\mathbf{\Xi}^l}}{\rm vec}(\mathcal{N})\big\|_F\Big) \label{proofp11} \\
	&\geq \frac{1}{\sqrt{k\! -\! c}}\Big(\big\|\ddot{\mathbf{D}}_{\mathbf{\Xi}\backslash\mathbf{\Xi}^l} \mathbf{P}^{\bot}_{\ddot{\mathbf{D}}_{\mathbf{\Xi}^l}} \big(\mathbf{D}_{n_{\mathbf{\Xi}_n\backslash\mathbf{\Xi}_n^l}}\otimes\cdots\otimes\mathbf{D}_{1_{\mathbf{\Xi}_1\backslash\mathbf{\Xi}_1^l}}\big)\nonumber\\
	&\hspace{5em}\times {\rm vec}\big(\mathcal{X}_{\mathbf{\Xi}\backslash\mathbf{\Xi}^l}\big)\big\|_F \nonumber\\
	&\hspace{4.5em}- \big\|\ddot{\mathbf{D}}_{\mathbf{\Xi}\backslash\mathbf{\Xi}^l} \mathbf{P}^{\bot}_{\ddot{\mathbf{D}}_{\mathbf{\Xi}^l}}{\rm vec}(\mathcal{N})\big\|_F\Big) \label{proofp12} \\
	&\geq \frac{1}{\sqrt{k\! -\! c}} \Big(\sigma_{\min} \Big(\ddot{\mathbf{D}}_{\mathbf{\Xi}\backslash\mathbf{\Xi}^l}\big(\mathbf{D}_{n_{\mathbf{\Xi}_n\backslash\mathbf{\Xi}_n^l}}\otimes\cdots\otimes\mathbf{D}_{1_{\mathbf{\Xi}_1\backslash\mathbf{\Xi}_1^l}}\big)\Big)\nonumber\\
	&\hspace{5em}\times\big\|\mathcal{X}_{\mathbf{\Xi}\backslash\mathbf{\Xi}^l}\big\|_F
	- \sigma_{\max} \ddot{\mathbf{D}}_{\mathbf{\Xi}\backslash\mathbf{\Xi}^l} \big\|\mathcal{N}\big\|_F\Big) \label{proofp13} \\
	&\geq \frac{1}{\sqrt{k\! -\! c}}\Big(\underline{W}^{\frac{1}{2}}_{\mathbf{\Upsilon},ls}\underline{W}^{\frac{1}{2}}_{\mathbf{\Upsilon},(k-c)^n}\big\|\mathcal{X}_{\mathbf{\Xi}\backslash\mathbf{\Xi}^l}\big\|_F\nonumber\\
	&\hspace{5em} - \overline{W}^{\frac{1}{2}}_{\mathbf{\Upsilon},ls}\|\mathcal{N}\|_F\Big), \label{proofp14}
\end{align}
where $c=|\mathbf{\Xi}\cap\mathbf{\Xi}^l|\geq l$,   $\underline{W}_{\mathbf{\Upsilon},k_n}$ and $\overline{W}_{\mathbf{\Upsilon},k_n}$ can be obtained by \textbf{Corollary \ref{Corollary6}},
(\ref{cauchy1}) follows from the Cauchy-Schwarz inequality, (\ref{proofp11}) is from (\ref{tensorCSmodel}), (\ref{proofp12}) is based on the triangle inequality, (\ref{proofp13}) is based on \cite[Lemma 5]{cai2011}, and (\ref{proofp14}) follows from \textbf{Corollary~\ref{Corollary6}}.

Denoting $\mathbf{\Psi}$ as the index set corresponding to the $s$ largest elements in $\big\{\big\|\mathcal{R}^l\times_1\mathbf{D}^{\rm H}_{1_{[i_1]}}\times_2\cdots\times_n\mathbf{D}^{\rm H}_{n_{[i_n]}}\big\|_F\big\}_{(i_1,\cdots,i_n)\in\mathbf{\Omega}\backslash(\mathbf{\Xi}\cup\mathbf{\Xi}^l)}$ and letting $q_s$ be the $s$th largest element in $\mathbf{\Psi}$, we have
\begin{align} % eq.197
	&q_s\nonumber\\
	 &\leq \frac{1}{\sqrt{s}}
	  \sqrt{\sum_{(i_1,\cdots,i_n)\in\mathbf{\Psi}} \big\|\mathcal{R}^l\times_1\mathbf{D}^{\rm H}_{1_{[i_1]}}\times_2\cdots\times_n\mathbf{D}^{\rm H}_{n_{[i_n]}}\big\|_F^2} \nonumber \\
	&= \frac{1}{\sqrt{s}}\big\|\ddot{\mathbf{D}}_{\mathbf{\Psi}} \mathbf{P}^{\bot}_{\ddot{\mathbf{D}}_{\mathbf{\Xi}^l}}{\rm vec}(\mathcal{Y})\big\|_F \nonumber \\
	&= \frac{1}{\sqrt{s}}\big\|\ddot{\mathbf{D}}_{\mathbf{\Psi}} \mathbf{P}^{\bot}_{\ddot{\mathbf{D}}_{\mathbf{\Xi}^l}}\big(\mathbf{D}_n\otimes\cdots\otimes\mathbf{D}_1\big) {\rm vec}(\mathcal{X})\nonumber\\
	 &\hspace{3em} + \ddot{\mathbf{D}}_{\mathbf{\Psi}} \mathbf{P}^{\bot}_{\ddot{\mathbf{D}}_{\mathbf{\Xi}^l}}{\rm vec}(\mathcal{N})\big\|_F \nonumber \\
	&\leq \frac{1}{\sqrt{s}}\Big(\big\|\ddot{\mathbf{D}}_{\mathbf{\Psi}} \mathbf{P}^{\bot}_{\ddot{\mathbf{D}}_{\mathbf{\Xi}^l}}\big(\mathbf{D}_{n_{\mathbf{\Xi}_n\backslash\mathbf{\Xi}_n^l}}\otimes\cdots\otimes\mathbf{D}_{1_{\mathbf{\Xi}_1\backslash\mathbf{\Xi}_1^l}}\big) \nonumber\\
	&\hspace{3em}\times{\rm vec}\big(\mathcal{X}_{\mathbf{\Xi}\backslash\mathbf{\Xi}^l}\big)\big\|_F + \big\|\ddot{\mathbf{D}}_{\mathbf{\Psi}} \mathbf{P}^{\bot}_{\ddot{\mathbf{D}}_{\mathbf{\Xi}^l}}{\rm vec}(\mathcal{N})\big\|_F\Big) \nonumber \\
	&\leq \frac{1}{\sqrt{s}}\Big(\big\|\ddot{\mathbf{D}}_{\mathbf{\Psi}} \mathbf{P}_{\ddot{\mathbf{D}}_{\mathbf{\Xi}^l}}
	  \big(\mathbf{D}_{n_{\mathbf{\Xi}_n\backslash\mathbf{\Xi}_n^l}}\otimes\cdots\otimes\mathbf{D}_{1_{\mathbf{\Xi}_1\backslash\mathbf{\Xi}_1^l}}\big) {\rm vec}\big(\mathcal{X}_{\mathbf{\Xi}\backslash\mathbf{\Xi}^l}\big)\big\|_F \nonumber \\
	& \qquad\quad + \big\|\ddot{\mathbf{D}}_{\mathbf{\Psi}}
	  \big(\mathbf{D}_{n_{\mathbf{\Xi}_n\backslash\mathbf{\Xi}_n^l}}\otimes\cdots\otimes\mathbf{D}_{1_{\mathbf{\Xi}_1\backslash\mathbf{\Xi}_1^l}}\big) {\rm vec}\big(\mathcal{X}_{\mathbf{\Xi}\backslash\mathbf{\Xi}^l}\big)\big\|_F\nonumber\\
	& \qquad\quad + \big\|\ddot{\mathbf{D}}_{\mathbf{\Psi}} \mathbf{P}^{\bot}_{\ddot{\mathbf{D}}_{\mathbf{\Xi}^l}} {\rm vec}(\mathcal{N})\big\|_F\Big), \label{thwo2last}
\end{align}
where (\ref{thwo2last}) is because $\mathbf{P}^{\bot}_{\ddot{\mathbf{D}}_{\mathbf{\Xi}^l}}=\mathbf{I}-\mathbf{P}_{\ddot{\mathbf{D}}_{\mathbf{\Xi}^l}}$. 
In the following, we derive upper bounds of the three terms in (\ref{thwo2last}), i.e., $\big\|\ddot{\mathbf{D}}_{\mathbf{\Psi}} \mathbf{P}_{\ddot{\mathbf{D}}_{\mathbf{\Xi}^l}}\big(\mathbf{D}_{n_{\mathbf{\Xi}_n\backslash\mathbf{\Xi}_n^l}}\otimes\cdots\otimes\mathbf{D}_{1_{\mathbf{\Xi}_1\backslash\mathbf{\Xi}_1^l}}\big) {\rm vec}\big(\mathcal{X}_{\mathbf{\Xi}\backslash\mathbf{\Xi}^l}\big)\big\|_F$, $\big\|\ddot{\mathbf{D}}_{\mathbf{\Psi}}
\big(\mathbf{D}_{n_{\mathbf{\Xi}_n\backslash\mathbf{\Xi}_n^l}}\otimes\cdots\otimes\mathbf{D}_{1_{\mathbf{\Xi}_1\backslash\mathbf{\Xi}_1^l}}\big) {\rm vec}\big(\mathcal{X}_{\mathbf{\Xi}\backslash\mathbf{\Xi}^l}\big)\big\|_F$ and $\big\|\ddot{\mathbf{D}}_{\mathbf{\Psi}} \mathbf{P}^{\bot}_{\ddot{\mathbf{D}}_{\mathbf{\Xi}^l}} {\rm vec}(\mathcal{N})\big\|_F$.

For the first term, note that
\begin{align} % eq.198
	& \Big\|\ddot{\mathbf{D}}_{\mathbf{\Psi}} \mathbf{P}_{\ddot{\mathbf{D}}_{\mathbf{\Xi}^l}} \big(\mathbf{D}_{n_{\mathbf{\Xi}_n\backslash\mathbf{\Xi}_n^l}}\otimes\cdots\otimes\mathbf{D}_{1_{\mathbf{\Xi}_1\backslash\mathbf{\Xi}_1^l}}\big) {\rm vec}\big(\mathcal{X}_{\mathbf{\Xi}\backslash\mathbf{\Xi}^l}\big)\Big\|_F \nonumber \\
	&  = \Big\|\ddot{\mathbf{D}}_{\mathbf{\Psi}} \ddot{\mathbf{D}}_{\mathbf{\Xi}^l} \big(\ddot{\mathbf{D}}_{\mathbf{\Xi}^l}^{{\rm H}}\ddot{\mathbf{D}}_{\mathbf{\Xi}^l}\big)^{-1} \ddot{\mathbf{D}}_{\mathbf{\Xi}^l}^{{\rm H}}\nonumber\\
	 &\hspace{1.5em}\times \big(\mathbf{D}_{n_{\mathbf{\Xi}_n\backslash\mathbf{\Xi}_n^l}}\otimes\cdots\otimes\mathbf{D}_{1_{\mathbf{\Xi}_1\backslash\mathbf{\Xi}_1^l}}\big) {\rm vec}\big(\mathcal{X}_{\mathbf{\Xi}\backslash\mathbf{\Xi}^l}\big)\Big\|_F \nonumber \\
	&  \leq \big\|\ddot{\mathbf{D}}_{\mathbf{\Psi}} \ddot{\mathbf{D}}_{\mathbf{\Xi}^l}\big\|_2 \big\|\big(\ddot{\mathbf{D}}_{\mathbf{\Xi}^l}^{{\rm H}}\ddot{\mathbf{D}}_{\mathbf{\Xi}^l}\big)^{-1}\big\|_2\nonumber\\
	  &\hspace{1em}\times\big\|\ddot{\mathbf{D}}_{\mathbf{\Xi}^l}^{{\rm H}} \big(\mathbf{D}_{n_{\mathbf{\Xi}_n\backslash\mathbf{\Xi}_n^l}}\otimes\cdots\otimes\mathbf{D}_{1_{\mathbf{\Xi}_1\backslash\mathbf{\Xi}_1^l}}\big)\big\|_2 \big\|\mathcal{X}_{\mathbf{\Xi}\backslash\mathbf{\Xi}^l}\big\|_F, \label{lasttwo}
\end{align}
where (\ref{lasttwo}) is from the sub-multiplicativity. Based on \textbf{Corollary~\ref{Corollary6}}, we have
\begin{align} % eqs.199,200
	&\big\|\ddot{\mathbf{D}}_{\mathbf{\Psi}} \ddot{\mathbf{D}}_{\mathbf{\Xi}^l}\big\|_2
	   \leq \max\Big\{s\varpi_{\mathbf{\Upsilon},n-1}^n\prod_{t=1}^{n}d_t,ls\varpi_{\mathbf{\Upsilon},n-1}^n\prod_{t=1}^{n}d_t\Big\}\nonumber\\
	& \hspace{5em} = l s \varpi_{\mathbf{\Upsilon},n-1}^n\prod_{t=1}^{n}d_t, \label{submultiplicativity1}\\
	&\Big\|\ddot{\mathbf{D}}_{\mathbf{\Xi}^l}^{{\rm H}}\big(\mathbf{D}_{n_{\mathbf{\Xi}_n\backslash\mathbf{\Xi}_n^l}}\otimes\cdots\otimes\mathbf{D}_{1_{\mathbf{\Xi}_1\backslash\mathbf{\Xi}_1^l}}\big)\Big\|_2\nonumber\\
	   &\leq \max \Big\{ls\varpi_{\mathbf{\Upsilon},n-1}^n\prod_{t=1}^{n}d_t,(k-c)^n\varpi_{\mathbf{\Upsilon},n-1}^n\prod_{t=1}^{n}d_t\Big\}. \label{submultiplicativity2}
\end{align}
It remains to derive an upper bound on $\big\|\big(\ddot{\mathbf{D}}_{\mathbf{\Xi}^l}^{{\rm H}}\ddot{\mathbf{D}}_{\mathbf{\Xi}^l}\big)^{-1}\big\|_2$. Denote $\ddot{\mathbf{D}}_{\mathbf{\Xi}^l}^{{\rm H}}\ddot{\mathbf{D}}_{\mathbf{\Xi}^l}$ as $\mathbf{I}+\mathbf{A}$ since the diagonal elements of $\ddot{\mathbf{D}}_{\mathbf{\Xi}^l}^{{\rm H}}\ddot{\mathbf{D}}_{\mathbf{\Xi}^l}$ are all 1, where $\mathbf{A}$ is a matrix of size $l s\prod_{t=1}^nd_t\times l s\prod_{t=1}^nd_t$ with blocks of size $\prod_{t=1}^nd_t\times\prod_{t=1}^nd_t$. Then, by using \textbf{Corollary \ref{Corollary1}}, we obtain
\begin{align} % eq.201
	\|\mathbf{A}\|_2 \leq \sum_{t=0}^{n-1}\underline{C}_t\tau_{\mathbf{\Upsilon},t}^{n-t}+\sum_{t=0}^{n-1}C_t\varpi_{\mathbf{\Upsilon},t}^n\prod^{n}_{i=1}d_i, \nonumber
\end{align}
where $\sum_{t=0}^{n-1}\underline{C}_t= \prod_{t=1}^n d_t-1$, $\sum_{t=0}^{n-1}C_t=ls-1$, and $\underline{C}_t$ and $C_t$ can be obtained by \textbf{Proposition \ref{prop2}}.

Suppose that $\sum_{t=0}^{n-1}\underline{C}_t\tau_{\mathbf{\Upsilon},t}^{n-t}+\sum_{t=0}^{n-1}C_t\varpi_{\mathbf{\Upsilon},t}^n\prod^{n}_{i=1}d_i<1$. Similar to (\ref{th8prof1}) to (\ref{th8prof5}), we have
\begin{align} % eq.202
	&\big\|\big(\ddot{\mathbf{D}}_{\mathbf{\Xi}^l}^{{\rm H}}\ddot{\mathbf{D}}_{\mathbf{\Xi}^l}\big)^{-1}\big\|_2\nonumber\\
	 & \leq \frac{1}{1-\sum_{t=0}^{n-1}\underline{C}_t\tau_{\mathbf{\Upsilon},t}^{n-t}-\sum_{t=0}^{n-1}C_t\varpi_{\mathbf{\Upsilon},t}^n\prod^{n}_{i=1}d_i}. \label{submultiplicativity3}
\end{align}
Combining (\ref{lasttwo}), (\ref{submultiplicativity1}), (\ref{submultiplicativity2}) and (\ref{submultiplicativity3}) leads to
\begin{align} % eq.203
	& \big\|\ddot{\mathbf{D}}_{\mathbf{\Psi}} \mathbf{P}_{\ddot{\mathbf{D}}_{\mathbf{\Xi}^l}}\big(\mathbf{D}_{n_{\mathbf{\Xi}_n\backslash\mathbf{\Xi}_n^l}}\otimes\cdots\otimes\mathbf{D}_{1_{\mathbf{\Xi}_1\backslash\mathbf{\Xi}_1^l}}\big) {\rm vec}\big(\mathcal{X}_{\mathbf{\Xi}\backslash\mathbf{\Xi}^l}\big)\big\|_F \nonumber \\
	&  \leq\big(ls\varpi_{\mathbf{\Upsilon},n-1}^n\prod_{t=1}^{n}d_t\big)\nonumber\\
	&\hspace{1em}\times\max\Big\{ls\varpi_{\mathbf{\Upsilon},n-1}^n\prod_{t=1}^{n}d_t,(k-c)^n\varpi_{\mathbf{\Upsilon},n-1}^n\prod_{t=1}^{n}d_t\Big\}\nonumber\\
	&\hspace{1em}\times\frac{1}{1-\sum_{t=0}^{n-1}\underline{C}_t\tau_{\mathbf{\Upsilon},t}^{n-t}-\sum_{t=0}^{n-1}C_t\varpi_{\mathbf{\Upsilon},t}^n\prod^{n}_{i=1}d_i}\|\mathcal{X}_{\mathbf{\Xi}\backslash\mathbf{\Xi}^l}\|_F\nonumber\\
	&  =\big(ls\varpi_{\mathbf{\Upsilon},n-1}^n\prod_{t=1}^{n}d_t\big)\nonumber\\
	&\hspace{1em}\times\max\Big\{ls\varpi_{\mathbf{\Upsilon},n-1}^n\prod_{t=1}^{n}d_t,(k-c)^n\varpi_{\mathbf{\Upsilon},n-1}^n\prod_{t=1}^{n}d_t\Big\}\nonumber\\
	&\hspace{1em}\times\frac{1}{\underline{W}_{\mathbf{\Upsilon},ls}}\|\mathcal{X}_{\mathbf{\Xi}\backslash\mathbf{\Xi}^l}\|_F.\label{submultiplicativitylast}
	\end{align}

For the second term, we have
\begin{align} % eq.204
	& \big\|\ddot{\mathbf{D}}_{\mathbf{\Psi}} \big(\mathbf{D}_{n_{\mathbf{\Xi}_n\backslash\mathbf{\Xi}_n^l}}\otimes\cdots\otimes\mathbf{D}_{1_{\mathbf{\Xi}_1\backslash\mathbf{\Xi}_1^l}}\big) {\rm vec}\big(\overline{\mathcal{X}}\big)\big\|_F\nonumber\\
	  &\qquad\leq \max\bigg\{s\varpi_{\mathbf{\Upsilon},n-1}^n\prod_{t=1}^{n}d_t,(k-c)^n\varpi_{\mathbf{\Upsilon},n-1}^n\prod_{t=1}^{n}d_t\bigg\}\nonumber\\
	  &\qquad\hspace{1em}\times\big\|\mathcal{X}_{\mathbf{\Xi}\backslash\mathbf{\Xi}^l}\big\|_F , \label{equu1}
\end{align}
where (\ref{equu1}) follows from \textbf{Corollary \ref{Corollary6}}.

For the third term, we have
\begin{align} % eq.205
	\big\|\ddot{\mathbf{D}}_{\mathbf{\Psi}} \mathbf{P}^{\bot}_{\ddot{\mathbf{D}}_{\mathbf{\Xi}^l}}{\rm vec}(\mathcal{N})\big\|_F
	&\leq \|\ddot{\mathbf{D}}_{\mathbf{\Psi}}\|_2 \|\mathcal{N}\|_F\nonumber\\
	& \leq
	  \overline{W}^{\frac{1}{2}}_{\mathbf{\Upsilon},s}\|\mathcal{N}\|_F. \label{noiselast}
\end{align}

Combining (\ref{thwo2last}), (\ref{submultiplicativitylast}), (\ref{equu1}) and (\ref{noiselast}), we have
\begin{align} % eq.206
	q_s \leq& \frac{1}{\sqrt{s}}\Bigg(\big(ls\varpi_{\mathbf{\Upsilon},n-1}^n\prod_{t=1}^{n}d_t\big)\nonumber\\
	&\qquad\times\max\Big\{ls\varpi_{\mathbf{\Upsilon},n-1}^n\prod_{t=1}^{n}d_t,(k-c)^n\varpi_{\mathbf{\Upsilon},n-1}^n\prod_{t=1}^{n}d_t\Big\}\nonumber\\
	&\qquad\times\frac{1}{\underline{W}_{\mathbf{\Upsilon},ls}}\|\mathcal{X}_{\mathbf{\Xi}\backslash\mathbf{\Xi}^l}\|_F \nonumber \\
	& \qquad + \max\Big\{s\varpi_{\mathbf{\Upsilon},n-1}^n\prod_{t=1}^{n}d_t,(k-c)^n\varpi_{\mathbf{\Upsilon},n-1}^n\prod_{t=1}^{n}d_t\Big\}\nonumber\\
	&\qquad\quad\times\|\mathcal{X}_{\mathbf{\Xi}\backslash\mathbf{\Xi}^l}\|_F
	  +\overline{W}^{\frac{1}{2}}_{\mathbf{\Upsilon},s}\|\mathcal{N}\|_F\Bigg). \label{upperqs}
\end{align}

Note that if $p_1>q_s$, $p_1$ belongs to the set of $s$ largest elements among all the elements in  $\big\{\big\|\mathcal{R}^l\times_1\mathbf{D}^{\rm H}_{1_{[i_1]}}\times_2\cdots\times_n\mathbf{D}^{\rm H}_{n_{[i_n]}}\big\|_F\big\}_{(i_1,\cdots,i_n)\in\mathbf{\Omega}\backslash\mathbf{\Xi}^l}$, then at least one correct index is selected in the $(l+1)$th iteration. Exploiting (\ref{proofp14}) and (\ref{upperqs}), $p_1 > q_s$ holds true if the following inequality is satisfied:
\begin{align} % eq.207
 	& \frac{1}{\sqrt{k-c}} \Big(\underline{W}^{\frac{1}{2}}_{\mathbf{\Upsilon},ls}\underline{W}^{\frac{1}{2}}_{\mathbf{\Upsilon},(k-c)^n}\big\|\mathcal{X}_{\mathbf{\Xi}\backslash\mathbf{\Xi}^l}\big\|_F - \overline{W}^{\frac{1}{2}}_{\mathbf{\Upsilon},ls}\|\mathcal{N}\|_F\Big) \nonumber \\
 	&  >
 	\frac{1}{\sqrt{s}}\Bigg(\big(ls\varpi_{\mathbf{\Upsilon},n-1}^n\prod_{t=1}^{n}d_t\big)\nonumber\\
 	&\hspace{3.5em}\times\max\Big\{ls\varpi_{\mathbf{\Upsilon},n-1}^n\prod_{t=1}^{n}d_t,(k-c)^n\varpi_{\mathbf{\Upsilon},n-1}^n\prod_{t=1}^{n}d_t\Big\} \nonumber \\
 	&\hspace{3.5em}\times\frac{1}{\underline{W}_{\mathbf{\Upsilon},ls}}\|\mathcal{X}_{\mathbf{\Xi}\backslash\mathbf{\Xi}^l}\|_F\nonumber \\
 	&\hspace{3.5em} + \max\Big\{s\varpi_{\mathbf{\Upsilon},n-1}^n\prod_{t=1}^{n}d_t,(k-c)^n\varpi_{\mathbf{\Upsilon},n-1}^n\prod_{t=1}^{n}d_t\Big\}\nonumber\\
 	&\hspace{4.5em}\times\|\mathcal{X}_{\mathbf{\Xi}\backslash\mathbf{\Xi}^l}\|_F
 	+\overline{W}^{\frac{1}{2}}_{\mathbf{\Upsilon},s}\|\mathcal{N}\|_F\Bigg). \label{p_1>q_s}
\end{align}
Note that
\begin{align} % eqs.208-210
	\frac{\|\overline{\mathcal{X}}\|_F}{\|\mathcal{N}\|_F} &\geq \frac{\sqrt{|\mathbf{\Xi}\backslash\mathbf{\Xi}^l|}\min\limits_{(i_1,\cdots,i_n)\in\mathbf{\Xi}}\|\mathcal{X}_{[i_1,\cdots,i_n]}\|_F}{\|\mathcal{N}\|_F} \nonumber\\
	 & = \frac{{\rm MAR}_*\sqrt{k-c}\|\mathcal{X}\|_F}{\sqrt{k}\|\mathcal{N}\|_F} \label{marsnrdef1} \\
	&\geq \frac{{\rm MAR}_*\sqrt{k-c}\|\mathcal{X}\times_1\mathbf{D}_1\times_2\cdots\times_n\mathbf{D}_n\|_F}{\sqrt{k\overline{W}_{\mathbf{\Upsilon},\prod_{t=1}^{n}k_t}}\|\mathcal{N}\|_F} \label{marsnrdef2} \\
	&= \frac{{\rm MAR}_*\sqrt{(k-c){\rm SNR}}}{\sqrt{k\overline{W}_{\mathbf{\Upsilon},\prod_{t=1}^{n}k_t}}}, \label{marsnrdef3}
\end{align}
where (\ref{marsnrdef1}) is from the definition of ${\rm MAR}^*$, (\ref{marsnrdef2}) is based on \textbf{Corollary \ref{Corollary6}}, and (\ref{marsnrdef3}) follows from the definition of SNR. Combining (\ref{p_1>q_s}) and (\ref{marsnrdef3}), we conclude that if
\begin{align} % eq.211
	&\sqrt{{\rm SNR}} \nonumber\\
	&> \Bigg(\frac{1}{\sqrt{s}}\overline{W}^{\frac{1}{2}}_{\mathbf{\Upsilon},s}+\frac{1}{\sqrt{k-c}}\overline{W}^{\frac{1}{2}}_{\mathbf{\Upsilon},ls}\Bigg) \sqrt{k}\overline{W}^{\frac{1}{2}}_{\mathbf{\Upsilon},\prod_{t=1}^{n}k_t} \nonumber \\
	& \hspace{1em}\times \Bigg(\! \bigg(\frac{1}{\sqrt{k\! -\! c}}\underline{W}^{\frac{1}{2}}_{\mathbf{\Upsilon},ls}\underline{W}^{\frac{1}{2}}_{\mathbf{\Upsilon},(k-c)^n} \nonumber\\
	  & \hspace{1em}- \frac{1}{\sqrt{s}}\Big(l s \varpi_{\mathbf{\Upsilon},n-1}^n \prod\limits_{t=1}^{n}d_t\Big)\nonumber\\
	  &\times \max\Big\{ls\varpi_{\mathbf{\Upsilon},n-1}^n\prod\limits_{t=1}^{n}d_t,(k\! -\! c)^n\varpi_{\mathbf{\Upsilon},n-1}^n\prod\limits_{t=1}^{n}d_t\Big\}\frac{1}{\underline{W}_{\mathbf{\Upsilon},ls}} \nonumber \\
	& -\frac{1}{\sqrt{s}}\max\Big\{s\varpi_{\mathbf{\Upsilon},n-1}^n\prod_{t=1}^{n}d_t,(k-c)^n\varpi_{\mathbf{\Upsilon},n-1}^n\prod_{t=1}^{n}d_t\Big\}\bigg)\, \nonumber\\
	&\times{\rm MAR}_* \sqrt{k-c} \Bigg)^{-1} \label{snrmin}
\end{align}
holds, the T-GBOMP chooses at least one correct index in the $l+1$th iteration.

\emph{c)}~So far, we have proved that under the condition (\ref{simpcontradictedcondition}) T-GBOMP succeeds in the first iteration, and under the condition (\ref{snrmin}) T-GBOMP succeeds in the general iteration. We now present a condition under which T-GBOMP succeeds in every iteration. Due to the fact that $1\leq k-c<k$, $1\leq l<k$ and $1\leq s\leq k$, (\ref{snrmin}) is transformed into
\begin{subequations}
	\begin{numcases}{}	% eqs.212a,212b	
		\sqrt{{\rm SNR}}>\Bigg(\frac{\sqrt{k}}{\sqrt{s}}\overline{W}^{\frac{1}{2}}_{\mathbf{\Upsilon},s}\overline{W}^{\frac{1}{2}}_{\mathbf{\Upsilon},k}+\sqrt{k}\overline{W}^{\frac{1}{2}}_{\mathbf{\Upsilon},ks}\overline{W}^{\frac{1}{2}}_{\mathbf{\Upsilon},k}\Bigg) \nonumber \\
		  \hspace{4em} \times \Bigg(\frac{1}{\sqrt{s}}\bigg(\frac{\sqrt{s}}{\sqrt{k}}\underline{W}^{\frac{1}{2}}_{\mathbf{\Upsilon},ks}\underline{W}^{\frac{1}{2}}_{\mathbf{\Upsilon},k}-k\varpi_{\mathbf{\Upsilon}}d_1\nonumber\\
		 \hspace{4em} -\frac{(ks\varpi_{\mathbf{\Upsilon}}d_1)^2}{\underline{W}_{\mathbf{\Upsilon},ks}}
		  \bigg) {\rm MAR}_*\Bigg)^{-1}\! ,  n=1, \label{transformeda} \\
		\sqrt{{\rm SNR}} > \Bigg(\frac{\sqrt{k}}{\sqrt{s}}\overline{W}^{\frac{1}{2}}_{\mathbf{\Upsilon},s}\overline{W}^{\frac{1}{2}}_{\mathbf{\Upsilon},\prod_{t=1}^{n}k_t}\nonumber\\
		\hspace{5em}+\sqrt{k}\overline{W}^{\frac{1}{2}}_{\mathbf{\Upsilon},ks}\overline{W}^{\frac{1}{2}}_{\mathbf{\Upsilon},\prod_{t=1}^{n}k_t}\Bigg) \nonumber \\
		 \hspace{1em} ~ \times \Bigg(\! \frac{1}{\sqrt{s}}\bigg(\frac{\sqrt{s}}{\sqrt{k}}\underline{W}^{\frac{1}{2}}_{\mathbf{\Upsilon},ks}\underline{W}^{\frac{1}{2}}_{\mathbf{\Upsilon},k^n}\! -\! k^n\varpi_{\mathbf{\Upsilon},n-1}^n\prod_{t=1}^{n}d_t\nonumber\\ 
		 \hspace{2em}- \frac{k^{n+1}s\varpi_{\mathbf{\Upsilon},n-1}^{2n} \big(\prod\limits_{t=1}^{n}d_t\big)^2}{\underline{W}_{\mathbf{\Upsilon},ks}} \bigg) {\rm MAR}_*\! \Bigg)^{-1}\!\! ,  n>1 . \label{transformedb}
	\end{numcases}
\end{subequations}
Since $\frac{\sqrt{k}}{\sqrt{s}}\geq\frac{\sqrt{s}}{\sqrt{k}}$, $\frac{1}{\sqrt{s}}\leq1$ and ${\rm MAR}_*\leq1$, it can be obtained that (\ref{transformeda}) and (\ref{transformedb}) are more restrictive than the bound in (\ref{simpcontradictedcondition}). Therefore, (\ref{transformeda}) and (\ref{transformedb}) guarantee that T-GBOMP chooses all correct indices in $k$ iterations. Simplifying (\ref{transformeda}) and (\ref{transformedb}) leads to (\ref{theo2maina}) and (\ref{theo2mainb}). This completes the proof.
\end{IEEEproof}

\subsection{Proof of \textbf{Theorem~\ref{theo3}}}\label{proofoftheo3} % Ap-K

\begin{IEEEproof}	
Denote $l^*$ as the number of iterations of T-GBOMP, where $l^*\leq k$. In the case that $s=1$, $l^*=k$. Based on \textbf{Theorem~\ref{theo2}}, we have $\hat{\mathbf{\Xi}}=\mathbf{\Xi}^k=\mathbf{\Xi}$ and $\hat{\mathcal{X}}=\mathcal{X}^k$, since $\hat{\mathcal{X}}=\arg\min\limits_{\mathcal{X}:{\rm supp}(\mathcal{X})=\mathbf{\Xi}}\big\|{\rm vec}(\mathcal{Y})-\sum_{(i_1,i_2,\cdots,i_n)\in\mathbf{\Xi}} \big(\mathbf{D}_{n_{[i_n]}}\otimes\cdots\otimes\mathbf{D}_{1_{[i_1]}}\big){\rm vec}(\mathcal{X})_{[i_x]}\big\|_F$ and $\hat{\mathcal{X}}_{\mathbf{\Omega}\backslash\mathbf{\Xi}}=\mathbf{0}$. Therefore,
\begin{align} % eqs.213-217
	\big\|\hat{\mathcal{X}}-\mathcal{X}\big\|_F &= \big\|\ddot{\mathbf{D}}_{\mathbf{\Xi}}^{\dagger}{\rm vec}(\mathcal{Y})-{\rm vec}(\mathcal{X}_{\mathbf{\Xi}})\big\|_F \nonumber\\
	 & = \big\|\ddot{\mathbf{D}}_{\mathbf{\Xi}}^{\dagger}({\rm vec}(\mathcal{Y})-\ddot{\mathbf{D}}_{\mathbf{\Xi}}{\rm vec}(\mathcal{X}_{\mathbf{\Xi}}))\big\|_F \label{theo3prof11} \\
	&= \big\|\ddot{\mathbf{D}}_{\mathbf{\Xi}}^{\dagger}{\rm vec}(\mathcal{N})\big\|_F \label{theo3prof1} \\
	&\leq \frac{\|\ddot{\mathbf{D}}_{\mathbf{\Xi}}\ddot{\mathbf{D}}_{\mathbf{\Xi}}^{\dagger}{\rm vec}(\mathcal{N})\|_F}{\underline{W}^{\frac{1}{2}}_{\mathbf{\Upsilon},k}} \label{theo3prof2} \\
	&= \frac{\|\mathbf{P}_{\ddot{\mathbf{D}}_{\mathbf{\Xi}}}{\rm vec}(\mathcal{N})\|_F}{\underline{W}^{\frac{1}{2}}_{\mathbf{\Upsilon},k}} \label{theo3prof3} \\
	&\leq \frac{\|\mathcal{N}\|_F}{\underline{W}^{\frac{1}{2}}_{\mathbf{\Upsilon},k}}, \label{theo3prof34}
\end{align}
where (\ref{theo3prof11}) is because $\ddot{\mathbf{D}}_{\mathbf{\Xi}}^{\dagger}\ddot{\mathbf{D}}_{\mathbf{\Xi}}=\mathbf{I}$, (\ref{theo3prof1}) follows from (\ref{tensorCSmodel}), (\ref{theo3prof2}) is due to \textbf{Corollary \ref{Corollary6}}, (\ref{theo3prof3}) is derived by $\mathbf{P}_{\ddot{\mathbf{D}}_{\mathbf{\Xi}}}=\ddot{\mathbf{D}}_{\mathbf{\Xi}}\ddot{\mathbf{D}}_{\mathbf{\Xi}}^{\dagger}$, and (\ref{theo3prof34}) is because $\|\mathbf{P}_{\ddot{\mathbf{D}}_{\mathbf{\Xi}}}\|_2\leq1$.

In the case that $s>1$, we have $\mathbf{\Xi}\subseteq\mathbf{\Xi}^k$. Letting $\mathcal{Q}^k$ be the best $k$ block-sparse approximation of $\mathcal{X}^k$, we obtain
\begin{align} % eqs.218-223
	&\|\mathcal{Q}^k-\mathcal{X}\|_F\nonumber\\
	 &= \|\mathcal{Q}^k-\mathcal{X}^k+\mathcal{X}^k-\mathcal{X}\|_F \nonumber\\
	 & \leq \|\mathcal{Q}^k-\mathcal{X}^k\|_F+\|\mathcal{X}^k-\mathcal{X}\|_F \label{theo3prof21} \\
	&\leq 2\|\mathcal{X}^k-\mathcal{X}\|_F \label{theo3prof22} \\
	&= 2\big\|\ddot{\mathbf{D}}_{\mathbf{\Xi}^k}^{\dagger}{\rm vec}(\mathcal{Y})-{\rm vec}(\mathcal{X}_{\mathbf{\Xi}^k})\big\|_F \label{theo3prof23} \\
	&= 2\big\|\ddot{\mathbf{D}}_{\mathbf{\Xi}^k}^{\dagger}{\rm vec}(\mathcal{X}_{\mathbf{\Xi}}\times_1\mathbf{D}_{1_{\mathbf{\Xi}_1}}\times_2\cdots\times_n\mathbf{D}_{n_{\mathbf{\Xi}_n}}+\mathcal{N})\nonumber\\
	&\hspace{2em}-{\rm vec}(\mathcal{X}_{\mathbf{\Xi}^k})\big\|_F \label{theo3prof24} \\
	&= 2\big\|\ddot{\mathbf{D}}_{\mathbf{\Xi}^k}^{\dagger}({\rm vec}(\mathcal{X}_{\mathbf{\Xi}^k}\times_1\mathbf{D}_{1_{\mathbf{\Xi}^k_1}}\times_2\cdots\times_n\mathbf{D}_{n_{\mathbf{\Xi}^k_n}}+\mathcal{N})\nonumber\\
	&\hspace{2em}-\mathbf{T}_k{\rm vec}(\mathcal{X}_{\mathbf{\Xi}^k}))\big\|_F \nonumber \\
	&= 2\big\|\ddot{\mathbf{D}}_{\mathbf{\Xi}^k}^{\dagger}{\rm vec}(\mathcal{N})\big\|_F 
	  \leq \frac{2\|\ddot{\mathbf{D}}_{\mathbf{\Xi}^k}\ddot{\mathbf{D}}_{\mathbf{\Xi}^k}^{\dagger}{\rm vec}(\mathcal{N})\|_F}{\underline{W}^{\frac{1}{2}}_{\mathbf{\Upsilon},sk}} \label{theo3prof25} \\
	&= \frac{2\|\mathbf{P}_{\ddot{\mathbf{D}}_{\mathbf{\Xi}^k}}{\rm vec}(\mathcal{N})\|_F}{\underline{W}^{\frac{1}{2}}_{\mathbf{\Upsilon},sk}} 
	  \leq\frac{2\|\mathcal{N}\|_F}{\underline{W}^{\frac{1}{2}}_{\mathbf{\Upsilon},sk}},\label{the3lower55}
\end{align}
where (\ref{theo3prof21}) is from the triangle inequality, (\ref{theo3prof22}) is because $\mathcal{Q}^k$ is the best $k$ block-sparse approximation to $\mathcal{X}^k$, (\ref{theo3prof23}) is because the nonzero tensor blocks of $\mathcal{X}^k$ are estimated by $\ddot{\mathbf{D}}_{\mathbf{\Xi}^k}^{\dagger}{\rm vec}(\mathcal{Y})$ and $\mathbf{\Xi}\subseteq\mathbf{\Xi}^k$, (\ref{theo3prof24}) is based on (\ref{tensorCSmodel}), and (\ref{theo3prof25}) is derived by \textbf{Corollary \ref{Corollary6}}.

From Appendix \ref{proofoftheo1}, we know that
\begin{align} % eq.224
  \|\mathcal{Q}^{k}-\mathcal{X}\|_F\geq\frac{\underline{W}^{\frac{1}{2}}_{\mathbf{\Upsilon},\prod_{t=1}^{n}k_t+k}\|\hat{\mathcal{X}}-\mathcal{X}\|_F-2\|\mathcal{N}\|_F}{\overline{W}^{\frac{1}{2}}_{\mathbf{\Upsilon},\prod_{t=1}^{n}k_t+k}} .\label{the3lower5}
\end{align}
The proof is completed by combining (\ref{the3lower55}) and (\ref{the3lower5}).
\end{IEEEproof}	

\subsection{Proof of \textbf{Lemma~\ref{lemma1}}}\label{proofoflemma1} % Ap-L

\begin{IEEEproof}	
From (\ref{proofoftheo522}), for any integer $c$ with $c\geq l$, we have
\begin{align} % eq.225
 	&\big\|\mathcal{R}^c\big\|^2_F - \big\|\mathcal{R}^{c+1}\big\|^2_F\nonumber\\
 	&	\geq \frac{\sum_{(i_1,i_2,\cdots,i_n)\in\mathbf{\Theta}^{c+1}}
 		\big\|\mathcal{R}^c\times_1\mathbf{D}^{\rm H}_{1_{[i_1]}}\times_2\cdots\times_n\mathbf{D}^{\rm H}_{n_{[i_n]}}\big\|^2_F}{\overline{W}_{\mathbf{\Upsilon},s}} . \label{rrres}
\end{align}
 	
Denote $\mathcal{Z}\in\mathbb{C}^{N_1\times_1\cdots\times_nN_n}$ as a tensor with ${\rm vec}(\mathcal{Z}_{\mathbf{\Xi}\cap\mathbf{\Xi}^l\cup\mathbf{\Lambda}^l_{\varrho}})={\rm vec}(\mathcal{X}_{\mathbf{\Xi}\cap\mathbf{\Xi}^l\cup\mathbf{\Lambda}^l_{\varrho}})$ and ${\rm vec}(\mathcal{Z}_{\mathbf{\Omega}\backslash(\mathbf{\Xi}\cap\mathbf{\Xi}^l\cup\mathbf{\Lambda}^l_{\varrho}})=\mathbf{0}$.
From \cite[C. 12]{jWang2016tsp}, we have
\begin{align} % eq.226
 	&\big\langle\big(\mathbf{D}^{\rm H}_n\otimes\cdots\otimes\mathbf{D}^{\rm H}_1\big) {\rm vec}(\mathcal{R}^c), {\rm vec}(\mathcal{Z})\big\rangle \nonumber\\
 	&\leq \bigg(\bigg\lceil\frac{|\mathbf{\Lambda}^l_{\varrho}|}{s}\bigg\rceil\bigg)^{\frac{1}{2}}\|\ddot{\mathbf{D}}_{\mathbf{\Theta}^{c+1}}^{\rm H}{\rm vec}(\mathcal{R}^c)\big\|_F \big\|{\rm vec}(\mathcal{Z}_{\mathbf{\Omega}\backslash\mathbf{\Xi}^c})\big\|_F . \label{jWang2016tspc12prof}
\end{align}
On the other hand,
\begin{align} % eqs.227-229
 	& \big\langle\big(\mathbf{D}^{\rm H}_n\otimes\cdots\otimes\mathbf{D}^{\rm H}_1\big) {\rm vec}(\mathcal{R}^c),{\rm vec}(\mathcal{Z})\big\rangle \nonumber\\
 		&\geq \big\|\big(\mathbf{D}_n\otimes\cdots\otimes\mathbf{D}_1\big) \big({\rm vec}(\mathcal{Z})-{\rm vec}\big(\hat{\mathcal{X}}^c\big)\big)\big\|_F \nonumber \\
 	& \quad \times \Big(\big\|{\rm vec}(\mathcal{R}^c)\big\|^2_F \nonumber\\
 	&\hspace{2.5em}- \big\|\big(\mathbf{D}_{n_{\mathbf{\Lambda}_n^l\backslash\mathbf{\Lambda}^l_{n_{{\varrho}}}}}\otimes\cdots\otimes\mathbf{D}_{1_{\mathbf{\Lambda}_1^l\backslash\mathbf{\Lambda}^l_{1_{\varrho}}}}\big) {\rm vec}\big(\mathcal{X}_{\mathbf{\Lambda}^l\backslash\mathbf{\Lambda}^l_{\varrho}}\big)\nonumber\\
 	&\hspace{2.5em} + {\rm vec}(\mathcal{N})\big\|^2_F \Big)^{\frac{1}{2}} \label{profoflemma411} \\
 	& \geq \underline{W}^{\frac{1}{2}}_{\mathbf{\Upsilon},|\mathbf{\Lambda}_{\varrho}^l\cup\mathbf{\Xi}^c|}\big\|{\rm vec}(\mathcal{Z})-{\rm vec}(\hat{\mathcal{X}}^c)\big\|_F \nonumber \\
 	& \quad \times \Big(\big\|{\rm vec}(\mathcal{R}^c)\big\|^2_F\nonumber\\
 	&\hspace{2.5em} - \big\|\big(\mathbf{D}_{n_{\mathbf{\Lambda}_n^l\backslash\mathbf{\Lambda}^l_{n_{{\varrho}}}}}\otimes\cdots\otimes\mathbf{D}_{1_{\mathbf{\Lambda}_1^l\backslash\mathbf{\Lambda}^l_{1_{\varrho}}}}\big) {\rm vec}\big(\mathcal{X}_{\mathbf{\Lambda}^l\backslash\mathbf{\Lambda}^l_{\varrho}}\big) \nonumber\\
 	&\hspace{2.5em}+ {\rm vec}(\mathcal{N})\big\|^2_F\Big)^{\frac{1}{2}} \label{profoflemma412} \\
 	& \geq \underline{W}^{\frac{1}{2}}_{\mathbf{\Upsilon},|\mathbf{\Lambda}_{\varrho}^l\cup\mathbf{\Xi}^c|} \big\|{\rm vec}\big(\mathcal{Z}_{\mathbf{\Omega}\backslash\mathbf{\Xi}^c}-\hat{\mathcal{X}}_{\mathbf{\Omega}\backslash\mathbf{\Xi}^c}^c\big)\big\|_F \nonumber \\
 	& \quad \times \Big(\big\|{\rm vec}(\mathcal{R}^c)\big\|^2_F \nonumber\\
 	&\hspace{2.5em}- \big\|\big(\mathbf{D}_{n_{\mathbf{\Lambda}_n^l\backslash\mathbf{\Lambda}^l_{n_{{\varrho}}}}}\otimes\cdots\otimes\mathbf{D}_{1_{\mathbf{\Lambda}_1^l\backslash\mathbf{\Lambda}^l_{1_{\varrho}}}}\big) {\rm vec}\big(\mathcal{X}_{\mathbf{\Lambda}^l\backslash\mathbf{\Lambda}^l_{\varrho}}\big) \nonumber\\
 	&\hspace{2.5em}+ {\rm vec}(\mathcal{N}) \big\|^2_F\Big)^{\frac{1}{2}} \nonumber \\
 	& \geq \underline{W}^{\frac{1}{2}}_{\mathbf{\Upsilon},|\mathbf{\Lambda}_{\varrho}^l\cup\mathbf{\Xi}^c|}\big\|{\rm vec}(\mathcal{Z}_{\mathbf{\Omega}\backslash\mathbf{\Xi}^c})\big\|_F \nonumber \\
 	& \quad \times \Big(\big\|{\rm vec}(\mathcal{R}^c)\big\|^2_F \nonumber\\
 	&\hspace{2.5em}- \big\|\big(\mathbf{D}_{n_{\mathbf{\Lambda}_n^l\backslash\mathbf{\Lambda}^l_{n_{{\varrho}}}}}\otimes\cdots\otimes\mathbf{D}_{1_{\mathbf{\Lambda}_1^l\backslash\mathbf{\Lambda}^l_{1_{\varrho}}}}\big) {\rm vec}\big(\mathcal{X}_{\mathbf{\Lambda}^l\backslash\mathbf{\Lambda}^l_{\varrho}}\big)\nonumber\\
 	&\hspace{2.5em} + {\rm vec}(\mathcal{N})\big\|^2_F\Big)^{\frac{1}{2}} , \label{profoflemma414}
 	\end{align}
where (\ref{profoflemma411}) is based on \cite[C. 13]{jWang2016tsp}, (\ref{profoflemma412}) is from \textbf{Corollary~\ref{Corollary6}}, and (\ref{profoflemma414}) is because ${\rm vec}(\hat{\mathcal{X}}_{\mathbf{\Omega}\backslash\mathbf{\Xi}^c}^c)=\mathbf{0}$. Combining (\ref{jWang2016tspc12prof}) and (\ref{profoflemma414}) yields
\begin{align} % eq.230
 	&\big\|\ddot{\mathbf{D}}_{\mathbf{\Theta}^{c+1}}^{\rm H}{\rm vec}(\mathcal{R}^c)\big\|_F \nonumber\\
 	&\geq \frac{\underline{W}^{\frac{1}{2}}_{\mathbf{\Upsilon},|\mathbf{\Lambda}_{\varrho}^l\cup\mathbf{\Xi}^c|}}{\Big(\Big\lceil\frac{|\mathbf{\Lambda}^l_{\varrho}|}{s}\Big\rceil\Big)^{\frac{1}{2}}}
 	 \bigg(\big\|{\rm vec}(\mathcal{R}^c)\big\|^2_F \nonumber \\
 	&\hspace{4em} -\big\|\big(\mathbf{D}_{n_{\mathbf{\Lambda}_n^l\backslash\mathbf{\Lambda}^l_{n_{{\varrho}}}}}\otimes\cdots\otimes\mathbf{D}_{1_{\mathbf{\Lambda}_1^l\backslash\mathbf{\Lambda}^l_{1_{\varrho}}}}\big) {\rm vec}\big(\mathcal{X}_{\mathbf{\Lambda}^l\backslash\mathbf{\Lambda}^l_{\varrho}}\big) \nonumber\\
 	&\hspace{4em}+ {\rm vec}(\mathcal{N})\big\|^2_F\bigg)^{\frac{1}{2}} . \label{comyields}
\end{align}
Based on (\ref{rrres}) and (\ref{comyields}) we have
\begin{align} % eq.231
	&\big\|\mathcal{R}^c\big\|^2_F - \big\|\mathcal{R}^{c+1}\big\|^2_F \nonumber\\
	&\geq \frac{\underline{W}_{\mathbf{\Upsilon},|\mathbf{\Lambda}_{\varrho}^l\cup\mathbf{\Xi}^c|}}{\Big\lceil\frac{|\mathbf{\Lambda}^l_{\varrho}|}{s}\Big\rceil\overline{W}_{\mathbf{\Upsilon},s}}
	  \bigg(\big\|{\rm vec}(\mathcal{R}^c)\big\|^2_F \nonumber \\
	& \hspace{4em}- \big\|\big(\mathbf{D}_{n_{\mathbf{\Lambda}_n^l\backslash\mathbf{\Lambda}^l_{n_{{\varrho}}}}}\otimes\cdots\otimes\mathbf{D}_{1_{\mathbf{\Lambda}_1^l\backslash\mathbf{\Lambda}^l_{1_{\varrho}}}}\big) {\rm vec}\big(\mathcal{X}_{\mathbf{\Lambda}^l\backslash\mathbf{\Lambda}^l_{\varrho}}\big)\nonumber\\
	&\hspace{4em} + {\rm vec}(\mathcal{N})\big\|^2_F\bigg) . \label{maincomyields}
\end{align}
Then, for $c^*\in\{c,\cdots,c+\Delta c-1\}$, we have
\begin{align} % eqs.232,233
	&\big\|\mathcal{R}^{c^*}\big\|^2_F - \big\|\mathcal{R}^{c^*+1}\big\|^2_F \nonumber\\
	&\geq \frac{\underline{W}_{\mathbf{\Upsilon},|\mathbf{\Lambda}_{\varrho}^l\cup\mathbf{\Xi}^{c^*}|}}{\Big\lceil\frac{|\mathbf{\Lambda}^l_{\varrho}|}{s}\Big\rceil\overline{W}_{\mathbf{\Upsilon},s}}
	  \bigg(\big\|\big({\rm vec}(\mathcal{R}^{c^*})\big\|^2_F \nonumber \\
	&\qquad\qquad -\big\|\big(\mathbf{D}_{n_{\mathbf{\Lambda}_n^l\backslash\mathbf{\Lambda}^l_{n_{{\varrho}}}}}\otimes\cdots\otimes\mathbf{D}_{1_{\mathbf{\Lambda}_1^l\backslash\mathbf{\Lambda}^l_{1_{\varrho}}}}\big) {\rm vec}\big(\mathcal{X}_{\mathbf{\Lambda}^l\backslash\mathbf{\Lambda}^l_{\varrho}}\big)\nonumber\\
	&\qquad\qquad + {\rm vec}(\mathcal{N})\big\|^2_F\bigg) \label{rrr1} \\
  \geq& \Bigg(1-\exp\bigg(-\frac{\underline{W}_{\mathbf{\Upsilon},|\mathbf{\Lambda}_{\varrho}^l\cup\mathbf{\Xi}^{c^*}|}}{\Big\lceil\frac{|\mathbf{\Lambda}^l_{\varrho}|}{s}\Big\rceil\overline{W}_{\mathbf{\Upsilon},s}}\bigg)\Bigg)
	  \bigg(\|{\rm vec}(\mathcal{R}^{c^*})\|^2_F \nonumber \\
	& -\big\|\big(\mathbf{D}_{n_{\mathbf{\Lambda}_n^l\backslash\mathbf{\Lambda}^l_{n_{{\varrho}}}}}\otimes\cdots\otimes\mathbf{D}_{1_{\mathbf{\Lambda}_1^l\backslash\mathbf{\Lambda}^l_{1_{\varrho}}}}\big) {\rm vec}\big(\mathcal{X}_{\mathbf{\Lambda}^l\backslash\mathbf{\Lambda}^l_{\varrho}}\big) \nonumber\\
	&+ {\rm vec}(\mathcal{N})\big\|^2_F\bigg) \nonumber \\
	\geq& \Bigg(1-\exp\bigg(-\frac{\underline{W}_{\mathbf{\Upsilon},|\mathbf{\Lambda}_{\varrho}^l\cup\mathbf{\Xi}^{c+\Delta c-1}|}}{\Big\lceil\frac{|\mathbf{\Lambda}^l_{\varrho}|}{s}\Big\rceil\overline{W}_{\mathbf{\Upsilon},s}}\bigg)\Bigg)
	  \bigg(\|{\rm vec}(\mathcal{R}^{c^*})\|^2_F\nonumber\\
	& -\big\|\big(\mathbf{D}_{n_{\mathbf{\Lambda}_n^l\backslash\mathbf{\Lambda}^l_{n_{{\varrho}}}}}\otimes\cdots\otimes\mathbf{D}_{1_{\mathbf{\Lambda}_1^l\backslash\mathbf{\Lambda}^l_{1_{\varrho}}}}\big) {\rm vec}\big(\mathcal{X}_{\mathbf{\Lambda}^l\backslash\mathbf{\Lambda}^l_{\varrho}}\big) \nonumber\\
	&+ {\rm vec}(\mathcal{N})\big\|^2_F\bigg) , \label{rrr3}
\end{align}
where (\ref{rrr1}) is from (\ref{maincomyields}), and (\ref{rrr3}) is due to the monotonic decreasing property of $\underline{W}_{\mathbf{\Upsilon},t}$ with respect to $t$. Subtract both sides of (\ref{rrr3}) by $\big\|{\rm vec}(\mathcal{R}^{c^*})\big\|^2_F - \big\|\big(\mathbf{D}_{n_{\mathbf{\Lambda}_n^l\backslash\mathbf{\Lambda}^l_{n_{{\varrho}}}}}\otimes\cdots\otimes\mathbf{D}_{1_{\mathbf{\Lambda}_1^l\backslash\mathbf{\Lambda}^l_{1_{\varrho}}}}\big) {\rm vec}\big(\mathcal{X}_{\mathbf{\Lambda}^l\backslash\mathbf{\Lambda}^l_{\varrho}}\big) + {\rm vec}(\mathcal{N})\big\|^2_F$, we obtain
\begin{align} % eq.234
	& \big\|\mathcal{R}^{c^*+1}\big\|^2_F\nonumber\\
	& - \big\|\big(\mathbf{D}_{n_{\mathbf{\Lambda}_n^l\backslash\mathbf{\Lambda}^l_{n_{{\varrho}}}}}\otimes\cdots\otimes\mathbf{D}_{1_{\mathbf{\Lambda}_1^l\backslash\mathbf{\Lambda}^l_{1_{\varrho}}}}\big) {\rm vec}\big(\mathcal{X}_{\mathbf{\Lambda}^l\backslash\mathbf{\Lambda}^l_{\varrho}}\big) + {\rm vec}(\mathcal{N})\big\|^2_F \nonumber\\
	&\leq \exp\bigg(-\frac{\underline{W}_{\mathbf{\Upsilon},|\mathbf{\Lambda}_{\varrho}^l\cup\mathbf{\Xi}^{c+\Delta c-1}|}}{\Big\lceil\frac{|\mathbf{\Lambda}^l_{\varrho}|}{s}\Big\rceil\overline{W}_{\mathbf{\Upsilon},s}}\bigg) \nonumber \\
	& \quad \times \Big(\big\|{\rm vec}(\mathcal{R}^{c^*})\big\|^2_F\nonumber\\
	 &\hspace{3em} - \big\|\big(\mathbf{D}_{n_{\mathbf{\Lambda}_n^l\backslash\mathbf{\Lambda}^l_{n_{{\varrho}}}}}\otimes\cdots\otimes\mathbf{D}_{1_{\mathbf{\Lambda}_1^l\backslash\mathbf{\Lambda}^l_{1_{\varrho}}}}\big) {\rm vec}\big(\mathcal{X}_{\mathbf{\Lambda}^l\backslash\mathbf{\Lambda}^l_{\varrho}}\big)\nonumber\\
	 &\hspace{3em} + {\rm vec}(\mathcal{N})\big\|^2_F\Big) . \label{plugg1}
\end{align}
Substituting $c^*=c,\cdots,c+\Delta c-1$ into (\ref{plugg1}), we have
\begin{align} % eqs.235-238
	& \big\|\mathcal{R}^{c+1}\big\|^2_F\nonumber\\
	& - \big\|\big(\mathbf{D}_{n_{\mathbf{\Lambda}_n^l\backslash\mathbf{\Lambda}^l_{n_{{\varrho}}}}}\otimes\cdots\otimes\mathbf{D}_{1_{\mathbf{\Lambda}_1^l\backslash\mathbf{\Lambda}^l_{1_{\varrho}}}}\big) {\rm vec}\big(\mathcal{X}_{\mathbf{\Lambda}^l\backslash\mathbf{\Lambda}^l_{\varrho}}\big) + {\rm vec}(\mathcal{N})\big\|^2_F \nonumber\\
	&\leq \exp\bigg(-\frac{\underline{W}_{\mathbf{\Upsilon},|\mathbf{\Lambda}_{\varrho}^l\cup\mathbf{\Xi}^{c+\Delta c-1}|}}{\Big\lceil\frac{|\mathbf{\Lambda}^l_{\varrho}|}{s}\Big\rceil\overline{W}_{\mathbf{\Upsilon},s}}\bigg) \nonumber \\
	& \quad \times\Big(\big\|{\rm vec}(\mathcal{R}^{c})\big\|^2_F\nonumber\\
	& \hspace{3em}  -\big\|\big(\mathbf{D}_{n_{\mathbf{\Lambda}_n^l\backslash\mathbf{\Lambda}^l_{n_{{\varrho}}}}}\otimes\cdots\otimes\mathbf{D}_{1_{\mathbf{\Lambda}_1^l\backslash\mathbf{\Lambda}^l_{1_{\varrho}}}}\big) {\rm vec}\big(\mathcal{X}_{\mathbf{\Lambda}^l\backslash\mathbf{\Lambda}^l_{\varrho}}\big) \nonumber\\
	&\hspace{3em} + {\rm vec}(\mathcal{N})\big\|^2_F\Big) , \label{plugg2} \\
	& \big\|\mathcal{R}^{c+2}\big\|^2_F \nonumber\\
	&- \big\|\big(\mathbf{D}_{n_{\mathbf{\Lambda}_n^l\backslash\mathbf{\Lambda}^l_{n_{{\varrho}}}}}\otimes\cdots\otimes\mathbf{D}_{1_{\mathbf{\Lambda}_1^l\backslash\mathbf{\Lambda}^l_{1_{\varrho}}}}\big) {\rm vec}\big(\mathcal{X}_{\mathbf{\Lambda}^l\backslash\mathbf{\Lambda}^l_{\varrho}}\big) + {\rm vec}(\mathcal{N})\big\|^2_F \nonumber\\
	&\leq \exp\bigg(-\frac{\underline{W}_{\mathbf{\Upsilon},|\mathbf{\Lambda}_{\varrho}^l\cup\mathbf{\Xi}^{c+\Delta c-1}|}}{\Big\lceil\frac{|\mathbf{\Lambda}^l_{\varrho}|}{s}\Big\rceil\overline{W}_{\mathbf{\Upsilon},s}}\bigg) \nonumber\\
	&\quad \times\Big(\big\|{\rm vec}(\mathcal{R}^{c+1})\big\|^2_F\nonumber\\
	 &\hspace{3em} - \big\|\big(\mathbf{D}_{n_{\mathbf{\Lambda}_n^l\backslash\mathbf{\Lambda}^l_{n_{{\varrho}}}}}\otimes\cdots\otimes\mathbf{D}_{1_{\mathbf{\Lambda}_1^l\backslash\mathbf{\Lambda}^l_{1_{\varrho}}}}\big) {\rm vec}\big(\mathcal{X}_{\mathbf{\Lambda}^l\backslash\mathbf{\Lambda}^l_{\varrho}}\big)\nonumber\\
	 &\hspace{3em}  + {\rm vec}(\mathcal{N})\big\|^2_F\Big) ,\label{plugg3} \\
	& \qquad\qquad\qquad \vdots \qquad\qquad , \nonumber \\
	& \big\|\mathcal{R}^{c+\Delta c}\big\|^2_F \nonumber\\
	&- \big\|\big(\mathbf{D}_{n_{\mathbf{\Lambda}_n^l\backslash\mathbf{\Lambda}^l_{n_{{\varrho}}}}}\otimes\cdots\otimes\mathbf{D}_{1_{\mathbf{\Lambda}_1^l\backslash\mathbf{\Lambda}^l_{1_{\varrho}}}}\big) {\rm vec}\big(\mathcal{X}_{\mathbf{\Lambda}^l\backslash\mathbf{\Lambda}^l_{\varrho}}\big) + {\rm vec}(\mathcal{N})\big\|^2_F\nonumber\\
	& \leq \exp\bigg(-\frac{\underline{W}_{\mathbf{\Upsilon},|\mathbf{\Lambda}_{\varrho}^l\cup\mathbf{\Xi}^{c+\Delta c-1}|}}{\Big\lceil\frac{|\mathbf{\Lambda}^l_{\varrho}|}{s}\Big\rceil\overline{W}_{\mathbf{\Upsilon},s}}\bigg) \nonumber\\
	& \quad \times \Big(\big\|{\rm vec}(\mathcal{R}^{c+\Delta c-1})\big\|^2_F \nonumber\\
	& \hspace{3em} - \big\|\big(\mathbf{D}_{n_{\mathbf{\Lambda}_n^l\backslash\mathbf{\Lambda}^l_{n_{{\varrho}}}}}\otimes\cdots\otimes\mathbf{D}_{1_{\mathbf{\Lambda}_1^l\backslash\mathbf{\Lambda}^l_{1_{\varrho}}}}\big) {\rm vec}\big(\mathcal{X}_{\mathbf{\Lambda}^l\backslash\mathbf{\Lambda}^l_{\varrho}}\big)\nonumber\\
		& \hspace{3em} + {\rm vec}(\mathcal{N})\big\|^2_F\Big) . \label{plugg4}
\end{align}
The desired result is obtained by combining the results of (\ref{plugg2}) to (\ref{plugg4}).
\end{IEEEproof}

\subsection{Proof of \textbf{Theorem~\ref{theo6}}}\label{proofoftheo6} % Ap-M

\begin{IEEEproof}	
We first present some definitions used in the proof. $\mathbf{\Lambda}^l=\mathbf{\Xi}\backslash\mathbf{\Xi}^l$ is the index set of the remaining support tensor blocks after $l$ iterations. Without loss of generality, assume that $\mathbf{\Lambda}^l=\{1,2,\cdots,|\mathbf{\Lambda}^l|\}$ corresponding to the nonzero blocks of ${\rm vec}(\mathcal{X})$, and ${\rm vec}(\mathcal{X})$ is arranged in descending order of their Frobenius norms, i.e., $\|{\rm vec}(\mathcal{X})_{[1]}\|_F\geq\|{\rm vec}(\mathcal{X})_{[2]}\|_F\geq\cdots\geq\|{\rm vec}(\mathcal{X})_{[|\mathbf{\Lambda}^l|]}\|_F$. The subset $\mathbf{\Lambda}^l_{\varrho}$ of $\mathbf{\Lambda}^l$ is given by
\begin{subequations} % eqs.239a,239b,239c
	\begin{numcases}{\mathbf{\Lambda}^l_{\varrho}=}		
		 \mathbf{\emptyset},\quad \varrho =0 , \label{lambdadef1} \\
		\{1,\cdots,\lceil \exp(\varrho-1)\rceil s\}, \nonumber\\ \qquad\varrho=1,\cdots,\max\Big\{0,\Big\lceil\log\frac{e|\mathbf{\Lambda}^l|}{s}\Big\rceil\Big\}, \label{lambdadef2} \\
		 \mathbf{\Lambda}^l, 
		 \hspace{0.55em}\varrho=\max\Big\{0,\Big\lceil\log\frac{e|\mathbf{\Lambda}^l|}{s}\Big\rceil\Big\}+1 . \label{lambdadef3}
	\end{numcases}
\end{subequations}
For a given set $\mathbf{\Lambda}^l$ and a constant $\eta\geq2$, let $E\in\Big\{1,2,\cdots,\max\Big\{0,\Big\lceil\log\frac{e|\mathbf{\Lambda}^l|}{s}\Big\rceil\Big\}+1\Big\}$ be a positive integer that
\begin{subequations} % eqs240a-240d
	\begin{numcases}{}	
		\big\|{\rm vec}(\mathcal{X}_{\mathbf{\Lambda}^l\backslash\mathbf{\Lambda}^l_0})\big\|^2_F < \eta \big\|{\rm vec}(\mathcal{X}_{\mathbf{\Lambda}^l\backslash\mathbf{\Lambda}^l_1})\big\|^2_F, \label{etadef1} \\
		\big\|{\rm vec}(\mathcal{X}_{\mathbf{\Lambda}^l\backslash\mathbf{\Lambda}^l_1})\big\|^2_F < \eta \big\|{\rm vec}(\mathcal{X}_{\mathbf{\Lambda}^l\backslash\mathbf{\Lambda}^l_2})\big\|^2_F, \label{etadef2} \\
		\qquad \vdots \qquad , \nonumber \\
		\big\|{\rm vec}(\mathcal{X}_{\mathbf{\Lambda}^l\backslash\mathbf{\Lambda}^l_{E-1}})\big\|^2_F \geq \eta \big\|{\rm vec}(\mathcal{X}_{\mathbf{\Lambda}^l\backslash\mathbf{\Lambda}^l_E})\big\|^2_F \label{etadef4} ,	
	\end{numcases}
\end{subequations}
It is worth noting that $E$ always exists according to \cite{jWang2016tsp}. Moreover, based on (\ref{etadef1})-(\ref{etadef4}), there exist two conclusions for $E\ge 2$ \cite{jWang2016tsp}:
\begin{align} % eqs.241,242
	&\big\|{\rm vec}\big(\mathcal{X}_{\mathbf{\Lambda}^l\backslash\mathbf{\Lambda}^l_{\varrho}}\big)\big\|^2_F \leq \eta^{E-1-\varrho}\big\|{\rm vec}\big(\mathcal{X}_{\mathbf{\Lambda}^l\backslash\mathbf{\Lambda}^l_{E-1}}\big)\big\|^2_F, \nonumber\\
	&\qquad\qquad\qquad\qquad\qquad\qquad \varrho =0,1,\cdots,E , \label{xiaoyu} \\
	&\hspace{5.5em}\big|\mathbf{\Lambda}^l\big| > \Big(\frac{2\eta-1}{2\eta-2}\Big)2^{E-2}s , \label{numberoflam}
\end{align}
where (\ref{numberoflam}) can be proved by considering the block structure in ${\rm vec}(\mathcal{X})$ and using the similar proof in \cite[Appendix B]{jWang2016tsp}.
Based on these preparations, we now turn to our proof of \textbf{Theorem~\ref{theo6}}, which is derived based on the mathematical induction. 

\emph{a)}~Consider the case in which $|\mathbf{\Lambda}^l|=0$. This indicates that all the support tensor blocks are selected such that
\begin{align} % eqs.243-247
	&\big\|\mathcal{R}^l\big\|_F \nonumber\\
	&= \big\|\mathcal{Y}-\hat{\mathcal{X}}^{l}\times_1\mathbf{D}_1\times_2\cdots\times_n\mathbf{D}_n\big\|_F \label{res1} \\
	&= \min\limits_{{\rm supp}(\mathcal{X})=\mathbf{\Xi}^{l}} \bigg\|{\rm vec}(\mathcal{Y}) - \sum_{(i_1,\cdots,i_n)\in\mathbf{\Xi}^{l}}\big(\mathbf{D}_{n_{[i_n]}}\otimes\cdots\otimes\mathbf{D}_{1_{[i_1]}}\big) \nonumber\\
	&\hspace{6em}\times{\rm vec}\big(\mathcal{X}_{[i_1,\cdots,i_n]}\big)\bigg \|_F\label{res2} \\
	&\leq \|\mathcal{Y}-\mathcal{X}\times_1\mathbf{D}_1\times_2\cdots\times_n\mathbf{D}_n\|_F \label{res3} \\
	&= \|\mathcal{N}\|_F \label{res4} \\
	&\leq \xi \|\mathcal{N}\|_F, \nonumber
\end{align}
where (\ref{res1}) is due to the definition of the residual, (\ref{res2}) is from Algorithm \ref{alg:T-GBOMP}, (\ref{res3}) is because $|\mathbf{\Lambda}^l|=0$, and (\ref{res4}) is based on (\ref{tensorCSmodel}).
	
\emph{b)}~Assume that the argument holds up to an integer $\theta-1$ $(\theta\geq1)$, i.e., whenever $|\mathbf{\Lambda}^l|\leq\theta-1$, the algorithm requires at most $l_E+\max\{|\mathbf{\Lambda}^l|-\lceil \exp(E-2)\rceil s\}$ additional iterations to choose all the support tensor blocks. 

\emph{c)}~Under this inductive assumption, we prove that if $|\mathbf{\Lambda}|=\theta$, the algorithm also needs at most $l_E+\max\{|\mathbf{\Lambda}^l|-\lceil \exp(E-2)\rceil s\}$ additional iterations to choose the remaining $\theta$ support tensor blocks in $\mathbf{\Xi}$.
	
Consider the case in which $E\geq2$. Observe that
\begin{align} % eqs.248-251
	&\|\mathcal{R}^{l_E}\|_F \nonumber\\
	&= \big\|\mathbf{P}^{\bot}_{\ddot{\mathbf{D}}_{\mathbf{\Xi}^{l_E}}} \big(\ddot{\mathbf{D}}_{\mathbf{\Xi}\backslash\mathbf{\Xi}^{l_E}}{\rm vec}\big(\mathcal{X}_{\mathbf{\Xi}\backslash\mathbf{\Xi}^{l_E}}\big) + {\rm vec}(\mathcal{N})\big)\big\|_F \nonumber \\
	&\geq \big\|\mathbf{P}^{\bot}_{\ddot{\mathbf{D}}_{\mathbf{\Xi}^{l_E}}}\ddot{\mathbf{D}}_{\mathbf{\Xi}\backslash\mathbf{\Xi}^{l_E}} {\rm vec}\big(\mathcal{X}_{\mathbf{\Xi}\backslash\mathbf{\Xi}^{l_E}}\big)\big\|_F\nonumber\\
	&\quad - \big\|\mathbf{P}^{\bot}_{\ddot{\mathbf{D}}_{\mathbf{\Xi}^{l_E}}}{\rm vec}(\mathcal{N})\big\|_F \label{theo5111} \\
	&\geq \sigma_{\min}\Big(\mathbf{P}^{\bot}_{\ddot{\mathbf{D}}_{\mathbf{\Xi}^{l_E}}}\ddot{\mathbf{D}}_{\mathbf{\Xi}\backslash\mathbf{\Xi}^{l_E}}\Big) \big\|{\rm vec}(\mathcal{X}_{\mathbf{\Xi}\backslash\mathbf{\Xi}^{l_E}})\big\|_F\nonumber\\
	&\quad - \big\|{\rm vec}(\mathcal{N})\big\|_F \label{theo512} \\
	&\geq \sigma_{\min}\big(\ddot{\mathbf{D}}_{\mathbf{\Xi}}\big) \big\|{\rm vec}\big(\mathcal{X}_{\mathbf{\Xi}\backslash\mathbf{\Xi}^{l_E}}\big)\big\|_F - \big\|{\rm vec}(\mathcal{N})\big\|_F \label{theo513} \\
	&\geq \underline{W}^{\frac{1}{2}}_{\mathbf{\Upsilon},k}\big\|{\rm vec}(\mathcal{X}_{\mathbf{\Lambda}^{l_E}})\big\|_F - \big\|{\rm vec}(\mathcal{N})\big\|_F , \label{theo514}
\end{align}
where (\ref{theo5111}) is based on the triangle inequality, (\ref{theo512}) is because $\big\|\mathbf{P}^{\bot}_{\ddot{\mathbf{D}}_{\mathbf{\Xi}^{l_E}}}{\rm vec}(\mathcal{N})\big\|_F\leq\|{\rm vec}(\mathcal{N})\|_F$, (\ref{theo513}) is from \cite[Lemma 5]{cai2011}, and (\ref{theo514}) is based on \textbf{Corollary \ref{Corollary6}}.

Let $l_0=l$ and $l_i=l+\alpha\sum_{\varrho=1}^{i}\Big\lceil\frac{|\mathbf{\Lambda}_{\varrho}^{l}|}{s}\Big\rceil$ for each of $i\in\{1,\cdots,E\}$. Based on \textbf{Lemma \ref{lemma1}} with $\varrho=i$, $c=l_{i-1}$ and $\Delta c=l_i=l_{i-1}$, we obtain
\begin{align} % eq.252
	& \big\|\mathcal{R}^{l_i}\big\|^2_F \nonumber\\
	&- \Big\|\big(\mathbf{D}_{n_{\mathbf{\Lambda}_n^l\backslash\mathbf{\Lambda}^l_{n_{{i}}}}}\otimes\cdots\otimes\mathbf{D}_{1_{\mathbf{\Lambda}_1^l\backslash\mathbf{\Lambda}^l_{1_{i}}}}\big) {\rm vec}\big(\mathcal{X}_{\mathbf{\Lambda}^l\backslash\mathbf{\Lambda}^l_{i}}\big) + {\rm vec}(\mathcal{N})\Big\|^2_F \nonumber \\
	&  \leq G_{\mathbf{\Upsilon},i,l_{i-1},l_i-l_{i-1}}
		\bigg(\big\|{\rm vec}\big(\mathcal{R}^{l_{i-1}}\big)\big\|^2_F \nonumber \\
	& \hspace{4em} -\Big\|\big(\mathbf{D}_{n_{\mathbf{\Lambda}_n^l\backslash\mathbf{\Lambda}^l_{n_{{i}}}}}\otimes\cdots\otimes\mathbf{D}_{1_{\mathbf{\Lambda}_1^l\backslash\mathbf{\Lambda}^l_{1_{i}}}}\big) {\rm vec}\big(\mathcal{X}_{\mathbf{\Lambda}^l\backslash\mathbf{\Lambda}^l_{i}}\big)\nonumber\\
	&\hspace{4em} + {\rm vec}(\mathcal{N})\Big\|^2_F\bigg) , \label{ginequ}
\end{align}
for each of $i\in\{1,\cdots,E\}$. Note that
\begin{align} % eq.253
	G_{\mathbf{\Upsilon},i,l_{i-1},l_i-l_{i-1}} &= \exp\bigg(-\frac{(l_i-l_{i-1})\underline{W}_{\mathbf{\Upsilon},|\mathbf{\Lambda}_{\varrho}^l\cup\mathbf{\Xi}^{l_i-1}|}}{\Big\lceil\frac{|\mathbf{\Lambda}^l_{\varrho}|}{s}\Big\rceil\overline{W}_{\mathbf{\Upsilon},s}}\bigg)\nonumber\\
	&	= \exp\Bigg(-\frac{\Big\lceil\frac{\alpha|\mathbf{\Lambda}^l_{\varrho}|}{s}\Big\rceil\underline{W}_{\mathbf{\Upsilon},|\mathbf{\Lambda}_{\varrho}^l\cup\mathbf{\Xi}^{l_i-1}|}}{\Big\lceil\frac{|\mathbf{\Lambda}^l_{\varrho}|}{s}\Big\rceil\overline{W}_{\mathbf{\Upsilon},s}}\Bigg) \nonumber \\
	&\leq \exp\bigg(-\frac{\alpha\underline{W}_{\mathbf{\Upsilon},|\mathbf{\Lambda}_{\varrho}^l\cup\mathbf{\Xi}^{l_i-1}|}}{\overline{W}_{\mathbf{\Upsilon},s}}\bigg), \nonumber
\end{align}
for each of $i \in \{1,2,\cdots,E\}$. Since
\begin{align} % eqs.254-256
	|\mathbf{\Lambda}_{\varrho}^l\cup\mathbf{\Xi}^{l_i-1}| &\leq |\mathbf{\Lambda}^l\cup\mathbf{\Xi}^{l_i-1}| \label{setlarge1} \\
	&\leq |\mathbf{\Xi}\cup\mathbf{\Xi}^{l_i-1}| \label{setlarge2} \\
	&\leq |\mathbf{\Xi}\cup\mathbf{\Xi}^{l_E}| , \label{setlarge3}
\end{align}
where (\ref{setlarge1}) and (\ref{setlarge2}) are from the definitions of $\mathbf{\Lambda}_{\varrho}^l$ and $\mathbf{\Lambda}^l$, respectively, and (\ref{setlarge3}) is because $l_i-1\leq l_E$.
Thus, we have
\begin{align} % eq.257
	G_{\mathbf{\Upsilon},i,l_{i-1},l_i-l_{i-1}} \leq \exp\bigg(-\frac{\alpha\underline{W}_{\mathbf{\Upsilon},|\mathbf{\Xi}\cup\mathbf{\Xi}^{l_E}|}}{\overline{W}_{\mathbf{\Upsilon},s}}\bigg). \label{gup}
\end{align}

From (\ref{lambdadef1}) to (\ref{lambdadef3}), we know that $|\mathbf{\Lambda}_{\varrho}^l|\leq \lceil \exp(\varrho-1)\rceil s$ for $\varrho=1,2,\cdots,E$. Hence we obtain
\begin{align} % eqs.258-262
	l_E &= l+\alpha\sum_{\varrho=1}^{E}\bigg\lceil\frac{|\mathbf{\Lambda}_{\varrho}^{l}|}{s}\bigg\rceil \label{llabel1} \\
	&\leq l+\alpha\sum_{\varrho=1}^{E}\big\lceil \exp(\varrho-1)\big\rceil \label{llabel2} \\
	&\leq l+ \alpha\sum_{\varrho=1}^{E}(\exp(\varrho-1)+1) \nonumber\\
	&  = l+\alpha\Big(\frac{\exp(E)-1}{e-1}+ E\Big) 
	  \leq l+\lceil \exp(E-2)\rceil s \label{llabel5} \\
	&\leq l+|\mathbf{\Lambda}^l| \label{llabel6} \\
	&\leq l+\theta, \label{llabel7}
\end{align}
where (\ref{llabel1}) follows from the definition of $l_E$, (\ref{llabel2}) is from (\ref{lambdadef1}) to (\ref{lambdadef3}), (\ref{llabel5}) is because $\alpha={\rm inf}\bigg\{x\in\Big\{1,\cdots,\max\Big\{0,\Big\lceil\log\frac{e|\mathbf{\Lambda}^l|}{s}\Big\rceil\Big\}+1\Big\},\frac{\exp(x-2) s (e-1)}{\exp(x-1) -1+(e-1)x}\bigg\}$, and (\ref{llabel6}) is from (\ref{lambdadef1}) to (\ref{lambdadef3}).
 Combining (\ref{gup}) and (\ref{llabel6}), we have
\begin{align} % eq.263
	G_{\mathbf{\Upsilon},i,l_{i-1},l_i-l_{i-1}} &\leq \exp\bigg(-\frac{\alpha\underline{W}_{\mathbf{\Upsilon},|\mathbf{\Xi}\cup\mathbf{\Xi}^{ l+\theta}|}}{\overline{W}_{\mathbf{\Upsilon},s}}\bigg)\nonumber\\
	 & \leq \exp\bigg(-\frac{\alpha\underline{W}_{\mathbf{\Upsilon},sl+s\theta+\theta}}{\overline{W}_{\mathbf{\Upsilon},s}}\bigg), \nonumber
\end{align}
since $|\mathbf{\Xi}\cup\mathbf{\Xi}^{ l+\theta}|=|\mathbf{\Xi}^{l+\theta}|+|\mathbf{\Lambda}^{l+\theta}|\leq s(l+\theta)+|\mathbf{\Lambda}^{l}|\leq sl+s\theta+\theta$.

Letting $\beta=\exp\Big(-\frac{\alpha\underline{W}_{\mathbf{\Upsilon},sl+s\theta+\theta}}{\overline{W}_{\mathbf{\Upsilon},s}}\Big)$, the following inequality holds:
\begin{align} % eqs.264-272
	&\big\|\mathcal{R}^{l_E}\big\|^2_F \nonumber\\
	&\leq \beta^E\big\|{\rm vec}(\mathcal{R}^{l_E})\big\|^2_F \nonumber \\
	& \quad+ (1-\beta) \sum_{\varrho=1}^{E} \beta^{E-\varrho} \Big\|\big(\mathbf{D}_{n_{\mathbf{\Lambda}_n^l\backslash\mathbf{\Lambda}^l_{n_{{\varrho}}}}}\otimes\cdots\otimes\mathbf{D}_{1_{\mathbf{\Lambda}_1^l\backslash\mathbf{\Lambda}^l_{1_{\varrho}}}}\big)\nonumber\\
	&\hspace{9em}\times {\rm vec}\big(\mathcal{X}_{\mathbf{\Lambda}^l\backslash\mathbf{\Lambda}^l_{\varrho}}\big) + {\rm vec}(\mathcal{N})\Big\|^2_F \label{rrrr1} \\
	&\leq \beta^E \big\|\big(\mathbf{D}_{n_{\mathbf{\Lambda}_n^l}}\otimes\cdots\otimes\mathbf{D}_{1_{\mathbf{\Lambda}_1^l}}\big) {\rm vec}(\mathcal{X}_{\mathbf{\Lambda}^l})+{\rm vec}(\mathcal{N})\big\|^2_F \nonumber \\
	&\quad + (1-\beta ) \sum_{\varrho=1}^{E} \beta^{E-\varrho} \Big\|\big(\mathbf{D}_{n_{\mathbf{\Lambda}_n^l\backslash\mathbf{\Lambda}^l_{n_{{\varrho}}}}}\otimes\cdots\otimes\mathbf{D}_{1_{\mathbf{\Lambda}_1^l\backslash\mathbf{\Lambda}^l_{1_{\varrho}}}}\big) \nonumber\\
	&\hspace{9em}\times{\rm vec}\big(\mathcal{X}_{\mathbf{\Lambda}^l\backslash\mathbf{\Lambda}^l_{\varrho}}\big) + {\rm vec}(\mathcal{N})\Big\|^2_F \label{rrrr2} \\
	&\leq \beta^E \Big((1+\gamma) \big\|\big(\mathbf{D}_{n_{\mathbf{\Lambda}_n^l}}\otimes\cdots\otimes\mathbf{D}_{1_{\mathbf{\Lambda}_1^l}}\big) {\rm vec}(\mathcal{X}_{\mathbf{\Lambda}^l})\big\|^2_F \nonumber\\
	&\quad+ (1+\gamma^{-1})\big\|{\rm vec}(\mathcal{N})\big\|^2_F\Big) 
	 + (1-\beta) \sum_{\varrho=1}^{E} \beta^{E-\varrho} \nonumber\\
	 &\quad\times\bigg( (1+\gamma)\big\|\big(\mathbf{D}_{n_{\mathbf{\Lambda}_n^l\backslash\mathbf{\Lambda}^l_{n_{{\varrho}}}}}\otimes\cdots\otimes\mathbf{D}_{1_{\mathbf{\Lambda}_1^l\backslash\mathbf{\Lambda}^l_{1_{\varrho}}}}\big) \nonumber\\
	&\hspace{2.5em}\times{\rm vec}\big(\mathcal{X}_{\mathbf{\Lambda}^l\backslash\mathbf{\Lambda}^l_{\varrho}}\big)\big\|_F + (1+\gamma^{-1}) \beta^{E-\varrho} \|{\rm vec}(\mathcal{N})\|^2_F\bigg) \label{rrrr3} \\
	&\leq \beta^E\Big((1+\gamma)\overline{W}_{\mathbf{\Upsilon},|\mathbf{\Lambda}^l|^n}\|{\rm vec}(\mathcal{X}_{\mathbf{\Lambda}^l})\|^2_F\nonumber\\
	&\hspace{3em}+(1+\gamma^{-1})\|{\rm vec}(\mathcal{N})\|^2_F\Big) \nonumber \\
	&\hspace{1em} + (1-\beta) \sum_{\varrho=1}^{E}\beta^{E-\varrho}\Big((1+\gamma)\overline{W}_{\mathbf{\Upsilon},|\mathbf{\Lambda}^l|^n}\|{\rm vec}(\mathcal{X}_{\mathbf{\Lambda}^l\backslash\mathbf{\Lambda}^l_{\varrho}})\|_F\nonumber\\
	 &\hspace{8em} + (1+\gamma^{-1})\beta^{E-\varrho}\|{\rm vec}(\mathcal{N})\|^2_F\Big) \label{rrrr4} \\
	=& \bigg(\beta^E(1+\gamma)\overline{W}_{\mathbf{\Upsilon},|\mathbf{\Lambda}^l|^n}\|{\rm vec}(\mathcal{X}_{\mathbf{\Lambda}^l})\|^2_F\nonumber\\
	&\quad+(1-\beta)\sum_{\varrho=1}^{E}\beta^{E-\varrho}(1+\gamma)\overline{W}_{\mathbf{\Upsilon},|\mathbf{\Lambda}^l|^n}\|{\rm vec}(\mathcal{X}_{\mathbf{\Lambda}^l\backslash\mathbf{\Lambda}^l_{\varrho}})\|_F\bigg) \nonumber \\
	& + \bigg(\beta^E(1+\gamma^{-1})\nonumber\\
	&\qquad+(1-\beta)\sum_{\varrho=1}^{E}\beta^{E-\varrho}(1+\gamma^{-1})\bigg)\|{\rm vec}(\mathcal{N})\|^2_F \label{rrrr5} \\
	&\leq \bigg(\eta^{E-1}\beta^E+(1-\beta)\sum_{\varrho=1}^{E}\eta^{E-1-\varrho}\beta^{E-\varrho}\bigg)\nonumber\\
	&\quad\times(1+\gamma)\overline{W}_{\mathbf{\Upsilon},|\mathbf{\Lambda}^l|^n}\|{\rm vec}(\mathcal{X}_{\mathbf{\Lambda}^l\backslash\mathbf{\Lambda}^l_{E-1}})\|_F \nonumber \\
	& \quad+ \bigg(\beta^E+(1-\beta)\sum_{\varrho=1}^{E}\beta^{E-\varrho}\bigg)(1+\gamma^{-1})\|{\rm vec}(\mathcal{N})\|^2_F \label{rrrr6} \\
	&< \bigg((\eta\beta)^{E}+\sum_{\varrho=0}^{E-1}(\eta\beta)^{\varrho}\bigg)\nonumber\\
	&\quad\times\eta^{-1}(1+\gamma)\overline{W}_{\mathbf{\Upsilon},|\mathbf{\Lambda}^l|^n}\|{\rm vec}(\mathcal{X}_{\mathbf{\Lambda}^l\backslash\mathbf{\Lambda}^l_{E-1}})\|_F \nonumber \\
	&\quad + \bigg(\beta^E+\sum_{\varrho=0}^{E-1}\beta^{\varrho}\bigg)(1+\gamma^{-1})\|{\rm vec}(\mathcal{N})\|^2_F \label{rrrr7} \\
  &< \bigg(\sum_{\varrho=0}^{\infty}(\eta\beta)^{\varrho}\bigg)\eta^{-1}(1+\gamma)\overline{W}_{\mathbf{\Upsilon},|\mathbf{\Lambda}^l|^n}\|{\rm vec}(\mathcal{X}_{\mathbf{\Lambda}^l\backslash\mathbf{\Lambda}^l_{E-1}})\|_F \nonumber \\
	&\quad + \bigg(\sum_{\varrho=0}^{\infty}\beta^{\varrho}\bigg)(1+\gamma^{-1})\|{\rm vec}(\mathcal{N})\|^2_F \label{rrrr8} \\
	&= \frac{1}{\eta(1-\eta\beta)}(1+\gamma)\overline{W}_{\mathbf{\Upsilon},\theta^n}\|{\rm vec}(\mathcal{X}_{\mathbf{\Lambda}^l\backslash\mathbf{\Lambda}^l_{E-1}})\|^2_F\nonumber\\
	 &\quad + \frac{1}{1-\beta}(1+\gamma^{-1})\|{\rm vec}(\mathcal{N})\|^2_F . \label{rrrr9}
\end{align}
Here (\ref{rrrr1}) is derived by first substituting $\beta$ into (\ref{ginequ}) to obtain
\begin{align} % eq.273
	&\big\|\mathcal{R}^{l_E}\big\|^2_F \leq \beta\big\|{\rm vec}(\mathcal{R}^{l})\big\|^2_ F\nonumber\\
	&+ (1-\beta)\big\|\big(\mathbf{D}_{n_{\mathbf{\Lambda}_n^l\backslash\mathbf{\Lambda}^l_{n_{{\varrho}}}}}\otimes\cdots\otimes\mathbf{D}_{1_{\mathbf{\Lambda}_1^l\backslash\mathbf{\Lambda}^l_{1_{\varrho}}}}\big) {\rm vec}\big(\mathcal{X}_{\mathbf{\Lambda}^l\backslash\mathbf{\Lambda}^l_{\varrho}}\big)\nonumber\\
	& + {\rm vec}(\mathcal{N})\big\|^2_F, \label{rrrr1because}
\end{align}
for each $i\in\{1,\cdots,E\}$, and (\ref{rrrr1}) is obtained after some additional calculations on (\ref{rrrr1because}), while (\ref{rrrr2}) is from \cite[Proposition 1]{jWang2016tsp}, (\ref{rrrr3}) is due to the fact that for $\gamma>0$, $\|\mathbf{x}+\mathbf{y}\|^2_F\leq(1+\gamma)\|\mathbf{x}\|^2_F+(1+\gamma^{-1})\|\mathbf{y}\|^2_F$ and $\mathbf{x}$, $\mathbf{y}$ are vectors, (\ref{rrrr4}) is based on \textbf{Corollary \ref{Corollary6}}, (\ref{rrrr5}) is from direct simplification, (\ref{rrrr6}) is because (\ref{xiaoyu}), (\ref{rrrr7}) is from $\beta<1$, (\ref{rrrr8}) is similar to the derivation of \cite[(49)]{jWang2016tsp}, and (\ref{rrrr9}) holds because $\eta\beta<1$.

Combining (\ref{theo514}) and (\ref{rrrr9}) yields
\begin{align} % eq.274
	&\|{\rm vec}(\mathcal{X}_{\mathbf{\Lambda}^{l_E}})\|_F \nonumber\\
	&\leq
	  \bigg(\frac{(1+\gamma)\overline{W}_{\mathbf{\Upsilon},\theta^n}}{\eta(1-\eta\beta)\underline{W}_{\mathbf{\Upsilon},k}}\bigg)^{\frac{1}{2}}\big\|{\rm vec}\big(\mathcal{X}_{\mathbf{\Lambda}^l\backslash\mathbf{\Lambda}^l_{E-1}}\big)\big\|_F \nonumber \\
	&\quad +\frac{1}{\underline{W}^{\frac{1}{2}}_{\mathbf{\Upsilon},k}}\bigg(1+\Big(\frac{1}{1-\beta}\big(1+\gamma^{-1}\big)\Big)^{\frac{1}{2}}\bigg)\|{\rm vec}(\mathcal{N})\|_F. \nonumber
\end{align}
By choosing an appropriate $\gamma$, we have $\Big(\frac{(1+\gamma)\overline{W}_{\mathbf{\Upsilon},\theta^n}}{\eta(1-\eta\beta)\underline{W}_{\mathbf{\Upsilon},k}}\Big)^{\frac{1}{2}}<1$.

Then, consider the following two cases. First, if
\begin{align} % eq.275
	&\frac{1}{\underline{W}^{\frac{1}{2}}_{\mathbf{\Upsilon},k}}\bigg(1+\Big(\frac{1}{1\! -\! \beta}\big(1+\gamma^{-1}\big)\Big)^{\frac{1}{2}}\bigg)\|{\rm vec}(\mathcal{N})\|_F\nonumber\\
	&<\Bigg(1-\bigg(\frac{(1\! +\! \gamma)\overline{W}_{\mathbf{\Upsilon},\theta^n}}{\eta(1\! -\! \eta\beta)\underline{W}_{\mathbf{\Upsilon},k}}\bigg)^{\frac{1}{2}}\Bigg)\|{\rm vec}(\mathcal{X}_{\mathbf{\Lambda}^l\backslash\mathbf{\Lambda}^l_{E-1}})\|_F, \nonumber
\end{align}
then we directly have
\begin{align} % eq.276
	\big\|\mathcal{X}_{\mathbf{\Lambda}^{l_E}}\big\|^2_F \leq \big\|\mathcal{X}_{\mathbf{\Lambda}^l\backslash\mathbf{\Lambda}^l_{E-1}}\big\|^2_F. \label{indicatelast}
\end{align}
Recall that $\mathbf{\Lambda}^{l_E}=\mathbf{\Xi}\backslash\mathbf{\Xi}^{l_E}$ and $l_E=l+\alpha\sum_{\varrho=1}^{E}\big\lceil\frac{|\mathbf{\Lambda}_{\varrho}^{l}|}{s}\big\rceil$. Then, (\ref{indicatelast}) indicates that $|\mathbf{\Lambda}^{l_E}|\leq|\mathbf{\Lambda}^l\backslash\mathbf{\Lambda}^l_{E-1}|=\theta-\lceil \exp(E-2)\rceil s$,
due to the similar analysis in \cite[Sec. V-B]{Kim2020tit}. Thus, there remain at most $\theta-\lceil \exp(E-2)\rceil s$ support tensor blocks after the $l_E$th iteration. Then, based on the induction hypothesis, we have
\begin{align} % eq.277
	\big\|\mathcal{R}^{l_E+\theta-\lceil \exp(E-2)\rceil s}\big\|_F \leq \xi\|\mathcal{N}\|_F. \label{rrxiaoyu1}
\end{align}
Note that
\begin{align} % eqs.278
	&l_E+\theta-\lceil \exp(E-2)\rceil s \nonumber\\
	&= l+\alpha \sum_{\varrho=1}^{E}\bigg\lceil\frac{|\mathbf{\Lambda}_{\varrho}^{l}|}{s}\bigg\rceil+\theta-\lceil \exp(E-2)\rceil s \nonumber \\ 
	&\leq l+\alpha\sum_{\varrho=1}^{E}\big\lceil \exp(\varrho-1) \big\rceil+\theta-\lceil \exp(E-2)\rceil s \nonumber \\
	&\leq l+\alpha\Big(\frac{1-\exp(E-1)}{1-e}+E\Big)+\theta-\lceil \exp(E-2)\rceil s\nonumber\\
	&  \leq l+\theta, \label{inf1}
\end{align}
where (\ref{inf1}) follows from that $\alpha={\rm inf}\bigg\{x\in\Big\{1,\cdots,\max\Big\{0,\Big\lceil\log\frac{e|\mathbf{\Lambda}^l|}{s}\Big\rceil\Big\}+1\Big\},\frac{\exp(x-2) s (e-1)}{\exp(x-1) -1+(e-1)x}\bigg\}$, and hence $\alpha\big(\frac{1-\exp(E-1)}{1-e}+E\big)-\lceil \exp(E-2)\rceil s\leq0$.
Thus, we have
\begin{align} % eqs.279,280
	\big\|\mathcal{R}^{l+\theta}\big\|_F &\leq \big\|\mathcal{R}^{l_E+\theta-\lceil \exp(E-2)\rceil s}\big\|_F \label{sest1} \\
	&\leq \xi\|\mathcal{N}\|_F. \label{sest2}
\end{align}
where (\ref{sest1}) is because the residual is non-increasing, and (\ref{sest2}) is from (\ref{rrxiaoyu1}).

Second, if
\begin{align} % eq.281
	&\frac{1}{\underline{W}^{\frac{1}{2}}_{\mathbf{\Upsilon},k}}\bigg(1+\Big(\frac{1}{1\! -\! \beta}\big(1+\gamma^{-1}\big)\Big)^{\frac{1}{2}}\bigg)\|{\rm vec}(\mathcal{N})\|_F\nonumber\\
	&\geq\Bigg(1-\bigg(\frac{(1\! +\! \gamma)\overline{W}_{\mathbf{\Upsilon},\theta^n}}{\eta(1\! -\! \eta\beta)\underline{W}_{\mathbf{\Upsilon},k}}\bigg)^{\frac{1}{2}}\Bigg)\|{\rm vec}(\mathcal{X}_{\mathbf{\Lambda}^l\backslash\mathbf{\Lambda}^l_{E-1}})\|_F,\label{sec1}
\end{align}
then we have
\begin{align} % eqs.282-285
	&\big\|\mathcal{R}^{l+\theta}\big\|_F \nonumber\\
	&\leq \big\|\mathcal{R}^{l_E}\big\|_F \label{rrrrr1} \\
	&\leq \bigg(\frac{1}{\eta(1-\eta\beta)}(1+\gamma)\overline{W}_{\mathbf{\Upsilon},\theta^n}\bigg)^{\frac{1}{2}}\big\|{\rm vec}\big(\mathcal{X}_{\mathbf{\Lambda}^l\backslash\mathbf{\Lambda}^l_{E-1}}\big)\big\|_F\nonumber\\
	 &\quad + \bigg(\frac{1}{1-\beta}(1+\gamma^{-1})\bigg)^{\frac{1}{2}}\|{\rm vec}(\mathcal{N})\|_F \label{rrrrr2} \\
	&\leq \Bigg(\frac{\Big(\frac{1}{\eta(1-\eta\beta)}(1+\gamma)\overline{W}_{\mathbf{\Upsilon},\theta^n}\Big)^{\frac{1}{2}}\Big(1+\Big(\frac{1}{1-\beta}\big(1+\gamma^{-1}\big)\Big)^{\frac{1}{2}}\Big)}{\underline{W}^{\frac{1}{2}}_{\mathbf{\Upsilon},k}\bigg(1-\Big(\frac{(1+\gamma)\overline{W}_{\mathbf{\Upsilon},\theta^n}}{\eta(1-\eta\beta)\underline{W}_{\mathbf{\Upsilon},k}}\Big)^{\frac{1}{2}}\bigg)} \nonumber \\ 
	& \quad\quad +\bigg(\frac{1}{1-\beta}(1+\gamma^{-1})\bigg)^{\frac{1}{2}}\Bigg)\| {\rm vec}(\mathcal{N})\|_F \label{rrrrr3} \\
	&\leq \xi\|\mathcal{N}\|_F, \label{eq.ApMa3}
\end{align}
where (\ref{rrrrr1}) is from (\ref{llabel7}), (\ref{rrrrr2}) is due to (\ref{rrrr9}), and (\ref{rrrrr3}) is because of (\ref{sec1}).

Now consider the case in which $E=1$. Similar to (\ref{theo514}), we have
\begin{align} % eq.286
	&\big\|\mathcal{R}^{l+1}\big\|_F\nonumber\\
	 &= \Big\|\mathbf{P}^{\bot}_{\ddot{\mathbf{D}}_{\mathbf{\Xi}^{l+1}}}\Big(\ddot{\mathbf{D}}_{\mathbf{\Xi}\backslash\mathbf{\Xi}^{l+1}} {\rm vec}\big(\mathcal{X}_{\mathbf{\Xi}\backslash\mathbf{\Xi}^{l+1}}\big) + {\rm vec}(\mathcal{N})\Big)\Big\|_F \nonumber \\
	&\geq \Big\|\mathbf{P}^{\bot}_{\ddot{\mathbf{D}}_{\mathbf{\Xi}^{l+1}}}\ddot{\mathbf{D}}_{\mathbf{\Xi}\backslash\mathbf{\Xi}^{l+1}} {\rm vec}\big(\mathcal{X}_{\mathbf{\Xi}\backslash\mathbf{\Xi}^{l+1}}\big)\Big\|_F - \Big\|\mathbf{P}^{\bot}_{\ddot{\mathbf{D}}_{\mathbf{\Xi}^{l+1}}}{\rm vec}(\mathcal{N}))\Big\|_F \nonumber \\
	&\geq \sigma_{\min}\Big(\mathbf{P}^{\bot}_{\ddot{\mathbf{D}}_{\mathbf{\Xi}^{l+1}}}\ddot{\mathbf{D}}_{\mathbf{\Xi}\backslash\mathbf{\Xi}^{l+1}}\Big)\big\|{\rm vec}\big(\mathcal{X}_{\mathbf{\Xi}\backslash\mathbf{\Xi}^{l+1}}\big)\big\|_F - \|{\rm vec}(\mathcal{N}))\|_F \nonumber \\
	&\geq \sigma_{\min}\big(\ddot{\mathbf{D}}_{\mathbf{\Xi}}\big) \big\|{\rm vec}\big(\mathcal{X}_{\mathbf{\Xi}\backslash\mathbf{\Xi}^{l+1}}\big)\big\|_F - \|{\rm vec}(\mathcal{N}))\|_F \nonumber \\
	&\geq \underline{W}^{\frac{1}{2}}_{\mathbf{\Upsilon},k}\big\|{\rm vec}\big(\mathcal{X}_{\mathbf{\Lambda}^{l+1}}\big)\big\|_F - \|{\rm vec}(\mathcal{N}))\|_F . \label{theo51115}
\end{align}
Then, from (\ref{maincomyields}), we obtain
\begin{align} % eq.287
	&\big\|\mathcal{R}^l\big\|^2_F - \big\|\mathcal{R}^{l+1}\big\|^2_F \nonumber\\
	&\geq \frac{\underline{W}_{\mathbf{\Upsilon},|\mathbf{\Lambda}_{\varrho}^l\cup\mathbf{\Xi}^l|}}{\Big\lceil\frac{|\mathbf{\Lambda}^l_{\varrho}|}{s}\Big\rceil\overline{W}_{\mathbf{\Upsilon},s}}
	  \bigg(\big\|{\rm vec}(\mathcal{R}^l)\big\|^2_F \nonumber \\
	& \quad-\Big\|\big(\mathbf{D}_{n_{\mathbf{\Lambda}_n^l\backslash\mathbf{\Lambda}^l_{n_{{\varrho}}}}}\otimes\cdots\otimes\mathbf{D}_{1_{\mathbf{\Lambda}_1^l\backslash\mathbf{\Lambda}^l_{1_{\varrho}}}}\big) {\rm vec}\big(\mathcal{X}_{\mathbf{\Lambda}^l\backslash\mathbf{\Lambda}^l_{\varrho}}\big)\nonumber\\
	&\hspace{3em} + {\rm vec}(\mathcal{N})\Big\|^2_F\bigg) \nonumber \\
  &= \frac{\underline{W}_{\mathbf{\Upsilon},|\mathbf{\Lambda}_{\varrho}^l\cup\mathbf{\Xi}^l|}}{\overline{W}_{\mathbf{\Upsilon},s}}
	  \bigg(\big\|{\rm vec}(\mathcal{R}^l)\big\|^2_F \nonumber \\
	&\quad -\Big\|\big(\mathbf{D}_{n_{\mathbf{\Lambda}_n^l\backslash\mathbf{\Lambda}^l_{n_{{\varrho}}}}}\otimes\cdots\otimes\mathbf{D}_{1_{\mathbf{\Lambda}_1^l\backslash\mathbf{\Lambda}^l_{1_{\varrho}}}}\big) {\rm vec}\big(\mathcal{X}_{\mathbf{\Lambda}^l\backslash\mathbf{\Lambda}^l_{\varrho}}\big) \nonumber\\
	&\hspace{3em}+ {\rm vec}(\mathcal{N})\Big\|^2_F\bigg), \label{theo5146}
\end{align}
where (\ref{theo5146}) is because $\Big\lceil\frac{|\mathbf{\Lambda}^l_{\varrho}|}{s}\Big\rceil\leq1$. Thus, for the same $\gamma$ as given in (\ref{rrrr3}), we have
\begin{align} % eqs.288-293
	&\big\|\mathcal{R}^{l+1}\big\|^2_F \nonumber\\
	&\leq \Big(1-\frac{\underline{W}_{\mathbf{\Upsilon},|\mathbf{\Lambda}_{1}^l\cup\mathbf{\Xi}^l|}}{\overline{W}_{\mathbf{\Upsilon},s}}\Big)\big\|\mathcal{R}^{l}\big\|^2_F \nonumber \\
	&\quad +\frac{\underline{W}_{\mathbf{\Upsilon},|\mathbf{\Lambda}_{\varrho}^l\cup\mathbf{\Xi}^l|}}{\overline{W}_{\mathbf{\Upsilon},s}}\Big\|\big(\mathbf{D}_{n_{\mathbf{\Lambda}_n^l\backslash\mathbf{\Lambda}^l_{n_{{\varrho}}}}}\otimes\cdots\otimes\mathbf{D}_{1_{\mathbf{\Lambda}_1^l\backslash\mathbf{\Lambda}^l_{1_{1}}}}\big)\nonumber\\
	&\hspace{8em}\times {\rm vec}\big(\mathcal{X}_{\mathbf{\Lambda}^l\backslash\mathbf{\Lambda}^l_{1}}\big) + {\rm vec}(\mathcal{N})\Big\|^2_F \label{232} \\
	&\leq \Big(1-\frac{\underline{W}_{\mathbf{\Upsilon},|\mathbf{\Lambda}_{1}^l\cup\mathbf{\Xi}^l|}}{\overline{W}_{\mathbf{\Upsilon},s}}\Big) \Big\|\big(\mathbf{D}_{n_{\mathbf{\Lambda}_n^l}}\otimes\cdots\otimes\mathbf{D}_{1_{\mathbf{\Lambda}_1^l}}\big) \nonumber\\
	&\hspace{10em}\times{\rm vec}\big(\mathcal{X}_{\mathbf{\Lambda}^l}\big) + {\rm vec}(\mathcal{N})\Big\|^2_F \nonumber \\
	&\quad +\frac{\underline{W}_{\mathbf{\Upsilon},|\mathbf{\Lambda}_{1}^l\cup\mathbf{\Xi}^l|}}{\overline{W}_{\mathbf{\Upsilon},s}} \Big\|\big(\mathbf{D}_{n_{\mathbf{\Lambda}_n^l\backslash\mathbf{\Lambda}^l_{n_{{1}}}}}\otimes\cdots\otimes\mathbf{D}_{1_{\mathbf{\Lambda}_1^l\backslash\mathbf{\Lambda}^l_{1_{1}}}}\big)\nonumber\\
	&\hspace{8em}\times {\rm vec}\big(\mathcal{X}_{\mathbf{\Lambda}^l\backslash\mathbf{\Lambda}^l_{1}}\big) + {\rm vec}(\mathcal{N})\Big\|^2_F \label{233} \\
	&\leq \Big(1-\frac{\underline{W}_{\mathbf{\Upsilon},|\mathbf{\Lambda}_{1}^l\cup\mathbf{\Xi}^l|}}{\overline{W}_{\mathbf{\Upsilon},s}}\Big)\nonumber\\
	&\quad\times \Big((1+\gamma) \big\|\big(\mathbf{D}_{n_{\mathbf{\Lambda}_n^l}}\otimes\cdots\otimes\mathbf{D}_{1_{\mathbf{\Lambda}_1^l}}\big) {\rm vec}\big(\mathcal{X}_{\mathbf{\Lambda}^l}\big)\big\|^2_F \nonumber\\
	&\qquad\quad+ (1+\gamma^{-1})\|{\rm vec}(\mathcal{N})\|^2_F\Big) \nonumber \\
	& \quad+\frac{\underline{W}_{\mathbf{\Upsilon},|\mathbf{\Lambda}_{1}^l\cup\mathbf{\Xi}^l|}}{\overline{W}_{\mathbf{\Upsilon},s}}\nonumber\\
	&\!\!\!\times \Big((1+\gamma)\big\|\big(\mathbf{D}_{n_{\mathbf{\Lambda}_n^l\backslash\mathbf{\Lambda}^l_{n_{{\varrho}}}}}\otimes\cdots\otimes\mathbf{D}_{1_{\mathbf{\Lambda}_1^l\backslash\mathbf{\Lambda}^l_{1_{1}}}}\big) {\rm vec}\big(\mathcal{X}_{\mathbf{\Lambda}^l\backslash\mathbf{\Lambda}^l_{1}}\big)\big\|^2_F \nonumber \\
	&\quad +(1+\gamma^{-1})\|{\rm vec}(\mathcal{N})\|^2_F\Big) \label{234} \\
	&\leq \Big(1-\frac{\underline{W}_{\mathbf{\Upsilon},|\mathbf{\Lambda}_{1}^l\cup\mathbf{\Xi}^l|}}{\overline{W}_{\mathbf{\Upsilon},s}}\Big)
	\times\Big((1+\gamma)\overline{W}_{\mathbf{\Upsilon},\theta^n}\|{\rm vec}(\mathcal{X}_{\mathbf{\Lambda}^l})\|^2_F\nonumber\\
	&\hspace{11em}+(1+\gamma^{-1})\|{\rm vec}(\mathcal{N})\|^2_F\Big) \nonumber \\
	& +\frac{\underline{W}_{\mathbf{\Upsilon},|\mathbf{\Lambda}_{1}^l\cup\mathbf{\Xi}^l|}}{\overline{W}_{\mathbf{\Upsilon},s}}\Big((1+\gamma)\overline{W}_{\mathbf{\Upsilon},\theta^n}\|{\rm vec}(\mathcal{X}_{\mathbf{\Lambda}^l\backslash\mathbf{\Lambda}^l_{1}})\|^2_F\nonumber\\
	&\hspace{7em}+(1+\gamma^{-1})\|{\rm vec}(\mathcal{N})\|^2_F\Big) \label{235} \\
  \leq& \Big(1-\frac{\underline{W}_{\mathbf{\Upsilon},|\mathbf{\Lambda}_{1}^l\cup\mathbf{\Xi}^l|}}{\overline{W}_{\mathbf{\Upsilon},s}}\Big)\Big((1+\gamma)\overline{W}_{\mathbf{\Upsilon},\theta^n}\|{\rm vec}(\mathcal{X}_{\mathbf{\Lambda}^l})\|^2_F\nonumber\\
  &\hspace{9em}+(1+\gamma^{-1})\|{\rm vec}(\mathcal{N})\|^2_F\Big) \nonumber \\
	& +\frac{\underline{W}_{\mathbf{\Upsilon},|\mathbf{\Lambda}_{1}^l\cup\mathbf{\Xi}^l|}}{\overline{W}_{\mathbf{\Upsilon},s}}\Big((1+\gamma)\overline{W}_{\mathbf{\Upsilon},\theta^n}\eta^{-1}\|{\rm vec}(\mathcal{X}_{\mathbf{\Lambda}^l})\|^2_F\nonumber\\
	&\hspace{7em}+(1+\gamma^{-1})\|{\rm vec}(\mathcal{N})\|^2_F\Big) \label{236} \\
	=& \bigg(\Big(1-\frac{\underline{W}_{\mathbf{\Upsilon},|\mathbf{\Lambda}_{1}^l\cup\mathbf{\Xi}^l|}}{\overline{W}_{\mathbf{\Upsilon},s}}\Big)(1+\gamma)\overline{W}_{\mathbf{\Upsilon},\theta^n}\nonumber\\
	&\quad+\frac{\underline{W}_{\mathbf{\Upsilon},|\mathbf{\Lambda}_{1}^l\cup\mathbf{\Xi}^l|}}{\overline{W}_{\mathbf{\Upsilon},s}}(1+\gamma)\overline{W}_{\mathbf{\Upsilon},\theta^n}\eta^{-1}\bigg)\|{\rm vec}(\mathcal{X}_{\mathbf{\Lambda}^l})\|^2_F\nonumber\\
	& +(1+\gamma^{-1})\|{\rm vec}(\mathcal{N})\|^2_F , \label{237}
\end{align}
where (\ref{232}) is from direct simplification, (\ref{233}) is derived based on \cite[Proposition 1]{jWang2016tsp}, (\ref{234}) is similar to the proof of (\ref{rrrr3}), (\ref{235}) is from \textbf{Corollary \ref{Corollary6}}, and (\ref{236}) is due to (\ref{etadef4}).

Combining (\ref{theo51115}) and (\ref{237}) yields
\begin{align} % eq.294
	&\big\|{\rm vec}(\mathcal{X}_{\mathbf{\Lambda}^{l+1}})\big\|_F \nonumber\\
	&\leq \Bigg(\Big(1-\frac{\underline{W}_{\mathbf{\Upsilon},|\mathbf{\Lambda}_{1}^l\cup\mathbf{\Xi}^l|}}{\overline{W}_{\mathbf{\Upsilon},s}}\Big)(1+\gamma)\overline{W}_{\mathbf{\Upsilon},\theta^n}\nonumber\\
	&\qquad+\frac{\underline{W}_{\mathbf{\Upsilon},|\mathbf{\Lambda}_{1}^l\cup\mathbf{\Xi}^l|}}{\overline{W}_{\mathbf{\Upsilon},s}}(1+\gamma)\overline{W}_{\mathbf{\Upsilon},\theta^n}\Bigg)^{\frac{1}{2}}\frac{1}{\underline{W}^{\frac{1}{2}}_{\mathbf{\Upsilon},k}}\|{\rm vec}(\mathcal{X}_{\mathbf{\Lambda}^l})\|_F \nonumber \\
	&\quad +\frac{1+(1+\gamma^{-1})^{\frac{1}{2}}}{\underline{W}^{\frac{1}{2}}_{\mathbf{\Upsilon},k}}\|{\rm vec}(\mathcal{N})\|_F. \label{zong1}
\end{align}
For an appropriate $\gamma$, we have
\begin{align} % eq.295
	&\Bigg(\Big(1-\frac{\underline{W}_{\mathbf{\Upsilon},|\mathbf{\Lambda}_{1}^l\cup\mathbf{\Xi}^l|}}{\overline{W}_{\mathbf{\Upsilon},s}}\Big)(1+\gamma)\overline{W}_{\mathbf{\Upsilon},\theta^n}\nonumber\\
	&\quad+\frac{\underline{W}_{\mathbf{\Upsilon},|\mathbf{\Lambda}_{1}^l\cup\mathbf{\Xi}^l|}}{\overline{W}_{\mathbf{\Upsilon},s}}(1+\gamma)\overline{W}_{\mathbf{\Upsilon},\theta^n}\Bigg)^{\frac{1}{2}}\frac{1}{\underline{W}^{\frac{1}{2}}_{\mathbf{\Upsilon},k}}<1 . \nonumber
\end{align}

Then, consider the following two cases. On one hand, if
\begin{align} % eq.296
	& \frac{1+(1+\gamma^{-1})^{\frac{1}{2}}}{\underline{W}^{\frac{1}{2}}_{\mathbf{\Upsilon},k}}\|{\rm vec}(\mathcal{N})\|_F \nonumber \\
	&  < \Bigg(1-\Bigg(\Big(1-\frac{\underline{W}_{\mathbf{\Upsilon},|\mathbf{\Lambda}_{1}^l\cup\mathbf{\Xi}^l|}}{\overline{W}_{\mathbf{\Upsilon},s}}\Big)(1+\gamma)\overline{W}_{\mathbf{\Upsilon},\theta^n}\nonumber\\
	&\quad+\frac{\underline{W}_{\mathbf{\Upsilon},|\mathbf{\Lambda}_{1}^l\cup\mathbf{\Xi}^l|}}{\overline{W}_{\mathbf{\Upsilon},s}}(1+\gamma)\overline{W}_{\mathbf{\Upsilon},\theta^n}\Bigg)^{\frac{1}{2}}\frac{1}{\underline{W}^{\frac{1}{2}}_{\mathbf{\Upsilon},k}}\Bigg)\|{\rm vec}(\mathcal{X}_{\mathbf{\Lambda}^l}) \|_F\nonumber
\end{align}
holds, then (\ref{zong1}) reveals that
\begin{align} % eq.297
	\|\mathcal{X}_{\mathbf{\Lambda}^{l+1}}\|^2_F \leq \|\mathcal{X}_{\mathbf{\Lambda}^l}\|^2_F. \nonumber
\end{align}
This indicates that $|\mathbf{\Lambda}^{l+1}|<\theta$. Similarly, there remain at most $\theta-1$ support tensor blocks after the $(l+1)$th iteration.
Since
\begin{align} % eq.298
	l+1+\theta-1 = l+\theta, \nonumber
\end{align}
we have
\begin{align} % eq.299
	\big\|\mathcal{R}^{l+1+\theta-1}\big\|_F& = \big\|\mathcal{R}^{l+\theta}\big\|_F
	  \leq \xi\|\mathcal{N}\|_F.\nonumber
\end{align}
On the other hand, if
\begin{align} % eq.300
	& \frac{1+(1+\gamma^{-1})^{\frac{1}{2}}}{\underline{W}^{\frac{1}{2}}_{\mathbf{\Upsilon},k}}\|{\rm vec}(\mathcal{N})\|_F \nonumber \\
	 &\geq\Bigg(1-\Bigg(\Big(1-\frac{\underline{W}_{\mathbf{\Upsilon},|\mathbf{\Lambda}_{1}^l\cup\mathbf{\Xi}^l|}}{\overline{W}_{\mathbf{\Upsilon},s}}\Big)(1+\gamma)\overline{W}_{\mathbf{\Upsilon},\theta^n}\nonumber\\
	 &\quad+\frac{\underline{W}_{\mathbf{\Upsilon},|\mathbf{\Lambda}_{1}^l\cup\mathbf{\Xi}^l|}}{\overline{W}_{\mathbf{\Upsilon},s}}(1+\gamma)\overline{W}_{\mathbf{\Upsilon},\theta^n}\eta^{-1}\Bigg)^{\frac{1}{2}}\frac{1}{\underline{W}^{\frac{1}{2}}_{\mathbf{\Upsilon},k}}\Bigg)\|{\rm vec}(\mathcal{X}_{\mathbf{\Lambda}^l})\|_F \label{if2}
\end{align}
holds, then we have
\begin{align} % eqs.301-305
	&\big\|\mathcal{R}^{l+\theta}\big\|_F 
	\leq \big\|\mathcal{R}^{l+1}\big\|_F \label{rmin1} \\
	&\leq \bigg(\Big(1-\frac{\underline{W}_{\mathbf{\Upsilon},|\mathbf{\Lambda}_{1}^l\cup\mathbf{\Xi}^l|}}{\overline{W}_{\mathbf{\Upsilon},s}}\Big)(1+\gamma)\overline{W}_{\mathbf{\Upsilon},\theta^n}\nonumber\\
	&\qquad+\frac{\underline{W}_{\mathbf{\Upsilon},|\mathbf{\Lambda}_{1}^l\cup\mathbf{\Xi}^l|}}{\overline{W}_{\mathbf{\Upsilon},s}}(1+\gamma)\overline{W}_{\mathbf{\Upsilon},\theta^n}\eta^{-1}\bigg)^{\frac{1}{2}}\|{\rm vec}(\mathcal{X}_{\mathbf{\Lambda}^l})\|_F \nonumber \\
  & \quad +(1+\gamma^{-1})^{\frac{1}{2}}\|{\rm vec}(\mathcal{N})\|_F \label{rmin2} \\
	&\leq \Bigg(\bigg(\Big(1-\frac{\underline{W}_{\mathbf{\Upsilon},|\mathbf{\Lambda}_{1}^l\cup\mathbf{\Xi}^l|}}{\overline{W}_{\mathbf{\Upsilon},s}}\Big)(1+\gamma)\overline{W}_{\mathbf{\Upsilon},\theta^n}\nonumber\\
	&\qquad+\frac{\underline{W}_{\mathbf{\Upsilon},|\mathbf{\Lambda}_{1}^l\cup\mathbf{\Xi}^l|}}{\overline{W}_{\mathbf{\Upsilon},s}}(1+\gamma)\overline{W}_{\mathbf{\Upsilon},\theta^n}\eta^{-1}\bigg)^{\frac{1}{2}}(1+(1+\gamma^{-1})^{\frac{1}{2}})\nonumber\\
	&\times\underline{W}^{-\frac{1}{2}}_{\mathbf{\Upsilon},k}\Bigg(1-\Bigg(\bigg(\Big(1-\frac{\underline{W}_{\mathbf{\Upsilon},|\mathbf{\Lambda}_{1}^l\cup\mathbf{\Xi}^l|}}{\overline{W}_{\mathbf{\Upsilon},s}}\Big)(1+\gamma)\overline{W}_{\mathbf{\Upsilon},\theta^n}\nonumber\\
	&\qquad+\frac{\underline{W}_{\mathbf{\Upsilon},|\mathbf{\Lambda}_{1}^l\cup\mathbf{\Xi}^l|}}{\overline{W}_{\mathbf{\Upsilon},s}}(1+\gamma)\overline{W}_{\mathbf{\Upsilon},\theta^n}\eta^{-1}\bigg)\frac{1}{\underline{W}_{\mathbf{\Upsilon},k}}\Bigg)^{\frac{1}{2}}\Bigg)^{-1} \nonumber \\
	& \quad +(1+\gamma^{-1})^{\frac{1}{2}}\Bigg)\|{\rm vec}(\mathcal{N})\|_F \label{rmin3} \\
	&\leq \Bigg(\bigg(\Big(1-\frac{\underline{W}_{\mathbf{\Upsilon},sl+s+\theta}}{\overline{W}_{\mathbf{\Upsilon},s}}\Big)(1+\gamma)\overline{W}_{\mathbf{\Upsilon},\theta^n}\nonumber\\
	&\qquad+\frac{\underline{W}_{\mathbf{\Upsilon},sl+s+\theta}}{\overline{W}_{\mathbf{\Upsilon},s}}(1+\gamma)\overline{W}_{\mathbf{\Upsilon},\theta^n}\eta^{-1}\bigg)^{\frac{1}{2}}(1+(1+\gamma^{-1})^{\frac{1}{2}})\nonumber\\
	&\quad\times\underline{W}^{-\frac{1}{2}}_{\mathbf{\Upsilon},k}\Bigg(1-\Bigg(\bigg(\Big(1-\frac{\underline{W}_{\mathbf{\Upsilon},sl+s+\theta}}{\overline{W}_{\mathbf{\Upsilon},s}}\Big)(1+\gamma)\overline{W}_{\mathbf{\Upsilon},\theta^n}\nonumber\\
	&\qquad+\frac{\underline{W}_{\mathbf{\Upsilon},sl+s+\theta}}{\overline{W}_{\mathbf{\Upsilon},s}}(1+\gamma)\overline{W}_{\mathbf{\Upsilon},\theta^n}\eta^{-1}\bigg)\frac{1}{\underline{W}_{\mathbf{\Upsilon},k}}\Bigg)^{\frac{1}{2}}\Bigg)^{-1} \nonumber \\
  &\quad +(1+\gamma^{-1})^{\frac{1}{2}}\Bigg)\|{\rm vec}(\mathcal{N})\|_F \label{rmin4} \\
	&\leq \xi\|\mathcal{N}\|_F, \nonumber
\end{align}
where (\ref{rmin1}) is because the residual is non-increasing, (\ref{rmin2}) is based on (\ref{237}), (\ref{rmin3}) is derived by (\ref{if2}), and (\ref{rmin4}) is because $|\mathbf{\Lambda}^l_1\cup\mathbf{\Xi}^l|\leq|\mathbf{\Xi}\cup\mathbf{\Xi}^{l+1}|=|\mathbf{\Xi}^{l+1}|+|\mathbf{\Lambda}^{l+1}|$.
This completes the proof.
\end{IEEEproof}

\subsection{Proof of \textbf{Theorem~\ref{theo7}}}\label{proofoftheo7} % Ap-N

\begin{IEEEproof}
Observe that
\begin{align} % eqs.306-309
	&\|\mathcal{R}^k\|_F \nonumber\\
	&= \|\mathcal{Y}-\hat{\mathcal{X}}^{k}\times_1\mathbf{D}_1\times_2\cdots\times_n\mathbf{D}_n\|_F \label{proftheo61} \\
	&= \|\mathcal{X}\times_1\mathbf{D}_1\times_2\cdots\times_n\mathbf{D}_n+\mathcal{N}\nonumber\\
	&\hspace{1.5em}-\hat{\mathcal{X}}^{k}\times_1\mathbf{D}_1\times_2\cdots\times_n\mathbf{D}_n\|_F \label{proftheo62} \\
	&= \big\|\big(\mathcal{X}-\hat{\mathcal{X}}^{k}\big)\times_1\mathbf{D}_1\times_2\cdots\times_n\mathbf{D}_n+\mathcal{N}\big\|_F \nonumber \\
	&\geq \big\|\big(\mathcal{X}-\hat{\mathcal{X}}^{k}\big)\times_1\mathbf{D}_1\times_2\cdots\times_n\mathbf{D}_n\big\|_F - \|\mathcal{N}\|_F \label{proftheo64} \\
	&\geq \underline{W}^{\frac{1}{2}}_{\mathbf{\Upsilon},\prod_{t=1}^{n}k_t+ks}\big\|(\mathcal{X}-\hat{\mathcal{X}}^{k})\big\|_F - \|\mathcal{N}\|_F , \label{proftheo65}
\end{align}
where (\ref{proftheo61}) is from the definition of residual, (\ref{proftheo62}) is based on (\ref{tensorCSmodel}), (\ref{proftheo64}) is derived by the triangle inequality, and (\ref{proftheo65}) is from \textbf{Corollary \ref{Corollary6}}.

Thus, based on (\ref{proftheo65}) and \textbf{Corollary~\ref{Corollary7}}, we have
\begin{align} % eq.310
	\big\|\mathcal{X}-\hat{\mathcal{X}}^{k}\big\|_F &\leq \underline{W}^{-\frac{1}{2}}_{\mathbf{\Upsilon},\prod_{t=1}^{n}k_t+ks}\big(\big\|\mathcal{R}^k\big\|_F + \|\mathcal{N}\|_F\big) \nonumber\\
	&\leq\underline{W}^{-\frac{1}{2}}_{\mathbf{\Upsilon},\prod_{t=1}^{n}k_t+ks}(\xi^*+1)\|\mathcal{N}\|_F . \label{xixi1}
\end{align}
On the other hand, similar to the proof in Appendix \ref{proofoftheo1}, using $\hat{\mathcal{Q}}^{k}$ as the best $k$ block-sparse tensor approximation of $\hat{\mathcal{X}}^{k}$, we have
\begin{align} % eqs.311-313
	\big\|\hat{\mathcal{Q}}^{k}-\mathcal{X}\big\|_F &= \big\|\hat{\mathcal{Q}}^{k}-\hat{\mathcal{X}}^{k}+\hat{\mathcal{X}}^{k}-\mathcal{X}\big\|_F \label{xixi2} \\
	&\leq \big\|\hat{\mathcal{Q}}^{k}-\hat{\mathcal{X}}^{k}\big\|_F + \big\|\hat{\mathcal{X}}^{k}-\mathcal{X}\big\|_F \label{xixi3} \\
	&\leq 2\big\|\hat{\mathcal{X}}^{k}-\mathcal{X}\big\|_F \nonumber\\
	&  \leq 2\underline{W}^{-\frac{1}{2}}_{\mathbf{\Upsilon},\prod_{t=1}^{n}k_t+ks}(\xi^*+1)\|\mathcal{N}\|_F,\label{xixi5}
\end{align}
where (\ref{xixi2}) is because $\hat{\mathcal{Q}}^{k}$ is the best approximation, (\ref{xixi3}) is from the triangle inequality, and (\ref{xixi5}) is based on (\ref{xixi1}).

From (\ref{the1lower5}), we have
\begin{align} % eq.314
	\big\|\hat{\mathcal{Q}}^{k}-\mathcal{X}\big\|_F \geq \frac{\underline{W}^{\frac{1}{2}}_{\mathbf{\Upsilon},\prod_{t=1}^{n}k_t+k}\|(\hat{\mathcal{X}}-\mathcal{X})\|_F-2\|\mathcal{N}\|_F}{\overline{W}^{\frac{1}{2}}_{\mathbf{\Upsilon},\prod_{t=1}^{n}k_t+k}}. \label{theo1lemm}
\end{align}
Then, using (\ref{xixi1}) and (\ref{theo1lemm}), we have
\begin{align} % eq.315
	&\frac{\underline{W}^{\frac{1}{2}}_{\mathbf{\Upsilon},\prod_{t=1}^{n}k_t+k}\|(\hat{\mathcal{X}}-\mathcal{X})\|_F-2\|\mathcal{N}\|_F}{\overline{W}^{\frac{1}{2}}_{\mathbf{\Upsilon},\prod_{t=1}^{n}k_t+k}}\nonumber\\
	&\qquad\qquad\qquad\leq2\underline{W}^{-\frac{1}{2}}_{\mathbf{\Upsilon},\prod_{t=1}^{n}k_t+ks}(\xi^*+1)\|\mathcal{N}\|_F , \nonumber
\end{align}
i.e.,
\begin{align} % eq.316
	 &\|\hat{\mathcal{X}}-\mathcal{X}\|_F \nonumber\\
	 &\leq \frac{2\Big(\underline{W}^{-\frac{1}{2}}_{\mathbf{\Upsilon},\prod_{t=1}^{n}k_t+ks}(\xi^*+1)\overline{W}^{\frac{1}{2}}_{\mathbf{\Upsilon},\prod_{t=1}^{n}k_t+k}+1\Big)}{\underline{W}^{\frac{1}{2}}_{\mathbf{\Upsilon},\prod_{t=1}^{n}k_t+k}}\|\mathcal{N}\|_F . \nonumber
\end{align}
This completes the proof.
\end{IEEEproof}

\ifCLASSOPTIONcaptionsoff
  \newpage
\fi

%\bibliographystyle{ieeetran}
%\bibliography{SBOLS}

\begin{IEEEbiographynophoto}{Liyang Lu}
	(Member, IEEE) received the B.S. degree from the School of Information Engineering, Beijing University of Posts and Telecommunications (BUPT), China, in 2017, and the Ph.D. degree from the School of Artificial Intelligence, BUPT, in 2022. He is currently a postdoctoral fellow at Tsinghua University. His area of research interests include compressed sensing, cognitive radios, and MIMO communications.
\end{IEEEbiographynophoto}

\begin{IEEEbiographynophoto}{Zhaocheng Wang}
	(Fellow, IEEE) received the B.S.,
	M.S., and Ph.D. degrees from Tsinghua University in
	1991, 1993, and 1996, respectively.
	From 1996 to 1997, he was a Post-Doctoral Fellow with Nanyang Technological University, Singapore.
	From 1997 to 1999, he was a Research Engineer/a Senior Engineer with OKI Techno Centre (Singapore) Pte.
	Ltd., Singapore. From 1999 to 2009, he was a Senior
	Engineer/a Principal Engineer with Sony Deutschland
	GmbH, Germany. Since 2009, he has been a Professor
	with the Department of Electronic Engineering, Tsinghua University, where he is currently the Director of the Broadband Communication Key Laboratory, Beijing National Research Center for Information Science
	and Technology (BNRist). He has authored or coauthored two books, which have
	been selected by IEEE Press Series on Digital and Mobile Communication (Wiley
	IEEE Press). He has also authored/coauthored more than 200 peer-reviewed
	journal articles. He holds 60 U.S./EU granted patents (23 of them as the first
	inventor). His research interests include wireless communications, millimeter
	wave communications, and optical wireless communications.
	Prof. Wang is a fellow of the Institution of Engineering and Technology. He was
	a recipient of the ICC2013 Best Paper Award, the OECC2015 Best Student Paper
	Award, the 2016 IEEE Scott Helt Memorial Award, the 2016 IET Premium Award,
	the 2016 National Award for Science and Technology Progress (First Prize), the
	ICC2017 Best Paper Award, the 2018 IEEE ComSoc Asia–Pacific Outstanding
	Paper Award, and the 2020 IEEE ComSoc Leonard G. Abraham Prize. 
\end{IEEEbiographynophoto}
 
\begin{IEEEbiographynophoto}{Zhen Gao}
	(Member, IEEE) received the B.S. degree
	in information engineering from the Beijing Institute
	of Technology, Beijing, China, in 2011, and the
	Ph.D. degree in communication and signal processing from the Tsinghua National Laboratory for
	Information Science and Technology, Department of
	Electronic Engineering, Tsinghua University, China,
	in 2016. He is currently an Associate Professor
	with the Beijing Institute of Technology. His current
	research interests include wireless communications,
	with a focus on multi-carrier modulations, multiple antenna systems, and sparse signal processing. He was a recipient of
	the 2016 IEEE Broadcast Technology Society Scott Helt Memorial Award
	(Best Paper), an Exemplary Reviewer Award of IEEE COMMUNICATION
	LETTERS in 2016, IET Electronics Letters Premium Award (Best Paper) in
	2016, and the Young Elite Scientists Sponsorship Program from the China
	Association for Science and Technology from 2018 to 2021.
\end{IEEEbiographynophoto}

\begin{IEEEbiographynophoto}{Sheng Chen}
	(Life Fellow, IEEE) received his BEng degree from the East China Petroleum Institute, Dongying, China, in January 1982, and his PhD degree from the City University, London, in September 1986, both in control engineering. In August 2005, he was awarded the higher doctoral degree, Doctor of Sciences (DSc), from the University of Southampton, Southampton, UK. From 1986 to 1999, He held research and academic appointments at the Universities of Sheffield, Edinburgh and Portsmouth, all in UK. Since 1999, he has been with the School of Electronics and Computer Science, the University of Southampton, UK, where he holds the post of Professor in Intelligent Systems and Signal Processing. Dr Chen's research interests include adaptive signal processing, wireless communications, modeling and identification of nonlinear systems, neural network and machine learning, evolutionary computation methods and optimization. He has published over 700 research papers. Professor Chen has 20,000+ Web of Science citations with h-index 61 and 39,000+ Google Scholar citations with h-index 83. Dr. Chen is a Fellow of the United Kingdom Royal Academy of Engineering, a Fellow of Asia-Pacific Artificial Intelligence Association and a Fellow of IET. He is one of the original ISI highly cited researchers in engineering (March 2004).
\end{IEEEbiographynophoto}

\begin{IEEEbiographynophoto}{H.~Vincent~Poor}
	(Life Fellow, IEEE) received the Ph.D. degree in EECS from Princeton University in 1977.  From 1977 until 1990, he was on the faculty of the University of Illinois at Urbana-Champaign. Since 1990 he has been on the faculty at Princeton, where he is currently the Michael Henry Strater University Professor. During 2006 to 2016, he served as the dean of Princeton’s School of Engineering and Applied Science, and he has also held visiting appointments at several other universities, including most recently at Berkeley and Cambridge. His research interests are in the areas of information theory, machine learning and network science, and their applications in wireless networks, energy systems and related fields. Among his publications in these areas is the book Machine Learning and Wireless Communications. (Cambridge University Press, 2022). Dr. Poor is a member of the National Academy of Engineering and the National Academy of Sciences and is a foreign member of the Royal Society and other national and international academies. He received the IEEE Alexander Graham Bell Medal in 2017.
\end{IEEEbiographynophoto}

\end{document}